\documentclass{ws-procs975x65}

\newcommand{\apjl}{Astrophysical Journal Letters}
\newcommand{\apj}{Astrophysical Journal}
\newcommand{\aap}{Astronomy \& Astrophysics}
\newcommand{\prd}{Physical Review D}

\newcommand{\mnras}{Monthly Notices of the Royal Astronomical Society}

\begin{document}

\title{On Gamma-Ray Bursts}

\author{R. Ruffini,$^{1,2,3}$ M.G. Bernardini,$^{1,2}$ C.L. Bianco,$^{1,2,3}$ L. Caito,$^{1,2}$ P. Chardonnet,$^{1,3,4}$ C. Cherubini,$^{1,5}$ M.G. Dainotti,$^{1,2}$ F. Fraschetti,$^6$ A. Geralico,$^{1,2}$ R. Guida,$^{1,2}$ B. Patricelli,$^{1,2}$ M. Rotondo,$^{1,2}$ J.A. Rueda Hernandez,$^{1,2}$ G. Vereshchagin,$^{1,2,3}$ S.-S. Xue$^{1,3}$}

\address{
$^1$ ICRANet and ICRA, Piazzale della Repubblica 10, I-65122 Pescara, Italy.\\
$^2$ Dip. di Fisica, Universit\`a di Roma ``La Sapienza'', Piazzale Aldo Moro 5, I-00185 Roma, Italy.\\
$^3$ ICRANet, Universit\'e de Nice Sophia Antipolis, Grand Ch\^ateau, BP 2135, 28, avenue de Valrose, 06103 NICE CEDEX 2, France.\\
$^4$ Universit\'e de Savoie, LAPTH - LAPP, BP 110, F-74941 Annecy-le-Vieux Cedex, France.\\
$^5$ Universit\`a Campus Biomedico, Facolt\`a di Ingegneria, Via Alvaro del Portillo 21, 00128 Rome, Italy.\\
$^6$ CEA/Saclay, F-91191 Gif-sur-Yvette Cedex, Saclay, France.\\
E-mails: ruffini@icra.it, maria.bernardini@icra.it, bianco@icra.it, letizia.caito@icra.it, chardon@lapp.in2p3.fr, cherubini@icra.it, dainotti@icra.it, fraschetti@icra.it, geralico@icra.it, roberto.guida@icra.it, barbara.patricelli@icranet.org, michael.rotondo@icra.it, jorge.rueda@icra.it, veresh@icra.it, xue@icra.it.
}

\begin{abstract}
We show by example how the uncoding of Gamma-Ray Bursts (GRBs) offers unprecedented possibilities to foster new knowledge in fundamental physics and in astrophysics. After recalling some of the classic work on vacuum polarization in uniform electric fields by Klein, Sauter, Heisenberg, Euler and Schwinger, we summarize some of the efforts to observe these effects in heavy ions and high energy ion collisions. We then turn to the theory of vacuum polarization around a Kerr-Newman black hole, leading to the extraction of the blackholic energy, to the concept of dyadosphere and dyadotorus, and to the creation of an electron-positron-photon plasma. We then present a new theoretical approach encompassing the physics of neutron stars and heavy nuclei. It is shown that configurations of nuclear matter in bulk with global charge neutrality can exist on macroscopic scales and with electric fields close to the critical value near their surfaces. These configurations may represent an initial condition for the process of gravitational collapse, leading to the creation of an electron-positron-photon plasma: the basic self-accelerating system explaining both the energetics and the high energy Lorentz factor observed in GRBs. We then turn to recall the two basic interpretational paradigms of our GRB model: 1) the Relative Space-Time Transformation (RSTT) paradigm and 2) the Interpretation of the Burst Structure (IBS) paradigm. These paradigms lead to a ``canonical'' GRB light curve formed from two different components: a Proper-GRB (P-GRB) and an extended afterglow comprising a raising part, a peak, and a decaying tail. When the P-GRB is energetically predominant we have a ``genuine'' short GRB, while when the afterglow is energetically predominant we have a so-called long GRB or a ``fake'' short GRB. We compare and contrast the description of the relativistic expansion of the electron-positron plasma within our approach and within the other ones in the current literature. We then turn to the special role of the baryon loading in discriminating between ``genuine'' short and long or ``fake'' short GRBs and to the special role of GRB 991216 to illustrate for the first time the ``canonical'' GRB bolometric light curve. We then propose a spectral analysis of GRBs, and proceed to some applications: GRB 031203, the first spectral analysis, GRB 050315, the first complete light curve fitting, GRB 060218, the first evidence for a critical value of the baryon loading, GRB 970228, the appearance of ``fake'' short GRBs. We finally turn to the GRB-Supernova Time Sequence (GSTS) paradigm: the concept of induced gravitational collapse. We illustrate this paradigm by the systems GRB 980425 / SN 1998bw, GRB 030329 / SN 2003dh, GRB 031203 / SN 2003lw, GRB 060218 / SN 2006aj, and we present the enigma of the URCA sources. We then present some general conclusions.
\end{abstract}

\bodymatter

\section{Introduction}

After almost a century of possible observational evidences of general relativistic effects, all very weak and almost marginal to the field of physics, the direct observation of gravitational collapse and of black hole formation promises to bring the field of general relativity into the mainstream of fundamental physics, testing a vast arena of unexplored regimes and leading to the explanation of a large number of yet unsolved astrophysical problems.

There are two alternative procedures for observing the process of gravitational collapse: either by gravitational waves or by joint observation of electromagnetic radiation and high energy particles. The gravitational wave observations may lead to the understanding of the global properties of the gravitational collapse process, inferred from the time-varying component of the global quadrupole moment of the system. Their observation is also made difficult and at times marginal due to the weak coupling of gravitational waves with the detectors. The observation of the electromagnetic radiation in the X, $\gamma$, optical and radio bands, and of the associated high energy particle component, is carried out by an unprecedented observational effort in space, ground and underground observatories. This effort is offering the possibility of giving for the first time a detailed description of the gravitational collapse process all the way to the formation of the black hole horizon.

The Grossmann Meetings have been dedicated to foster the mathematical and physical developments of Einstein theories. They have grown in recent years due to remarkable progress both in fundamental physics and in astrophysics. In this sense, I will present here some highlights of recent progress on the theoretical understanding of Gamma-Ray Bursts (GRBs, for a review see e.g. Ref.~\refcite{2007AIPC..910...55R} and references therein), which nicely represents two complementary aspects of the problem: progress in probing fundamental theories in yet unexplored regimes, as well as understanding the astrophysical scenario underlying novel astrophysical phenomena.

\begin{figure}
\includegraphics[width=\hsize,clip]{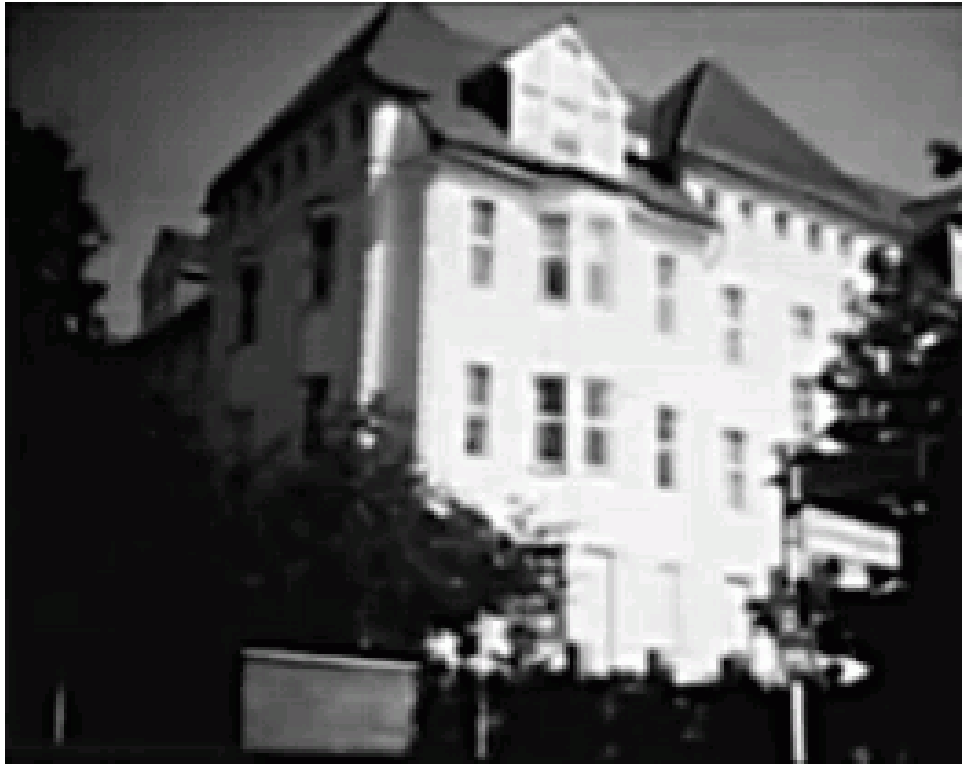}
\caption{Einstein's home in Dahlem}
\label{EinsteinHome}
\end{figure}

It is particularly inspiring that this MG11 takes place in Dahlem, close to where many of the fundamental breakthroughs and discoveries of modern physics have indeed occurred. A few hundred meters from here, in Ehrenbergstrasse 33, Albert Einstein once lived while introducing and developing the theory of general relativity (see Fig.~\ref{EinsteinHome}).

\begin{figure}
\centering
\includegraphics[width=\hsize,clip]{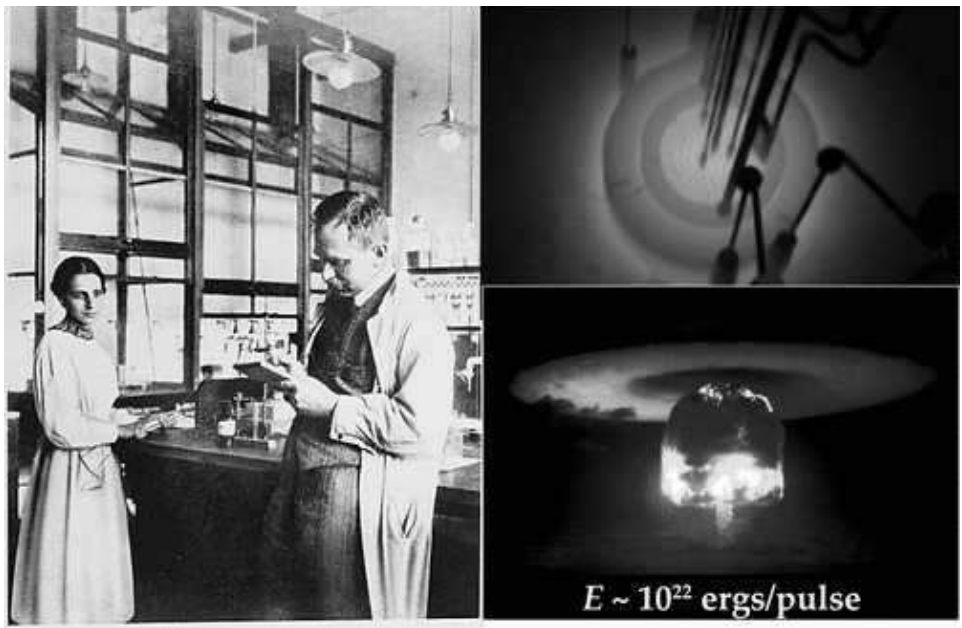}
\caption{{\bf Left:} Lise Meitner and Otto Hahn in the Kaiser-Wilhelm-Institut in Berlin. {\bf Right:} Picture of a nuclear reactor and an atomic bomb, both resulting from Uranium fission.}
\label{Meitner_Hahn}
\end{figure}

A few hundred meters from here there is also the Kaiser-Wilhelm-Institut, where Lise Meitner, Otto Hahn and Fritz Strassmann (see Fig.~\ref{Meitner_Hahn}) continued the work on neutron capture by Uranium initiated a few years before by Fermi\cite{1936PhRv...50..899A}. This work led to the revolutionary evidence for Uranium fission\cite{1939NW.....27...11H,mf39}. There is no need to stress the enormous consequences of this discovery following the work of Fermi\cite{1952AmJPh..20..536F}, Feynman, Metropolis \& Teller\cite{1949PhRv...75.1561F}, Oppenheimer \cite{OppenheimerCase}, Wigner\cite{WignerBook}, etc., and, just to keep symmetry, the work of Kurchatov, Sakharov, and Zel'dovich (see e.g. Ref.~\refcite{StalinBomb}). The message from the nuclear and thermonuclear work leads to nuclear reactors and to explosive events typically of $\sim 10^{22}$ ergs/pulse (see Fig.~\ref{Meitner_Hahn}).

\begin{figure}
\centering
\includegraphics[width=\hsize,clip]{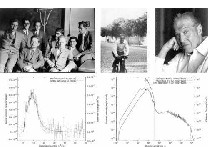}
\caption{{\bf Above Left:} A historical picture taken at the time of the Heisenberg-Euler work. Surrounding Heisenberg, counterclockwise, are some of the major interpreters of modern physics: Rudolf Peierls, George Placzek, Giovanni Gentile, Giancarlo Wick, Felix Bloch, Wicky Weisskopf and Fritz Sauter. {\bf Above Center:} H. Euler. {\bf Above Right:} Julian Schwinger. {\bf Below Left:} Observed light curve of GRB 980425 with our theoretical fit with a total energy of $10^{48}$ ergs. {\bf Below Right:} Observed light curve of GRB 050315 with our theoretical fit with a total energy of $10^{53}$ ergs.}
\label{Heisenberg-Group}
\end{figure}

What I would like to stress in this lecture is the possible role of a theoretical work, coeval to the work of Otto Hahn and Lise Meitner, developed close to Dahlem, at the University of Leipzig. Such a theoretical work may very well lead in the future to the understanding of yet bigger explosions and have an equally, if not more, fundamental role on the existence and the dynamics of our Universe. We refer here to the work pioneered by Klein\cite{1929ZPhy...53..157K}, Sauter\cite{1931ZPhy...69..742S}, Euler and Heisenberg\cite{1935NW.....23..246E,1936ZPhy...98..714H}, Weisskopf\cite{w36,1934ZPhy...89...27W} (see Fig.~\ref{Heisenberg-Group}). The work deals with the creation of electron-positron pairs out of the vacuum generated by an overcritical electric field. In the following we give evidences that this process is indeed essential to the extraction of energy from the black hole and occurs in the GRBs. The characteristic energy of these sources is typically on the order of $10^{49}$--$10^{54}$ ergs/pulse. I am giving here, as examples of these GRBs, the light curves of GRB 980425 (with a total energy of $\sim 10^{48}$ ergs) and GRB 050315 (with a total energy of $\sim 10^{53}$ ergs).

So much for fundamental science. From the astrophysical point of view, GRBs offers an equally rich scenario, being linked to processes like supernovae, coalescence of binary neutron stars, black holes and binaries in globular clusters, and possibly intermediate mass black holes. After reviewing some of the fundamental work on vacuum polarization, I will focus on a new class of 4 particularly weak GRBs which promises to clarify the special connection between GRBs and supernovae. I will then conclude on the very surprising aspect that GRBs can indeed originate from a very wide variety of different astrophysical sources, but their features can be explained within a unified theoretical model, which applies in the above-mentioned enormous range of energies. The reason of this ``uniqueness'' of the GRBs is strictly linked to the late phases of gravitational collapse leading to the formation of the black hole and to its theoretically expected ``uniqueness'': the black hole is uniquely characterized by mass, charge and angular momentum\cite{1971PhT....24a..30R}. The GRB phenomenon originates in the late phases of the process of gravitational collapse, when the formation of the horizon of the black hole is being reached. The phenomenon is therefore quite independent of the different astrophysical settings giving origin to the black hole formation.

\section{Vacuum polarization in a uniform electric field}

We recall some early work on pair creation following the introduction by Dirac of the relativistic field equation for the electron and leading to the classical results of Klein\cite{1929ZPhy...53..157K}, Heisenberg \& Euler\cite{1936ZPhy...98..714H}, Schwinger\cite{s51,s54a,s54b}.

\subsection{Klein and Sauter work}

\begin{figure}
\begin{picture}(350,350)
\put(60,50){\includegraphics[height=7.8cm,width=10.8cm]{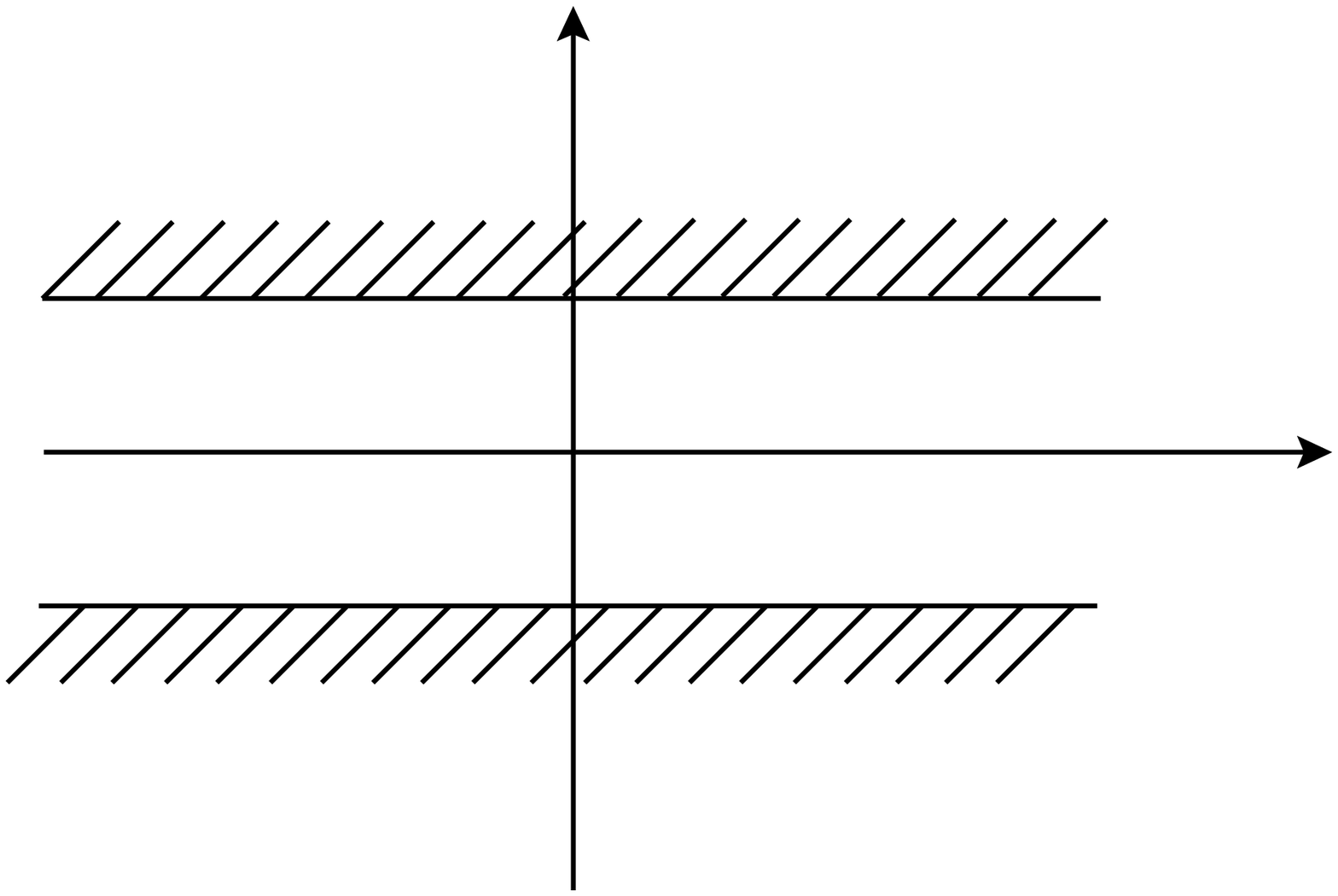}}
\put(350,170){$z$}
\put(200,260){${\mathcal E}$}
\put(30,230){positive continuum ${\mathcal E}_+>m_ec^2$}
\put(30,80){negative continuum ${\mathcal E}_-<m_ec^2$}
\put(40,200){$m_ec^2$}
\put(40,120){$-m_ec^2$}
\end{picture}
\caption{The mass-gap $2m_ec^2$  that separates the positive continuum spectrum ${\mathcal E}_+$
from the negative continuum spectrum ${\mathcal E}_-$.}%
\label{gap}%
\end{figure}

It is well known that every relativistic wave equation of a free relativistic 
particle of mass $m_e$, momentum ${\bf p}$ and energy ${\mathcal E}$,
admits symmetrically ``positive energy'' and ``negative energy'' solutions. Namely the wave-function 
\begin{equation}
\psi^{\pm}({\bf x},t)\sim e^{\frac{i}{\hbar}({\bf k}\cdot{\bf x}-{\mathcal E}_{\pm}t)}
\label{freeparticle}
\end{equation}
describes a relativistic particle, whose energy, mass and momentum must satisfy,
\begin{equation}
{\mathcal E}_{\pm}^2=m_e^2c^4 +c^2|{\bf p}|^2;\quad {\mathcal E}_\pm=\pm\sqrt{m_e^2c^4 +c^2|{\bf p}|^2},
\label{klein0} 
\end{equation}
this gives rise to the familiar positive and negative energy spectrum (${\mathcal E}_\pm$) of positive and negative energy states ($\psi^{\pm}({\bf x},t)$) of the relativistic particle, as represented in Fig.~\ref{gap}. 
In such free particle situation (flat space, no external field), 
all the quantum states are stable; that is, there is no possibility of ``positive'' (``negative'') energy states decaying into a ``negative'' (``positive'') energy states, since all negative energy states are fully filled and
there is an energy gap $2m_ec^2$ separating the negative energy spectrum from the positive energy spectrum. This is the 
view of Dirac theory on the spectrum of a relativistic particle \cite{Dir30,Dir33}. 

Klein studied a relativistic particle moving in an external {\it constant} potential $V$ and in this case 
Eq.~(\ref{klein0}) is modified as 
\begin{equation}
[{\mathcal E}-V]^2=m_e^2c^4 +c^2|{\bf p}|^2, ;\quad {\mathcal E}_\pm=V\pm\sqrt{m_e^2c^4 +c^2|{\bf p}|^2}.
\label{klein0bis} 
\end{equation}
He solved this relativistic wave equation by considering an incident free relativistic wave of positive energy states
scattered by the constant potential $V$, leading to reflected and transmitted waves. He found a paradox that in 
the case $V\ge {\mathcal E}+m_ec^2$, the reflected flux is larger than the incident flux $j_{\rm ref} >j_{\rm inc}$, 
although the total flux is conserved, i.e., $j_{\rm inc}=j_{\rm ref}+j_{\rm tran}$. This was known as the 
Klein paradox (see Ref.~\refcite{1929ZPhy...53..157K}). This implies that negative energy states have contributions to both the 
transmitted flux $j_{\rm tran}$ and reflected flux $j_{\rm ref}$.     

\begin{figure}
\includegraphics[width=\hsize,clip]{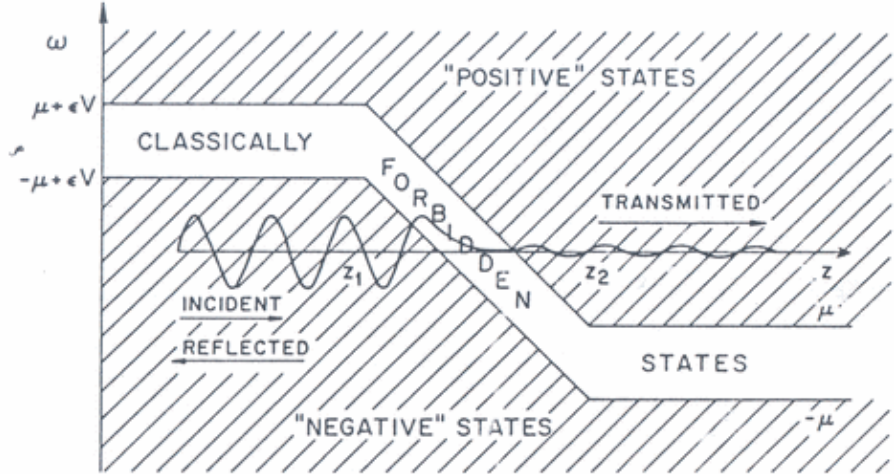}
\caption{In the presence of a strong enough electric field the boundaries of the classically allowed states 
(``positive'' or ``negative'') can also be so tilted that a ``negative'' is at the same level as a ``positive'' 
(level crossing). Therefore a ``negative'' wave-packet from the left will be partially transmitted,
after an exponential damping due to the tunneling through the classically forbidden states, as a ``positive''
wave-packet outgoing to the right. This figure
is reproduced from Fig.~II in Ref.~\refcite{demourwkb}, and $\mu=m_ec^2, \epsilon V=V(z), \omega={\mathcal E}$.}%
\label{demourgape}%
\end{figure}

Sauter studied this problem by considering an electric potential of an external constant 
electric field $E$ in the $\hat {\bf z}$ direction \cite{1931ZPhy...69..742S}. In this case the energy 
${\mathcal E}$ is shifted by the amount $V(z)=-e Ez$, where $e$ is the electron charge. 
In the case of the electric field $E$ uniform between $z_1$ and $z_2$ and null outside,
Fig.~\ref{demourgape} represents the corresponding sketch of allowed states.
The key point now, which is the essence of the Klein paradox \cite{1929ZPhy...53..157K}, is that the above mentioned stability of the ``positive energy'' states is lost for sufficiently strong electric fields. The same is true for ``negative energy'' states. Some ``positive energy'' and ``negative energy'' states have the same energy-levels, i.e., the crossing of
energy-levels occurs. Thus, these ``negative energy'' waves incident from the left will be both reflected back by the electric field and partly transmitted to the right as a ```positive energy'' wave, as shown in Fig.~\ref{demourgape} \cite{demourwkb}. This transmission is nothing else but a quantum tunneling of the wave function through the 
electric potential barrier, where classical states are forbidden. This is the same as the so-called the Gamow tunneling of the wave function through nuclear potential barrier (Gamow-wall)\cite{gamow-book}.

Sauter first solved the relativistic Dirac equation \cite{Dir30,Dir33}
in the presence of the constant electric field by the ansatz,
\begin{equation}
\psi_s({\bf x},t)= e^{\frac{i}{\hbar}(k_xx+k_yy-{\mathcal E}_\pm t)}\chi_{s_3}(z)
\label{Eparticle}
\end{equation}
Where the spinor function $\chi_{s_3}(z)$ obeys the following equation ($\gamma_0, \gamma_i $ are Dirac matrices)
\begin{equation}
\left[\hbar c\gamma_3\frac{d}{dz}+\gamma_0(V(z)-{\mathcal E}_\pm)+(m_ec^2+ic\gamma_2p_y
+ic\gamma_1 p_x)\right]\chi_{s_3}(z)=0,
\label{sauterfunction}
\end{equation}
and the solution $\chi_{s_3}(z)$ can be expressed in terms of hypergeometric functions \cite{1931ZPhy...69..742S}. 
Using this wave-function $\psi_s({\bf x},t)$ (\ref{Eparticle}) and the flux 
$ic\psi_s^\dagger\gamma_3\psi_s$, Sauter computed the transmitted flux of 
positive energy states, the incident and reflected
fluxes of negative energy states, as well as exponential decaying flux of classically forbidden states, 
as indicated in Fig.~\ref{demourgape}. Using continuous conditions of wave functions and fluxes 
at boundaries of the potential, Sauter found that         
the transmission coefficient $|T|^2$ of the wave through the electric potential 
barrier from the negative energy state to positive energy states:
\begin{equation}
|T|^2=\frac{|{\rm transmission\hskip0.2cm flux}|}{|{\rm incident\hskip0.2cm flux}|}
\sim e^{-\pi\frac{m_e^2c^3}{\hbar e E}}.
\label{transmission}       
\end{equation}
This is the probability of negative energy states decaying to positive energy states, caused by an external
electric field. The method that Sauter adopted to calculate the transmission coefficient $|T|^2$ is the same as
the one Gamow used at that time to calculate quantum tunneling of the wave function 
through nuclear potential barrier (Gamow-wall), leading to the $\alpha$-particle emission \cite{gamow-book}.

\subsection{Heisenberg-Euler-Weisskopf effective theory}

To be able to explain elastic light-light scattering \cite{1935NW.....23..246E},
Heisenberg and Euler\cite{1936ZPhy...98..714H} and Weisskopf\cite{w36,1934ZPhy...89...27W} proposed a  theory that
attributes to the vacuum certain
nonlinear electromagnetic properties, as if it were
a dielectric and
permeable medium\cite{1936ZPhy...98..714H,w36}.

Let ${\mathcal L}$ to be the Lagrangian density of electromagnetic fields ${\bf E},{\bf B}$;
a Legendre transformation
produces the Hamiltonian density:
\begin{equation}
{\mathcal H}=E_i{\frac{\delta {\mathcal L}}{ \delta E_i}} - {\mathcal L}.
\label{hlrelation}
\end{equation}
In  Maxwell's theory, the two densities are given by
\begin{equation}
{\mathcal L}_M={\frac{1}{8\pi}}({\bf  E}^2-{\bf B}^2),\hskip0.5cm {\mathcal H}_M
={\frac{1}{8\pi}}({\bf E}^2+{\bf B}^2).
\label{maxwell}
\end{equation}
To quantitatively describe nonlinear electromagnetic properties of the vacuum
based on the Dirac theory, the above authors
introduced the concept of an
 effective Lagrangian ${\mathcal L}_{\rm eff}$
of the vacuum state in the presence of electromagnetic fields, and an associated
Hamiltonian density
\begin{equation}
{\mathcal L}_{\rm eff}={\mathcal L}_M+\Delta {\mathcal L},\hskip0.5cm
{\mathcal H}_{\rm eff}={\mathcal H}_M + \Delta {\mathcal H}.
\label{effectivelh}
\end{equation}
From these one derives
induced fields $\bf  D,\bf  H$ as the derivatives
\begin{equation}
D_i=4\pi{\frac{\delta {\mathcal L}_{\rm eff}}{\delta E_i}},\hskip0.5cm H_i=-4\pi{\frac{\delta {\mathcal L}_{\rm eff}}{\delta B_i}}.
\label{dh}
\end{equation}
In Maxwell's theory, $\Delta {\mathcal L}\equiv 0$ in the vacuum, so that
$\bf  D={\bf E}$ and $\bf  H={\bf B}$.
In Dirac's theory, however, $\Delta {\mathcal L}$ is
a complex function of ${\bf E}$ and ${\bf B}$.
Correspondingly, the vacuum behaves as a dielectric and permeable medium \cite{1936ZPhy...98..714H,w36} in which,
\begin{equation}
D_i=\sum_k\epsilon_{ik}E_k,\hskip0.5cm H_i=\sum_k\mu_{ik}B_k,
\label{dh1}
\end{equation}
where complex $\epsilon_{ik}$ and $\mu_{ik}$ are the field-dependent dielectric and permeability tensors of the vacuum. 

The discussions on complex dielectric and permeability tensors ($\epsilon_{ik}$ and $\mu_{ik}$) 
can be found for example in Ref.~\refcite{landaumedium}.
The effective Lagrangian and Hamiltonian densities in such a medium is given by,
\begin{equation}
{\mathcal L}_{\rm eff}={\frac{1}{8\pi}}({\bf E}\cdot {\bf  D}-{\bf B}\cdot \bf  H),\hskip0.5cm {\mathcal H}_{\rm eff}
={\frac{1}{8\pi}}({\bf E}\cdot {\bf  D}+{\bf B}\cdot {\bf  H}).
\label{effmaxwell}
\end{equation}
In this medium, the conservation of electromagnetic energy
has the form
\begin{equation}
-{\rm div}{\bf  S}={\frac{1}{4\pi}}\left({\bf E}\cdot{\frac{\partial {\bf  D}}{\partial t}}+{\bf B}\cdot {\frac{\partial {\bf  H}}{\partial t}}\right),\hskip0.5cm \bf  S={\frac{c}{4\pi}}{\bf E}\times {\bf B},
\label{effcons}
\end{equation}
where $\bf  S$ is the Poynting vector  describing
the density of electromagnetic energy flux.
By considering electromagnetic fields complex and monochromatic
\begin{equation}
{\bf E}={\bf E}(\omega) \exp-i(\omega t);\quad 
{\bf B}={\bf B}(\omega) \exp-i(\omega t),
\label{monochromaticfields}
\end{equation} 
of frequency $\omega$, the dielectric and permeability tensors are frequency-dependent, i.e., $\epsilon_{ik}(\omega)$ 
and $\mu_{ik}(\omega)$. 
Substituting these fields and tensors into the r.h.s. of Eq.~(\ref{effcons}),
one obtains the dissipation of electromagnetic energy per time into the medium,
\begin{equation}
Q={\frac{\omega}{8\pi}}\left\{{\rm Im}\left[\epsilon_{ik}(\omega)\right]E_i E^*_k
+{\rm Im}\left[\mu_{ik}(\omega)\right]B_iB^*_k\right\}.
\label{dissipation}
\end{equation}
This is nonzero if $\epsilon_{ik}(\omega)$ and $\mu_{ik}(\omega)$ contain an imaginary part. The dissipation of electromagnetic 
energy is accompanied by heat production. In light of the third thermodynamical law of entropy increase,
the energy $Q$  of electromagnetic fields lost in the medium is always positive, i.e., $Q>0$. As a consequence,
${\rm Im}[\epsilon_{ik}(\omega)]>0$ and ${\rm Im}[\mu_{ik}(\omega)]>0$.  
The real parts of $\epsilon_{ik}(\omega)$ and $\mu_{ik}(\omega)$
represent an electric and magnetic polarizability of the vacuum and
lead, for example, to the refraction of light in an electromagnetic field, 
or to the elastic scattering of light from light \cite{1935NW.....23..246E}. 
The $n_{ij}(\omega)=\sqrt{\epsilon_{ik}(\omega)\mu_{kj}(\omega)}$ is the reflection index of the medium.
The field-dependence of $\epsilon_{ik}$ and $\mu_{ik}$ implies
nonlinear electromagnetic properties of the vacuum as a dielectric and permeable medium.

The effective Lagrangian density (\ref{effectivelh}) is a relativistically
invariant function of the
field strengths ${\bf E}$ and ${\bf B}$.
Since $({\bf E}^2-{\bf B}^2)$ and $({\bf E}\cdot {\bf B})^2$ are relativistic invariants, one can formally expand
$\Delta {\mathcal L}$ in powers of weak field strengths:
\begin{equation}
\Delta {\mathcal L} = \kappa_{20}
 ({\bf E}^2-{\bf B}^2)^2+\kappa_{02}
 ({\bf E}\cdot {\bf B})^2
+ \kappa _{30} ({\bf E}^2 -{\bf B}^2)^3 + \kappa _{12} ({\bf E}^2- {\bf B}^2)({\bf E}\cdot {\bf B})^2+\dots~ ,
\label{leffective}
\end{equation}
where $\kappa_{ij}$ are field-independent constants
whose subscripts indicate the powers of
$({\bf E}^2-{\bf B}^2)$ and
${\bf E}\cdot{\bf B}$, respectively.
Note that the invariant
${\bf E}\cdot{\bf B}$ appears only in even powers since it is odd under parity
and electromagnetism is
parity invariant.
The Lagrangian density (\ref{leffective}) corresponds, via
relation (\ref{hlrelation}), to
\begin{eqnarray}
\Delta {\mathcal H}&=&\kappa_{2,0}
 ({\bf E}^2-{\bf B}^2)(3{\bf E}^2+{\bf B}^2)+ \kappa _{0,2}
 ({\bf E}\cdot {\bf B})^2 \nonumber\\
&&+ \kappa _{3,0} ({\bf E}^2 -{\bf B}^2)^2(5{\bf E}^2+{\bf B}^2)+
 \kappa _{1,2}(3{\bf E}^2- {\bf B}^2)({\bf E}\cdot {\bf B})^2+\dots~ .
\label{heffective}
\end{eqnarray}
To obtain ${\mathcal H}_{\rm eff}$
in Dirac's theory,
one has to calculate
\begin{equation} \Delta {\mathcal H}=\sum_k\left\{\psi^*_k,\Big[
\mbox{\boldmath$\alpha$}\cdot(-i h c{\bf \nabla} + e{\bf A}\
)) +\beta m_ec^2\Big]\psi_k\right\},
\label{wcal}
\end{equation}
where $\alpha_i,\beta$ are Dirac matrices, ${\bf  A}$
is the vector potential, and
$\{\psi_k(x)\}$
are the wave functions
of the occupied negative-energy states.
When performing the sum,
one encounters infinities
which were removed
by Weisskopf\cite{w36,1934ZPhy...89...27W}, Dirac\cite{dirac1934}, Heisenberg\cite{heisenberg1934}
by a suitable subtraction.

Heisenberg \cite{heisenberg1934}
expressed the
Hamiltonian density in terms of the
density matrix
$ \rho (x,x')=\sum_k\psi^*_k(x)\psi_k(x')$ \cite{dirac1934}. Euler and Kockel\cite{1935NW.....23..246E}, and Heisenberg and Euler\cite{1936ZPhy...98..714H} calculated the coefficients $ \kappa _{ij}$.
They did so by solving the Dirac equation
in the presence of parallel electric and magnetic fields
${\bf E}$ and ${\bf B}$ in
a specific direction,
\begin{equation}
\psi_k(x)\rightarrow
\psi_{p_z,n,s_3}\equiv
e^{{\frac{i}{\hbar}}(zp_z-{\mathcal E}t)}u_{n}(y)\chi_{s_3}(x),
\hskip0.5cm  n=0,1,2,\dots~
\label{diracsolution}
\end{equation}
where $\{u_n(y)\}$ are the Landau states \cite{LanLif75a,LanLif75b} depending on the magnetic field and
$\chi_{s_3}(x)$
are the spinor functions
calculated
by Ref.~\refcite{1931ZPhy...69..742S}.
Heisenberg and Euler used the Euler-Maclaurin formula
to perform the sum over $ n$, and
obtained for the additional Lagrangian  in (\ref{effectivelh})
the integral representation,     
\begin{eqnarray}
 \Delta {\mathcal L}_{\rm eff}&=&\frac{e^2}{16\pi^2\hbar c}\int^\infty_0
e^{-s}\frac{ds}{s^3}\Big[is^2\,\bar E \bar B
\frac{\cos(s[\bar E^2-\bar B^2+2i(\bar E\bar B)]^{1/2})
+{\rm c.c.}}{\cos(s[\bar E^2-\bar B^2+2i(\bar E\bar B)]^{1/2})-{\rm c.c.}}\nonumber\\
&&+ \left(\frac{m_e^2c^3}{e\hbar}\right)^2
+\frac{s^2}{3}(|\bar B|^2-|\bar E|^2)\Big],
\label{effectiveint}
\end{eqnarray}
where
$\bar E,\bar B$ are the dimensionless reduced fields in the unit of the critical field $E_c$,
\begin{equation}
\bar E=\frac{|{\bf E}|}{E_c},~~~~\bar B=\frac{|{\bf B}|}{E_c};~~~~E_c\equiv \frac{m_e^2c^3}{e\hbar}.
\label{dimenlessEB}
\end{equation}
Expanding this in powers of $ \alpha $ up to $ \alpha ^3$
yields
the following values for the four constants:
\begin{equation}
\kappa_{2,0}=\frac{\alpha}{90\pi^2}E_c^{-2},\hskip0.3cm \kappa _{0,2} = 7\kappa_{2,0} ,\hskip0.3cm
\kappa _{3,0}= \frac{32\pi \alpha}{315}E_c^{-4},
\hskip0.3cm \kappa_{1,2} = {\frac{13}{2}} \kappa _{3,0} .
\label{lconstant}
\end{equation}
Weisskopf\cite{w36} adopted a simpler method.
He considered first
the special case in which ${\bf E}=0, {\bf B}\not=0$
and used the Landau states
to find
$\Delta {\mathcal H}$ of Eq.~(\ref{heffective}),
extracting from this
$\kappa_{2,0}$ and $ \kappa _{3,0}$.
Then he  added a
weak electric field ${\bf E}\not=0$
to calculate perturbatively
its contributions to $\Delta {\mathcal H}$
in the Born approximation (see for example
Landau and Lifshitz \cite{LanLif75a,LanLif75b}). This led again to
the coefficients (\ref{lconstant}).

The above results receive higher corrections in QED
and are correct only up to order $ \alpha ^2$.
Up to this order,
the field-dependent dielectric and permeability tensors $\epsilon_{ik}$
and $\mu_{ik}$ (\ref{dh1}) have the following real parts
for weak fields
\begin{eqnarray}
{\rm Re}(\epsilon_{ik})&=&\delta_{ik} +
\frac{4\alpha}{45}\big[2(\bar E^2-\bar B^2)\delta_{ik}+7\bar B_i\bar B_k\big]
+{\mathcal O}( \alpha ^2),
\nonumber\\
{\rm Re}(\mu_{ik})&=&\delta_{ik} +
\frac{4\alpha}{45}\big[2(\bar E^2-\bar B^2)\delta_{ik}+7\bar E_i\bar B_k\big]
+{\mathcal O}( \alpha ^2).
\label{dielectricity}
\end{eqnarray}

\subsection{Imaginary part of the effective Lagrangian}

Heisenberg and Euler \cite{1936ZPhy...98..714H} were the first
to  realize that for ${\bf E}\not=0$ the powers
series expansion (\ref{leffective}) is not convergent,
due to singularities of the integrand in (\ref{effectiveint}) at
$s=\pi/\bar E, 2\pi/\bar E,\dots~$. They concluded
that the powers series
expansion (\ref{leffective}) does not yield
all corrections to the Maxwell
Lagrangian, calling for a more careful evaluation
of the integral representation, see Eq.~(\ref{effectiveint}).
Selecting an integration path that avoids
these singularities, they found an imaginary term.
Motivated by Sauter's work \cite{1931ZPhy...69..742S}
on Klein paradox \cite{1929ZPhy...53..157K}, Heisenberg and Euler
estimated the size of the imaginary term  in the effective Lagrangian as
\begin{equation}
-{\frac{8i}{\pi}}{\bar E}^2 m_ec^2\left(\frac{m_ec}{h}\right)^3 e^{-\pi/\bar E},
\label{herate}
\end{equation}
and pointed out that it is associated with pair production by the electric field. This imaginary term
in the effective Lagrangian is related to the imaginary parts of field-dependent dielectric $\epsilon$ 
and permeability $\mu$ of the vacuum. 

In 1950's, Schwinger \cite{s51,s54a,s54b}  derived the
same formula (\ref{effectiveint}) once more
within the quantum field theory of 
Quantum Electromagnetics (QED),
\begin{equation} \!\!\!\!\!\!\!\!
\frac{ \tilde\Gamma }{V}
=\frac{\alpha E^2}{ \pi^2\hbar}\sum_{n=1}^\infty \frac{1}{ n^2}\exp
\left(-\frac{n\pi E_c}{ E}\right).
\label{probability1}
\end{equation}
and its Lorentz-invariant expression in terms of electromagnetic fields  ${\bf E}$ and ${\bf B}$,
\begin{equation}
\frac{ \tilde\Gamma }{V}=\frac{  \alpha   \varepsilon^2}{ \pi^2 }
\sum_{n=1}  \frac{1}{n^2}
\frac{ n\pi\beta / \varepsilon }
{\tanh {n\pi \beta/ \varepsilon}}\exp\left(-\frac{n\pi E_c}{ \varepsilon}\right),
\label{probabilityeh}
\end{equation}
where
\begin{eqnarray} \!\!\!\!\!\!\!\!\!\!\!\!
\left\{
{ \frac{\varepsilon}{\beta} }\right\}
 & \equiv\!&
\frac{1}{ \sqrt{2} }
 \sqrt{
 \sqrt{
 ({\bf E}^2-{\bf B}^2)^2+4( {\bf E}\cdot {\bf B})^2
}
\pm ({{\bf E}^2-{\bf B}^2})
}.
\label{fieldinvariant}
\end{eqnarray}

The exponential factor $e^{\pi\frac{m_e^2c^3}{\hbar e E}}$ in Eqs.~(\ref{transmission}) and (\ref{herate})
characterizes the transmission coefficient of quantum tunneling, 
Ref.~\refcite{1936ZPhy...98..714H} introduced the critical field strength $E_c=\frac{m_e^2c^3}{\hbar e}$ (\ref{dimenlessEB}). 
They compared
it with the field strength $E_e$ of an electron at its classical radius, $E_e=e/r_e^2$ where 
$r_e=\alpha \hbar/(m_ec)$ and $\alpha=1/137$. They found the field strength $E_e$ is 137 time larger than 
the critical field strength $E_c$, i.e., $E_e=\alpha ^{-1}E_c$. At a critical radius $r_c=\alpha^{1/2}\hbar/(m_ec)> r_e$, the
field strength of the electron would be equal to the critical field strength $E_c$. 

As shown in Fig.~\ref{gap}, the negative-energy
spectrum
of solutions of the Dirac equation
has energies ${\mathcal E}_-<-m_ec^2$, and  is separated from the
positive energy-spectrum ${\mathcal E}_+> m_ec^2$
by a gap $2m_ec^2\approx1.02$MeV.
The negative-energy states are all filled.
The energy gap
 is by a factor $4/ \alpha^2
\approx 10^5 $ larger
than the typical binding energy  of atoms
($\sim 13.6$eV).
In order to create an
electron-positron
pair, one must spend this large amount of energy.
The source of this energy can  be an external field.

If an electric field attempts to tear an electron out of the filled state
the gap energy must be gained over the distance of
two electron radii. The virtual particles
give an electron a radius of the order of the
Compton wavelength $ \lambda \equiv \hbar/m_ec$.
Thus we expect  a significant
creation
of electron-positron pairs
if the work done by the electric field $E$
over twice the Compton wave length $\hbar/m_ec$ is larger than $2m_ec^2$
\begin{equation}
eE\left({\frac{2\hbar}{ m_ec}}\right)>2m_ec^2 .
\nonumber
\end{equation}
This condition defines a critical electric field
\begin{equation}
E_c\equiv{\frac{m_e^2c^3}{ e\hbar}}\simeq 1.3\cdot 10^{16}\, {\rm V/cm},
\label{critical1}
\end{equation}
above which
pair creation becomes abundant.
To have an idea how large this critical electric field is, we
compare it with the value of the electric field required to ionize a hydrogen atom.
There the above inequality
holds for twice of the Bohr radius and the Rydberg energy
\begin{equation}
eE^{\rm ion}\left({\frac{2\hbar}{  {\alpha  m_ec}}}\right)> \alpha ^2{m_ec^2},
\nonumber
\end{equation}
so that $E_c\approx E^{\rm ion}_c/ \alpha^3 $ is about $10^6$
times as large,
a value that has so far
not been reached in a laboratory on Earth.

\section{Pair production in Coulomb potential of nuclei and heavy-ion collisions}

By far the major attention to build a critical electric field has occurred in the physics of heavy nuclei and in heavy ion collisions. We recall in the following some of the basic ideas, calculations, as well as experimental attempts to obtain the pair creation process in nuclear physics.

\subsection{The $Z=137$ catastrophe}

\begin{figure}
\centering 
\includegraphics[width=\hsize,clip]{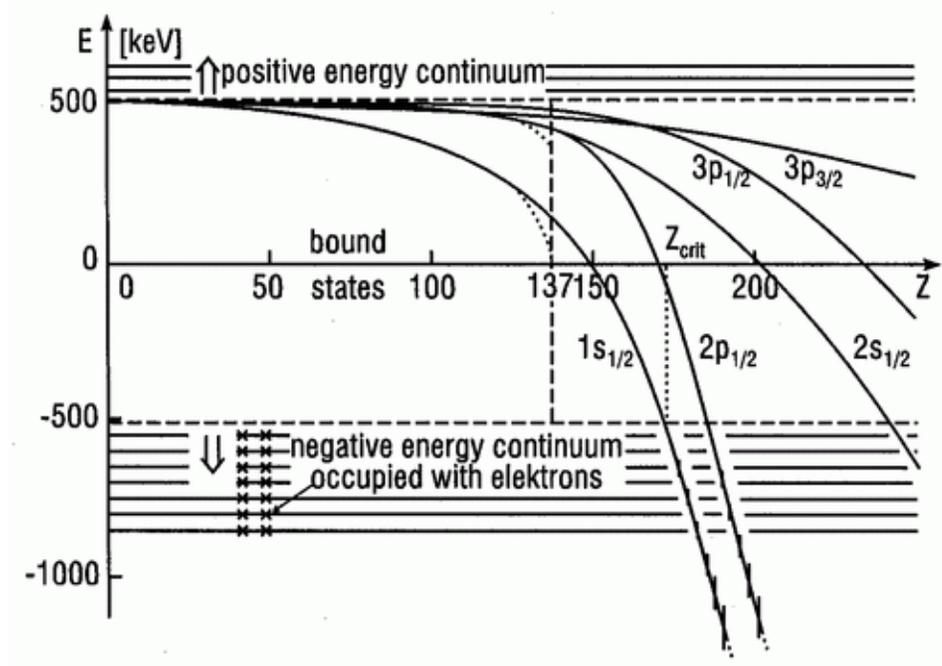}
\caption{Atomic binding energies as function of nuclear charge Z. This figure is reproduced from
Fig.~1 in Ref.~\refcite{grc98}.}%
\label{zspectrum}%
\end{figure}

Soon after the Dirac equation for a relativistic electron was discovered \cite{z4a,z4b,z5}, Gordon \cite{z7} (for all $Z< 137$) and Darwin \cite{z6} (for $Z=1$) found its solution in the point-like Coulomb potential $V(r)=-Z\alpha/r, \quad 0<r<\infty$. 
Solving the differential equations for the Dirac wave function, they obtained the well-known Sommerfeld's formula \cite{Sommerfeld-X} for the energy-spectrum, 
\begin{equation}
{\mathcal E}(n,j)=m_ec^2\left[1+\left(\frac{Z\alpha}{ n-|K|+(K^2-Z^2\alpha^2)^{1/2}}\right)^2\right]^{-1/2}.
\label{dirac}
\end{equation}
Here the principle quantum number $n=1,2,3,\cdot\cdot\cdot$ and 
\begin{equation}
K=\left\{\begin{array}{ll} -(j+1/2)= -(l+1), & {\rm if}\quad j=l+\frac{1}{2}, \quad l\ge 0 \\
 (j+1/2)= l, & {\rm if}\quad j=l-\frac{1}{2}, \quad l\ge 1
 \end{array}\right.
\label{dirac-k}
\end{equation}
where $l=0,1,2,\ldots$ is the orbital angular momentum corresponding to the upper component of Dirac bi-spinor, 
$j$ is the total angular momentum, and the states with 
$K=\mp 1,\mp 2,\mp 3,\cdot\cdot\cdot, \mp (n-1)$ are doubly degenerate, while the state $K=-n$ is a singlet \cite{z7,z6}. 
The integer values $n$ and $K$ label bound states whose energies are ${\mathcal E}(n,j)\in (0,m_ec^2)$. For the example, 
in the case of the lowest energy states, one has     
\begin{eqnarray}
{\mathcal E}(1S_{\frac{1}{2}})&=& \sqrt {1-(Z\alpha)^2},\label{dirac-k1}\\
{\mathcal E}(2S_{\frac{1}{2}})&=&{\mathcal E}(2P_{\frac{1}{2}})
= \sqrt{\frac{1+\sqrt{1-(Z\alpha)^2}}{2}},\label{dirac-k2}\\
{\mathcal E}(2P_{\frac{3}{2}})&=& \sqrt {1-\frac{1}{4}(Z\alpha)^2}.
\label{dirac-k3}
\end{eqnarray}
For all states of the discrete spectrum, the binding energy
$mc^2-{\mathcal E}(n,j)$ increases as the nuclear charge $Z$ increases, as shown in
Fig.~\ref{zspectrum}.  When $Z=137$, ${\mathcal E}(1S_{1/2})=0$, 
${\mathcal E}(2S_{1/2})={\mathcal E}(2P_{1/2})=1/\sqrt{2}$ 
and ${\mathcal E}(2S_{3/2})=\sqrt{3}/2$. 
Gordon noticed in his pioneer paper \cite{z7} that no regular solutions with $n=1,j=1/2,l=0,$ and $K=-1$ (the $1S_{1/2}$ ground state) are found beyond $Z=137$.
This phenomenon is the so-called ``$Z=137$ catastrophe'' and it is associated with the
assumption that the nucleus is point-like in calculating the electronic energy-spectrum.

\subsection{Semi-Classical description}\label{eff}

In order to have further understanding of this phenomenon, we study it
in the semi-classical scenario.
Setting the origin of spherical coordinates $(r,\theta,\phi)$ at the
point-like charge, we introduce the 
vector potential $A_\mu=({\bf A}, A_0)$, where ${\bf A}=0$ and $A_0$
is the Coulomb potential. The motion 
of a relativistic ``electron'' (scalar particle) with mass $m$ and
charge $e$ is described by its radial momentum $p_r$, angular momenta 
$p_\phi$ and the Hamiltonian,
\begin{eqnarray}
H_\pm &=& \pm
mc^2\sqrt{1+(\frac{p_r}{mc})^2+(\frac{p_\phi}{mcr})^2}-V(r),
\label{toth}
\end{eqnarray}
where the potential energy $V(r)=eA_0$, and $\pm$ corresponds for
positive and negative solutions.
The states corresponding to negative energy solutions are fully
occupied. 
The angular momentum $p_\phi$ is conserved, for the Hamiltonian is
spherically symmetric. 
For a given angular momentum $p_\phi$, the Hamiltonian (\ref{toth})
describes electron's radial motion in 
the following the effective potential
\begin{eqnarray}
E_\pm &=& \pm mc^2\sqrt{1+(\frac{p_\phi}{mcr})^2}-V(r).
\label{tote}
\end{eqnarray} 
The Coulomb potential energy $V(r)$ is given by
\begin{equation}
V(r)=\frac{Ze^2}{r},
\label{potenoutside}
\end{equation}
where  $Ze^2=|Qe|$. 
 
In the classical scenario, given different values of angular momenta
$p_\phi$, the stable circulating orbits (states) 
are determined by the minimum of the effective potential $E_+(r)$
(\ref{tote}).
Using $dE_+(r)/dr =0$, we obtain the stable 
orbit location at the radius $R_L$ in the unit of the Compton length
$\lambda=\hbar/mc$,
\begin{equation}
R_L(p_\phi)=Z\alpha\lambda\sqrt{1-\left(\frac{Z\alpha}{p_\phi/\hbar}\right)^2},
\label{stable}
\end{equation}
where $\alpha=e^2/\hbar c$ and $p_\phi> Z\alpha$.
Substituting Eq.~(\ref{stable}) into Eq.~(\ref{tote}), we find the
energy of the electron at each stable orbit,
\begin{equation}
{\mathcal E}(p_\phi)\equiv {\rm min}(E_+) =
mc^2\sqrt{1-\left(\frac{Z\alpha}{p_\phi/\hbar}\right)^2}.
\label{mtote}
\end{equation}
The last stable orbits (minimal energy) are given by   
\begin{equation}
p_\phi\rightarrow Z\alpha \hbar  + 0^+,
\quad R_L(p_\phi) \rightarrow 0^+,\quad {\mathcal E}(p_\phi) \rightarrow 0^+.
\label{mtote1}
\end{equation}
For stable orbits for $p_\phi\gg 1$, the radii $R_L\gg 1$ and energies
${\mathcal E}\rightarrow mc^2+0^-$; 
electrons in these orbits are critically bound since their banding energy
goes to zero. As the energy-spectrum 
(\ref{dirac}) (see Eqs.~(\ref{dirac-k1},\ref{dirac-k2},\ref{dirac-k3}),
Eq.~(\ref{mtote}) shows, only
positive or null energy solutions (states) exist in the presence of
a point-like nucleus.  
 
In the semi-classical scenario, the discrete values of angular momentum
$p_\phi$ are selected by the Bohr-Sommerfeld quantization rule
\begin{equation}
\int p_\phi d\phi \simeq h (l+\frac{1}{2}),\quad \Rightarrow\quad
p_\phi(l) \simeq \hbar(l+\frac{1}{2}),\quad l=0,1,2,3,\ldots
\label{angq}
\end{equation}
describing the semi-classical states of radius and energy
\begin{eqnarray}
R_L(l) &\simeq &
Z\alpha\lambda\sqrt{1-\left(\frac{2Z\alpha}{2l+1}\right)^2},\label{rlq}\\
{\mathcal E}(l) &\simeq &
mc^2\sqrt{1-\left(\frac{2Z\alpha}{2l+1}\right)^2}\label{elq}.
\end{eqnarray}
Other values of angular momentum $p_\phi$, radius $R_L$ and energy
${\mathcal E}$ given by Eqs.~(\ref{stable},\ref{mtote}) 
in the classical scenario are not allowed. 
When these semi-classical states are not occupied as required by the
Pauli Principle, 
the transition from one state 
to another with different discrete values ($l_1,l_2$ and $\Delta
l=l_2-l_1=\pm 1$) is made by emission or absorption of a spin-1
($\hbar$) 
photon. Following the energy and angular-momentum conservations, photon
emitted or absorbed in the transition have angular momenta 
$p_\phi(l_2)-p_\phi(l_1)=\hbar (l_2-l_1)=\pm\hbar$ and energy 
${\mathcal E}(l_2)-{\mathcal E}(l_1)$. As required by the Heisenberg
indeterminacy principle 
$\Delta\phi\Delta p_\phi \simeq 4\pi p_\phi(l) \gtrsim h$,
the absolute ground state for minimal energy and angular momentum is
given by the $l=0$ state, $p_\phi\sim \hbar/2$, $R_L\sim
Z\alpha\lambda\sqrt{1-(2Z\alpha)^2}>0$
and ${\mathcal E} \sim  mc^2\sqrt{1-(2Z\alpha)^2}>0$ for $Z\alpha \le
1/2$. Thus the stability of all semi-classical states 
$l>0$ is guaranteed by the Pauli principle. In contrast for $Z\alpha > 1/2$,
there is not an absolute ground state
in the semi-classical scenario.
 
We see now how the lowest energy states are selected by the quantization
rule in the semi-classical scenario
out of the last stable orbits (\ref{mtote1}) in the classical scenario.
For the case of $Z\alpha\le 1/2$, equating Eq.~(\ref{mtote1}) to
$p_\phi=\hbar(l+ 1/2)$ (\ref{angq}), we find the selected state $l=0$ is
only possible solution so that the ground state $l=0$ 
in the semi-classical scenario corresponds to the last stable orbits
(\ref{mtote1}) in the classical scenario. On the other hand for the case 
$Z\alpha > 1/2$, equating Eq.~(\ref{mtote1}) to
$p_\phi=\hbar(l+ 1/2)$ (\ref{angq}), we find the selected state
$l=\tilde l\equiv (Z\alpha-1)/2>0$ 
in the semi-classical scenario corresponds to the last stable orbits
(\ref{mtote1}) in the classical scenario. This state $l=\tilde l>0$
is not protected by the Heisenberg indeterminacy principle from
quantum-mechanically decaying in $\hbar$-steps to the states 
with lower angular momentum and energy (correspondingly smaller radius
$R_L$ (\ref{rlq})) via photon emissions. This clearly shows
that the ``$Z=137$-catastrophe'' corresponds to $R_L\rightarrow 0$,
falling to the center of the Coulomb potential 
and all semi-classical states ($l$) are unstable. 

\subsection{The critical value of the nuclear charge $Z_{cr}=173$.}

A very different situation is encountered when considering the fact the nucleus is not point-like and has an extended charge distribution \cite{g1a,g1b,g1c,g1d,z10,z11a,z11b,z12,z}. When doing so, the $Z=137$ catastrophe disappears and the energy-levels ${\mathcal E}(n,j)$ of the bound states $1S$, $2P$ and $2S$, $\cdot\cdot\cdot$ smoothly continue to drop toward the negative energy continuum as $Z$ increases to values larger than $137$, as shown in Fig.~\ref{zspectrum}. The reason is that the finite size $R$ of the nucleus charge distribution provides a cutoff for the boundary condition at the origin $r\rightarrow 0$ and the energy-levels ${\mathcal E}(n,j)$ of the Dirac equation are shifted due to this cutoff. In order to determine the critical value $Z_{cr}$ when the negative energy continuum (${\mathcal E}<- m_ec^2$) is encountered (see Fig.~\ref{zspectrum}), Zel'dovich and Popov\cite{z10,z11a,z11b,z12,z} solved the Dirac equation corresponding to a nucleus of finite extended charge distribution, i.e., the Coulomb potential is modified as
\begin{equation}
V(r)=\left\{\begin{array}{ll} -\frac{Ze^2}{ r}, &  r>R, \\
 -\frac{Ze^2}{ R}f\left(\frac{r}{ R}\right), &  r<R,
 \end{array}\right.
\label{extpotential}
\end{equation}
where $R\sim 10^{-12}$cm is the size of the nucleus. The form of the cutoff function $f(x)$ depends on the
distribution of the electric charge over the volume of the nucleus $(x=r/R, 0<x<1$, with $f(1)=1)$.
Thus, $f(x)=(3-x^2)/2$ corresponds to a constant volume density of charge. 

Solving the Dirac equation with the modified Coulomb potential (\ref{extpotential}) and calculating
the corresponding perturbative shift $\Delta {\mathcal E}_R$ of the lowest energy level (\ref{dirac-k1})
one obtains\cite{z10,z}
\begin{equation}
\Delta {\mathcal E}_R= m_ec^2\frac{(\xi)^2(2\xi e^{-\Lambda})^{2\gamma_z}}{\gamma_z (1+2\gamma_z)}
\left[1-2\gamma_z\int_0^1f(x)x^{2\gamma_z} dx\right],
\label{zshift}
\end{equation}
where $\xi=Z\alpha$, $\gamma_z=\sqrt{1-\xi^2}$ and $\Lambda=\ln(\hbar / m_ecR)\gg 1$ is a logarithmic parameter 
in the problem under consideration.
The asymptotic expressions for the $1S_{1/2}$ energy that were obtained are\cite{z12,z} 
\begin{equation}
{\mathcal E}(1S_{1/2})=m_ec^2\left\{\begin{array}{ll} \sqrt {1-\xi^2}\coth(\Lambda\sqrt {1-\xi^2}), &  0<\xi<1, \\
 \Lambda^{-1}, &  \xi=1,\\
 \sqrt {\xi^2-1}\cot(\Lambda\sqrt {\xi^2-1}), &  \xi>1.
 \end{array}\right.
\label{easymptotic}
\end{equation}
As a result, the ``$Z=137$ catastrophe'' in Eq.~(\ref{dirac}) disappears and 
${\mathcal E}(1S_{1/2})=0$ gives
\begin{equation}
\xi_0=1+\frac{\pi^2}{8\Lambda}+{\mathcal O}(\Lambda^{-4});
\label{0xi}
\end{equation}
the state $1S_{1/2}$ energy continuously
goes down to the negative energy continuum since $Z\alpha >1$, 
and ${\mathcal E}(1S_{1/2})=-1$ gives
\begin{equation}
\xi_{cr}=1+\frac{\pi^2}{2\Lambda(\Lambda +2)}+{\mathcal O}(\Lambda^{-4})
\label{0xicr}
\end{equation}
as shown in Fig.~\ref{zspectrum}. 
In Ref.~\refcite{z10,z} it is found that the critical value $\xi_c^{(n)}= Z_c\alpha$
for the energy-levels $nS_{1/2}$ and $nP_{1/2}$ to reach the negative energy continuum
is equal to
\begin{equation}
\xi_c^{(n)}= 1+ \frac{n^2\pi^2}{ 2\Lambda^2}+{\mathcal O}(\Lambda^{-3}).
\label{criticalxi}
\end{equation}
The critical value increases rapidly with
increasing $n$. As a result, it is found that
$Z_{cr}\simeq 173$ is a critical value at which the lowest energy-level of the bound state
$1S_{1/2}$ encounters the negative energy continuum, while other bound states encounter the negative energy continuum at $Z_{cr}>173$ (see also Ref.~\refcite{g1c} for a numerical estimation of the same spectrum).
We refer the readers to Ref.~\refcite{z10,z11a,z11b,z12,z,popov2001} for mathematical and numerical details.

When $Z>Z_{cr}=173$, the lowest energy-level of the bound state $1S_{1/2}$ enters the negative energy continuum.
Its energy-level can be estimated as follows
\begin{equation}
{\mathcal E}(1S_{1/2})=m_ec^2 - \frac{Z \alpha}{\bar r}<-m_ec^2,
\label{1S}
\end{equation}
where $\bar r $ is the average radius of the $1S_{1/2}$ state's orbit, and the binding energy of this
state satisfies $Z\alpha/\bar r > 2 m_ec^2$. If this bound state is unoccupied, 
the bare nucleus gains a binding energy $Z\alpha/\bar r$ larger than 
$2m_ec^2$, and becomes unstable against the production of an electron-positron pair. Assuming this 
pair-production occurs around the radius $\bar r$, we have energies for the electron ($\epsilon_-$) and positron ($\epsilon_+$) given by
\begin{equation}
\epsilon_-=\sqrt{|c{\bf p}_-|^2+m_e^2c^4}-\frac{Z \alpha}{\bar r};\ \quad \epsilon_+=\sqrt{|c{\bf p}_+|^2+m_e^2c^4}+\frac{Z \alpha}{\bar r},
\label{eofep}
\end{equation}
where ${\bf p}_\pm$ are electron and positron momenta, and ${\bf p}_-=-{\bf p}_+$. 
The total energy required for the production of a pair is
\begin{equation}
\epsilon_{-+}=\epsilon_-+\epsilon_+=2\sqrt{|c{\bf p}_-|^2+m_e^2c^4},
\label{totaleofep}
\end{equation}
which is independent of the potential $V(\bar r)$. The potential energies $\pm eV(\bar r)$ of the electron 
and positron cancel 
each other out and do not contribute to the total energy (\ref{totaleofep}) required for pair production. 
This energy (\ref{totaleofep}) is acquired from the binding energy ($Z\alpha/\bar r > 2 m_ec^2$)
by the electron filling into the bound state $1S_{1/2}$. A part of the binding energy becomes 
the kinetic energy of positron that goes out.  
This is analogous to the familiar case that a proton ($Z=1$)
catches an electron into the ground state $1S_{1/2}$, and a photon is emitted with the energy not less than
13.6 eV.   
In the same way, 
more electron-positron pairs are produced, when $Z\gg Z_{cr}=173$ the energy-levels of 
the next bound states $2P_{1/2},2S_{3/2},\ldots$ 
enter the negative energy continuum, provided these bound states of the bare nucleus are unoccupied.  

\subsection{Positron production}

Ref.~\refcite{GZ69,GZ70} proposed that when $Z>Z_{cr}$, the bare nucleus spontaneously produces pairs of electrons and positrons: the two positrons\footnote{Hyperfine structure of $1S_{1/2}$ state: single and triplet.} 
go off to infinity and the effective charge of the bare nucleus decreases
by two electrons, which corresponds exactly to filling the K-shell.\footnote{The supposition was made 
in Ref.~\refcite{GZ69,GZ70} that the electron density of $1S_{1/2}$ state, as well as the vacuum polarization density, 
is delocalized at $Z\rightarrow Z_{cr}$. Later it was proven to be incorrect \cite{z11a,z11b,z}.}
A more detailed investigation was made for the solution of
the Dirac equation at $Z\sim Z_{cr}$, when the lowest electron level $1S_{1/2}$ merges with the
negative energy continuum, by Ref.~\refcite{z10,z11a,z11b,z12,z65}. This further clarified the situation, showing that at
$Z\gtrsim Z_{cr}$, an imaginary resonance energy of Dirac equation appears
\begin{equation}
\epsilon = \epsilon_0 - i\frac{\Gamma}{2},
\label{zimaginary}
\end{equation}
where
\begin{eqnarray}
\epsilon_0 &=&-m_ec^2- a(Z-Z_{cr}),
\label{epsilon0}\\
\Gamma &\sim& \theta(Z-Z_{cr})\exp \left(-b\sqrt\frac{Z_{cr}}{ Z-Z_{cr}}\right),
\label{zprobability}
\end{eqnarray}
and $a,b$ are constants, depending on the cutoff $\Lambda$ (for example, $b=1.73$ for $Z=Z_{cr}=173$, see Ref.~\refcite{z11a,z11b,z}). The
energy and momentum of the emitted positrons are $|\epsilon_0|$ and $|{\bf p}|=\sqrt{|\epsilon_0|-m_ec^2}$.

The kinetic energy of the two positrons at infinity is given by
\begin{equation}
\varepsilon_p = |\epsilon_0| - m_ec^2 = a(Z-Z_{cr})+\ldots,
\label{zkinetic}
\end{equation}
which is proportional to $Z-Z_{cr}$ (as long as $(Z-Z_{cr})\ll Z_{cr}$) and tends to zero as $Z\rightarrow Z_{cr}$.
The pair-production resonance at the energy (\ref{zimaginary}) is extremely narrow
and practically all positrons are emitted with almost same kinetic energy for $Z\sim Z_{cr}$, i.e., nearly
monoenergetic spectra (sharp line structure).
Apart from a pre-exponential factor, $\Gamma$ in Eq.~(\ref{zprobability}) coincides with the probability of positron production, i.e., the penetrability of the Coulomb barrier.
The related problems of vacuum charge density due to electrons filling into the K-shell and charge renormalization due to the change of wave function of electron states are discussed by Ref.~\refcite{z20,z21,z22,z23,z24}.
An extensive and detailed review on this theoretical issue can be found in Ref.~\refcite{grc98,z,popov2001,Greinerbook}.

On the other hand, some theoretical work has been done studying the possibility that
pair production due to bound states encountering the negative energy continuum is
prevented from occurring by higher order processes of quantum
field theory, such as charge renormalization, electron self-energy and nonlinearities in electrodynamics and even the Dirac field itself \cite{g3a,g3b,g3c,g3d,grc98-5a,grc98-5b,grc98-6}.
However, these studies show that various effects modify $Z_{cr}$ by a few percent, but have no way to prevent the binding energy from increasing to $2m_ec^2$ as $Z$ increases, without simultaneously contradicting the existing precise experimental data on stable
atoms \cite{g}.

Following this, special attention has been given to understand the process of creating a nucleus with $Z > Z_{cr}$ by collision of two nuclei of charge $Z_1$ and $Z_2$ such that $Z=Z_1+Z_2 \ge Z_{cr}$\cite{z65,GZ69,GZ70,greiner1972a,greiner1972b}. To observe the emission of positrons coming from pair production occurring near an overcritical nucleus
temporarily formed by two nuclei, the following necessary conditions have to be fullfilled:
(i) the atomic number of an overcritical nucleus must be larger than $Z_{cr}=173$;
(ii) the lifetime (the sticking time of two-nuclear collisions) of the overcritical
nucleus must be much longer than the characteristic time $(\hbar/m_ec^2)$ of pair production;
(iii) inner shells (K-shell) of the overcritical nucleus should be unoccupied.

\begin{figure}
\centering 
\includegraphics[width=\hsize,clip]{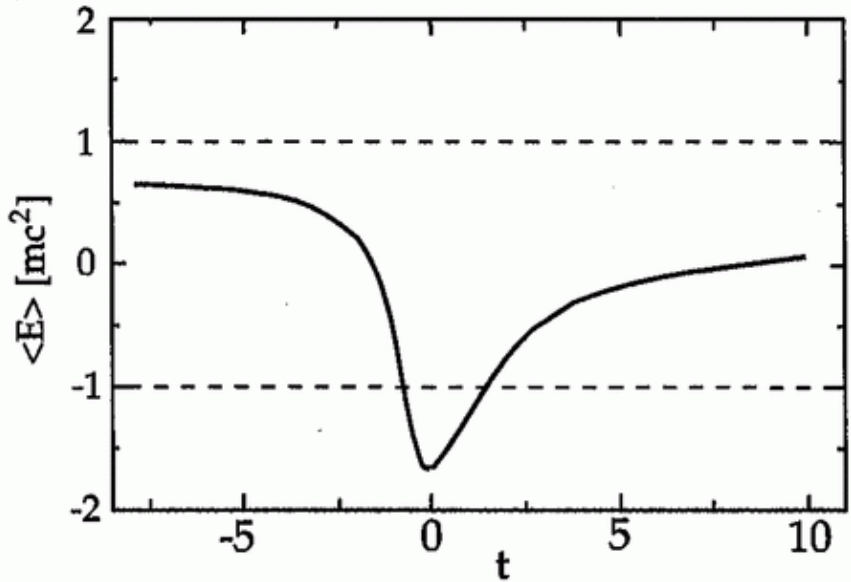}
\caption{Energy expectation values of the $1s\sigma$ state in a U+U collision at 10 GeV/nucleon.
The unit of time is $\hbar/m_ec^2$. This figure is reproduced from Fig.~4 in Ref.~\refcite{grc98}.}%
\label{example}%
\end{figure}

When in the course of a heavy-ion collision the two nuclei come into contact, some deep-inelastic reactions have been claimed to exist for a certain time $\Delta t_s$. The duration $\Delta t_s$ of this contact (sticking time) is expected to depend on the nuclei involved in the reaction and on beam energy. For very heavy nuclei, the Coulomb interaction is the dominant force between the nuclei, so that the sticking times $\Delta t_s$ are typically much shorter and on the average
probably do not exceed $1\sim 2\cdot 10^{-21}$ sec \cite{grc98}. Accordingly the calculations (see
Fig.~\ref{example})
also show that the time when the binding energy is overcritical is very short,
about $1.2\cdot 10^{-21}$ sec. 
Theoretical attempts have been proposed to study the nuclear aspects
of heavy-ion collisions at energies very close to the Coulomb barrier and search
for conditions, which would serve as a trigger for prolonged nuclear reaction times,
(the sticking time $\Delta t_s$) to enhance the amplitude of pair production \cite{grc98,g,g5,g6,grc98-30}.
Up to now no conclusive theoretical or experimental
evidence exists for long nuclear delay times in very heavy collision systems.

It is worth noting that several other dynamical processes contribute to
the production of positrons in undercritical as well as in overcritical collision systems \cite{g3a,g3b,g3c,g3d}.
Due to the time-energy uncertainty relation (collision broadening),
the energy-spectrum of such positrons has a rather broad and
oscillating structure, considerably different from a sharp line structure
that we would expect from pair-production positron emission alone.

\subsection{Experiments}

As already remarked, if the sticking time $\Delta t_s$ could be prolonged, the probability of pair production in vacuum around the super heavy 
nucleus would be enhanced. As a consequence, the spectrum of emitted positrons is expected to develop a sharp line structure, indicating the
spontaneous vacuum decay caused by the overcritical electric field of a forming super heavy nuclear system with $Z\ge Z_{cr}$. If the sticking time $\Delta t_s$ is not long enough
and the sharp line of pair production positrons has not yet well-developed, in the observed
positron spectrum it is difficult to distinguish the pair production
positrons from positrons created through other different mechanisms.
Prolonging the sticking time and identifying pair production positrons among all other
particles \cite{g10,g11} created in the collision process are important experimental tasks \cite{g7a,g7b,g7c,g8,g9,g18-22a,g18-22b,g18-22c}.

\begin{figure}
\centering 
\includegraphics[width=\hsize,clip]{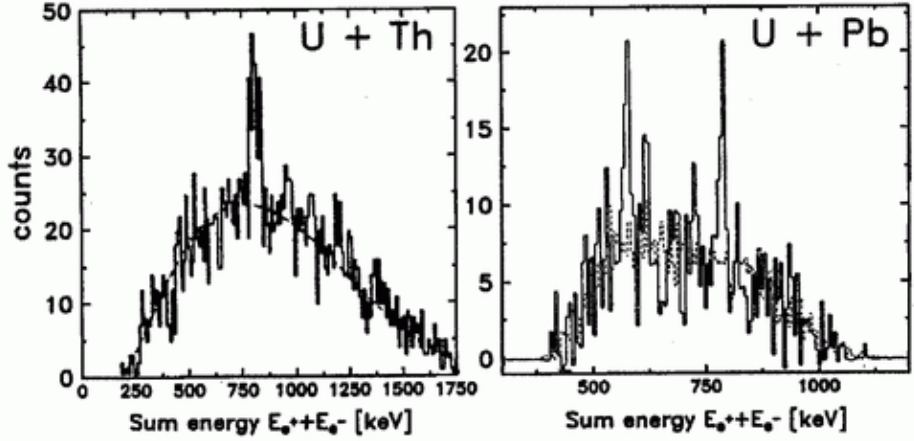}
\caption{Two typical example of coincident electron-positron spectra measured by the
Epose group in the system U+Th (left) and by the Orange group in U+Pb collisions (right).
When plotted as a function of the total energy of the electron and positron, very narrow line structures were observed.
This figure is reproduced from Fig.~7 in Ref.~\refcite{grc98}.}%
\label{linestructure}%
\end{figure}

For nearly 20 years the study of atomic excitation processes and in particular of positron creation in heavy-ion collisions has been a major research topic at GSI (Darmstadt) \cite{zexp1,G...96,L...97,zexp4,H...98}.
The Orange and Epos groups at GSI (Darmstadt) discovered narrow line structures
(see Fig.~\ref{linestructure})
of unexplained origin, first in the single positron energy spectra and later in coincident
electron-positron pair emission. Studying more collision systems with a wider range of the
combined nuclear charge $Z=Z_1+Z_2$ they found that narrow line structures are essentially
independent of $Z$. This rules out the explanation of a pair-production positron,
since the line would be expected at the position of the $1s\sigma$ resonance, i.e., at a
kinetic energy given by Eq.~(\ref{zkinetic}), which is strongly $Z$ dependent.
Attempts to link this positron line to spontaneous pair production have failed.
Other attempts to explain this positron line in term of atomic
physics and new particle scenario were not successful as well \cite{grc98}.

The anomalous positron line problem has perplexed experimentalists and theorists alike for more than a decade.
Moreover, later results
obtained by the Apex collaboration at Argonne National Laboratory showed no statistically significant positron line structures \cite{A...95,grc98-52b}. This is in strong contradiction with
the former results obtained by the Orange and Epos groups. However, the analysis of
Apex data was challenged in the comment by Ref.~\refcite{grc98-53a,grc98-53b} for the Apex measurement
would have been less sensitive to extremely narrow positron lines. A new generation of
experiments (Apex at Argonne and the new Epos and Orange setups at GSI) with much improved
counting statistics has failed to reproduce the earlier results \cite{grc98}.

To overcome the problem posed by the short time scale of
pair production ($10^{-21}$ sec), hopes rest on the idea to select collision systems
in which a nuclear reaction with sufficient sticking time occurs. Whether such a situation
can be realized still is an open question \cite{grc98}. In addition, the anomalous positron line
problem and its experimental contradiction overshadow on the field of studying
the pair production in heavy ion collisions.
In summary, clear experimental signals for electron-positron pair production in heavy ion 
collisions are still missing \cite{grc98} at the present time.

\section{The extraction of blackholic energy from a black hole by vacuum polarization processes}

We recall here the basic steps leading to the study of a critical electric field in a Kerr-Newman black hole. We recall the theoretical framework to apply the Schwinger process in general relativity in the field of a Kerr-Newman geometry as well as the process of extraction of the ``blackholic'' energy. We then recall the basic concepts of dyadosphere and dyadotorus leading to a photon-electron-positron plasma surrounding the black hole.

\subsection{The mass-energy formula of black holes}

The same effective potential technique (see Landau and Lifshitz \cite{ll2}) which allowed the analysis of circular orbits around the black hole was crucial in reaching the equally interesting discovery of the reversible and irreversible transformations of black holes by Christodoulou and Ruffini \cite{cr71}, which in turn led to the mass-energy formula of the black hole
\begin{equation} 
E_{BH}^2 = M^2c^4 = \left(M_{\rm ir}c^2 + \frac{Q^2}{2\rho_+}\right)^2+\frac{L^2c^2}{\rho_+^2}\, ,
\label{em} 
\end{equation} 
with 
\begin{equation} 
\frac{1}{\rho_+^4}\left(\frac{G^2}{c^8}\right)\left(Q^4+4L^2c^2\right)\leq 1\, , 
\label{s1}
\end{equation} 
where 
\begin{equation} 
S=4\pi\rho_+^2=4\pi(r_+^2+\frac{L^2}{c^2M^2})=16\pi\left(\frac{G^2}{c^4}\right) M^2_{\rm ir}\, ,
\label{sa} 
\end{equation} 
is the horizon surface area, $M_{\rm ir}$ is the irreducible mass, $r_{+}$ is the horizon radius and $\rho_+$ is the quasi-spheroidal cylindrical coordinate of the horizon evaluated at the equatorial plane. Extreme black holes satisfy the equality in Eq.~(\ref{s1}).

From Eq.~(\ref{em}) follows that the total energy of the black hole $E_{BH}$ can be split into three different parts: rest mass, Coulomb energy and rotational energy. In principle both Coulomb energy and rotational energy can be extracted from the black hole (Christodoulou and Ruffini \cite{cr71}). The maximum extractable rotational energy is 29\% and the maximum extractable Coulomb energy is 50\% of the total energy, as clearly follows from the upper limit for the existence of a black hole, given by Eq.~(\ref{s1}). We refer in the following to both these extractable energies as the blackholic energy. We outline how the extraction of the blackholic energy is indeed made possible by electron-positron pair creation. We also introduce the concept of a ``dyadosphere'' or ``dyadotorus'' around a black hole and we will the outline how the evolution of the electron-positron plasma will naturally lead to a self-acceleration process creating the very high Lorentz gamma factor observed in GRBs.

\subsection{Vacuum polarization in Kerr-Newman geometries}\label{ruffini}

We already discussed the phenomenon of
electron-positron pair production in a strong electric field in a flat
space-time. We study the same phenomenon
occurring around a black hole with electromagnetic structure (EMBH).
For simplicity and in order to give the fundamental energetic estimates, we postulate that the collapse has already occurred and
has led to the formation of an EMBH. Clearly, this is done only in order to give an estimate of the transient phase occurring during the gravitational collapse. In reality, an EMBH will never be formed because the vacuum polarization process will carry away its electromagnetic energy during the last phase of gravitational collapse. Indeed being interested in this transient phenomenon, we can estimate its energetics by the conceptual analysis of an already formed EMBH, and this is certainly valid for the estimate of the energetics which we will encounter in a realistic phase of gravitational collapse.

The spacetime around the EMBH is
described by the Kerr-Newman geometry whose metric we rewrite here for convenience
in Boyer-Lindquist coordinates $(t,r,\theta,\phi)$
\begin{equation}
ds^{2}={\frac{\Sigma}{\Delta}}dr^{2}+\Sigma d\theta^{2}+{\frac{\Delta}{\Sigma
}}(dt-a\sin^{2}\theta d\phi)^{2}+{\frac{\sin^{2}\theta}{\Sigma}}\left[
(r^{2}+a^{2})d\phi-adt\right]  ^{2},\label{kerrnewmannBL}%
\end{equation}
where $\Delta=r^{2}-2Mr+a^{2}+Q^{2}$ and $\Sigma=r^{2}+a^{2}\cos^{2}\theta$,
as before and as usual $M$ is the mass, $Q$ the charge and $a$ the angular
momentum per unit mass of the EMBH. We recall that the Reissner-Nordstr{\o}m
geometry is the particular case $a=0$ of a non-rotating black hole.  Natural units $G=\hbar=c=1$ will be adopted in this 
section.

The
electromagnetic vector potential around the Kerr-Newman black hole is given
in Boyer-Lindquist coordinates by
\begin{equation}
\mathbf{A}=-Q\Sigma^{-1}r(dt-a\sin^{2}\theta d\phi).
\label{potentialcurve}
\end{equation}
The electromagnetic field tensor is then
\begin{align}
\mathbf{F}= &\  d\mathbf{A}=   2Q\Sigma^{-2}[(r^{2}-a^{2}\cos^{2}\theta)dr\wedge
dt-2a^{2}r\cos\theta\sin\theta d\theta\wedge dt\nonumber\\
& -a\sin^{2}\theta(r^{2}-a^{2}\cos^{2}\theta)dr\wedge d\phi+2ar(r^{2}%
+a^{2})\cos\theta\sin\theta d\theta\wedge d\phi].
\end{align}

After some preliminary work in Refs.~\refcite{dr12z,dr12s,dr12unruh}, the occurrence of pair production in a Kerr-Newman geometry was addressed by Deruelle \cite{dr12mg1}.
In a Reissner-Nordstr\"om geometry, QED pair production has been studied
by Zaumen\cite{dr13zaumen} and Gibbons\cite{dr13gibbons}. The corresponding problem of QED pair production in the Kerr-Newman geometry was addressed by Damour and Ruffini\cite{dr75}, who obtained the rate of pair production with particular emphasis on:
\begin{itemize}
\item the limitations imposed by pair production on the strength of the
electromagnetic field of a black hole \cite{r70};
\item the efficiency of extracting rotational and Coulomb energy (the ``blackholic'' energy) from a black hole by pair production;
\item the possibility of having observational consequences of astrophysical interest.
\end{itemize}
The third point was in fact a far-reaching prevision of possible energy
sources for gamma ray bursts that are now one of the most important phenomena under current
theoretical and observational study. In the following, we recall the
main results of the work by Damour and Ruffini.

In order to study the pair production in the Kerr-Newman geometry, they introduced
at each event $(t,r,\theta,\phi)$ a local Lorentz frame associated with a
stationary observer ${\mathcal{O}}$ at the event $(t,r,\theta,\phi)$. A
convenient frame is defined by the following orthogonal tetrad \cite{dr15}
\begin{align}
\boldsymbol{\omega}^{(0)} &  =(\Delta/\Sigma)^{1/2}(dt-a\sin^{2}\theta
d\phi),\label{tetrad1}\\
\boldsymbol{\omega}^{(1)} &  =(\Sigma/\Delta)^{1/2}dr,\label{tetrad2}\\
\boldsymbol{\omega}^{(2)} &  =\Sigma^{1/2}d\theta,\label{tetrad3}\\
\boldsymbol{\omega}^{(3)} &  =\sin\theta\Sigma^{-1/2}((r^{2}+a^{2}%
)d\phi-adt).\label{tetrad4}%
\end{align}
In this Lorentz frame, the electric potential $A_{0}$, the electric
field ${\bf E}$ and the magnetic field ${\bf B}$ are given by the following
formulas (c.e.g. Ref.~\refcite{mtw73}),
\begin{align*}
A_{0} &  =\boldsymbol{\omega}_{a}^{(0)}A^{a},\\
{\bf E}^{\alpha} &  =\boldsymbol{\omega}_{\beta}^{(0)}F^{\alpha\beta},\\
{\bf B}^{\beta} &  ={\frac{1}{2}}\boldsymbol{\omega}_{\gamma}^{(0)}%
\epsilon^{\alpha\gamma\delta\beta}F_{\gamma\delta}.
\end{align*}
We then obtain
\begin{equation}
A_{0}=-Qr(\Sigma\Delta)^{-1/2},\label{gaugepotential}%
\end{equation}
while the electromagnetic fields ${\bf E}$ and ${\bf B}$ are parallel to the
direction of $\boldsymbol{\omega}^{(1)}$ and have strengths given by
\begin{align}
E_{(1)} &  =Q\Sigma^{-2}(r^{2}-a^{2}\cos^{2}\theta),\label{e1}\\
B_{(1)} &  =Q\Sigma^{-2}2ar\cos\theta,\label{b1}%
\end{align}
respectively. The maximal strength $E_{\mathrm{\max}}$ of the electric field
is obtained in the case $a=0$ at the horizon of the EMBH: $r=r_{+}$. We have
\begin{equation}
E_{\max}=Q^{2}/r_{+}^{2}\label{emax2}.%
\end{equation}
Equating the maximal electric field strength (\ref{emax2}) to the critical
value (\ref{critical1}), one obtains the maximal black hole mass
$M_{\mathrm{max}}\simeq7.2\cdot10^{6}M_{\odot}$ for pair production to occur.
For any black hole with mass smaller than $M_{\mathrm{max}}$, the pair
production process can drastically modify its electromagnetic structure.

Both the gravitational and the electromagnetic background fields of the
Kerr-Newman black hole are stationary when considering the quantum field of the
electron, which has mass $m_e$ and charge $e$. If $m_eM\gg1$, then the spatial
variation scale $GM/c^{2}$ of the background fields is much larger than the
typical wavelength $\hbar/m_ec$ of the quantum field. As far as
purely QED phenomena such as pair production are concerned, it is possible to consider the
electric and magnetic fields defined by Eqs.~(\ref{e1},\ref{b1}) as constants
in the neighborhood of a few wavelengths around any events $(r,\theta,\phi,t)$.
Thus, the analysis and discussion on the Sauter-Euler-Heisenberg-Schwinger
process over a flat space-time can be locally applied to the case of the
curved Kerr-Newman geometry, based on the equivalence principle.

The rate of pair production around a Kerr-Newman black hole can be obtained from the Schwinger 
formula (\ref{probabilityeh}) for parallel electromagnetic fields 
$\varepsilon =E_{(1)}$ and $\beta= B_{(1)}$ as:
\begin{equation}
\frac{\tilde\Gamma}{V}={\frac{e^2E_{(1)}B_{(1)}}{4\pi^{2}}}%
\sum_{n=1}^{\infty}{\frac{1}{n}}%
\coth\left(  {\frac{n\pi B_{(1)}}{E_{(1)}}}\right)  \exp\left(
-{\frac{n\pi E_{\mathrm{c}}}{E_{(1)}}}\right)  .\label{drw}%
\end{equation}
The total number of pairs
produced in a region $D$ of the space-time is
\begin{equation}
N=\int_{D}d^{4}x\sqrt{-g}\frac{\tilde\Gamma}{V},\label{drn}%
\end{equation}
where $\sqrt{-g}=\Sigma\sin\theta$. In Ref.~\refcite{dr75}, it was assumed that for
each created pair the particle (or antiparticle) with the same sign of charge
as the black hole is expelled to infinity with charge $e$, energy
$\omega$ and angular momentum $l_{\phi}$
while the antiparticle is absorbed by the black hole. This implies the
decrease of charge, mass and angular momentum of the black hole and a
corresponding extraction of all three quantities. The rates of change of the three
quantities are then determined by the rate of pair production (\ref{drw}) and
by the conservation laws of total charge, energy and angular momentum
\begin{align}
\dot{Q} &  =-Re,\nonumber\\
\dot{M} &  =-R\langle\omega\rangle\label{damour1},\\
\dot{L} &  =-R\langle l_{\phi}\rangle,\nonumber
\end{align}
where $R=\dot{N}$ is the rate of pair production and $\langle\omega\rangle$ and
$\langle l_{\phi}\rangle$ represent some suitable mean values for the energy
and angular momentum carried by the pairs.

Supposing the maximal variation of black hole charge to be $\Delta Q=-Q$, one
can estimate the maximal number of pairs created and the maximal mass-energy
variation. It was concluded in Ref.~\refcite{dr75} that
the maximal mass-energy variation in the pair production process is larger than
$10^{41}$erg and up to $10^{58}$erg, depending on the black hole mass.
They concluded at the time {\itshape ``this work naturally leads to a most simple model for the explanation of the recently discovered $\gamma$-ray bursts''}.

\subsection{The ``Dyadosphere''}

We first recall the three theoretical results which provide the foundation of the EMBH theory.

In 1971 in the article {\itshape ``Introducing the Black Hole''}\cite{1971PhT....24a..30R}, the theorem was advanced that the most general black hole is characterized uniquely by three independent parameters: the mass-energy $M$, the angular momentum $L$ and the charge  $Q$ making it an EMBH. Such an ansatz, which came to be known as the ``uniqueness theorem'' has turned out to be one of the most difficult theorems to be proven in all of physics and mathematics. The progress in the proof has been authoritatively summarized by Ref.~\refcite{c97}. The situation can be considered satisfactory from the point of view of the physical and astrophysical considerations. Nevertheless some fundamental mathematical and physical issues concerning the most general perturbation analysis of an EMBH are still the topic of active scientific discussion\cite{bcjr02}. 
 
In 1971 Christodoulou and Ruffini\cite{cr71} obtained the mass-energy formula of a Kerr-Newman black hole, see Eqs.(\ref{em}--\ref{sa}). As we just recalled, in 1975 there was the Damour and Ruffini work \cite{dr75} on the vacuum polarization of a Kerr-Newman geometry. The key point of this work was the possibility that energy on the order of $10^{54}$ ergs could be released almost instantaneously by the vacuum polarization process of a black hole. At the time, however, nothing was known about the distance of GRB sources or their energetics. The number of theories trying to explain GRBs abounded, as mentioned by Ruffini in the Kleinert Festschrift\cite{ruKl}.

After the discovery in 1997 of the afterglow of GRBs\cite{1997Natur.387..783C} and the determination of the cosmological distance of their sources, we noticed the coincidence between their observed energetics and the one theoretically predicted by Damour and Ruffini\cite{dr75}. We therefore returned to these theoretical results with renewed interest developing some additional basic theoretical concepts \cite{rukyoto,prx02,prx98,1999A&A...350..334R,2000A&A...359..855R} such as the dyadosphere and, more recently, the dyadotorus.

As a first simplifying assumption we have developed our considerations in the absence of rotation using spherically symmetric models. The space-time is then described by the Reissner-Nordstr\"{o}m geometry, whose spherically symmetric metric is given by
\begin{equation}
d^2s=g_{tt}(r)d^2t+g_{rr}(r)d^2r+r^2d^2\theta +r^2\sin^2\theta
d^2\phi ~,
\label{s}
\end{equation}
where $g_{tt}(r)= - \left[1-{2GM\over c^2r}+{Q^2G\over c^4r^2}\right] \equiv - \alpha^2(r)$ and $g_{rr}(r)= \alpha^{-2}(r)$.

The first result we obtained is that the pair creation process does not occur at the horizon of the EMBH: it extends over the entire region outside the horizon in which the electric field exceeds the value $E^\star$ of the order of magnitude of the critical value given by Eq.~(\ref{critical1}). The pair creation process, as we recalled in the previous sections (see e.g. Fig.\ref{demourgape}), is a quantum tunnelling between the positive and negative energy states, which needs only a level crossing but can occur, of course, also for $E^\star < E_c$ if the extent of the field is large enough, although with decreasing intensity. Such a process occurs also for $E^\star > E_c$. In order to give a scale of the phenomenon, and for definiteness, in Ref.~\refcite{prx98} we first consider the case of $E^\star \equiv E_c$. Since the electric field in the Reissner-Nordstr\"{o}m geometry has only a radial component given by\cite{r78}
\begin{equation}
{\cal E}\left(r\right)=\frac{Q}{r^2}\, ,
\label{edir}
\end{equation}
this region extends from the horizon radius
\begin{eqnarray}
r_{+}&=&1.47 \cdot 10^5\mu (1+\sqrt{1-\xi^2})\hskip0.1cm {\rm cm}
\label{r+}
\end{eqnarray}
out to an outer radius\cite{rukyoto}
\begin{equation}
r^\star=\left({\hbar\over mc}\right)^{1\over2}\left({GM\over
c^2}\right)^{1\over2} \left({m_{\rm p}\over m}\right)^{1\over2}\left({e\over
q_{\rm p}}\right)^{1\over2}\left({Q\over \sqrt{G}M}\right)^{1\over2}=1.12\cdot 10^8\sqrt{\mu\xi} \hskip0.1cm {\rm cm},
\label{rc}
\end{equation}
where we have introduced the dimensionless mass and charge parameters $\mu={M\over M_{\odot}}$, $\xi={Q\over (M\sqrt{G})}\le 1$, see Fig.~\ref{dyaon}.

The second result has been to realize that the local number density of electron and positron pairs created in this region as a function of radius is given by
\begin{equation}
n_{e^+e^-}(r) = {Q\over 4\pi r^2\left({\hbar\over
mc}\right)e}\left[1-\left({r\over r^\star}\right)^2\right] ~,
\label{nd}
\end{equation}
and consequently the total number of electron and positron pairs in this region is
\begin{equation}
N^\circ_{e^+e^-}\simeq {Q-Q_c\over e}\left[1+{
(r^\star-r_+)\over {\hbar\over mc}}\right],
\label{tn}
\end{equation}
where $Q_c = E_c r_+^2$.

\begin{figure}
\includegraphics[width=10cm,clip]{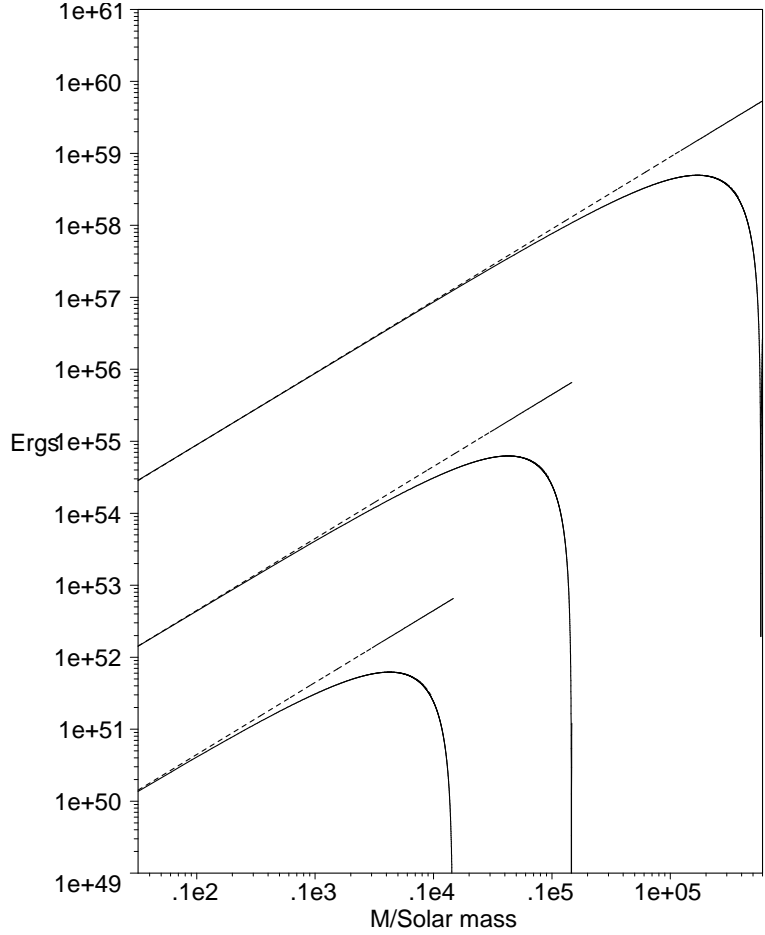}
\caption{The energy extracted by the process of vacuum polarization is plotted (solid lines) as a function of the mass $M$ in solar mass units for selected values of the charge parameter $\xi=1,0.1,0.01$ (from top to bottom) for an EMBH, the case $\xi=1$ reachable only as a limiting process. For comparison we have also plotted the maximum energy extractable from an EMBH (dotted lines) given by Eq.~(\ref{em}). Details in Ref.~\refcite{prx01}.}
\label{prep}
\end{figure}

\begin{figure}
\includegraphics[width=10cm,clip]{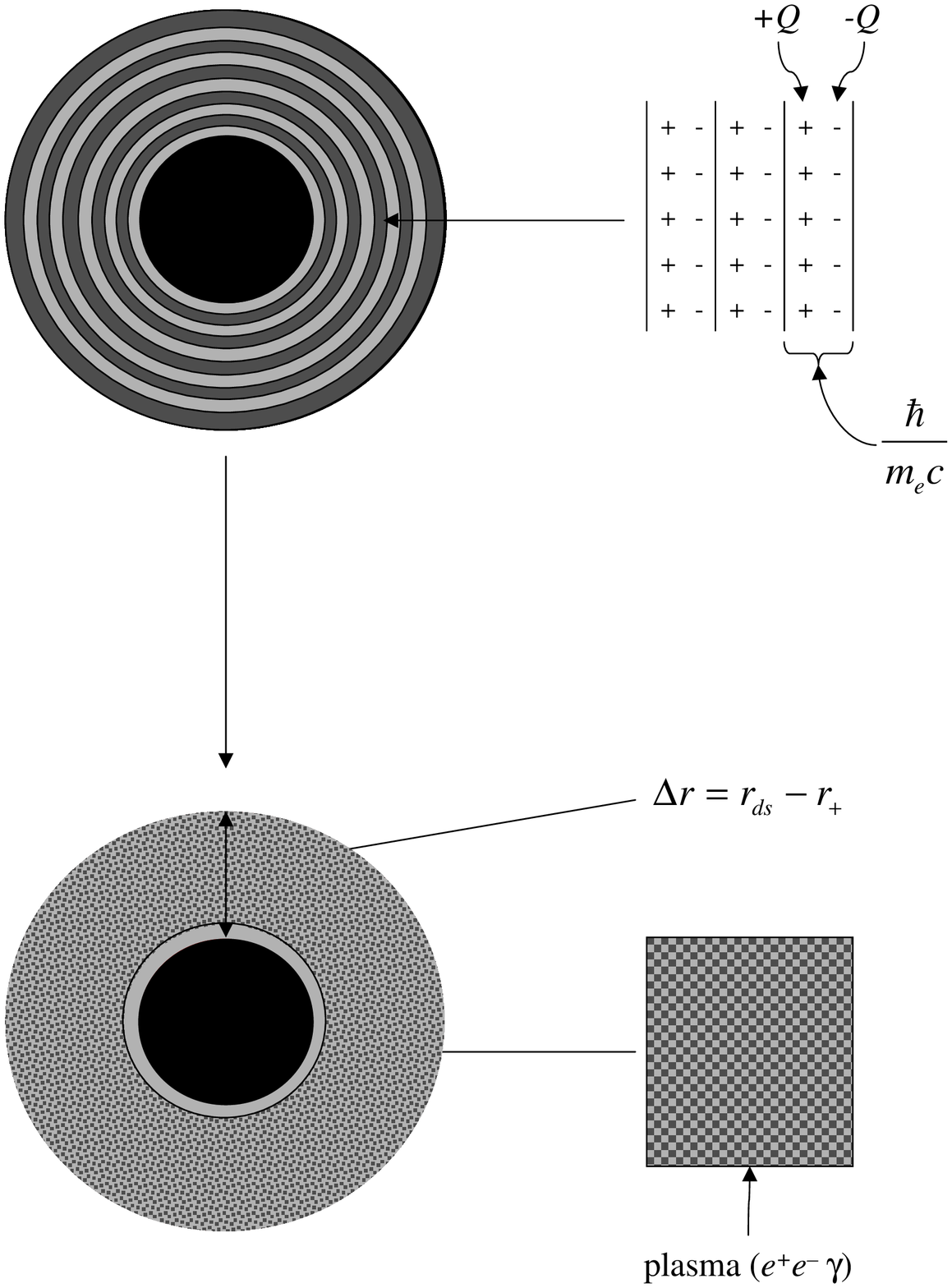}
\caption{The dyadosphere of a Reissner-Nordstr\"{o}m black hole can be represented as equivalent to a concentric set of capacitor shells, each one of thickness $\hbar/m_ec$ and producing a number of $e^+e^-$ pairs of the order of $\sim Q/e$ on a time scale of $10^{-21}$ s, where $Q$ is the EMBH charge. The shells extend in a region of thickness $\Delta r$, from the horizon $r_{+}$ out to the dyadosphere outer radius $r_{\rm ds}$ (see text). The system evolves to a thermalised plasma configuration.}
\label{dyaon}
\end{figure}

The total number of pairs is larger by an enormous factor $r^{\star}/\left(\hbar/mc\right) > 10^{18}$ than the value $Q/e$ which a naive estimate of the discharge of the EMBH would have predicted. Due to this enormous amplification factor in the number of pairs created, the region between the horizon and $r^{\star}$ is dominated by an essentially high density neutral plasma of electron-positron pairs. We have defined this region as the dyadosphere of the EMBH from the Greek duas, duadsos for pairs. Consequently we have called $r^\star$ the dyadosphere radius $r^\star \equiv r_{\rm ds}$\cite{rukyoto,prx02,prx98}. The vacuum polarization process occurs as if the entire dyadosphere is subdivided into a concentric set of shells of capacitors each of thickness $\hbar/m_ec$ and each producing a number of $e^+e^-$ pairs on the order of $\sim Q/e$ (see Fig.~\ref{dyaon}). The energy density of the electron-positron pairs is given by
\begin{equation}
\epsilon(r) = {Q^2 \over 8 \pi r^4} \biggl(1 - \biggl({r \over
r_{\rm ds}}\biggr)^4\biggr) ~, \label{jayet}
\end{equation}
(see Figs.~2--3 of Ref.~\refcite{prx02}). The total energy of pairs converted from the static electric energy and deposited within the dyadosphere is then
\begin{equation}
E_{\rm dya}={1\over2}{Q^2\over r_+}\left(1-{r_+\over r_{\rm ds}}\right)\left[1-\left({r_+\over r_{\rm ds}}\right)^4\right] ~.
\label{tee}
\end{equation}

\begin{figure}
\centering
\includegraphics[width=0.49\hsize,clip]{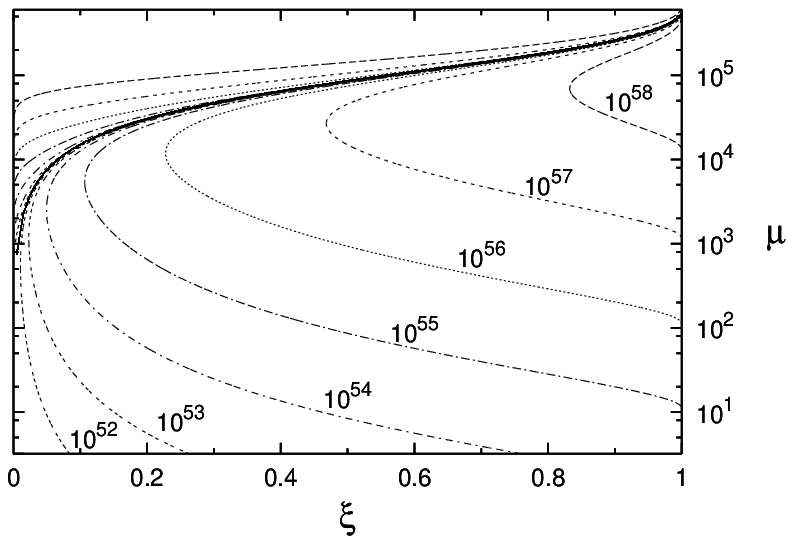}
\includegraphics[width=0.49\hsize,clip]{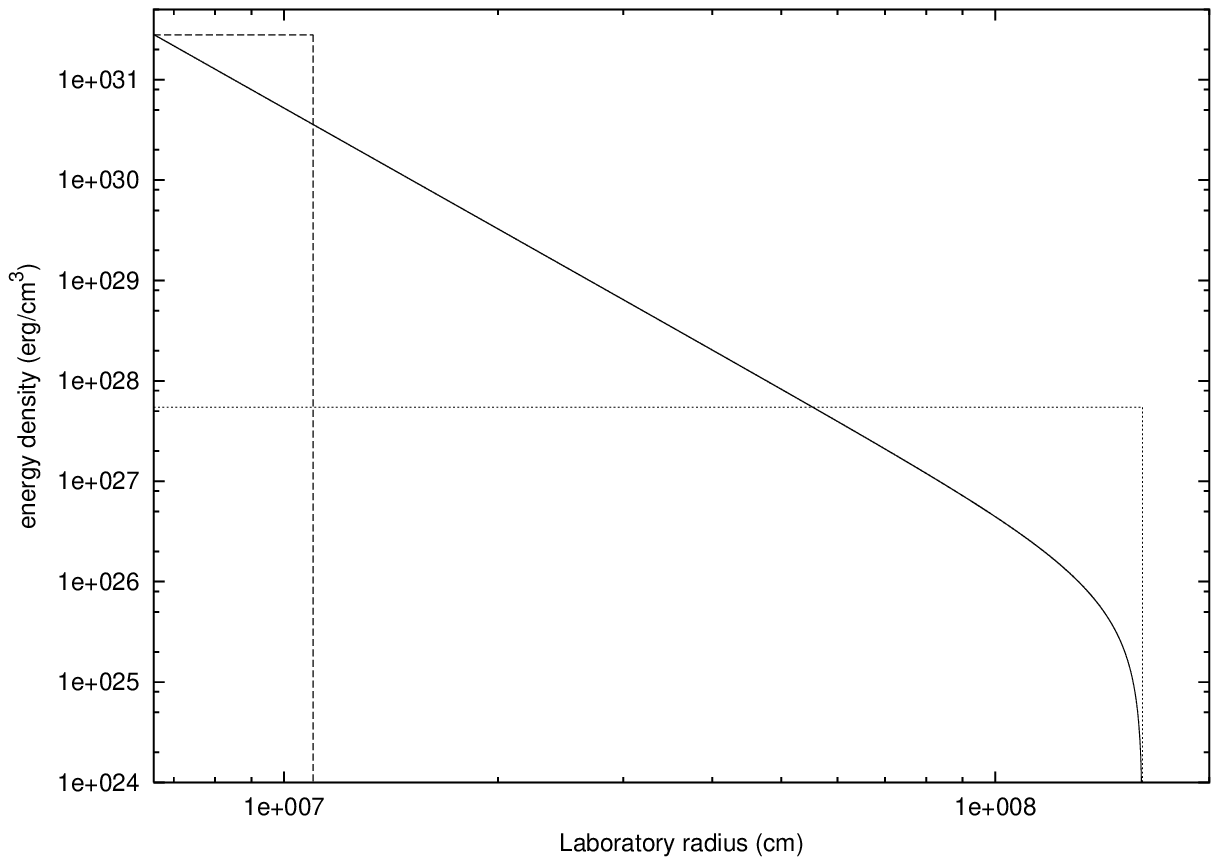}
\caption{{\bf Left)} Selected lines corresponding to fixed values of the $E_{dya}$ are given as a function of the two parameters $\mu$ $\xi$, only the solutions below the continuous heavy line are physically relevant.The configurations above the continuous heavy lines correspond to unphysical solutions with $r_{\rm ds} < r_+$. {\bf Right)} Two different approximations for the energy density profile inside the dyadosphere. The first one (dashed line) fixes the energy density equal to its peak value, and computes an ``effective'' dyadosphere radius accordingly. The second one (dotted line) fixes the dyadosphere radius to its correct value, and assumes an uniform energy density over the dyadosphere volume. The total energy in the dyadosphere is of course the same in both cases. The solid curve represents the real energy density profile.}
\label{muxi}
\label{dens}
\end{figure}

As we will see in the following this is one of the two fundamental parameters of the EMBH theory (see Fig.~\ref{muxi}). In the limit ${r_+\over r_{\rm ds}}\rightarrow 0$, Eq.~(\ref{tee}) leads to $E_{\rm dya}\rightarrow {1\over2}{Q^2\over r_+}$, which coincides with the energy extractable from EMBHs by reversible processes ($M_{\rm ir}={\rm const.}$), namely $E_{BH}-M_{\rm ir}={1\over2}{Q^2\over r_+}$\cite{cr71}, see Fig.~\ref{prep}. Due to the very large pair density given by Eq.~(\ref{nd}) and to the sizes of the cross-sections for the process $e^+e^-\leftrightarrow \gamma+\gamma$, the system has been assumed to thermalize to a plasma configuration for which
\begin{equation}
n_{e^+}=n_{e^-} \sim n_{\gamma} \sim n^\circ_{e^+e^-},
\label{plasma}
\end{equation}
where $n^\circ_{e^+e^-}$ is the total number density of $e^+e^-$-pairs created in the dyadosphere\cite{prx02,prx98}. This assumption has been in the meantime rigorously proven by Aksenov, Ruffini and Vereshchagin\cite{arv07}.

The third result which we have introduced again for simplicity is that for a given $E_{dya}$ we have assumed either a constant average energy density over the entire dyadosphere volume, or a more compact configuration with energy density equal to its peak value. These are the two possible initial conditions for the evolution of the dyadosphere (see Fig.~\ref{dens}).

The above theoretical results permit a good estimate of the general energetics processes originating in the dyadosphere, assuming an already formed EMBH. In reality, if the GRB data become accurate enough, the full dynamical description of the dyadosphere formation will be needed in order to explain the observational data by the general relativistic effects and characteristic time scales of the approach to the EMBH horizon\cite{crv02,rv02a,rv02b,rvx03sep}.

\subsection{The ``Dyadotorus''}

We turn now to examine how the presence of rotation modifies
the geometry of the surface containing the region where electron-positron
pairs are created as well as the conditions for the existence of such a surface. Due to the axial symmetry of the problem, we have called this region the ``dyadotorus'' \cite{Jorge_IC4}.

Following Damour \cite{damour,damourotros} we introduce at each point of the
spacetime the orthogonal Carter tetrad
\begin{align*}
&\omega^{(0)}=(\Delta/\Sigma)^{1/2}(dt-a \sin^2\theta d\varphi)\, ,\\
&\omega^{(1)}=(\Sigma/\Delta)^{1/2}dr\, ,\quad \omega^{(2)}=\Sigma^{1/2}d\theta\, ,\\
&\omega^{(3)}=\sin\theta \Sigma^{-1/2} [(r^2+a^2)d\varphi-a dt]\, ,
\end{align*}
where $\Sigma=r^2+a^2\cos ^2 \theta$ and $\Delta=r^2-2Mr+a^2+Q^2$, $M$ being
the mass, $Q$ the electric charge and $a$ the angular momentum per unit mass of
the black hole.  Thus, in the Lorentz frame defined by the above tetrad the
Kerr--Newman metric reads
\begin{equation}\label{metric}
ds^2=-(\omega^{(0)})^2+(\omega^{(1)})^2+(\omega^{(2)})^2+(\omega^{(3)})^2\, ,
\end{equation}
and the outer horizon of the black hole is localized at
$r_+=M+\sqrt{M^2-(a^2+Q^2)}$.
In this frame
the electric and magnetic field obtained from the potential
\begin{equation}
A=-Q r (\Sigma \Delta)^{-1/2} \omega^{(0)}\, ,
\end{equation}
are parallel, i.e.,
\begin{align}\label{fields}
&E_{(1)}=Q \Sigma^{-2} (r^2-a^2 \cos^2\theta)\, ,\\
&B_{(1)}=2 a Q \Sigma^{-2} r \cos\theta\, .
\end{align}

The rate $R$ of pair creation using the Schwinger approach is \cite{dr75}
\begin{equation}
R=\int 2 \textrm{Im}\,{\cal L} \sqrt{|g|} d^4 x\, ,
\end{equation}
where
\begin{align}
2 \textrm{Im}\,{\cal L} &=(4 \pi )^{-1} (E_{(1)} {\epsilon} / \pi)^2
\sum_{n=1}^{\infty} n^{-2} (n \pi B_{(1)}/E_{(1)}) \nonumber \\
&\coth (n \pi B_{(1)}/E_{(1)}) \exp (-n \pi m^2_e /E_{(1)})\, ,
\end{align}
where $\epsilon=m^2_e/E_c$.

We define the dyadotorus by the condition $|{\bf E}|=\kappa E_c$, where
$10^{-1} \leq \kappa \leq 10$.  Solving for $r$ and introducing the
dimensionless quantities $\xi=Q/M$, $\alpha=a/M$, $\mu=M/M_\odot$, ${\cal
E}=\kappa\,E_c\,M_\odot\approx 1.873\times10^{-6}$ and $\tilde{r}=r/M$ (with
$M_\odot \approx 1.477\times10^{5}$ cm) we get
\begin{equation}\label{dyadosurf}
\left(\frac{r^d_\pm}{M}\right)^2=\frac{\xi}{2\mu {\cal
E}}-\alpha^2\cos^2\theta\pm \sqrt{\frac{\xi^2}{4\mu^2 {\cal E}^2}
-\frac{2\xi}{\mu {\cal E}}\alpha^2\cos^2\theta}\, ,
\end{equation}
where the $\pm$ signs correspond to the two different parts of the surface.

\begin{figure}
\centering
$\begin{array}{cc}
\includegraphics[width=0.45\hsize]{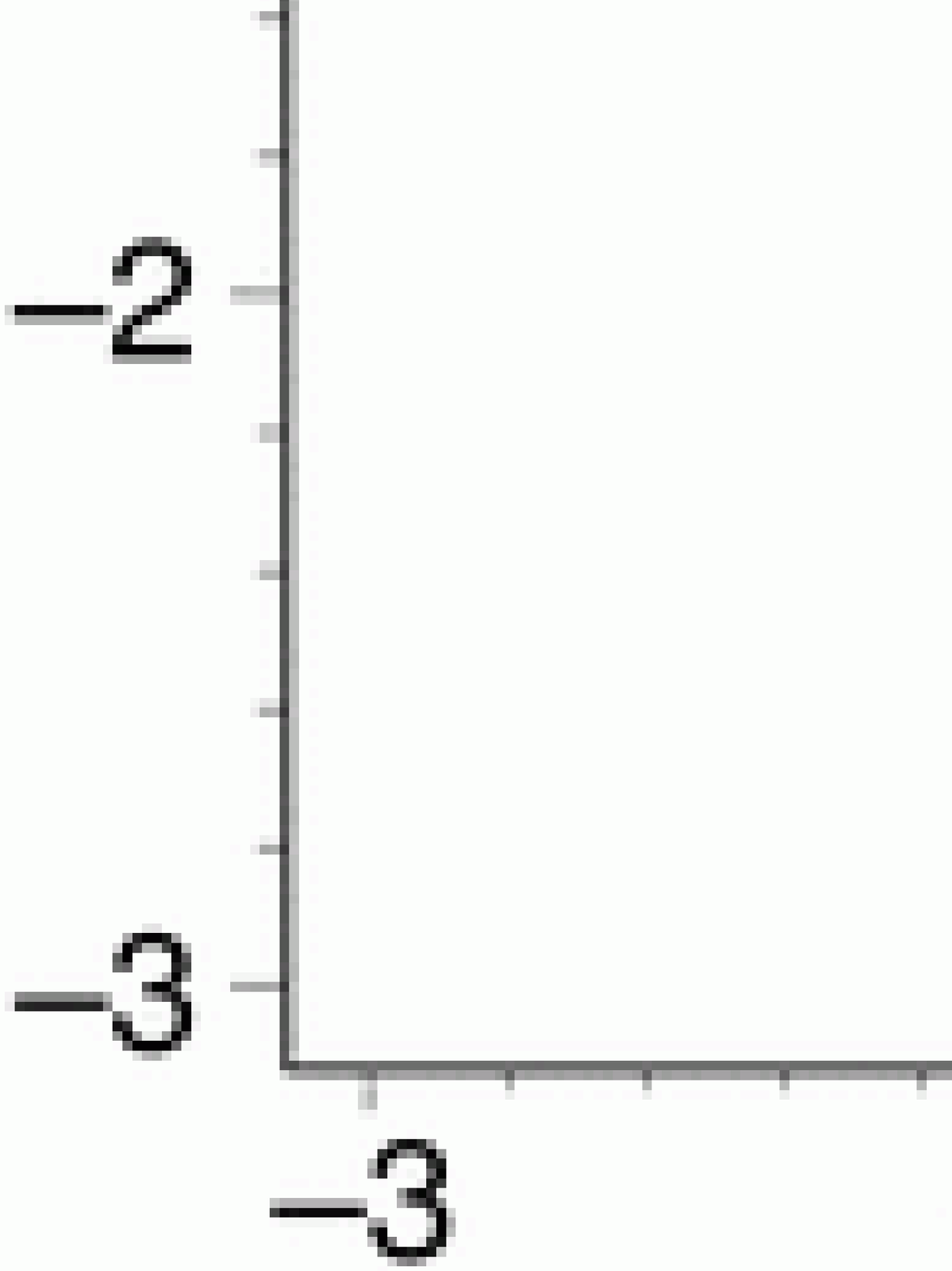}&\qquad
\includegraphics[width=0.45\hsize]{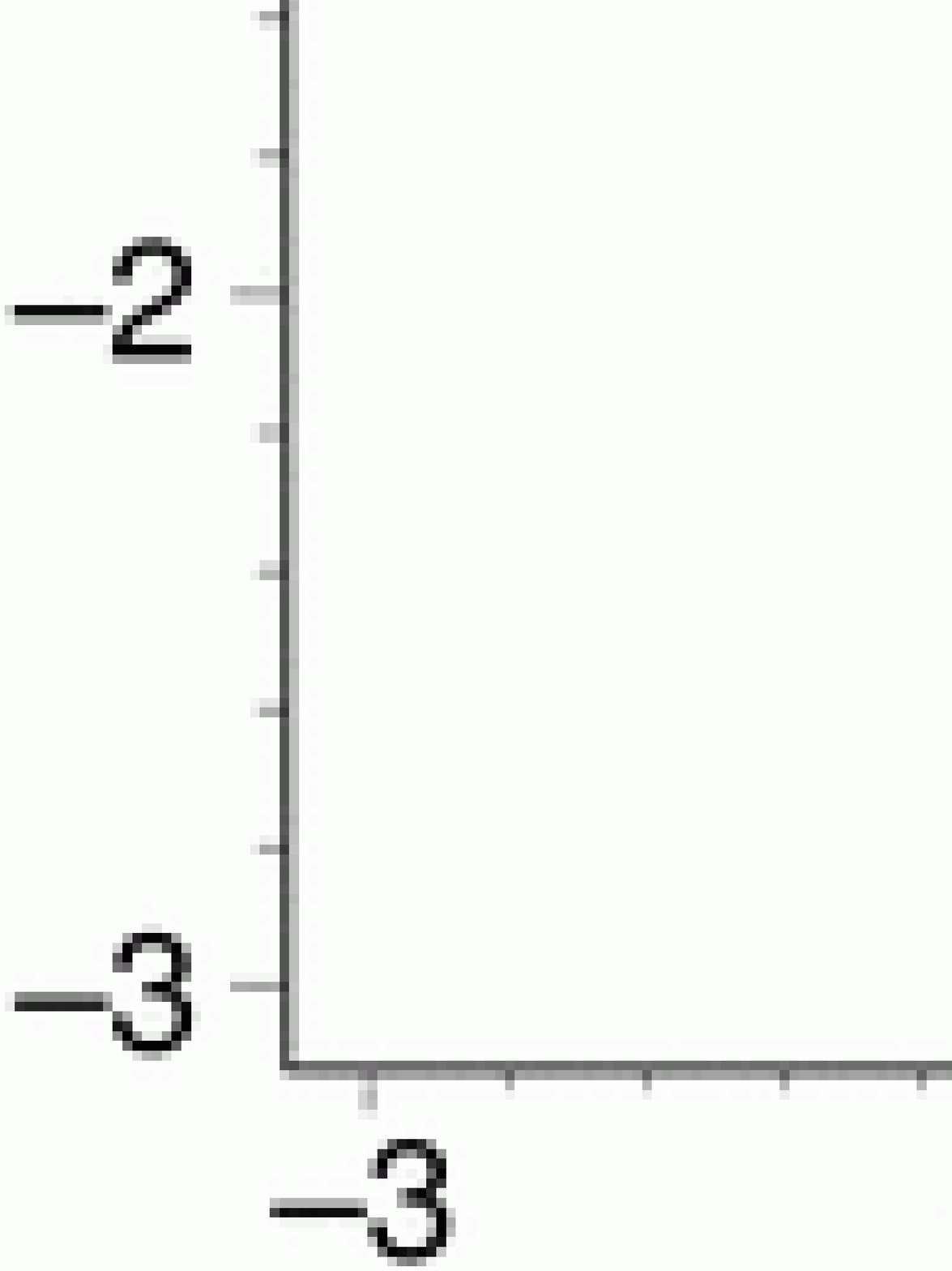}\\[0.4cm]
\mbox{(a)} & \mbox{(b)}\\[0.6cm]
\includegraphics[width=0.45\hsize]{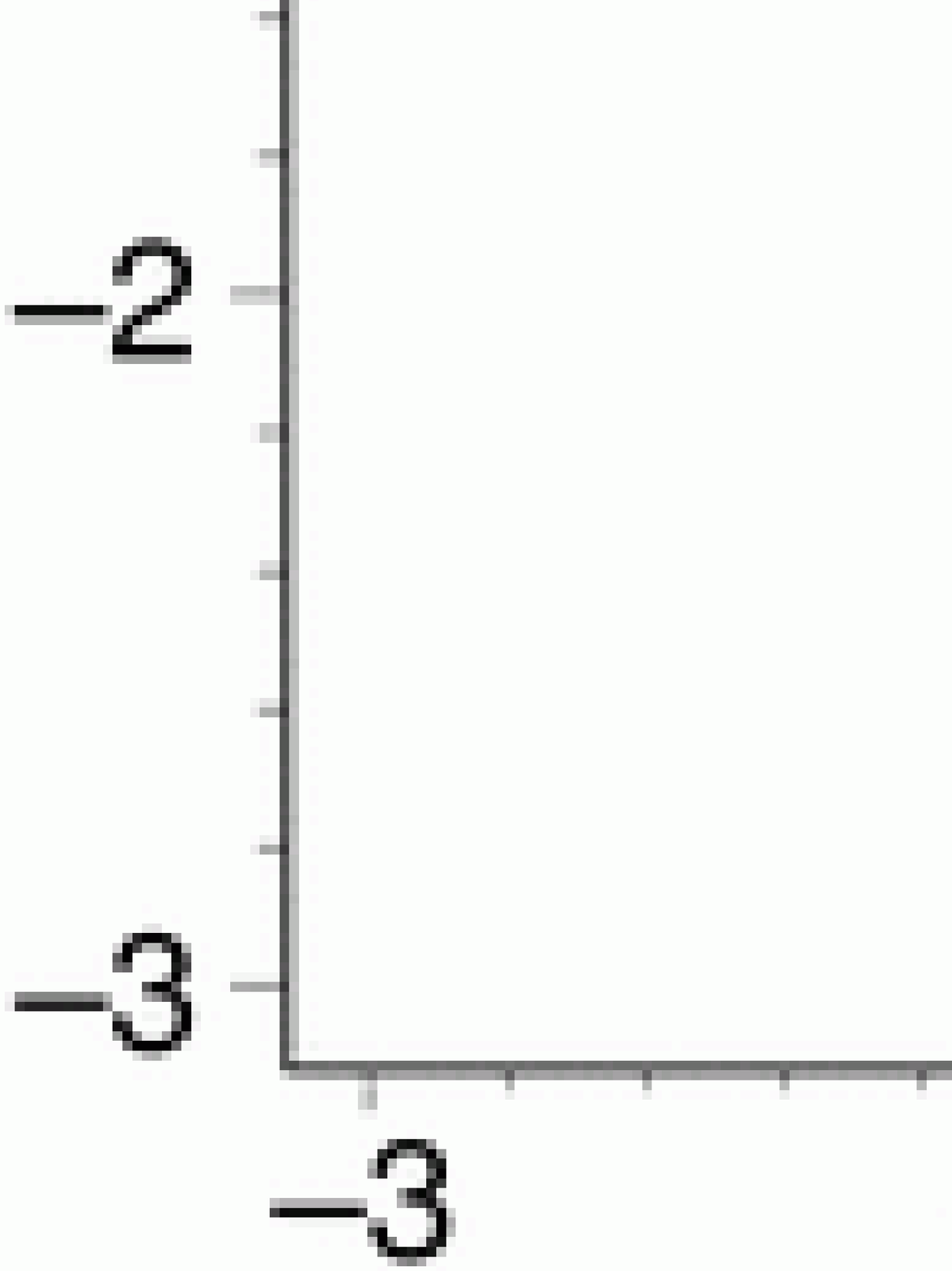}&\qquad
\includegraphics[width=0.45\hsize]{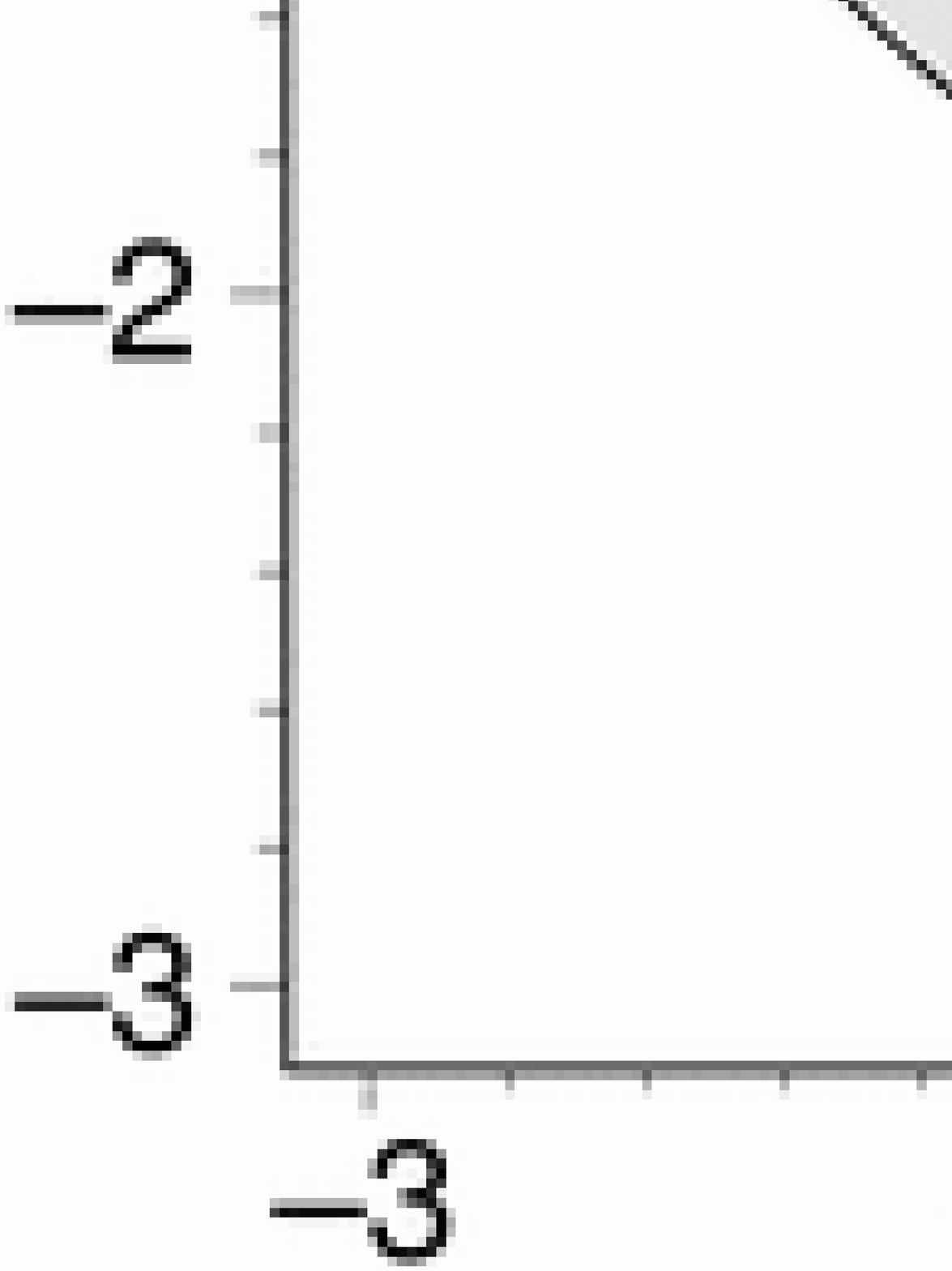}\\[0.4cm]
\mbox{(c)} & \mbox{(d)}\\
\end{array}$
\caption{The projection of the dyadotorus on the $X-Z$ plane ($X=r\sin \theta$,
$Z=r\cos \theta$ are Cartesian-like coordinates built up simply using the
Boyer-Lindquist radial and angular coordinates) is shown for an extreme
Kerr-Newman black hole with $\mu=10$ and different values of the charge
parameter $\xi=[1,1.3,1.49,1.65]\times10^{-4}$ (from (a) to (d) respectively).
 The black circle represents the black hole horizon.}\label{figure1}
\end{figure}

The two parts of the surface join at the particular values $\theta^*$ and
$\pi-\theta^*$ of the polar angle where
$\theta^*=\arccos\left(\frac{1}{2\sqrt{2}\alpha}\sqrt{\frac{\xi}{\mu{\cal
E}}}\right)$.
The requirement that $\cos\theta^*\leq1$ can be solved for instance for the
charge parameter $\xi$, giving a range of values of $\xi$ for which the
dyadotorus takes one of the shapes (see fig.\ref{figure1})
\begin{equation}
\textrm{surface}=
\begin{cases}
\textrm{ellipsoid--like} & \textrm{if}\,\, \xi \geq \xi_{*}\\
\textrm{thorus--like} & \textrm{if}\,\, \xi < \xi_{*}
\end{cases}
\end{equation}
where $\xi_{*}=8 \mu {\cal E} \alpha^2$.

In Fig.~\ref{figure1} we show some examples of the dyadotorus geometry
for different sets of parameters for an extreme Kerr--Newman black hole
($M^2=a^2+Q^2$), we can see the transition from a toroidal geometry to an
ellipsoidal one depending on the value of the black hole charge.

\begin{figure}
\centering
\includegraphics[scale=0.45]{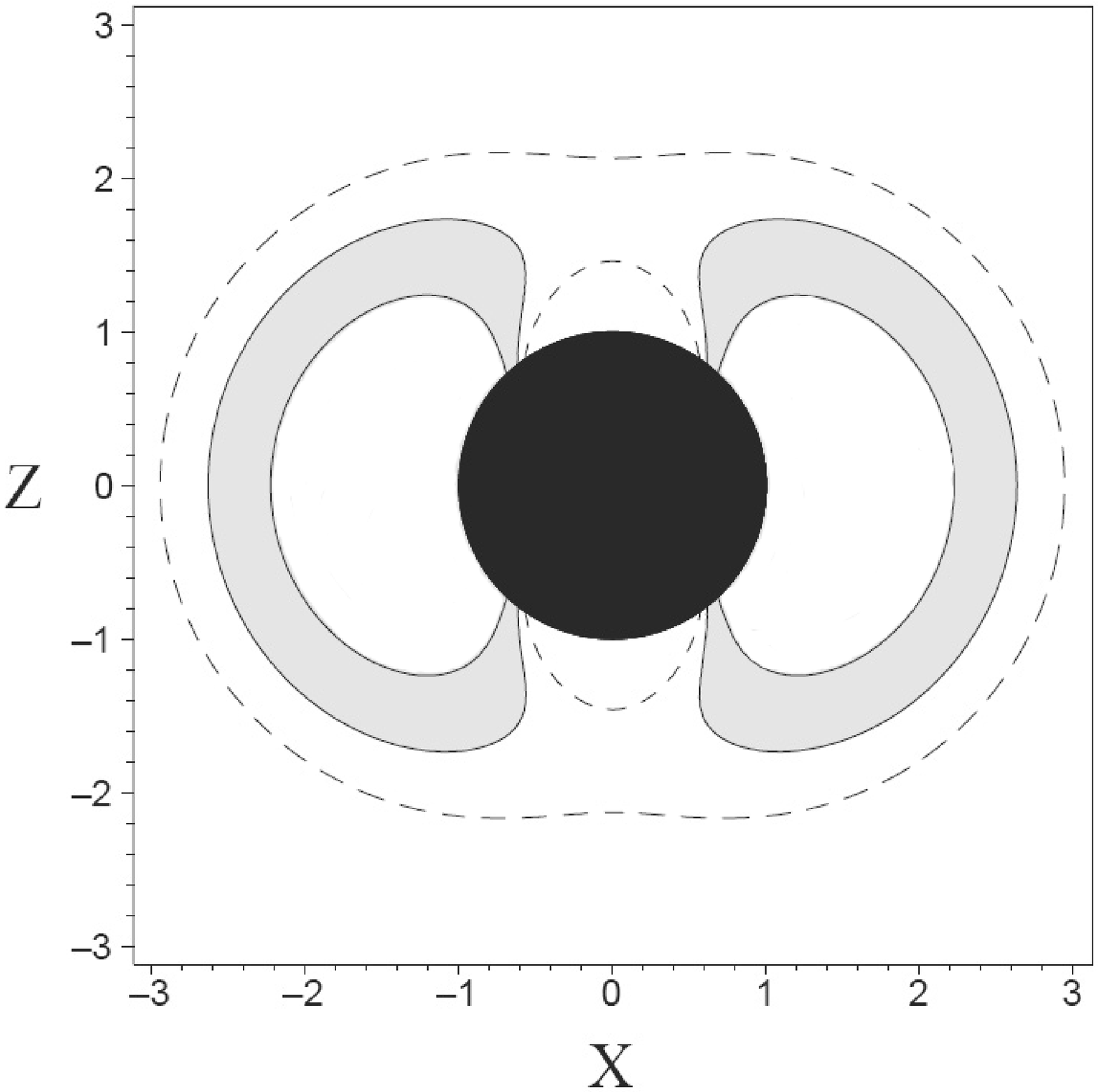}
\caption{The projections of the surfaces corresponding to different values of
the ratio $|{\bf E}|/E_c\equiv k$ are shown for the same choice of parameters
as in Fig.~\ref{figure1} (b), as an example.  The gray shaded region is part of
the ``dyadotorus'' corresponding to the case $\kappa=1$ as plotted in Fig.
\ref{figure1} (b).  The region delimited by dashed curves corresponds to
$\kappa=0.8$, i.e., to a value of the strength of the electric field smaller
than the critical one, and contains the dyadotorus; the latter in turn contains
the white region corresponding to $\kappa=1.4$, i.e., to a value of the strength
of the electric field greater than the critical one. } \label{figure2}
\end{figure}

Fig.~\ref{figure2} shows instead the projections of the surfaces
corresponding to different values of the ratio $|{\bf E}|/E_c\equiv \kappa$ for
the same choice of parameters as in Fig.~\ref{figure1} (b), as an example. We
see that the region enclosed by such surfaces shrinks for increasing values of
$\kappa$.

\section{On the observability of electron-positron pairs created by vacuum polarization in Earth-bound experiments and in astrophysics}

In summary, from the considerations we have presented in the previous sections three different earth-bound experiments and one astrophysical observation
have been proposed for identifying the polarization of the electronic vacuum
due to a supercritical electric field postulated by
Sauter-Heisenberg-Euler-Schwinger (see Ref.~\refcite{1931ZPhy...69..742S,1936ZPhy...98..714H,s51,n70}):

\begin{enumerate}
\item In collisions of heavy ions near the Coulomb barrier, as first
proposed in Ref.~\refcite{GZ69,GZ70} (see also Ref.~\refcite{PR71,z65,z}). Despite some
apparently encouraging results (see Ref.~\refcite{S...83}), such efforts have failed so
far due to the small contact time of the colliding ions \cite{A...95,G...96,L...97,B...95,H...98}. Typically the electromagnetic energy
involved in the collisions of heavy ions with impact parameter $%
l_{1}\sim10^{-12}$cm is $E_{1}\sim10^{-6}$erg and the lifetime of the
diatomic system is $t_{1}\sim10^{-22}$s.
\item In collisions of an electron beam with optical laser pulses: a signal
of positrons above background has been observed in collisions of a 46.6 GeV
electron beam with terawatt pulses of optical laser in an experiment at the
Final Focus Test Beam at SLAC \cite{B...97}; it is not clear if this
experimental result is an evidence for the vacuum polarization phenomenon.
The energy of the laser pulses was $E_{2}\sim10^{7}$erg, concentrated in a
space-time region of spacial linear extension (focal length) $l_{2}\sim
10^{-3}$cm and temporal extension (pulse duration) $t_{2}\sim10^{-12}$s \cite{B...97}.
\item At the focus of an X-ray free electron laser (XFEL) (see Ref.~\refcite{R01,AHRSV01,RSV02} and references therein). Proposals for this experiment
exist at the TESLA collider at DESY and at the LCLS facility at SLAC \cite{R01}. Typically the electromagnetic energy at the focus of an XFEL can be $%
E_{3}\sim10^{6}$erg, concentrated in a space-time region of spacial linear
extension (spot radius) $l_{3}\sim10^{-8}$cm and temporal extension
(coherent spike length) $t_{3}\sim10^{-13}$s \cite{R01}.
\end{enumerate}

and from astrophysics

\begin{enumerate}
\item Around an electromagnetic black hole (black hole) \cite{dr75,prx98,prx02}, giving rise to the observed phenomenon of GRBs \cite{2001ApJ...555L.107R,2001ApJ...555L.113R,2001ApJ...555L.117R,2002ApJ...581L..19R}. The electromagnetic
energy of an black hole of mass $M\sim10M_{\odot}$ and charge $Q\sim0.1M/%
\sqrt{G}$ is $E_{4}\sim10^{54}$erg and it is deposited in a space-time
region of spacial linear extension $l_{4}\sim10^{8}$cm \cite{prx98,rv02a}
and temporal extension (collapse time) $t_{4}\sim10^{-2}$s \cite{rvx03}.
\end{enumerate}

As we will see in the following, the creation of an electron-positron plasma is indeed essential to explain not only the GRB energetics but also the unprecedentedly large Lorentz gamma factor observed in GRBs.

\section{Electrodynamics for nuclear matter in bulk}

We have seen how critical fields may be generated in heavy ion collisions, in collisions of electron beams with optical laser pulses and in X-ray free electron lasers. The explanation of GRBs leads to a theoretical framework postulating the existence of critical and overcritical fields in black holes, in order to extract their blackholic energy. It becomes natural, then, to ask if there is any mechanism which can lead to the existence of a critical field not only on the above mentioned microscopic scale, but also on macroscopic scales, possibly to be encountered at the onset of the process of gravitational collapse. It is clear that such processes should be common to a large variety of initial conditions occurring in the gravitational collapse either of a neutron star or of two binary neutron stars or again of a binary system formed by a neutron star and a white dwarf or, finally, in the general formation of an intermediate mass black hole. It is already clear from the work of Damour and Ruffini that the gravitational collapse giving birth to the black holes in active galactic nuclei with masses much larger that $10^6 M_\odot$ cannot give rise to instantaneous vacuum polarization processes leading to GRBs. For this reason we present here an alternative treatment of the electrodynamics for nuclear matter in bulk, presenting for the first time a unified approach to neutron star physics and nuclear physics, covering the range of atomic number from $A \sim 10^2$ to $A\sim 10^{57}$.

\subsection{The Thomas-Fermi equations for heavy ions}

It is well know that the Thomas-Fermi equation is the exact theory for atoms, molecules and solids as $Z\rightarrow\infty$ \cite{liebsimon}. We show in this section that the relativistic Thomas-Fermi theory developed to study atoms for heavy nuclei with $Z \simeq 10^6$ \cite{g1a,ruffinistella81} gives important basic new information about the state of nuclear matter in bulk in the limit of $N \simeq (m_{\rm Planck}/m_n)^3$ nucleons of mass $m_n$ and about its electrodynamic properties. 

The analysis of bulk nuclear matter in neutron stars composed of a degenerate gas of neutrons, protons and electrons 
has traditionally been approached by microscopically implementing the charge neutrality condition  by requiring the electron density $n_e(x)$ to coincide with the proton density $n_p(x)$
\begin{eqnarray}
n_e(x)=n_p(x).
\label{localnp}
\end{eqnarray}
It is clear, however, that especially when conditions close to gravitational collapse occur, there is an ultra-relativistic component of degenerate electrons whose confinement requires the existence of very strong 
electromagnetic fields, in order to guarantee the overall charge neutrality of the neutron star. Under these conditions Eq.~(\ref{localnp}) will necessarily be violated. We will show here that they will develop electric fields close to the critical value $E_c$ introduced by Ref.~\refcite{1931ZPhy...69..742S,1936ZPhy...98..714H,s51,s54a,s54b}:
\begin{eqnarray}
E_c=\frac{m^2c^3}{e\hbar}.
\label{ec2}
\end{eqnarray}

Special attention to the existence of critical electric fields and the possible condition for electron-positron ($e^+e^-$)
pair creation out of the vacuum  in the case of heavy bare nuclei with the atomic number $Z\geq 173$ has been 
given by Ref.~\refcite{g1a,GZ70,z11b,z,g,muller72}.
They analyzed the specific pair creation process of an electron-positron pair around both a point-like 
and extended bare nucleus by direct integration of Dirac equation.
These considerations have been extrapolated to much heavier nuclei $Z\gg 1600$, implying the creation of a large number of  $e^+e^-$ pairs by using a statistical approach based on the relativistic Thomas-Fermi equation by Ref.~\refcite{muller75,migdal76}. 
Using substantially the same statistical approach based on the relativistic Thomas-Fermi equation, Ref.~\refcite{ruffinistella80,ruffinistella81} have analyzed the electron densities around an extended nucleus in a neutral 
atom all the way  up to $Z\simeq 6000$. They have shown the effect of the penetration of the electron orbitals well 
inside the nucleus, leading to a screening of the nuclei positive charge and to the concept of an ``effective'' nuclear charge distribution.

All of this work assumes for the radius of the extended nucleus the semi-empirical formula \cite{segrebook},
\begin{eqnarray}
R_c\approx r_0 A^{1/3},\quad r_0=1.2\cdot 10^{-13}{\rm cm},
\label{dn}
\end{eqnarray}
where the mass number $A=N_n+N_p$, $N_n$ and $N_p$ are the neutron and proton numbers.  
The approximate relation between $A$ and the atomic number $Z=N_p$ 
\begin{eqnarray}
Z \simeq \frac{A}{2}
\label{z2a}
\end{eqnarray}
was adopted in Ref.~\refcite{muller75,migdal76}, or the empirical formula
\begin{eqnarray}
Z &\simeq & [\frac{2}{A}+\frac{3}{200}\frac{1}{A^{1/3}}]^{-1}
\label{zae}
\end{eqnarray}
was adopted in Ref.~\refcite{ruffinistella80,ruffinistella81}.

\subsection{Electroweak equilibrium in Nuclear Matter in Bulk}
 
We outline an alternative approach of the description of nuclear matter in bulk: it generalizes the above treatments, already developed and tested for the study of heavy nuclei, to the case of $N \simeq (m_{\rm Planck}/m_n)^3$ nucleons. 
This more general approach differs in many aspects from the ones in the current literature and reproduces the above treatments in the limiting case of $A$ 
smaller than $10^6$,. We will look for a solution implementing the condition of overall charge neutrality of the star as given by 
\begin{eqnarray}
N_e=N_p,
\label{golbalnp}
\end{eqnarray}
which significantly modifies Eq.~(\ref{localnp}), since now $N_e (N_p)$ is the total number of electrons (protons) of the equilibrium configuration.

We present here only a
simplified prototype of this approach, outlining the essential relative role
of the four fundamental interactions present in the neutron star physics: the gravitational, weak, strong and
electromagnetic interactions. In addition, we also implement the fundamental role of Fermi-Dirac statistics   
and the phase space blocking due to the Pauli principle in the degenerate configuration. The new results essentially depend from the coordinated action of the above five theoretical components and cannot be obtained if any one of them is neglected.

Let us first recall the role of gravity.
In the case of neutron stars, unlike the case of nuclei where its effects can be neglected, gravitation has the fundamental role of defining the  basic parameters of the equilibrium configuration. As pointed out by Ref.~\refcite{gamow-book} at a Newtonian level and by Ref.~\refcite{ov39} in general relativity, configurations of equilibrium exist at approximately one solar mass and at an average density around the nuclear density. This result is obtainable considering only the gravitational interaction of a system of Fermi degenerate self-gravitating neutrons, neglecting all other particles and interactions. This situation can be formulated within a Thomas-Fermi self-gravitating model (see e.g. Ref.~\refcite{ruffiniphd}). 

In the present case of our simplified prototype model directed at revealing new electrodynamic properties, the role of gravity is taken into account simply by considering in line with the generalization of the above results a mass-radius relation for the baryonic core
\begin{eqnarray}
R^{NS}=R_c\approx \frac{\hbar}{m_\pi c}\frac{m_{\rm Planck}}{m_n} .
\label{dnns}
\end{eqnarray}
This formula generalizes the one given by Eq.~(\ref{dn}) extending its validity  to  $N\approx (m_{\rm Planck}/m_n)^3$, 
leading to a baryonic core radius $R_c\approx  10$km.
We also recall that a more detailed analysis of nuclear matter in bulk in neutron stars (see e.g. Ref.~\refcite{sato1970,cameron1970}) shows that at mass densities larger than the ``melting" density of 
\begin{eqnarray}
\rho_c=4.34 \cdot 10^{13} g/cm^3,
\label{melting}
\end{eqnarray}
all nuclei disappear. In the description of nuclear matter in bulk we have to consider then the three Fermi degenerate gases of neutrons, protons and electrons. In turn this naturally leads to considering the role of strong and weak interactions among the nucleons. In the nucleus, the role of the strong and weak interaction, with a short range of one Fermi, is to bind the nucleons, with a binding energy of 8 MeV, in order to balance the Coulomb repulsion of the protons. In the neutron star case we have seen that the neutron confinement is due to gravity. We still assume that an essential role of the strong interactions is to balance the effective Coulomb repulsion due to the protons, partly screened by the electron distribution inside the neutron star core. We shall verify, for self-consistency, the validity of this assumption for the final equilibrium solution we will obtain.

We now turn to the essential weak interaction role in establishing the relative balance between neutrons, protons and electrons via the direct and inverse 
$\beta$-decay
\begin{eqnarray}
p+ e  &\longrightarrow & n + \nu_e ,
\label{beta}\\
n  &\longrightarrow & p +e + \bar\nu_e.
\label{ibeta}
\end{eqnarray} 
Since neutrinos escape from the star and the Fermi energy of the electrons is zero, as we will show below, the only non-vanishing terms in the equilibrium condition given by the weak interactions are 
\begin{eqnarray}
[(P_n^Fc)^2+M^2_nc^4]^{1/2}-M_nc^2=  [(P_p^Fc)^2+M^2_pc^4]^{1/2}-M_pc^2  + |e|V^p_{\rm coul},  
\label{neq}
\end{eqnarray}
where $P_n^F$ and $P_p^F$  are respectively, the neutron and proton Fermi momenta, and $V^p_{\rm coul}$ is the Coulomb potential of the protons. At this point, having fixed all these physical constraints, the main task is to find the electron distributions satisfying not only the Dirac-Fermi statistics but also the electrostatic Maxwell equations. The condition of equilibrium for the  Fermi degenerate electrons implies a zero value for the Fermi energy
\begin{eqnarray}
[(P_e^Fc)^2+m^2c^4]^{1/2}-mc^2  + eV_{\rm coul}(r)=0,
\label{eeq2}
\end{eqnarray}
where $P_e^F$ is the electron Fermi momentum and $V_{\rm coul}(r)$ is the Coulomb potential.

\subsection{Relativistic Thomas-Fermi Equation for Nuclear Matter in Bulk}

In line with the procedure already followed for heavy atoms  
\cite{ruffinistella80,ruffinistella81} we adopt here the relativistic Thomas-Fermi Equation
\begin{eqnarray}
\frac {1}{x}\frac {d^2\chi(x)}{dx^2}= - 4\pi \alpha\left\{\theta(x-x_c)
- \frac {1}{3\pi^2}\left[\left(\frac {\chi(x)}{x}+\beta\right)^2-\beta^2\right]^{3/2}\right\},
\label{eqless}
\end{eqnarray}
where $\alpha=e^2/(\hbar c)$, $\theta(x-x_c)$ represents the normalized proton density distribution, the variables $x$  and $\chi$  are related to the radial coordinate and the electron Coulomb potential $V_{\rm coul}$ by 
\begin{eqnarray}
x=\frac {r}{R_c}\left(\frac {3N_p}{4\pi}\right)^{1/3};\quad eV_{\rm coul}(r)\equiv \frac {\chi(r)}{r},
\label{dless}
\end{eqnarray} 
and the constants $x_c (r=R_c)$ and $\beta$ are respectively
\begin{eqnarray}
x_c\equiv\left(\frac {3N_p}{4\pi}\right)^{1/3};\quad \beta\equiv  
\frac {mcR_c}{\hbar}\left(\frac{4\pi}{3N_p}\right)^{1/3}.
\label{dbeta}
\end{eqnarray}
The solution has the boundary conditions
\begin{eqnarray}
\chi(0)=0;\quad \chi(\infty)=0,
\label{bchi}
\end{eqnarray} 
with the continuity of the function $\chi$ and its first derivative $\chi'$ at the boundary of the core $R_c$.
The crucial point is the determination of the eigenvalue of the first derivative at the center 
\begin{eqnarray}
\chi'(0)={\rm const}. ,
\label{bchi1}
\end{eqnarray} 
which has to be determined by satisfying the above boundary conditions (\ref{bchi}) and constraints given by 
Eq.~(\ref{neq}) and Eq.~(\ref{golbalnp}).

The difficulty of the integration of the Thomas-Fermi equations is certainly one of the most celebrated chapters in theoretical physics and mathematical physics, still challenging a proof of the existence and uniqueness of the solution and strenuously avoiding the occurrence of exact analytic solutions. We recall after the original papers of Ref.~\refcite{Thomas,Fermi}, the works of Ref.~\refcite{scorza28,scorza29,sommerfeld,Miranda} all the way to the many hundred papers reviewed in the classical articles of Ref.~\refcite{liebsimon,Lieb,Spruch}. The situation here is more difficult since we are working on the special relativistic generalization of the Thomas-Fermi equation. 
We must therefore proceed by numerical integration in this case as well. The difficulty of this numerical task is further enhanced by a consistency check in order to satisfy all the various constraints. 

\begin{figure}
\centering 
\includegraphics[width=\hsize,clip]{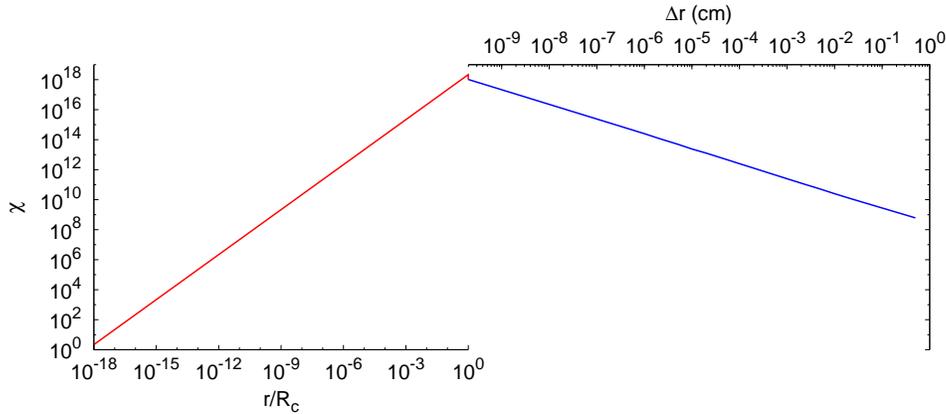}
\caption{The solution $\chi$ of the relativistic Thomas-Fermi equation for $A=10^{57}$ and core radius $R_c=10$km is plotted 
as a function of radial coordinate. The left red line corresponds to the internal solution and it is plotted as a function of radial coordinate in units of $R_c$ in logarithmic scale. The right blue line corresponds to the solution external to the core and it is plotted as function of the distance $\Delta r$ from the surface in a logarithmic scale in centimeters.}%
\label{chif}%
\end{figure}

\begin{figure}
\centering 
\includegraphics[width=\hsize,clip]{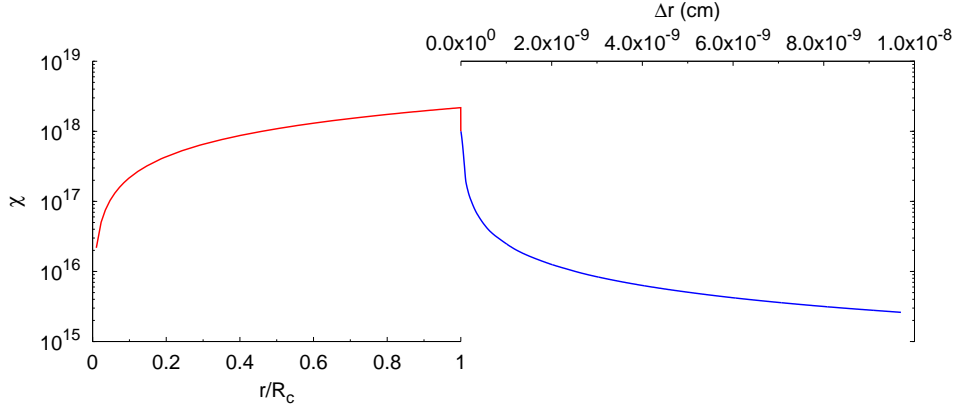}
\caption{The same as Fig.~(\ref{chif}): enlargement around the core radius $R_c$ showing explicitly the continuity of function $\chi$ and its 
derivative $\chi'$ from the internal to the external solution. }%
\label{chircf}%
\end{figure}

We start the computations by assuming a total number of protons and a value of the core radius $R_c$. We integrate the Thomas-Fermi equation and determine the number of neutrons from the Eq.~(\ref{neq}). We iterate the procedure until a value of $A$ is reached consistent 
with our choice of the core radius. The paramount difficulty of the problem is the numerical determination of the 
eigenvalue in Eq.~(\ref{bchi1}) which already for $A \approx 10^{4}$ had presented remarkable numerical difficulties \cite{ruffinistella80}. In the present context we have been faced for a few months with an  apparently insurmountable numerical task: the determination of the eigenvalue seemed to necessitate a  significant number of decimal places in the first derivative (\ref{bchi1}) comparable to the number of the electrons in the problem! 
We shall discuss elsewhere the way we overcame this difficulty by splitting the problem on the basis of the physical interpretation of the solution \cite{rrx06}. The solution is given in 
Fig.~(\ref{chif}) and Fig.~(\ref{chircf}).

\begin{figure}
\centering 
\includegraphics[width=\hsize,clip]{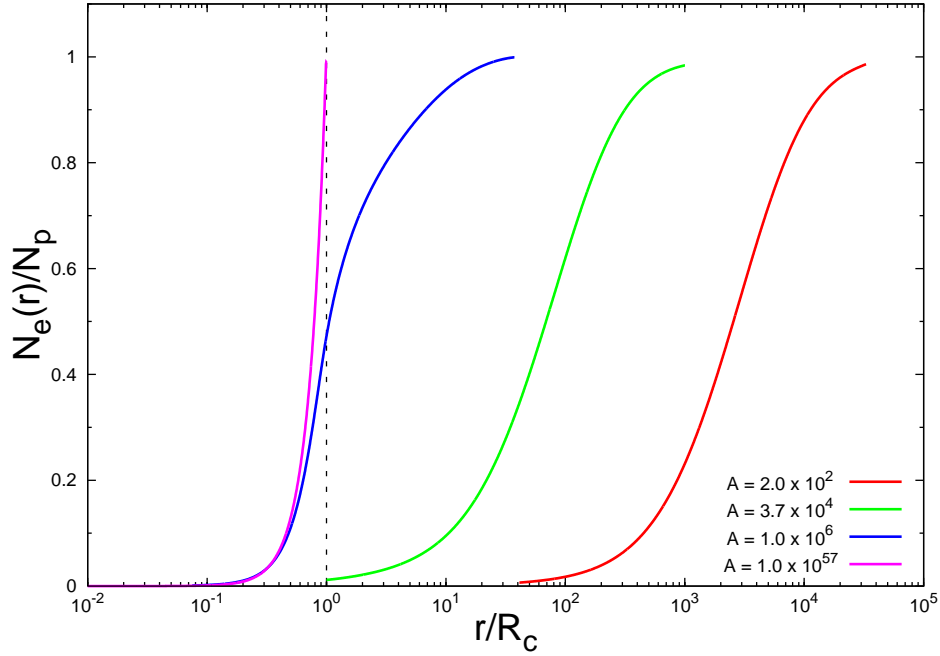}
\caption{The electron number (\ref{tein}) in units of the total proton number $N_p$ is given as function of radial distance in units of the core radius $R_c$, for selected values of 
$A$, again in a logarithmic scale. It is clear that by increasing the value of $A$, the penetration of electrons inside the core increases. The detail shown in Fig.~(\ref{efieldf}) and Fig.~(\ref{enumberf1}) demonstrates how  for $N \simeq (m_{\rm Planck}/m_n)^3$ a relatively small
tail of electrons outside the core exists and generates on the baryonic core surface an electric field close to the critical value. A significant electron density outside the core is found.
}%
\label{enumberf}%
\end{figure}

\begin{figure}
\centering 
\includegraphics[width=\hsize,clip]{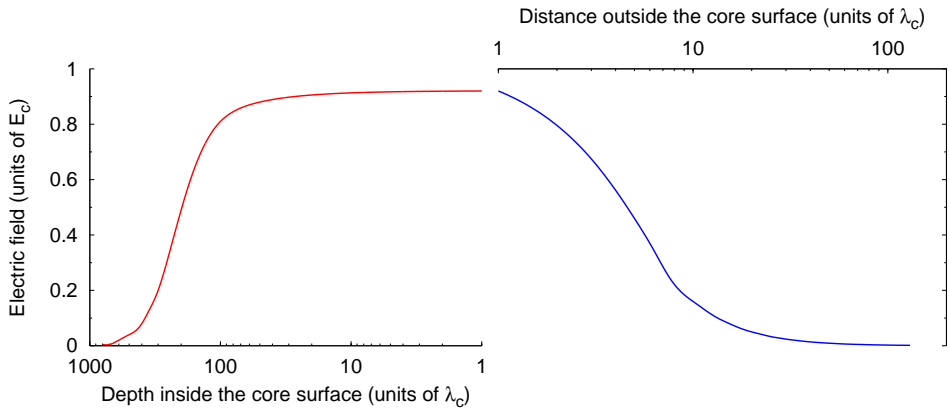}
\caption{The electric field in units of the critical field $E_c$ is plotted around the core radius $R_c$. The left (right) diagram in the red (blue) refers the region just inside (outside) the core radius plotted logarithmically. By increasing the density of the star the field approaches the critical field. }%
\label{efieldf}%
\end{figure}

\begin{figure}
\centering 
\includegraphics[width=\hsize,clip]{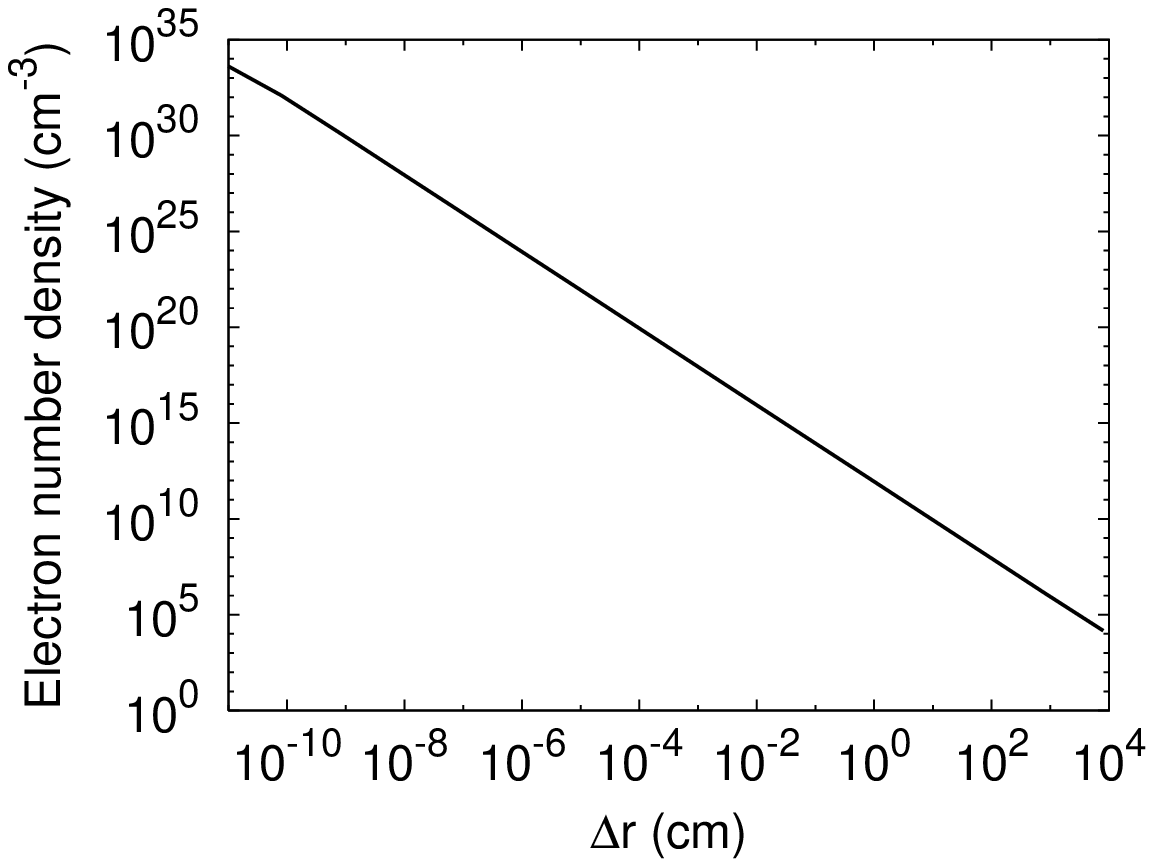}
\caption{ The density of electrons for $A=10^{57}$ in the region outside the core; both scales are logarithmic.
}%
\label{enumberf1}%
\end{figure}

A relevant quantity for exploring the physical significance of the solution is given by the number of electrons within a given radius $r$
\begin{eqnarray}
N_e(r)=\int_0^{r} 4\pi (r')^2 n_e(r')dr'.
\label{tein}
\end{eqnarray}
This allows the determination of the distribution of the electrons inside and outside the core for selected values of the $A$ parameter, and follows the progressive penetration of the electrons in the core as $A$ increases [ see Fig.~(\ref{enumberf})]. 
Then we can evaluate the net charge inside the core
\begin{eqnarray}
N_{\rm net} = N_p-N_e(R_c) < N_p,
\label{net}
\end{eqnarray} 
generalizing the results in Ref.~\refcite{ruffinistella80,ruffinistella81}
and consequently determine the electric field at the core surface, as well as inside and outside the core 
[see Fig.~(\ref{efieldf})] and evaluate as well the  Fermi degenerate electron distribution outside the core 
[see Fig.~(\ref{enumberf1})].

It is interesting to explore the solution of the problem under the same conditions and constraints imposed by the fundamental interactions and the quantum statistics and imposing the corresponding Eq.~(\ref{golbalnp}) instead of Eq.~(\ref{localnp}). Indeed a solution exists and is much simpler  
\begin{eqnarray}
n_n(x)=n_p(x)=n_e(x)=0,\quad \chi=0.
\label{trivial}
\end{eqnarray}


\subsection{The energetic stability of the solution}

Before drawing our conclusions we should check the theoretical consistency of the solution. We obtain an overall neutral configuration for the nuclear matter in bulk, with a positively charged baryonic core with  
\begin{equation}
N_{\rm net}= 0.92\left(\frac{m}{m_\pi}\right)^2\left(\frac{e}{m_n\sqrt{G}}\right)^2\left(\frac{1}{\alpha}\right)^2 ,
\label{nete}
\end{equation}
and an electric field at the baryonic core surface (see Fig.~(\ref{efieldf}) )
\begin{equation}
\frac{E}{E_c}=0.92.
\label{esurface}
\end{equation} 
The corresponding Coulomb repulsive energy per nucleon is given by 
\begin{equation}
U^{\rm max}_{\rm coul}= \frac{1}{2\alpha}\left(\frac{m}{m_\pi}\right)^3mc^2\approx 1.78\cdot 10^{-6}({\rm MeV}),
\label{coul1}
\end{equation}
well below the nucleon binding energy per nucleon. It is also important to verify that this charge core is gravitationally stable.
We have in fact 
\begin{equation}
\frac{Q}{\sqrt{G}M}=\alpha^{-1/2}\left(\frac{m}{m_\pi}\right)^2\approx 1.56\cdot 10^{-4}.
\label{nuclb}
\end{equation}
The electric field of the baryonic core is screened to infinity by an electron distribution given in Fig.~(\ref{enumberf1}).

As has been the case previously, any new solution for a Thomas-Fermi system has relevance and finds its justification in the domain of theoretical and mathematical physics. We expect that as in the case of other solutions that have appeared in the literature of the relativistic Thomas-Fermi equations, this new one presented here will find important applications in physics and astrophysics.
There are a variety of new effects that such a generalized approach naturally leads to: 
(1) the energetics of the global neutrality solution is greatly different from the one obtained from the condition of local neutrality; 
(2) the formation process for a neutron star can also have specific new signatures, due to reaching a more tightly bound system; 
(3) we expect important 
consequences on the initial conditions in the physics of gravitational collapse of the baryonic core as soon as the protons and neutrons become relativistic and the critical mass for gravitational collapse to a black hole is reached. The consequent collapse to a black hole will have very different energetics properties, since the initial conditions will imply the existence of a critical electric field. Such a field will naturally lead to very strong processes of pair creation during the following phases of gravitational collapse. This research is ongoing.

We now turn to the interpretation of the GRB data within the above theoretical framework and recall some basic interpretational paradigms that we have introduced in order to reach a systematic understanding of these sources.

\section{The first paradigm: The Relative Space-Time Transformation (RSTT) paradigm}

The ongoing dialogue between our work and that of others who model GRBs still rests on some elementary considerations presented by Einstein in his classic article of 1905\cite{e05}. These considerations are quite general and even precede Einstein's derivation of the Lorentz transformations from first principles. We recall here Einstein's words: ``We might, of course, content ourselves with time values determined by an observer stationed together with the watch at the origin of the coordinates, and coordinating the corresponding positions of the hands with light signals, given out by every event to be timed, and reaching him through empty space. But this coordination has the disadvantage that it is not independent of the standpoint of the observer with the watch or clock, as we know from experience."

\begin{figure} 
\centering 
\includegraphics[width=\hsize,clip]{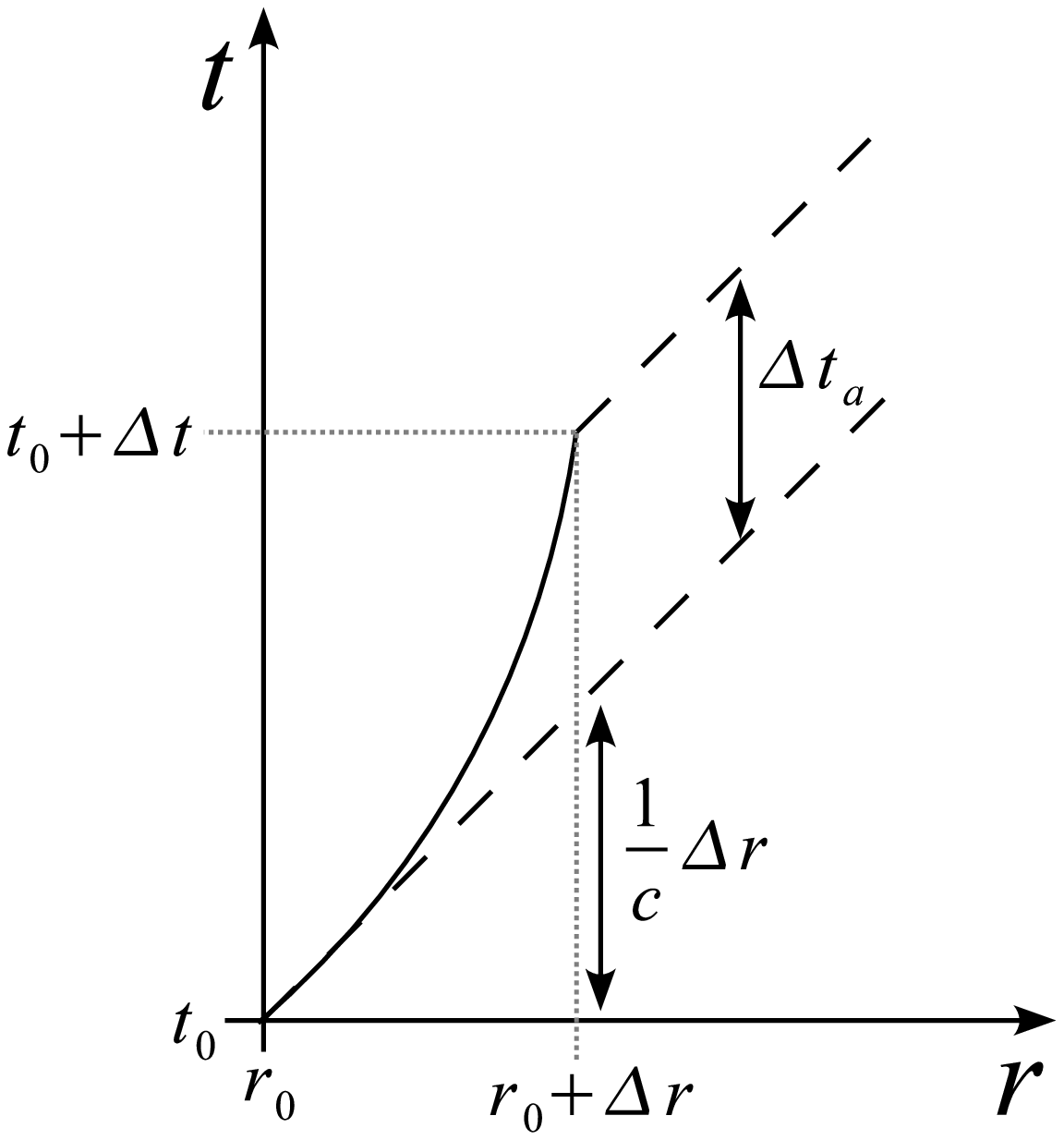} 
\caption{Relation between the arrival time $t_a$ and the laboratory time $t$. Details in Ref.~\refcite{2001ApJ...555L.107R,2003AIPC..668...16R}.} 
\label{ttasch_new_bn} 
\end{figure} 

Einstein's message is simply illustrated in Fig.~\ref{ttasch_new_bn}. If we consider in an inertial frame a source (solid line) moving with high speed and emitting light signals (dashed lines) along the direction of its motion, a far away observer will measure a delay $\Delta t_a$ between the arrival time of two signals respectively emitted at the origin and after a time interval $\Delta t$ in the laboratory frame, which in our case is the frame where the black hole is at rest. The real velocity of the source is given by 
\begin{equation} 
v = \frac{\Delta r}{\Delta t} 
\label{v} 
\end{equation} 
and the apparent velocity is given by: 
\begin{equation} 
v_{app} = \frac{\Delta r}{\Delta t_a}\, , 
\label{vapp} 
\end{equation} 
As pointed out by Einstein, the adoption of coordinating light signals simply by their arrival time as in Eq.~(\ref{vapp}), without an adequate definition of synchronization, is incorrect and leads to insurmountable difficulties as well as to apparently ``superluminal" velocities as soon as motions close to the speed of light are considered.

The use of $\Delta t_a$ as a time coordinate, often tacitly adopted by astronomers, should be done, if at all, cautiously. The relation between $\Delta t_a$ and the correct time parameterization in the laboratory frame has to be taken into account
\begin{equation} 
\Delta t_a = \Delta t - \frac{\Delta r}{c} = \Delta t - 
\frac{1}{c}\int_{t_\circ}^{t_\circ + \Delta t}{v\left(t'\right) dt'}\, . 
\label{tadef} 
\end{equation} 
In other words, the relation between the arrival time and the laboratory time cannot be done without a knowledge of the speed along the \emph{entire} world line of the source. In the case of GRBs, such a world line starts at the moment of gravitational collapse. It is of course clear that the parameterization in the laboratory frame has to take into account the cosmological redshift $z$ of the source. We then have at the detector
\begin{equation}
\Delta t_a^d = \left(1+z\right) \Delta t_a\, .
\label{taddef}
\end{equation}

In the current GRB literature, Eq.~(\ref{tadef}) has been systematically neglected by addressing only the afterglow description and neglecting the previous history of the source. Often the integral equation has been approximated by a clearly incorrect instantaneous value 
\begin{equation} 
\Delta t_a \simeq \frac{\Delta t}{2\gamma^2}\, . 
\label{taapp} 
\end{equation}
The approach has been adopted to consider the afterglow part of the GRB phenomenon separately without knowledge of the entire equation of motion of the source. 

This point of view has reached its most extreme expression in the work reviewed by Ref.~\refcite{1999PhR...314..575P,p00}, where the so-called ``prompt radiation'', lasting on the order of $10^2$ s, is considered as a burst emitted by the prolonged activity of an ``inner engine." In these models, generally referred to as the ``internal shock model," the emission of the afterglow is assumed to follow the ``prompt radiation'' phase\cite{rm94,px94,sp97,f99,fcrsyn99}.
As we outline in the following sections, such an extreme point of view originates from the inability to obtain the time scale of the ``prompt radiation'' from a burst structure. These authors consequently appeal to the existence of an ``ad hoc'' inner engine in the GRB source to solve this problem.

We show in the following sections how this difficulty has been overcome in our approach by interpreting the ``prompt radiation'' as an integral part of the afterglow and {\em not} as a burst. This explanation can be reached only through a relativistically correct theoretical description of the entire afterglow (see next sections). Within the framework of special relativity we show that it is not possible to describe a GRB phenomenon by disregarding the knowledge of the entire past world line of the source. We show that at $10^2$ seconds the emission occurs from a region of dimensions of approximately $10^{16}$ cm, well within the region of activity of the afterglow. This point was not appreciated in the current literature due to the neglect of the apparent superluminal effects implied by the use of the ``pathological'' parametrization of the GRB phenomenon by the arrival time of light signals.

We now turn to the first paradigm, the relative space-time transformation (RSTT) paradigm,\cite{2001ApJ...555L.107R} which emphasizes the importance of a global analysis of the GRB phenomenon encompassing both the optically thick and the afterglow phases. Since all the data are received in terms of the detector arrival time, it is essential to know the equations of motion of all relativistic phases of the GRB sources with $\gamma > 1$ in order to reconstruct the corresponding time coordinate in the laboratory frame, see Eq.~\eqref{tadef}. Contrary to other phenomena in nonrelativistic physics or astrophysics, where every phase can be examined separately from the others, in the case of GRBs all the phases are inter-related by their signals received in the arrival time $t_a^d$. In order to describe the physics of the source, there is the need to derive the laboratory time $t$ as a function of the arrival time $t_a^d$ along the entire past world line of the source using Eq.~\eqref{taddef}.

An additional difference, also linked to special relativity, between our treatment and others in the current literature relates to the assumption of the existence of scaling laws in the afterglow phase: the power law dependence of the Lorentz gamma factor on the radial coordinate is usually systematically assumed. From the proper use of the relativistic transformations and by the direct numerical and analytic integration of the special relativistic equations of motion we demonstrate (see next sections) that no simple power-law relation can be derived for the equations of motion of the system. This situation is not new for workers in relativistic theories: scaling laws exist in the extreme ultrarelativistic regimes and in the Newtonian ones but not in the intermediate fully relativistic regimes (see e.g. Ref.~\refcite{r70}).

\section{The second paradigm: The Interpretation of the Burst Structure (IBS) paradigm}

We turn now to the second paradigm, which is more complex since it deals with all the different phases of the GRB phenomenon. We first address the dynamical phases following the dyadosphere formation.

After the vacuum polarization process around a black hole, one of the topics of the greatest scientific interest is the analysis of the dynamics of the electron-positron plasma formed in the dyadosphere. This issue was addressed by us in a collaboration with Jim Wilson at Livermore. The numerical simulations of this problem were developed at Livermore, while the semi-analytic approach was developed in Rome (see Ruffini et al.\cite{1999A&A...350..334R,2000A&A...359..855R} and next sections).
The corresponding treatment in the framework of the Cavallo, Rees et al.\ analysis was performed by Piran et al.\cite{1993MNRAS.263..861P} also using a numerical approach, by Bisnovatyi-Kogan \& Murzina\cite{bm95} using an analytic approach and by M\'esz\'aros et al.\cite{1993ApJ...415..181M} using a numerical and semi-analytic approach.

Although some similarities exist between these treatments, they are significantly different in the theoretical details and in the final results (see Ref.~\refcite{2007arXiv0705.2411B} and next sections). Since the final result of the GRB model is extremely sensitive to any departure from the correct treatment, it is indeed very important to detect at every step the appearance of possible fatal errors.

\subsection{The optically thick phase of the fireshell}

A conclusion common to all these treatments is that the electron-positron plasma is initially optically thick and expands till transparency reaching very high values of the Lorentz gamma 
factor. A second point, which is also common, is the discovery of a clearly new feature: the plasma shell expands but the Lorentz contraction is such that its width in the laboratory frame appears to be constant. This self acceleration of the thin shell is the distinguishing factor of GRBs, conceptually very different from the physics of a fireball developed by the inner pressure of an atomic bomb explosion in the Earth's atmosphere. In the case of GRBs the region interior to the shell is inert and with pressure totally negligible: the entire dynamics occurs on the shell itself. For this reason, we refer in the following to the self accelerating shell as the ``fireshell.''

\begin{figure}
\begin{minipage}{\hsize}
\includegraphics[width=\hsize,clip]{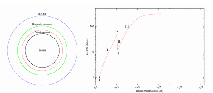}
\includegraphics[width=\hsize,clip]{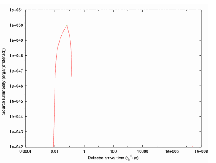}
\end{minipage}
\caption{{\bf Above:} The optically thick phase of the fireshell evolution are qualitatively represented in this diagram. There are clearly recognizable 1) the PEM pulse phase, 2) the impact on the baryonic remnant, 3) the PEMB pulse phase and the final approach to transparency with the emission of the P-GRB. Details in Ref.~\refcite{2003AIPC..668...16R}. {\rm Below:} The P-GRB emitted at the transparency point at a time of arrival $t_a^d$ which has been computed following the prescriptions of Eq.~\eqref{tadef}. Details in Ref.~\refcite{2001ApJ...555L.113R,2003AIPC..668...16R}.}
\label{cip2}
\end{figure}

There is a major difference between our approach and those of Piran, M\'esz\'aros and Rees in that we assume the dyadosphere to be initially filled only with an electron-positron plasma. Such a plasma expands in substantial agreement with the results presented in the work of Ref.~\refcite{bm95}. In our model the fireshell of electron-positron pairs and photons (PEM pulse)\cite{1999A&A...350..334R} evolves and encounters the remnant of the star progenitor of the newly formed black hole. The fireshell is then loaded with baryons. A new fireshell is formed of electron-positron-photons and baryons (PEMB pulse)\cite{2000A&A...359..855R} which expands all the way until transparency is reached. At transparency the emitted photons give origin to what we define as the Proper-GRB (P-GRB, see Ref.~\refcite{2001ApJ...555L.113R} and Fig.~\ref{cip2}).

In our approach, the baryon loading is measured by a dimensionless quantity
\begin{equation}
B = \frac{M_B c^2}{E_{dya}}\, ,
\label{Bdef}
\end{equation}
which gives direct information about the mass $M_B = N_B m_p$ of the remnant, where $m_p$ is the proton mass. The corresponding treatment done by Piran and collaborators\cite{1990ApJ...365L..55S,1993MNRAS.263..861P} and by Ref.~\refcite{1993ApJ...415..181M} differs in one important respect: the baryonic loading is assumed to occur from the beginning of the electron-positron pair formation and no relation to the mass of the remnant of the collapsed progenitor star is attributed to it.

A further difference also exists between our description of the rate equation for the electron-positron pairs and the ones by those authors. While our results are comparable with the ones obtained by Piran under the same initial conditions, the set of approximations adopted by Ref.~\refcite{1993ApJ...415..181M} appears to be too radical and leads to very different results violating energy and momentum conservation (see next sections and Ref.~\refcite{2007arXiv0705.2411B}).

\begin{figure}
\includegraphics[width=\hsize,clip]{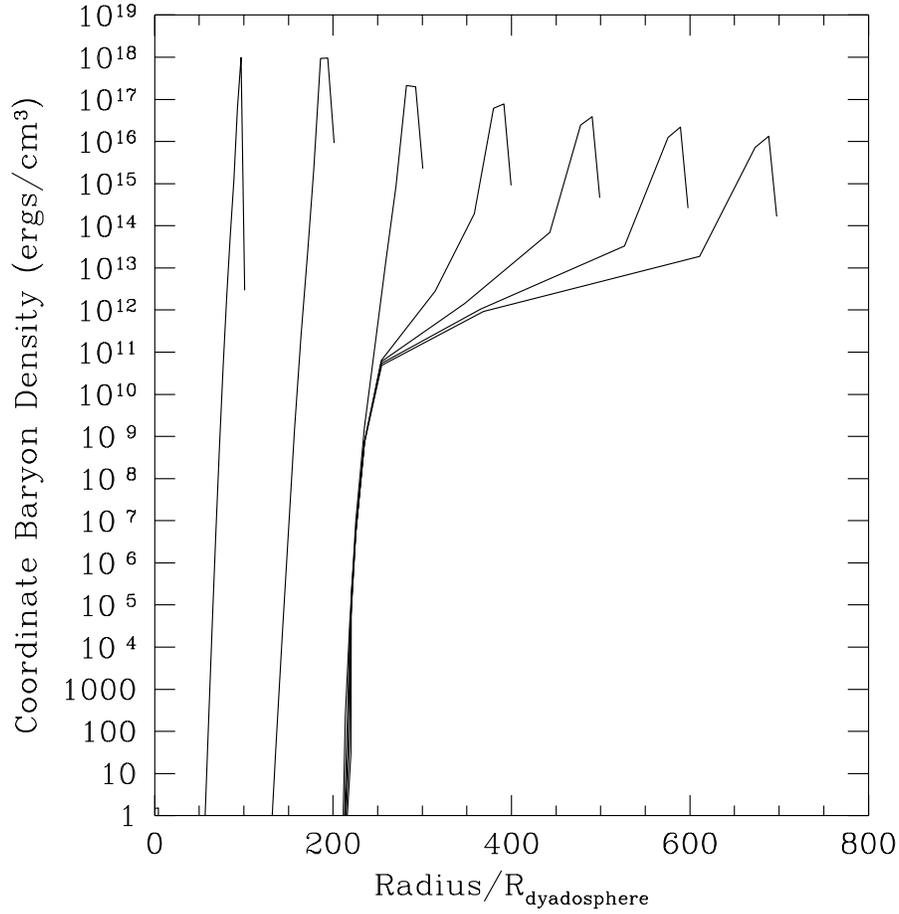}
\caption{A sequence of snapshots of the coordinate baryon energy density is shown from the one dimensional hydrodynamic calculations of the Livermore code. The radial coordinate is given in units of dyadosphere radii ($r_{ds}$). At $r \simeq 100 r_{ds}$ there is located a baryonic matter shell corresponding to a baryon loading $B = 1.3\times 10^{-2}$. For this baryon shell mass we see a significant departure from the constant thickness solution for the fireshell dynamics and a clear instability occurs. Details in Ref.~\refcite{2000A&A...359..855R}. As we will see, this result, peculiar to our treatment, will play a major role in the theoretical interpretation of GRBs.}
\label{B10-2}
\end{figure}

From our analysis\cite{2000A&A...359..855R} it also becomes clear that such an expanding dynamical evolution can only occur for values of $B \le 10^{-2}$ (see Fig.~\ref{B10-2}). This prediction, as we will show shortly in the many GRB sources considered, is very satisfactorily confirmed by observations and is indeed essential in order to reach the high values of the Lorentz gamma factor observed in GRBs.

From the value of the $B$ parameter, related to the mass of the remnant, it therefore follows that the collapse to a black hole leading to a GRB is drastically different from the collapse to a neutron star. While in the case of a neutron star collapse a very large amount of matter is expelled, in many instances well above the mass of the neutron star itself, in the case of black holes leading to a GRB only a very small fraction of the initial mass ($\sim 10^{-2}$ or less) is expelled. The collapse to a black hole giving rise to a GRB appears to be much smoother than any collapse process considered until today: almost 99.9\% of the star has to collapse at once to form the black hole!

We summarize in Fig.~\ref{cip2} the optically thick phase of the fireshell evolution: we start from a given dyadosphere of energy $E_{dya}$; the fireshell self-accelerates outward; an abrupt decrease in the value of the Lorentz gamma factor occurs due to the engulfment of the baryonic loading followed by a further self-acceleration until the fireshell becomes transparent.

The photon emission at this transparency point is the P-GRB. An accelerated beam of baryons with an initial Lorentz gamma factor $\gamma_\circ$ starts to interact with the interstellar medium at typical distances from the black hole of $r_\circ \sim 10^{14}$ cm and at a photon arrival time at the detector on the Earth's surface of $t_a^d \sim 0.1$ s. These values determine the initial conditions of the afterglow.

\subsection{Hydrodynamics of the pair plasma}

\label{appendix2}

We give a systematic derivation of the main equations, present a critical review of existing models for isotropic relativistic fireballs, and compare and contrast these models, following Ref.~\refcite{2007arXiv0705.2411B}. In the next section, following Ref.~\refcite{Vereshchagin07} we derive basic equations and describe the approximations involved. Then we present the model \cite{1999A&A...350..334R,2000A&A...359..855R} which differs
from other models in the literature in that it describes the dynamics of the
fireshell taking into account the \emph{rate equations} for electron-positron
pairs. Then we compare and contrast the above mentioned models.

\subsubsection{Local, global and average conservation laws}

\paragraph{Particle number.}

The first relevant equation represents continuity of relativistic flux and
reads\footnote{Greek indices denote four-dimensional components and run from 0
to 3 while Latin indices run from 1 to 3. General relativistic effects are
neglected, which is a good approximation, but we use the general definition
of energy-momentum conservation to take into account the most general
coordinate system.}
\begin{equation}
(nU^{\mu})_{;\mu}=\frac{1}{\sqrt{-g}}\frac{\partial\left(  \sqrt{-g}\,nU^{\mu
}\right)  }{\partial x^{\mu}}=0, \label{consnum}%
\end{equation}
where $n$ is the number density of relativistic fluid and $U^{\mu}$ is its
velocity field. Defining the particle number by
\begin{equation}
N=\int_{V}\sqrt{-g}\,nU^{0}dV, \label{num}%
\end{equation}
we see that
\begin{equation}
\frac{dN}{dt}=-\int_{V}\sqrt{-g}\,nU^{i}dV=-\oint_{\Sigma}\sqrt{-g}%
\,nU^{i}dS_{i}, \label{connum}%
\end{equation}
where we have used the Ostrogradsky-Gauss theorem. Thus if particles do not
cross the surface $\Sigma$ bounding the volume $V$ considered under consideration, the total number of
particles is constant during the system evolution.

Now assume spherical symmetry\footnote{The only nonvanishing components of the
energy-momentum tensor are $T^{00}$, $T^{01}$, $T^{10}$, $T^{11}$, $T^{22}$
, $T^{33}$. The factor $\sqrt{-g}=r^{2}\sin\vartheta$ in all expressions above
becomes simply a volume measure and the differentials are $dV=drd\vartheta
d\mathcal{\varphi}$, $dS=d\vartheta d\mathcal{\varphi}$, so the differential
laboratory volume can be written as $d\mathcal{V}\equiv\sqrt{-g}dV=r^{2}%
\sin\vartheta drd\vartheta d\mathcal{\varphi}$.}, which is usually done for
fireball descriptions. With spherical spatial coordinates $x^{i}%
=\{r,\vartheta,\mathcal{\varphi}\}$ the interval is
\begin{equation}
ds^{2}=-dt^{2}+dr^{2}+r^{2}d\vartheta^{2}+r^{2}\sin^{2}\vartheta d\varphi^{2}.
\end{equation}
Assuming the absence of fluxes through the boundary $\Sigma$ we rewrite
(\ref{consnum})

\begin{equation}
\frac{\partial\left(  n\gamma\right)  }{\partial t}+\frac{1}{r^{2}}%
\frac{\partial}{\partial r}\left(  r^{2}n\sqrt{\gamma^{2}-1}\right)  =0.
\label{continuityeq}%
\end{equation}
Integrating this equation over the volume from some initial $r_{i}(t)$ to a radius
$r_{e}(t)$ which we assume to be comoving with the fluid%
\begin{equation}
\frac{dr_{i}(t)}{dt}=\beta(r_{i},t),\qquad\frac{dr_{e}(t)}{dt}=\beta(r_{e},t),
\end{equation}
and ignoring a factor $4\pi$ we have%
\begin{gather}%
{\displaystyle\int\limits_{r_{i}}^{r_{e}}}
\frac{\partial\left(  n\gamma\right)  }{\partial t}r^{2}dr+%
{\displaystyle\int\limits_{r_{i}}^{r_{e}}}
\frac{\partial}{\partial r}\left(  r^{2}n\sqrt{\gamma^{2}-1}\right)
dr=\label{nslabcon}\\
\frac{\partial}{\partial t}%
{\displaystyle\int\limits_{r_{i}}^{r_{e}}}
\left(  n\gamma\right)  r^{2}dr-\frac{dr_{e}}{dt}n(r_{e},t)\gamma
(r_{e},t)r_{e}^{2}+\frac{dr_{i}}{dt}n(r_{i},t)\gamma(r_{i},t)r_{i}%
^{2}+\nonumber\\
+r_{e}^{2}n(r_{e},t)\sqrt{\gamma^{2}(r_{e},t)-1}-r_{i}^{2}n(r_{i}%
,t)\sqrt{\gamma^{2}(r_{i},t)-1}=\nonumber\\
=\frac{d}{dt}%
{\displaystyle\int\limits_{r_{i}}^{r_{e}}}
\left(  n\gamma\right)  r^{2}dr=0,\nonumber
\end{gather}
Since we deal with arbitrary comoving boundaries, this means that the number
of particles in each shell between the boundaries is conserved as well as the
total number of particles integrated over all shells, in other words,%

\begin{equation}
N=4\pi%
{\displaystyle\int\limits_{0}^{R(t)}}
n\gamma r^{2}dr=\mathrm{const}, \label{Ncons}%
\end{equation}
where $R(t)$ is the external radius of the fireshell.

Following Ref.~\refcite{1993MNRAS.263..861P} one can transform (\ref{continuityeq})
from the variables $(t,r)$ to the new variables $(s=t-r,r)$ and then show that%
\begin{equation}
\frac{1}{r^{2}}\frac{\partial}{\partial r}\left(  r^{2}n\sqrt{\gamma^{2}%
-1}\right)  =-\frac{\partial}{\partial s}\left(  \frac{n}{\gamma+\sqrt
{\gamma^{2}-1}}\right)  .
\end{equation}

Now assume the expansion velocity is ultrarelativistic,%
\begin{equation}
\gamma\gg1. \label{gll1}%
\end{equation}
In this approximation, therefore,%
\begin{equation}
dN=4\pi n\gamma r^{2}dr\approx\mathrm{const}. \label{dNcons}%
\end{equation}
Relations (\ref{dNcons}) and (\ref{Ncons}) then imply%
\begin{equation}
4\pi%
{\displaystyle\int\limits_{r_{i}}^{r_{e}}}
\left(  n\gamma r^{2}\right)  dr=4\pi\left[  n(r,t)\gamma(r,t)r^{2}\right]
{\displaystyle\int\limits_{r_{i}}^{r_{e}}}
dr=4\pi\left(  n\gamma r^{2}\right)  \Delta\approx\mathrm{const},
\end{equation}
where the first argument of the functions $n(r,t)$ and $\gamma(r,t)$ is restricted
to the interval $r_{i}<r<r_{e}$ and
\begin{equation}
\Delta\equiv r_{e}-r_{i}\approx\mathrm{const}. \label{Delta}%
\end{equation}
This means that the fluid shell does not broaden, but rather has a constant
thickness. This fact proves the constant thickness approximation, adopted in
\refcite{1999A&A...350..334R,2000A&A...359..855R}.

The volume element measured by the observer outside the fireshell (to be
referred to as the lab frame in what follows), for which it appears to be moving
with velocity $\beta$ is just%
\begin{equation}
d\mathcal{V}=4\pi r^{2}dr, \label{dVlab}%
\end{equation}
while the volume element comoving with the fireshell, for which the fluid is
at rest, is%
\begin{equation}
dV=4\pi\gamma r^{2}dr, \label{dVcom}%
\end{equation}
with the conversion of the volumes%
\begin{equation}
dV=\gamma d\mathcal{V}. \label{Vconv}%
\end{equation}
Then the average value of the Lorentz factor is defined as follows%
\begin{equation}
\left\langle \gamma\right\rangle \equiv\frac{4\pi%
{\displaystyle\int}
\gamma r^{2}dr}{4\pi%
{\displaystyle\int}
r^{2}dr}=\frac{V}{\mathcal{V}}. \label{gammaav}%
\end{equation}

Now we can formulate the conservation law for the average value of the number
density in the lab frame%
\begin{equation}
\left\langle n\right\rangle _{\mathrm{lab}}\equiv\frac{N}{\mathcal{V}}%
=\frac{4\pi%
{\displaystyle\int\limits_{r_{i}}^{r_{e}}}
n\gamma r^{2}dr}{4\pi%
{\displaystyle\int\limits_{r_{i}}^{r_{e}}}
r^{2}dr}. \label{naverage}%
\end{equation}
Assuming $r\gg\Delta$ we then obtain%
\begin{equation}
\left\langle n\right\rangle _{\mathrm{lab}}\simeq\frac{4\pi n\gamma
r^{2}\Delta}{4\pi r^{2}\Delta}=n(r,t)\gamma(r,t)\propto r^{-2}. \label{nav}%
\end{equation}
Therefore, the average number density in the lab frame scales like $r^{-2}$.

At the same time, recalling the expression for the divergence of the
four-velocity%
\begin{equation}
U^{\mu}{}_{;\mu}=\frac{1}{V}\frac{dV}{d\tau}, \label{Vcom}%
\end{equation}
where $\tau$ is the proper time, and remembering that $U^{\mu}\frac{\partial
}{\partial x^{\mu}}=\frac{d}{d\tau}$, from (\ref{consnum}) we get%
\begin{gather}
(nU^{\mu})_{;\mu}=U^{\mu}n_{;\mu}+nU^{\mu}{}_{;\mu}=\frac{dn}{d\tau}+\frac
{n}{V}\frac{dV}{d\tau}=0,\nonumber\\
d\ln n+d\ln V=0.
\end{gather}
This means that the number of particles is conserved along the flow lines of
the fluid. The solution of this equation provides the definition for the
comoving average number density%
\begin{equation}
\left\langle n\right\rangle _{\mathrm{com}}\equiv\frac{N}{V}=\frac{4\pi%
{\displaystyle\int\limits_{r_{i}}^{r_{e}}}
n\gamma r^{2}dr}{4\pi%
{\displaystyle\int\limits_{r_{i}}^{r_{e}}}
\gamma r^{2}dr}=\frac{\left\langle n\right\rangle _{\mathrm{lab}}%
}{\left\langle \gamma\right\rangle }. \label{ncom}%
\end{equation}

Clearly, the condition (\ref{ncom}) gives a link between the description of
the fireshell evolution in terms of local functions entering
(\ref{continuityeq}) on one side of the equation, and global quantities (\ref{gammaav})
and (\ref{nav}) on the other. The presence of the global conservation
(\ref{Ncons}) in both these cases ensures equivalence of the local
(\ref{continuityeq}) and the average (\ref{naverage}) descriptions for the
fireshell, unless its detailed structure is considered.

\paragraph{Energy-momentum conservation.}

The basis of the description of a relativistic fireshell is the energy-momentum
principle. It allows one to obtain the relativistic hydrodynamic equations, or
equations of motion for the fireshell, the energy and momentum conservation
equations which are used extensively to describe interaction of relativistic
baryons of the fireshell with the interstellar matter, and the boundary conditions
which are used to understand shock wave propagation in the decelerating
baryons and in the outer medium. 

Consider energy-momentum conservation in the
most general form:%

\begin{equation}
\left(  T_{\mu}~^{\nu}\right)  {}_{;\nu}=\frac{\partial(\sqrt{-g}\,T_{\mu
}~^{\nu}{})}{\partial x^{\nu}}+\sqrt{-g}\,\Gamma_{\nu\lambda}^{\mu}%
T^{\nu\lambda}=0, \label{ce}%
\end{equation}
where $\Gamma_{\nu\lambda}^{\mu}$ are the Christoffel symbols and $g$ is the
determinant of the metric tensor. Integrating over the entire three-dimensional
volume we obtain
\begin{equation}
\int_{V}T_{\mu}~^{\nu}{}{}_{;\nu}dV=0. \label{cev}%
\end{equation}
Integrating over the entire four-dimensional volume and applying the divergence
theorem we get \cite{1948PhRv...74..328T}
\begin{equation}
\int_{t}\int_{V}T_{\mu}~^{\nu}{}{}_{;\nu}dVdt=\oint_{V}T_{\mu}~^{\nu}{}%
\lambda_{\nu}dV=0, \label{ceom}%
\end{equation}
where $\lambda_{\alpha}$ are covariant components of the outer normal
to the three-dimensional hypersurface (volume $V$) enclosing the spacetime region.

Now suppose that there is a discontinuity on the fluid flow. Taking the volume
to be a spherical shell and choosing the coordinate system in which the
discontinuity is at rest so that in (\ref{ceom}) for normal vectors to the
discontinuity hypersurface $\lambda_{\alpha}$, we have
\begin{equation}
\lambda_{\alpha}\lambda^{\alpha}=1,\quad\quad\lambda_{0}=0.
\end{equation}
Let the radius of the shell $R_{s}$ be very large and the shell thickness $\Delta$
very small. With $R_{s}\rightarrow\infty$ and $\Delta\rightarrow0$ from
(\ref{ceom}) we get
\begin{equation}
\left[  T^{\alpha i}\right]  =0,
\end{equation}
where the brackets mean that the quantity inside is the same on both sides of
the discontinuity surface. This equation together with continuity condition
for particle density flux $[nU^{i}]=0$ was used by Ref.~\refcite{1948PhRv...74..328T} to obtain the relativistic Rankine-Hugoniot equations.
These equations govern shock wave dynamics which are supposed to appear
during the collision of the baryonic material left from the fireshell with the
interstellar medium \cite{1976PhFl...19.1130B}. The origin of the afterglow
could be connected \cite{1992MNRAS.258P..41R,1992ApJ...395L..83N,1994ApJ...422..248K} to the conversion of kinetic energy into radiative energy
in these shocks.

Consider now the energy-momentum tensor of the perfect fluid in the lab frame
(where the fluid was initially at rest)
\begin{equation}
T^{\mu\nu}=p\,g^{\mu\nu}+\omega U^{\mu}U^{\nu},
\end{equation}
where $\omega=\rho+p$ is the proper enthalpy, $p$ is the proper pressure and $\rho$ is the
proper internal energy density.

Rewrite (\ref{ce}) in the spherically symmetric case
\begin{align}
\frac{\partial T_{0}~^{0}{}}{\partial t}+\frac{1}{r^{2}}\frac{\partial
}{\partial r}\left(  r^{2}T_{0}~^{1}{}\right)   &  =0,\\
\frac{\partial T_{1}~^{0}{}}{\partial t}+\frac{1}{r^{2}}\frac{\partial
}{\partial r}\left(  r^{2}T_{1}~^{1}{}\right)  -\frac{1}{r}\left(  T_{2}%
~^{2}{}+T_{3}~^{3}{}\right)   &  =0,
\end{align}
arriving at the equations of motion of a relativistic fireshell
\cite{1976PhFl...19.1130B,1993MNRAS.263..861P,1999A&A...350..334R}
\begin{align}
\frac{\partial(\gamma^{2}\omega)}{\partial t}-\frac{\partial p}{\partial
t}+\frac{1}{r^{2}}\frac{\partial}{\partial r}\left(  r^{2}\gamma^{2}%
\beta\omega\right)   &  =0,\label{conseq1}\\
\frac{\partial(\gamma^{2}\beta\omega)}{\partial t}+\frac{1}{r^{2}}%
\frac{\partial}{\partial r}\left[  r^{2}(\gamma^{2}-1)\omega\right]
+\frac{\partial p}{\partial r}  &  =0, \label{conseq2}%
\end{align}
where the four-velocity and the relativistic Lorentz factor are defined as
follows\footnote{Throughout this chapter we set the speed of light equal to
1.}
\begin{equation}
U^{\mu}=(\gamma,\gamma\beta,0,0),\quad\quad\gamma\equiv(1-\beta^{2})^{-1/2},
\end{equation}
and $\beta$ is the radial velocity.

The total momentum of the spherically symmetric expanding shell vanishes. However,
from the local conservation equations (\ref{conseq1}) one finds that the
radial component of the four-momentum vector does not vanish. In analogy with
the continuity equation (\ref{continuityeq}) we integrate the first equation
in (\ref{conseq1}) over the volume starting from some internal radius $r_{i}(t)$
up to some external radius $r_{e}(t)$, and ignoring a factor $4\pi$ we obtain
\begin{gather}
\int_{r_{i}}^{r_{e}}\frac{\partial(\gamma^{2}\omega)}{\partial t}r^{2}%
dr-\int_{r_{i}}^{r_{e}}\frac{\partial p}{\partial t}r^{2}dr+\int_{r_{i}%
}^{r_{e}}\frac{1}{r^{2}}\,\frac{\partial}{\partial r}\,(r^{2}\gamma^{2}%
\beta\omega)r^{2}dr=\nonumber\\
\frac{\partial}{\partial t}\int_{r_{i}}^{r_{e}}\gamma^{2}\omega r^{2}%
dr+r_{i}^{2}\gamma^{2}(r_{i})\omega(r_{i})\beta(r_{i})-r_{e}^{2}\gamma
^{2}(r_{e})\omega(r_{e})\beta(r_{e})-\\
\frac{\partial}{\partial t}\int_{r_{i}}^{r_{e}}pr^{2}dr+r_{i}^{2}%
p(r_{i})-r_{e}^{2}p(r_{e})+r_{e}^{2}\gamma^{2}(r_{e})\omega(r_{e})\beta
(r_{e})-\\
r_{i}^{2}\gamma^{2}(r_{i})\omega(r_{i})\beta(r_{i})=0.\nonumber
\end{gather}
If the boundaries $r_{i}(t)$ and $r_{e}(t)$ are comoving with the fluid we
have%
\begin{equation}
\frac{d}{dt}\int_{r_{i}}^{r_{e}}\left(  \gamma^{2}\omega-p\right)
r^{2}dr=r_{e}^{2}p(r_{e})-r_{i}^{2}p(r_{i}). \label{dEdt}%
\end{equation}
Further, if one assumes (\ref{gll1}), one gets the following result%
\begin{equation}
E=4\pi\int_{0}^{R(t)}\gamma^{2}\omega r^{2}dr=\mathrm{const}. \label{Econs}%
\end{equation}

The differential conservation law follows from the same arguments which lead
to (\ref{dNcons}), so we also have%
\begin{equation}
dE=4\pi\gamma^{2}\omega r^{2}dr\approx\mathrm{const}. \label{dEcons}%
\end{equation}

Analogously to (\ref{naverage}) we introduce the average energy density in the
lab frame%
\begin{equation}
\left\langle \rho\right\rangle _{_{\mathrm{lab}}}\equiv\frac{E}{\mathcal{V}%
}=\frac{4\pi%
{\displaystyle\int\limits_{r_{i}}^{r_{e}}}
(\gamma^{2}\omega)r^{2}dr}{4\pi%
{\displaystyle\int\limits_{r_{i}}^{r_{e}}}
r^{2}dr}, \label{Eaverage}%
\end{equation}
Taking the polytropic equation of state with the thermal index%
\begin{equation}
\Gamma\equiv1+\frac{p}{\rho}, \label{thermalgamma}%
\end{equation}
and requiring also $r\gg\Delta$ and (\ref{gll1}) we find from (\ref{Eaverage})%
\begin{equation}
\left\langle \rho\right\rangle _{\mathrm{lab}}\simeq\rho(r)\gamma
^{2}(r)\propto r^{-2}. \label{Eav}%
\end{equation}

The radial momentum equation follows from (\ref{conseq2})%

\begin{gather}
\int_{r_{i}}^{r_{e}}\frac{\partial(\gamma^{2}\beta\omega)}{\partial t}%
r^{2}dr+\int_{r_{i}}^{r_{e}}\frac{1}{r^{2}}\,\frac{\partial}{\partial
r}\,\left[  r^{2}\left(  \gamma^{2}-1\right)  \omega\right]  r^{2}%
dr+\int_{r_{i}}^{r_{e}}\frac{\partial p}{\partial r}r^{2}dr\nonumber\\
=\frac{\partial}{\partial t}\int_{r_{i}}^{r_{e}}\gamma^{2}\beta\omega
r^{2}dr+r_{i}^{2}\gamma^{2}(r_{i})\omega(r_{i})\beta^{2}(r_{i})-r_{e}%
^{2}\gamma^{2}(r_{e})\omega(r_{e})\beta^{2}(r_{e})+\\
+r_{e}^{2}\left(  \gamma^{2}(r_{e})-1\right)  \omega(r_{e})-r_{i}^{2}\left(
\gamma^{2}(r_{i})-1\right)  \omega(r_{i})+\int_{r_{i}}^{r_{e}}\frac{\partial
p}{\partial r}r^{2}dr=\nonumber\\
\frac{\partial}{\partial t}\int_{r_{i}}^{r_{e}}\left(  \gamma^{2}\beta
\omega\right)  r^{2}dr+\int_{r_{i}}^{r_{e}}\frac{\partial p}{\partial r}%
r^{2}dr=0.\nonumber
\end{gather}
This leads to%
\begin{equation}
\frac{d}{dt}\int_{r_{i}}^{r_{e}}\left(  \gamma^{2}\beta\omega\right)
r^{2}dr=2\int_{r_{i}}^{r_{e}}prdr+r_{i}^{2}p(r_{i})-r_{e}^{2}p(r_{e}).
\label{dPdt}%
\end{equation}
For the radial momentum we have%
\begin{equation}
\frac{d\mathbf{P}_{tot}}{dt}=\frac{d}{dt}\int_{0}^{R(t)}4\pi\left(  \gamma
^{2}\beta\omega\right)  r^{2}dr=8\pi\int_{0}^{R(t)}prdr. \label{prslabcon}%
\end{equation}
The left hand side of this equation is the time derivative of the radial
momentum, i.e., the radial ``force". The right hand side is the integral of the
pressure over all the shells, so it is clear that unless the pressure in the
fireshell is zero, it experiences self-acceleration due to internal pressure.

\paragraph{Entropy conservation.}

Yet another relevant equation is entropy conservation which may be obtained
from (\ref{ce}) by projection along the flow line%
\begin{gather}
U^{\mu}\left(  T_{\mu}~^{\nu}\right)  {}_{;\nu}=\left(  U^{\mu}T_{\mu}~^{\nu
}\right)  {}_{;\nu}-T_{\mu}~^{\nu}\left(  U^{\mu}\right)  {}_{;\nu}=\\
=-\left(  \rho U^{\mu}\right)  {}_{;\mu}-\omega U^{\nu}\left(  U_{\mu}U^{\mu
}{}_{;\nu}\right)  -pU^{\mu}{}_{;\mu}=0.\nonumber
\end{gather}
The second term on the last line vanishes since $U_{\mu}U^{\mu}=-1$, so we
have another conservation equation%
\begin{equation}
-U^{\mu}\left(  T_{\mu}~^{\nu}\right)  {}_{;\nu}=\left(  \rho U^{\mu}\right)
{}_{;\mu}+pU^{\mu}{}_{;\mu}=0. \label{econ}%
\end{equation}

This conservation law corresponds to another conserved quantity, the entropy.
In fact, Eq.~(\ref{econ}) can be rewritten as%
\begin{equation}
\left(  \rho U^{\mu}\right)  {}_{;\mu}+pU^{\mu}{}_{;\mu}=\left(  \omega
U^{\mu}\right)  {}_{;\mu}-U^{\mu}p{}_{;\mu}=0.
\end{equation}
Now using the continuity equation (\ref{consnum}) and the identity $\omega U^{\mu
}=nU^{\mu}\left(  \frac{\omega}{n}\right)  $ we find%
\begin{equation}
\left(  \omega U^{\mu}\right)  {}_{;\mu}-U^{\mu}p{}_{;\mu}=nU^{\mu}\left[
\left(  \frac{\omega}{n}\right)  _{;\mu}-\frac{1}{n}p{}_{;\mu}\right]  =0.
\end{equation}
But the functions inside the brackets are scalars, and therefore covariant
derivatives can be replaced by ordinary derivatives. Then we recall the second
law of thermodynamics \cite{1987per..book.....L}%
\begin{equation}
d\left(  \frac{\omega}{n}\right)  =Td\left(  \frac{\sigma}{n}\right)
+\frac{1}{n}dp, \label{dentropy}%
\end{equation}
and finally obtain%
\begin{equation}
nTU^{\mu}\left(  \frac{\sigma}{n}\right)  _{;\mu}=0,
\end{equation}
which can be rewritten using (\ref{consnum}) as%
\begin{equation}
\left(  \sigma U^{\mu}\right)  _{;\mu}=0. \label{entcons}%
\end{equation}
This is the continuity equation for the entropy. Since it has exactly the same
form as (\ref{consnum}), all the conservation equations such as (\ref{dNcons}) and
(\ref{Ncons}) hold for the entropy as well,%
\begin{align}
d\sigma &  =4\pi\left(  \sigma\gamma\right)  r^{2}dr\approx\mathrm{const}%
,\label{dScons}\\
\mathbf{S}  &  =%
{\displaystyle\int\limits_{0}^{R(t)}}
d\sigma=\mathrm{const.} \label{Scons}%
\end{align}
Assuming (\ref{thermalgamma}) we find from (\ref{econ}) and (\ref{Vcom}) the
following result%
\begin{gather}
U^{\mu}{}\rho_{;\mu}+\Gamma\rho U^{\mu}{}_{;\mu}=\nonumber\\
d\ln\rho+\Gamma d\ln V=0,\label{entrcon}\\
\left\langle \rho\right\rangle _{\mathrm{com}}V^{\Gamma}=\mathrm{const}%
.\nonumber
\end{gather}

Finally, due to the similarity of Eqs.~(\ref{dNcons}) and (\ref{dScons}), the
average entropy can be defined in the same manner as (\ref{naverage}).

\paragraph{Analogy with a Friedmann Universe.}

It is easy to show that the conservation equations (\ref{dNcons}),(\ref{dEcons})
and (\ref{dScons}) imply an analogy between the fireball and the Friedmann
Universe, noticed first by Shemi and Piran \cite{1990ApJ...365L..55S,1993MNRAS.263..861P}. In fact, this analogy is valid for a polytropic
equation of state (\ref{thermalgamma}) and the ultrarelativistic expansion
condition (\ref{gll1}). First, using (\ref{thermalgamma}) and the integral
form of (\ref{dentropy}) we obtain%
\[
\sigma=\frac{\omega}{T},
\]
which leads to%
\begin{equation}
\rho\propto\sigma^{\Gamma},
\end{equation}
we rewrite the above mentioned conservation equations using
(\ref{thermalgamma})
\begin{align}
n\gamma r^{2}  &  =\mathrm{const},\nonumber\\
\rho^{\frac{1}{\Gamma}}\gamma r^{2}  &  =\mathrm{const},\label{dcons}\\
\rho\gamma^{2}r^{2}  &  =\mathrm{const}.\nonumber
\end{align}
From these equations we then easily find%
\begin{align}
\gamma &  \propto r^{\frac{2\left(  \Gamma-1\right)  }{2-\Gamma}},\nonumber\\
n  &  \propto r^{-\frac{2}{2-\Gamma}},\label{gensol}\\
\rho &  \propto r^{-\frac{2\Gamma}{2-\Gamma}}.\nonumber
\end{align}
Taking the ultrarelativistic equation of state with $\Gamma=4/3$ we immediately
obtain%
\begin{align}
\gamma &  \propto r,\nonumber\\
n  &  \propto r^{-3},\label{scalingsUR}\\
\rho &  \propto r^{-4},\nonumber
\end{align}
as opposed to the nonrelativistic equation of state with $\Gamma=1$ with
different scaling%
\begin{align}
\gamma &  =\mathrm{const},\nonumber\\
n  &  \propto r^{-2},\label{scalingsNR}\\
\rho &  \propto r^{-2}.\nonumber
\end{align}

Actually, scaling laws (\ref{scalingsUR}) take
place for the homogeneous isotropic radiation-dominated Universe
\cite{1990ApJ...365L..55S,1993MNRAS.263..861P}. This fact allowed the
authors of Ref.~\refcite{1993MNRAS.263..861P} to speak about the frozen-pulse profile
for $\gamma\gg1$ where number, energy and entropy density is conserved within
each differential shell with thickness $dr$, although the radial distribution of
matter and energy can be inhomogeneous.

Although for observer \emph{inside} the radiation-dominated fireshell it looks
indistinguishable from a portion of radiation-dominated Universe, for the
observer \emph{outside} it looks drastically different. In fact, the validity of
differential conservation laws (\ref{dNcons}),(\ref{dEcons}) and
(\ref{dScons}) together with integral ones (\ref{Ncons}),(\ref{Econs}) and
(\ref{Scons}) implies constant thickness approximation assumed in
\refcite{1999A&A...350..334R,2000A&A...359..855R}.

Clearly, if the condition (\ref{gll1}) is satisfied, also
\begin{equation}
\Delta\ll R(t)
\end{equation}
is valid. Given the scalings (\ref{scalingsUR}) we then find%
\begin{equation}
V=4\pi%
{\displaystyle\int_{R(t)-\Delta}^{R(t)}}
\gamma r^{2}dr\simeq4\pi%
{\displaystyle\int_{R(t)-\Delta}^{R(t)}}
\delta r^{3}dr\simeq4\pi R^{3}, \label{Vcom1}%
\end{equation}
where we put $\gamma=\delta r$, $\delta$ is a constant. At the same time,%
\begin{equation}
\mathcal{V}=4\pi%
{\displaystyle\int_{0}^{R(t)}}
r^{2}dr=\frac{4\pi}{3}R^{3}. \label{Vlab1}%
\end{equation}
Equality of (\ref{Vcom1}) and (\ref{Vlab1}) up to a numerical factor suggests
that the initially homogeneous energy and particle number distribution looks
highly compressed in the lab frame expanding with ultrarelativistic velocity
and compression factor $\gamma$.

\subsubsection{Self-acceleration of the fireshell}

For a fireshell which is initially optically thick the total energy is conserved.
Assume that the fireshell consists of relativistic electrons, positrons and
photons, and also that some admixture of a plasma in the form of photons and
electrons is present such that the total charge is zero. While electrons are
relativistic, the protons are not. The equation of state for pairs and electrons in
such a case is given to a good
approximation by that of an ultrarelativistic fluid: $p_{e^{\pm},\gamma}=\rho_{e^{\pm},\gamma}/3$. At the same time
for protons we have $p_{p}\simeq0$. Therefore positrons and electrons together
with photons can be considered to be one fluid with $p_{r}=\rho_{r}/3$ since they
are strongly coupled since the medium is optically thick. Instead protons have
low pressure and little internal energy compared to their rest mass energy.

According to (\ref{Econs}) we find
\begin{equation}%
{\displaystyle\int\limits_{0}^{R(t)}}
(\gamma^{2}\omega-p)r^{2}dr=%
{\displaystyle\int\limits_{0}^{R(t)}}
\gamma^{2}\rho_{p}r^{2}dr+\frac{4}{3}%
{\displaystyle\int\limits_{0}^{R(t)}}
\gamma^{2}\rho_{r}r^{2}dr. \label{enconsme}%
\end{equation}
These two terms are the rest mass energy of protons $M_{B}$ and the energy of the
ultrarelativistic fluid $E$ correspondingly, so we arrive at a simple result,
expressing the total energy of the expanding relativistic shell in the lab frame
\begin{equation}
\gamma(E+M)=\mathrm{const}, \label{enm}%
\end{equation}
which reads simply as $E+M=\mathrm{const}$ in the comoving frame taking into
account the conversion of volumes (\ref{Vconv}).

For homogeneous distributions of matter, energy density and pressure the
integrals (\ref{nslabcon}), (\ref{dEdt}) and (\ref{dPdt}) reduce to%

\begin{align}
n\gamma\mathcal{V}  &  =\mathrm{const},\nonumber\\
\left[  \gamma^{2}\left(  \rho+p\right)  \right]  \mathcal{V}  &
=\mathrm{const},
\end{align}
while in the comoving frame instead we would have%

\begin{align}
nV  &  =\mathrm{const},\\
\rho V  &  =\mathrm{const},
\end{align}
which means the energy and number of particles do not change.

From the above we have%

\begin{align}
nU_{com}^{0}V  &  =nV=\mathrm{const}=nU_{lab}^{0}\mathcal{V}=n\gamma
\mathcal{V},\\
T_{com}^{00}V  &  =\rho V=\mathrm{const}=T_{lab}^{00}\mathcal{V}=\left[
\gamma^{2}\left(  \rho+p\right)  \right]  \mathcal{V},
\end{align}
remembering that all quantities $n,\rho,p$ are always defined as comoving ones.

Energy conservation (\ref{enconsme}) for (\ref{gll1}) implies%

\begin{equation}
\gamma=\gamma_{0}\sqrt{\frac{\rho_{p}^{0}+\Gamma\rho_{0}\mathcal{V}_{0}}%
{\rho_{p}+\Gamma\rho\mathcal{V}}}. \label{gamma}%
\end{equation}

Clearly all the equations given above can be written for the average values of
the number and energy densities.

\subsubsection{Quasi-analytic model of GRBs}

\label{qam}

The first detailed models for the expansion of a relativistic fireball were
suggested in the beginning of nineties \cite{1990ApJ...365L..55S,1993MNRAS.263..861P,1993ApJ...415..181M}. Independent
calculations were performed in Ref.~\refcite{1999A&A...350..334R} and
\refcite{2000A&A...359..855R}. The main difference of these last two articles from
the other models in the literature is that initially not photons but pairs are
created by an overcritical electric field, and these pairs produce photons later.
This plasma referred to as the pair-electro-magnetic (PEM) pulse expands
initially into the vacuum surrounding the black hole reaching 
relativistic velocities very quickly. Then the collision with the baryonic remnant of the
collapsed star takes place and the PEM pulse becomes a
pair-electro-magnetic-baryonic (PEMB) pulse, see Ref.~\refcite{2003AIPC..668...16R}
for details. This difference is not large, since it was shown that the final
gamma factor does not depend on the distance from the baryonic remnant or on the
parameters of the black hole. The only crucial parameters are again the
initial energy $E_{0}$ and baryonic admixture $B$.

The model is based on the numerical integration of the relativistic energy-momen\-tum
conservation equations (\ref{conseq1},\ref{conseq2}) together with the
baryonic number conservation equation (\ref{consnum}). However, the most
important point distinct from all previous models is that the \emph{rate
equation} for electron-positron pairs is added to the model and integrated
simultaneously in order to give a more detailed description of the transparency.
This latter fact leads to quantitative differences in predictions of the model
with respect to the simplified models in the literature.

Here we concentrate on the simple quasi-analytical treatment presented in
\refcite{1999A&A...350..334R,2000A&A...359..855R}, see also
\refcite{2003AIPC..668...16R}. The PEMB pulse is supposed to contain a finite
number of shells each with a flat density profile. Their dynamics is governed by
the set of equations derived in the previous subsections. We collect (\ref{ncom}%
),(\ref{entrcon}) and (\ref{gamma}) together (omitting brackets for brevity)%

\begin{align}
\frac{n_{B}^{0}}{n_{B}}  &  =\frac{V}{V_{0}}=\frac{\mathcal{V}}{\mathcal{V}%
_{0}}\frac{\gamma}{\gamma_{0}},\label{eqn_rr_1}\\
\frac{\rho_{0}}{\rho}  &  =\left(  \frac{V}{V_{0}}\right)  ^{\Gamma}=\left(
\frac{\mathcal{V}}{\mathcal{V}_{0}}\right)  ^{\Gamma}\left(  \frac{\gamma
}{\gamma_{0}}\right)  ^{\Gamma},\label{eqn_rr_2}\\
\frac{\gamma}{\gamma_{0}}  &  =\sqrt{\frac{\rho_{p}^{0}+\Gamma\rho
_{0}\mathcal{V}_{0}}{\rho_{p}+\Gamma\rho\mathcal{V}},} \label{eqn_rr_3}%
\end{align}
where subscript ``0" denotes initial values, and all quantities are assumed
to be averaged over a finite distribution of shells with constant width and
density profiles. All components such as photons, electrons, positrons and
plasma ions give a contribution to the energy density and pressure. This set of
equations is equivalent to (\ref{eqn_sp_adi}) and (\ref{eqn_sp_gamma}) (see
below). The next step is to take into account the rate equation for positrons
and elections, accounting for non-instant transparency:%

\begin{equation}
(n_{e^{\pm}}U^{\mu})_{;\mu}=\overline{\sigma\mathfrak{v}}(n_{e^{\pm}}%
^{2}(T)-n_{e^{\pm}}^{2}),
\end{equation}
or, integrating over the volume%

\begin{equation}
\frac{\partial}{\partial t}N_{e^{\pm}}=-N_{e^{\pm}}\frac{1}{\mathcal{V}}%
\frac{\partial\mathcal{V}}{\partial t}+\overline{\sigma\mathfrak{v}}\frac
{1}{\gamma^{2}}(N_{e^{\pm}}^{2}(T)-N_{e^{\pm}}^{2}), \label{eqn_rr_4}%
\end{equation}
where $\sigma$ is the mean pair annihilation-creation cross section
and $\mathfrak{v}$ is the thermal velocity of $e^{\pm}$-pairs. The coordinate
number density of $e^{\pm}$-pairs in equilibrium is $N_{e^{\pm}}(T)=\gamma
n_{e^{\pm}}(T)$ and the coordinate number density of $e^{\pm}$-pairs is
$N_{e^{\pm}}=\gamma n_{e^{\pm}}$. For $T>m_{e}c^{2}$ we have $n_{e^{\pm}%
}(T)\simeq n_{\gamma}(T)$, i.e., the number densities of pairs and photons are
nearly equal. The pair number densities are given by appropriate Fermi
integrals with zero chemical potential at the equilibrium temperature $T$.
For an infinitesimal expansion of the coordinate volume from $\mathcal{V}_{0}
$ to $\mathcal{V}$ in the coordinate time interval $t-t_{0}$ one can
discretize the last differential equation for numerical computations.

The most important results of the analysis performed in
\refcite{2000A&A...359..855R} are the following:

\begin{itemize}
\item the appropriate model for the geometry of an expanding fireshell (PEM-pulse) is
given by the constant width approximation (this conclusion is achieved by
comparing results obtained using (\ref{conseq1}),(\ref{conseq2}) and the
simplified treatment described above),

\item there is a bound on parameter $B$ which comes from violation of the constant
width approximation, $B\leq10^{-2}$ ($\eta\geq10^{2}$).
\end{itemize}

In a previous subsection we proved the applicability of the constant
thickness approximation for the fireshell. The second conclusion appears to be
crucial, since it shows that there is a critical loading of baryons at which
their presence produce a turbulence in the outflow from the fireshell, and its
motion becomes very complicated and the fireshell evolution does not lead in
general to a GRB.

Exactly because of this reason, the optically thick fireshell never reaches
a radius as large as $r_{b}=r_{0}\eta^{2}$ which is discussed in
\refcite{1993ApJ...415..181M}, see section (\ref{lmr}), since to do this the
baryonic fraction should exceed the critical value $B_{c}=10^{-2}$. For
larger values of $B_{c}$ the theory reviewed here does not apply. This means
in particular that all the conclusions of Ref.~\refcite{1993ApJ...415..181M} obtained for
$r>r_{b}$ are invalid. In fact, for $B<B_{c}$ the gamma factor even does not
reach saturation.

Notice that another way to obtain the constraint $B<B_{c}$ is to require the
optical depth of the emitting region to be smaller than 1, leading to the
requirement that the Lorentz factor be greater than $\gamma\geq10^{2}$, see
the introduction. At the same time, there is a simple relation between the Lorentz
factor and the baryonic loading parameter $B=\gamma^{-1}$ in the region
$10^{-2}<B<10^{-4}$, see Fig.~\ref{fig3_2}, which leads to $B\leq10^{-2}$.

\begin{figure}
\includegraphics[width=\hsize]{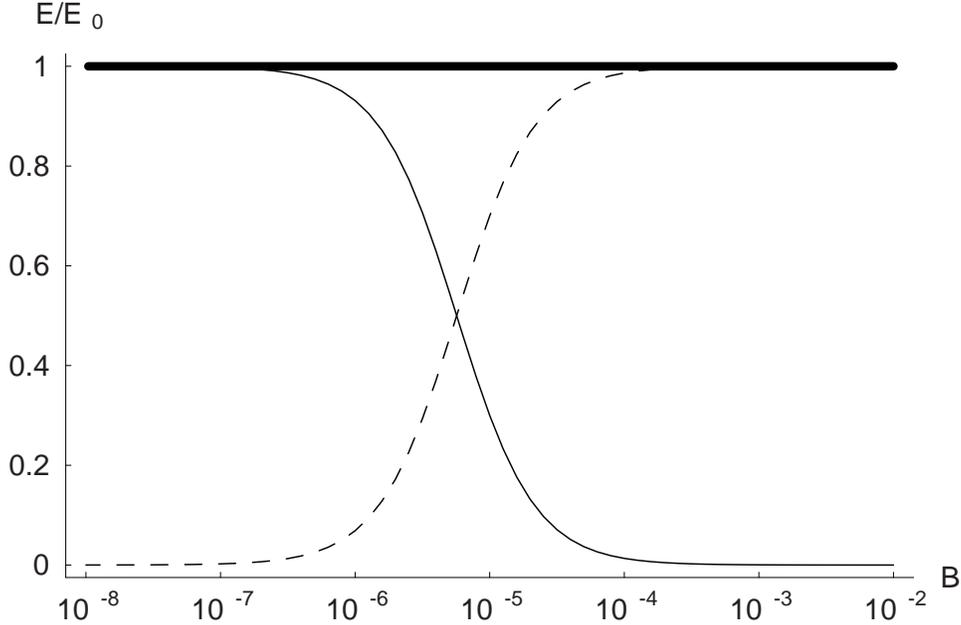}
\caption{Relative energy release in the form of photons emitted at the
transparency point (solid line) and the kinetic energy of the plasma (dashed line)
of the baryons in terms of the initial energy of the fireball depending on
parameter $B$ obtained on the basis of quasi-analytic model. The thick line denotes
the total energy of the system in terms of the initial energy.}
\label{fig3_1}
\end{figure}

\begin{figure}
\includegraphics[width=\hsize]{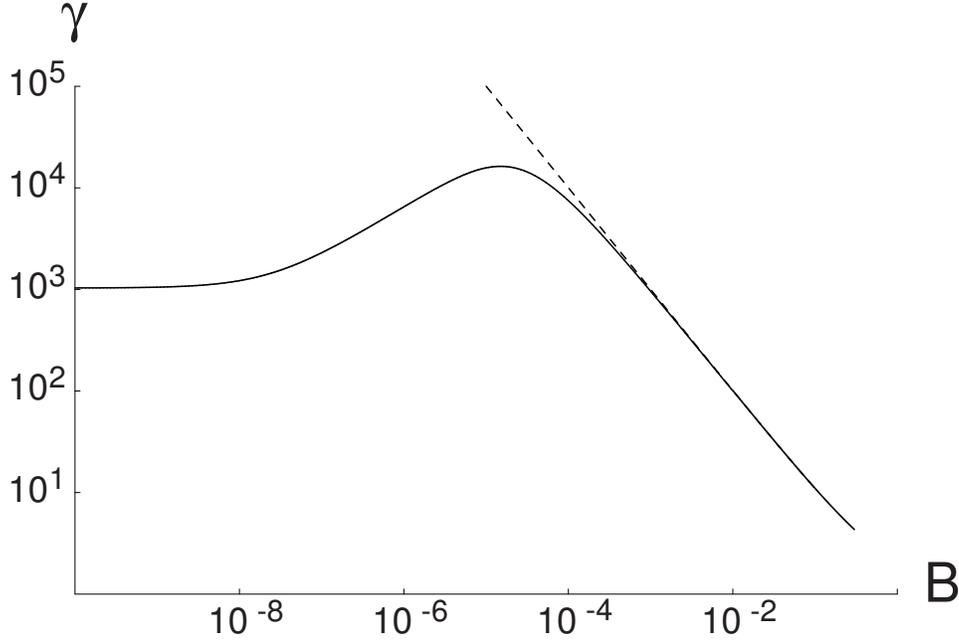}
\caption{Relativistic gamma factor of the fireball when it reaches transparency
depending on the value of parameter $B$. The dashed line gives the asymptotic value.}
\label{fig3_2}
\end{figure}

The fundamental result coming from this model are the diagrams shown in
Fig.~\ref{fig3_1} and Fig.~\ref{fig3_2}. The first one shows basically which
portion of the initial energy is emitted in the form of gamma rays $E_{\gamma}$
when the fireshell reaches the transparency condition $\tau\simeq1$ and how much
energy gets converted into the kinetic energy $E_{k}$ of the baryons left after
pair annihilation and the photons escape.

The second diagram gives the value of gamma factor at the moment when the systems
reaches transparency.

Energy conservation holds, namely
\begin{equation}
E_{0}=E_{\gamma}+E_{k}.
\end{equation}
Clearly when the baryon abundance is low, most energy is emitted when the
fireshell becomes transparent. It is remarkable that almost all initial energy is
converted into the kinetic energy of the baryons already in the region of validity of the
constant thickness approximation $B<10^{-2}$, so the region $10^{-8}%
<B<10^{-2}$ is the most interesting from this point of view.

\subsubsection{Alternative models}

\paragraph{Shemi and Piran model.}

\label{sp}

In this section we discuss the model proposed by Ref.~\refcite{1990ApJ...365L..55S}. This quantitative model gives a rather good general
picture of relativistic fireballs.
Shemi and Piran found that the temperature at which the fireball becomes
optically thin is determined to be%

\begin{equation}
\mathcal{T}_{esc}=\min(\mathcal{T}_{g},\mathcal{T}_{p}),
\end{equation}
where $\mathcal{T}_{g}$ and $\mathcal{T}_{p}$\ is the temperature when it
reaches transparency with respect to the gas (plasma) or the pairs:%

\begin{align}
\mathcal{T}_{g}^{2}  &  \simeq\frac{45}{8\pi^{3}}\frac{m_{p}}{m_{e}}\frac
{1}{\alpha^{2}g_{0}^{\frac{1}{3}}}\frac{1}{\mathcal{T}_{0}^{2}\mathcal{R}%
0}\eta,\label{eqn_sp_Tg}\\
\mathcal{T}_{p}  &  \simeq0.032,
\end{align}
where $m_{p}$ is the proton mass, $g_{0}=\frac{11}{4},$ $\alpha=\frac{1}%
{137},$ and the dimensionless temperature $\mathcal{T}$ and radius $\mathcal{R}$ of
the fireball are measured in units of $\frac{m_{e}c^{2}}{k_{B}}$ and
$\lambda_{e}\equiv\frac{\hbar}{m_{e}c}$ correspondingly, and the subscript ``0"
denotes initial values. The temperature at the transparency point in the case when the
plasma admixture is unimportant is nearly a constant for a range of parameters
of interest and it almost equals%

\begin{equation}
T_{p}=15\,\,\mathrm{keV}. \label{eqn_Tp}%
\end{equation}

Adiabatic expansion of the fireball implies%

\begin{equation}
\frac{\boldsymbol{E}}{\boldsymbol{E}_{0}}=\frac{\mathcal{T}}{\mathcal{T}_{0}%
}=\frac{\mathcal{R}_{0}}{\mathcal{R}}, \label{eqn_sp_adi}%
\end{equation}
where $\boldsymbol{E}=\frac{E}{m_{e}c^{2}}$ is a radiative energy. From the
energy conservation (\ref{ce}), assuming the fluid to be pressureless and its
energy density profile to be constant, we have in the coordinate frame%

\begin{equation}
\int T_{0}~^{0}d\mathcal{V}=\gamma E_{tot}=\mathrm{const}.
\end{equation}

Assuming at initial moment $\gamma_{0}=1$ and remembering that
$\boldsymbol{E}_{tot}=\boldsymbol{E}+Mc^{2}$ we arrive at the following
fundamental expression for the relativistic gamma factor $\gamma$ at the transparency point:%

\begin{equation}
\gamma=\frac{\boldsymbol{E}_{0}+\mathcal{M}c^{2}}{\boldsymbol{E}%
+\mathcal{M}c^{2}}=\frac{\eta+1}{(\frac{\mathcal{T}_{esc}}{\mathcal{T}_{0}%
})\eta+1}, \label{eqn_sp_gamma}%
\end{equation}
where $\mathcal{M}=\frac{M}{m_{e}}.$

One can use this relation to get such important characteristics of the GRB as the
observed temperature and observed energy. In fact, they can be expressed as follows:%

\begin{align}
\mathcal{T}_{obs}  &  =\gamma\mathcal{T}_{esc},\\
\boldsymbol{E}_{obs}  &  =\boldsymbol{E}_{0}\frac{\mathcal{T}_{obs}%
}{\mathcal{T}_{esc}}.
\end{align}

\begin{figure}
\includegraphics[width=\hsize]{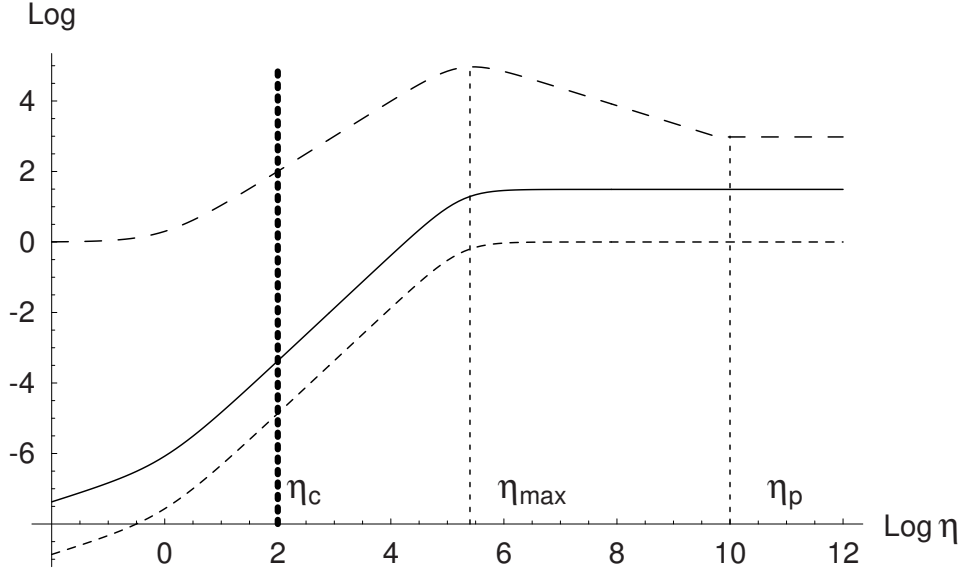}
\caption{The relativistic gamma factor (upper dashed line), the observed
temperature (solid line), and the ratio of observed energy to the initial
energy of the fireball (lower dashed line) as a function of baryonic loading
parameter, see Ref.~\refcite{1990ApJ...365L..55S}. The values of parameters are the
same as in the cited paper. The thick dashed line denotes the limiting value of
the baryonic loading. Its values when the gamma factor reaches a maximum and becomes
constant are also shown.}
\label{fig3_3}
\end{figure}

These results are shown in Fig.~\ref{fig3_3}. In the limit of small $\eta$
we have $\gamma=(1+\eta)$, while for very large $\eta$ the value of the gamma
factor at the transparency point is $\gamma=\mathcal{T}_{0}/\mathcal{T}_{esc}$,
and it has a maximum at intermediate values of $\eta$. We denote by a dashed
thick line the limiting value of the $\eta$ parameter $\eta_{c}\equiv B_{c}^{-1}$.
For $\eta<\eta_{c}$ the approximations used to construct the model do not
hold. It is clear that because of the presence of bound $\eta_{c}$ the value
$\gamma=\eta$ can be reached only asymptotically. In effect, the value
$\eta_{c}$ cuts the region where saturation of the gamma factor happens before
the moment when the fireball becomes transparent.%

It was found that for relatively large $\eta\geq10^{5}$ the photons emitted
when the fireball becomes transparent carry most of the initial energy.
However, since the observed temperature in GRBs is smaller than the initial
temperature of the fireball, one may assume that a large part of the initial
energy is converted to kinetic energy of the plasma.

\paragraph{Shemi, Piran and Narayan model.}

\refcite{1993MNRAS.263..861P} presents a generalization of
this model to arbitrary initial density profile of the fireball. These authors
performed numerical integrations of energy-momentum relativistic
conservation equations (\ref{conseq1}),(\ref{conseq2}) and baryon number
conservation equation (\ref{connum}). They were mainly interested in the
evolution of the observed temperature, gamma factor and other quantities as the
the radius increases. Their study results in a number of important conclusions, namely:

\begin{itemize}
\item The expanding fireball has two basic phases: a radiation dominated phase
and a matter-dominated phase. In the former, the gamma factor grows linearly
with the radius of the fireball: $\gamma\varpropto r$, while in the latter the
gamma factor reaches the asymptotic value $\gamma\simeq\eta+1$.

\item The numerical solutions are reproduced with good accuracy by the
frozen-pulse approximation when the pulse width is given by initial radius of
the fireball.
\end{itemize}

The last conclusion is important, since the fireball becomes a fireshell, and the
volume $V$ can be calculated as%

\begin{equation}
V=4\pi R^{2}\Delta, \label{volume}%
\end{equation}
where $\Delta\simeq R_{0}$ is the width of the leading shell with constant
energy density profile and $R$ is the radius of the fireball.

They also present the following scaling solution:%

\begin{align}
R  &  =R_{0}\left(  \frac{\gamma_{0}}{\gamma}D^{3}\right)  ^{1/2},\\
\frac{1}{D}  &  \equiv\frac{\gamma_{0}}{\gamma}+\frac{3\gamma_{0}}{4\gamma
\eta}-\frac{3}{4\eta},
\end{align}
where the subscript ``0" denotes some initial time when $\gamma\gtrsim$ 2, which
can be inverted to give $\gamma(R)$.

\paragraph{M\'{e}sz\'{a}ros, Laguna and Rees model.}

\label{lmr}

The next step in developing this model was made by Ref.~\refcite{1993ApJ...415..181M}.
In order to reconcile the model with observations, these authors proposed a
generalization to the anisotropic (jet) case. Nevertheless, their analytic results
apply to the case of homogeneous isotropic fireballs and we will follow their
analytical isotropic model in this section.

Starting from the same point as Shemi and Piran, consider (\ref{eqn_sp_adi})
and (\ref{eqn_sp_gamma}). The analytic part of the paper describes the geometry of
the fireball, the gamma factor behavior and the final energy balance between
radiation and kinetic energy. Magnetic field effects are also considered, but
we are not interested in this part here.

Three basic regimes are found in Ref.~\refcite{1993ApJ...415..181M} for the evolution of
the fireball. In two first regimes there is a correspondence between the
analysis in the paper and results of Ref.~\refcite{1993MNRAS.263..861P}, so the
constant thickness approximation holds. It is claimed in
\refcite{1993ApJ...415..181M} that when the radius of the fireball reaches very
large values comparable to $R_{b}=R_{0}\eta^{2}$, a noticeable departure from
the constant width of the fireball occurs. However, it is important to note that
the fireball becomes transparent much earlier and this effect never becomes
important (see section \ref{qam}).

The crucial quantity presented in the paper is $\Gamma_{m}$---the maximum
possible bulk Lorentz factor achievable for a given initial radiation energy
$E_{0}$ deposited within a given initial radius $R_{0}$:%

\begin{align}
\Gamma_{m}  &  \equiv\eta_{m}=\left(  \tau_{0}\eta\right)  ^{1/3}=\left(
\Sigma_{0}\kappa\eta\right)  ^{1/3},\label{eqn_mlr_Gm}\\
\Sigma_{0}  &  =\frac{M}{4\pi R_{0}^{2}},\quad\quad\kappa=\frac{\sigma_{T}%
}{m_{p}},
\end{align}
where $\Sigma_{0}$ is initial baryon (plasma) mass surface density.

All subsequent calculations in the paper Ref.~\refcite{1993ApJ...415..181M} involve
this quantity. It is evident from (\ref{eqn_mlr_Gm}) that a linear
dependence between the gamma factor $\Gamma$ and parameter $\eta$ is assumed.
However, this is certainly not true as can be seen from Fig.~\ref{fig3_2}. We
will come back to this point in the following section.

Another important quantity is given in this paper, namely%

\begin{equation}
\Gamma_{p}=\frac{T_{0}}{T_{p}}.
\end{equation}
This is just the asymptotic behavior of the gamma factor at Fig.~\ref{fig3_3}
for very large $\eta$. Using it, the authors calculate the value of the $\eta$
parameter above which the pair dominated regime occurs:%

\begin{equation}
\eta_{p}=\frac{\Gamma_{m}^{3}}{\Gamma_{p}^{2}}.
\end{equation}
This means that above $\eta_{p}$ the presence of baryons in the fireball is
insufficient to keep the fireball opaque after pairs are annihilated and
almost all initial energy deposited in the fireball is emitted immediately.

\begin{figure}
\includegraphics[width=\hsize]{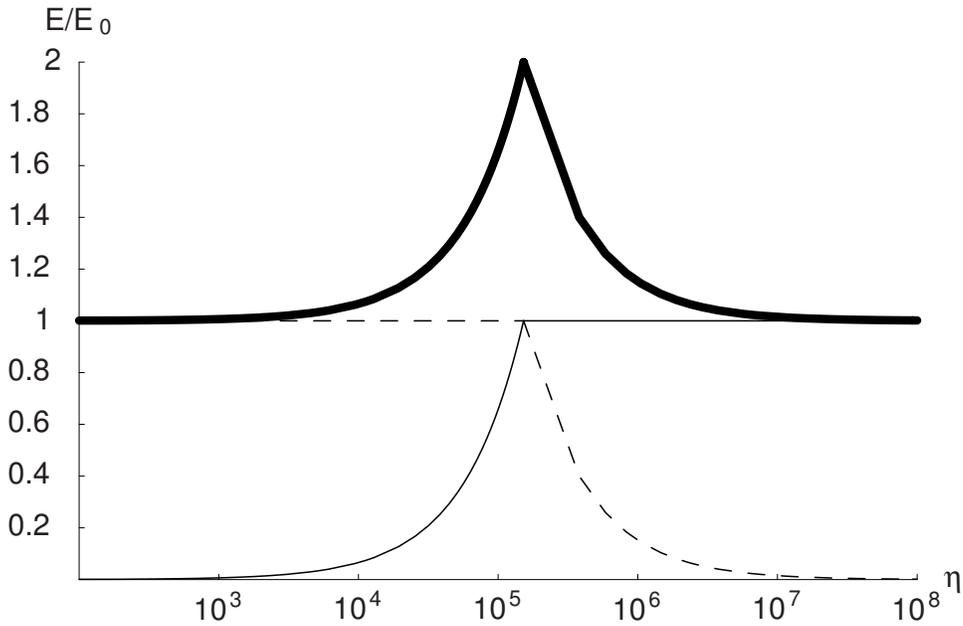}
\caption{The ratios of radiation and kinetic energy to the initial energy of
the fireball predicted by the M\'{e}sz\'{a}ros, Laguna and Rees model. The thick line
denotes the total energy of the system in terms of the initial energy. Energy
conservation does not hold.}
\label{fig3_4}
\end{figure}

The estimate of the final radiation to kinetic energy ratio
\cite{1993ApJ...415..181M} is incorrect, because kinetic and radiation
energies do not sum up to the initial energy of the fireball thus violating energy
conservation. This is illustrated in Fig.~\ref{fig3_4}. The correct analytic
diagram is instead presented in Fig.~\ref{fig3_1}.%

\paragraph{Approximate results.}

\label{approx}

All models for isotropic fireballs are based on the following points:

\begin{enumerate}
\item flat space-time,

\item relativistic energy-momentum principle,

\item baryonic number conservation.
\end{enumerate}

Although the model \cite{1999A&A...350..334R,2000A&A...359..855R}
starts with Reissner-Nord\-str\"{o}m geometry, the numerical code is written for
the case of flat space-time simply because curved space-time effects become
insignificant soon after the fireshell reaches relativistic expansion
velocities. The presence of the rate equation in the model
\cite{1999A&A...350..334R,2000A&A...359..855R} has a strong physical
basis and its absence in the other treatments means incompleteness of their
models. Indeed, the number density of pairs influences the speed of expansion
of the fireshell. However, in this section we neglect the rate equation and
discuss those points common among all the models being considered.

First of all, let us return to Fig.~\ref{fig3_2}. For almost all values of the
$\eta$ parameter the gamma factor is determined by the gas (i.e., plasma or
baryons) admixture according to (\ref{eqn_sp_Tg}); consider this case in what follows.
For given initial energy and radius this temperature depends only on $\eta$, so one can write:%

\begin{equation}
\gamma=\frac{\eta+1}{(\frac{\mathcal{T}_{g}}{\mathcal{T}_{0}})\eta+1}%
=\frac{\eta+1}{a\eta^{\frac{3}{2}}+1}, \label{eqn_an}%
\end{equation}
where%

\begin{equation}
a =2.1\cdot10^{3}\mathcal{T}_{0}^{-2}\mathcal{R}_{0}^{-0.5}.
\end{equation}
From this formula one has immediately the two asymptotic regimes, namely:%

\begin{equation}
\gamma=\left\{
\begin{array}
[c]{cc}%
\eta+1, & \eta<\eta_{\max},\\
\frac{1}{a \sqrt{\eta}}, & \eta>\eta_{\max}%
\end{array}
\right.  . \label{eqn_g_as}%
\end{equation}

Notice, that the constant $a$ is an extremely small number, so that after
obtaining a precise value of $\eta_{\max}$ by equating to zero the derivative of
function (\ref{eqn_an}), one can expand the result in a Taylor series and get to
the lowest order in $a$  the result:%

\begin{align}
\eta_{\max}  &  \simeq\left(  \frac{2}{a}\right)  ^{\frac{2}{3}}-2,\\
\gamma_{\max}  &  \equiv\gamma(\eta_{\max})\simeq\frac{1}{3}\left[  1+\left(
\frac{2}{a}\right)  ^{\frac{2}{3}}\right]  .
\end{align}

In particular, in the case shown in Fig.~\ref{fig3_3} one has $\eta_{\max
}=2.8\cdot10^{5}$, $\gamma_{\max}=9.3\cdot10^{4}$ while according to
(\ref{eqn_mlr_Gm}), $\Gamma_{m}=\eta_{\max}=1.75\cdot10^{5}$. Clearly, our
result is much more accurate. Actually, the value $\Gamma_{m}$ in
(\ref{eqn_mlr_Gm}) is obtained from equating asymptotes in (\ref{eqn_g_as})
and there exists the following relation:%

\begin{align}
a=(\tau_{0}\eta)^{-1/2}.
\end{align}

Now we are ready to explain why the observed temperature (and consequently the
observed energy) does not depend on $\eta$ in the region $\eta_{max}<\eta
<\eta_{p}$. From the second line in (\ref{eqn_g_as}) it follows that the gamma
factor in this region behaves like $\gamma\propto\eta^{-1/2}$, while
$\mathcal{T}_{esc}\propto\eta^{1/2}$. These two exactly compensate each other
leading to the independence of the observed quantities on $\eta$ in this region.
This remains the same for $\eta>\eta_{p}$ also, since here $\mathcal{T}%
_{esc}=\mathcal{T}_{p}=$const and from (\ref{eqn_an}) $\gamma=$const.

\paragraph{Nakar, Piran and Sari revision.}

Recently a revision of the fireball model was made in Ref.~\refcite{2005ApJ...635..516N}. These authors presented a new diagram for the final
Lorentz gamma factor and for the energy budget of the fireball. Their work was
motivated by the observation of giant flares with the subsequent afterglow
spreading up to the radio region with a thermal spectrum. They concluded that the
fireball has to be loaded by either baryons or magnetic field, and cannot be
only a pure $e^{\pm},\gamma$ plasma in order to have $10^{-3}$ of the total
energy radiated in the giant flare.

In analogy with cosmology the authors define the number density of pairs which
survive because the expansion rate becomes larger than the annihilation
rate\footnote{This effect is accounted for automatically in our approach where
rate equations for pairs include an expansion term.} which gives the condition
\begin{equation}
n_{\pm}\approx\frac{1}{\sigma_{T}R_{0}}.
\end{equation}
Then, recalling (\ref{eqn_sp_adi}), if we want to estimate the number of pairs it
turn out to be
\begin{equation}
N_{\pm}=\frac{4\pi R_{0}ct}{\sigma_{T}}\,\left(  \frac{T_{0}}{T_{\pm}}\right)
^{2},
\end{equation}
where we identify $\Delta=ct$ in (\ref{volume}). In Ref.~\refcite{2005ApJ...635..516N}
the authors obtained the third power of the ratio of temperatures which influences
all of their subsequent results.

Having reached the conclusion that the afterglow cannot be obtained as the result
of interaction of the $e^{\pm},\gamma$ plasma with the CircumBurst Medium (CBM), the authors turn to baryonic
loading considerations. They attempt to define critical values of the loading
parameter $\eta$ finding in general 4 such values\footnote{The last value
$\eta_{4}$ corresponds to the case of heavy loading where spreading of the
expanding shell is observed, and is not considered here.}, in particular:
\begin{align}
\eta_{1}=\frac{E_{0}\sigma_{T}}{4\pi R_{0} ct m_{e} c^{2}}\left(  \frac
{T_{\pm}}{T_{0}}\right)  ^{3},\\
\eta_{2}=\frac{E_{0}\sigma_{T}}{4\pi R_{0} ct m_{p} c^{2}}\left(  \frac
{T_{\pm}}{T_{0}}\right)  ^{3},\\
\eta_{3}=\left(  \frac{E_{0}\sigma_{T}}{4\pi R_{0} ct m_{p} c^{2}}\right)
^{1/4}.
\end{align}

We recall that the first two quantities are based on the formula for $N_{\pm}$
and should contain factors of $\left(  \frac{T_{\pm}}{T_{0}}\right)  ^{2}$ instead.

The first `critical' value, $\eta_{1}$, comes from the condition $N_{p}
m_{p}=N_{\pm}m_{e}$, where $N_{p}\equiv\frac{E_{0}}{m_{p} c^{2}\eta}$ is just
the number of protons in the plasma admixture. It does not correspond to any
critical change in the physics of the phenomena; for instance it cannot be
interpreted as the equality of masses (equal inertia) of pairs and baryons since
the former is mainly due to their total energy $E_{\pm}$, while the latter
to their rest mass $N_{p} m_{p}$. This value is, however, close to the one
defined above $\eta_{1}\approx\eta_{p}$.

The second `critical' value, $\eta_{2}$, corresponds to the condition
$N_{p}=N_{\pm}$, namely equality of the numbers of protons and pairs. It is also
incorrectly interpreted as equal contributions to the Thompson scattering. In
fact, the cross-section for Thompson scattering of protons contains an additional
factor $\left(  \frac{m_{e}}{m_{p}}\right)  ^{2}$ with respect to the usual
formula for electrons.

The definition of the third `critical' value, $\eta_{3}$, is not clear, but
an important feature is its closeness to the critical value $\eta_{c}$ quoted above.

Assuming adiabatic conditions (\ref{eqn_sp_adi}) the authors present the
new diagram for the final gamma factor and energy budget of the pair-baryonic
plasma at transparency. In fact, this diagram, shown in our Fig.~\ref{fig3_6}
by the dashed curve for the parameter $B$, is very similar to the one obtained in Ref.~\refcite{1998MNRAS.300.1158G}, which considered
the hydrodynamics of relativistic $e^{\pm},\gamma$ winds. That problem is very
different from ours, because of different boundary conditions\footnote{Note
that the authors of Ref.~\refcite{1998MNRAS.300.1158G} also use rate equations
describing decoupling plasma from photons.}. In particular, in the wind energy
conservation, (\ref{ce}) does not hold; the reason is that constant energy
(mass) supply takes place parametrized in Ref.~\refcite{1998MNRAS.300.1158G} by
$\dot{E}$ ($\dot{M}$). In that paper in fact authors present the diagram for the
asymptotic value of the Lorentz gamma factor depending on the ratio
$\frac{\dot{E}}{\dot{M}}$ which is very different from the quantity $\eta$.

Surprisingly, the fundamental result about the presence of a maximum in the
diagram for the gamma factor on Fig.~\ref{fig3_6} which was found by the same
authors previously in Ref.~\refcite{1990ApJ...365L..55S} (see Fig.~\ref{fig3_3}) that
comes from the energy conservation (\ref{eqn_sp_gamma}) is ignored in
\refcite{2005ApJ...635..516N}. It can be understood in the following way. For
small loading (small $B$) the more baryons are present in the plasma the
larger the number density of corresponding electrons becomes, and the larger
the optical depth is. Therefore, transparency is reached later, which gives a larger
gamma factor at transparency. On the other hand, for heavy baryon loading
(relatively large $B$) the more baryons are present, the more inertia the
plasma has, and by energy conservation, the less the final gamma factor has to be.

\subsubsection{Significance of the rate equation}

The rate equation describes the evolution of the number densities for electrons and
positrons. In analytic models it is assumed that pairs are annihilated
instantly when the transparency condition is fulfilled. Moreover, the dynamics of
expansion is influenced by the electron-positron energy density as can be seen
from (\ref{eqn_rr_1})--(\ref{eqn_rr_4}). Therefore, it is important to clarify whether neglecting the rate equation is a crude approximation or not.

\begin{figure}
\includegraphics[width=\hsize]{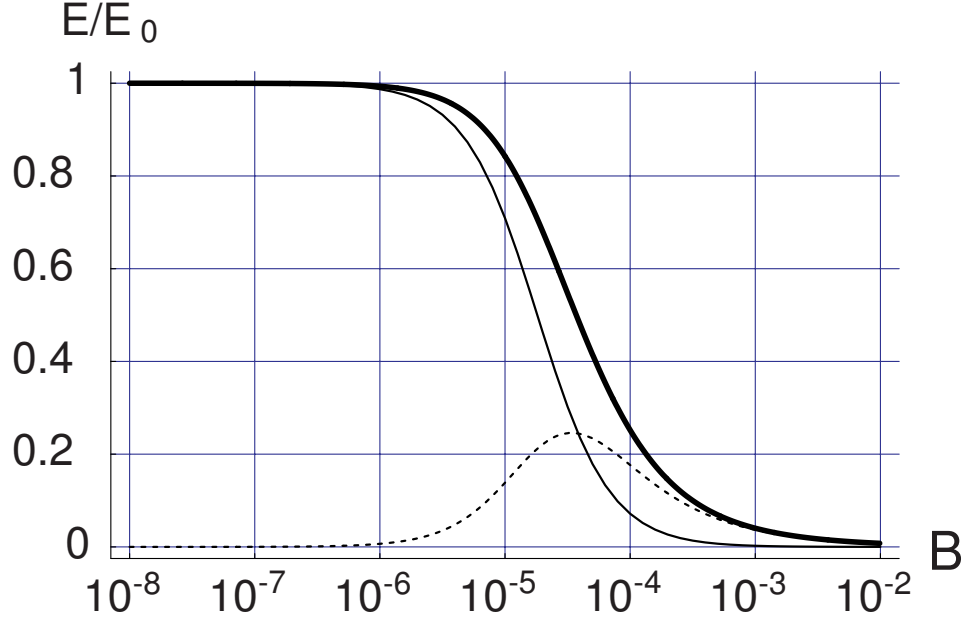}
\caption{Relative energy release in the form of photons emitted at the
transparency point of the GRB in terms of the initial energy of the fireball
depending on parameter $B$. Thick line represents numerical results and it is
the same as in Fig.~\ref{fig3_1}. The normal line shows the results for the analytic
model of Ref.~\refcite{1990ApJ...365L..55S}. The dashed line shows the
difference between the exact numerical and approximate analytical results.}
\label{fig3_5}
\end{figure}

Using Eq.~ (\ref{eqn_sp_gamma}) one can obtain the analytic dependence of the
energy emitted at the transparency point on parameter $B$ and we compare them in
Fig.~\ref{fig3_5}.%

We also show the difference between numerical results based on integration of
Eqs.~(\ref{eqn_rr_1}--\ref{eqn_rr_4}) and analytic results from the Shemi and Piran
model. The values of the parameters are: $\mu=10^{3}$ and $\xi=0.1$ (which
correspond to $E_{0}=2.87\cdot10^{54}\,$ergs and $R_{0}=1.08\cdot10^{9}\,$cm).
One can see that the difference peaks at intermediate values of $B$. The
crucial deviations, however, appear for large $B$, where analytical predictions
for the observed energy are about two orders of magnitude smaller than the
numerical ones. This is due to the difference in predictions of the radius of
the fireshell at the transparency moment. In fact, the analytical model
overestimates this value by about two orders of magnitude for $B=10^{-2}$. So
for large $B$ with correct treatment of pair dynamics the fireshell becomes
transparent \emph{earlier} comparing to the analytical treatment.

\begin{figure}
\includegraphics[width=\hsize]{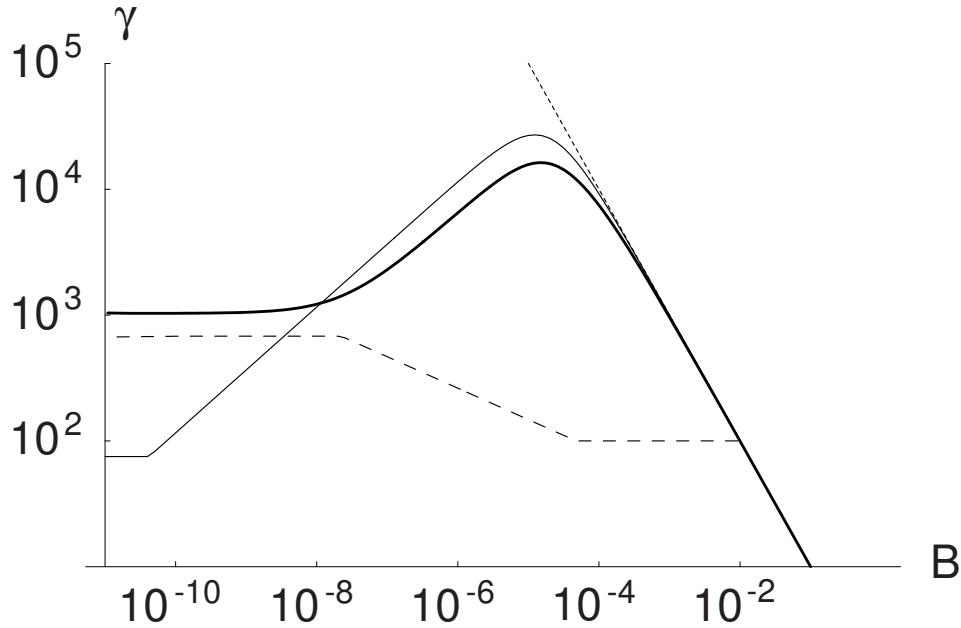}
\caption{Relativistic gamma factor when transparency is reached. The thick
line denotes exact numerical results, the normal line corresponds to
the analytical estimate from the  Shemi and Piran model, while the dotted line denotes the
asymptotic value of the baryonic loading parameter. The dashed line shows the
results of Nakar, Piran and Sari.}
\label{fig3_6}
\end{figure}

At the same time, the difference between numerical and analytical results for
gamma factor is significant for small $B$ as illustrated at Fig.~\ref{fig3_6}.
While both results coincide for $B>10^{-4}$ there is a constant difference for
the range of values $10^{-8}<B<10^{-4}$ and asymptotic constant values for the
gamma factor are also different. Besides, this asymptotic behavior takes place
for larger values of $B$ in disagreement with analytical expectations. Thus
the acceleration of the fireshell for small $B$ is larger if one accounts for
pair dynamics.

It is clear that the error coming from neglecting the rate equation is
significant. This implies that the simple analytic model of Shemi and Piran gives
only a qualitative picture of the fireshell evolution and in order to get the
correct description of the fireshell one cannot neglect the rate equation.

Moreover, the difference between the exact numerical model
\cite{1999A&A...350..334R,2000A&A...359..855R} and the approximate
analytical models \cite{1990ApJ...365L..55S} becomes apparent in various
physical aspects, namely in predictions of the radius of the shell when it
reaches transparency, the gamma factor at transparency and the ratio between
the energy released in the form of photons and that converted into kinetic
form. The last point is crucial. It is assumed in the literature that the
entire initial energy of the fireshell gets converted into kinetic energy of
the shell during adiabatic expansion. Indeed, taking a typical value of
the parameter $B$ like $10^{-3}$ we find that according to the Shemi and Piran model we
have only $0.2\%$ of the initial energy left in the form of photons. However,
exact numerical computations \cite{1999A&A...350..334R,2000A&A...359..855R} give $3.7\%$ for the energy of photons radiated
when the fireshell reaches transparency, which is a significant value and it
cannot be neglected.%

\begin{table}
\tbl{Comparison of different models for fireballs.}
{
\begin{tabular}{|c|c|c|c|c|}
\hline
& { Ruffini et al.} & { Shemi, Piran} & { Piran, et al.} &
{ M\'{e}sz\'{a}ros, et al.}\\\hline
{ conservation:} &  &  &  & \\
{ EM,} & { yes} & { yes} & { yes} &
{ yes}\\
{ bar. number,} & { yes} & { not consider} &
{ yes} & { yes}\\\hline
{ rate equation} & { yes} & { no} & { no} &
{ no}\\\hline
{ const width} & { justify} & { not consider} &
{ justify} & { in part}\\
{ approx.} &  &  &  & \\\hline
{ model }$\gamma(r)$ & { num./analyt.} & { no} &
{ num./analyt.} & { num./analyt.}\\\hline
{ model }$\gamma(\eta)$ & { num.} & { analyt.} &
{ not consider} & { analytic}\\\hline
\end{tabular}
}
\label{tab3}
\end{table}%

To summarize the above discussion, we present the results of this survey in the
Table \ref{tab3}. It is important to notice again that comparing to the simplified
analytic treatment, accounting for the rate of change of electron-positron
pairs densities gives quantitatively different results for the ratio of kinetic
versus photon energies produced in the GRB and the gamma factor at the
transparency moment, which in turn leads to different afterglow properties.
Therefore, although analytical models presented in sections \ref{sp} and
\ref{approx} agree and give the correct qualitative description of the fireshell,
one should use the numerical approach described in section \ref{qam} in order to
compare the theory and observations.

\paragraph{Reaching transparency.}%

\begin{figure}
\includegraphics[width=\hsize]{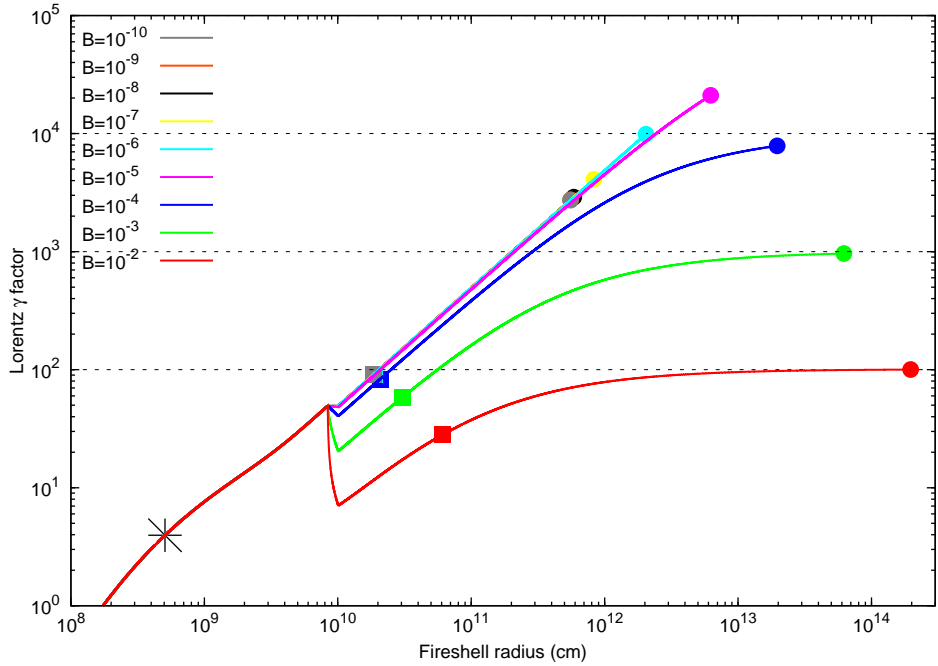}
\caption{The Lorentz factor as a function of the radius for selected values of the
baryonic loading parameter. Filled circles show the radius at the moment of
transparency. Squares show the moment of departure from thermal distributions
of electron-positron pairs. The star denotes the moment when the temperature
of the fireshell equals $511 $ keV.}
\label{fig:gamma}
\end{figure}

\begin{figure}
\includegraphics[width=\hsize]{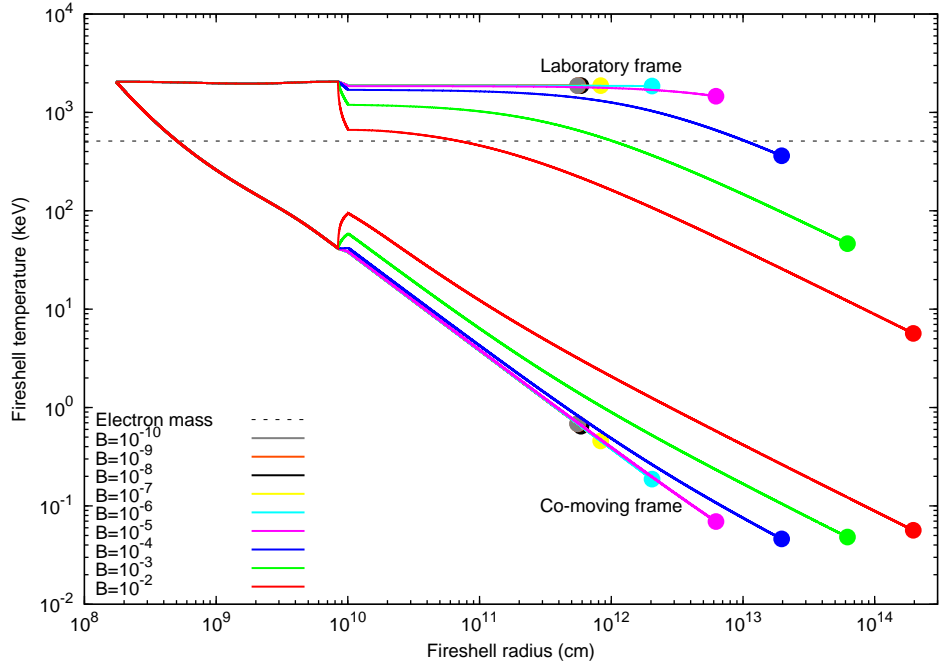}
\caption{The temperature of the fireshell as a function of the radius for selected
values of the baryonic loading parameter. Filled circles show the radius at
the moment of transparency.}
\label{fig:temperature}
\end{figure}

\begin{figure}
\includegraphics[width=\hsize]{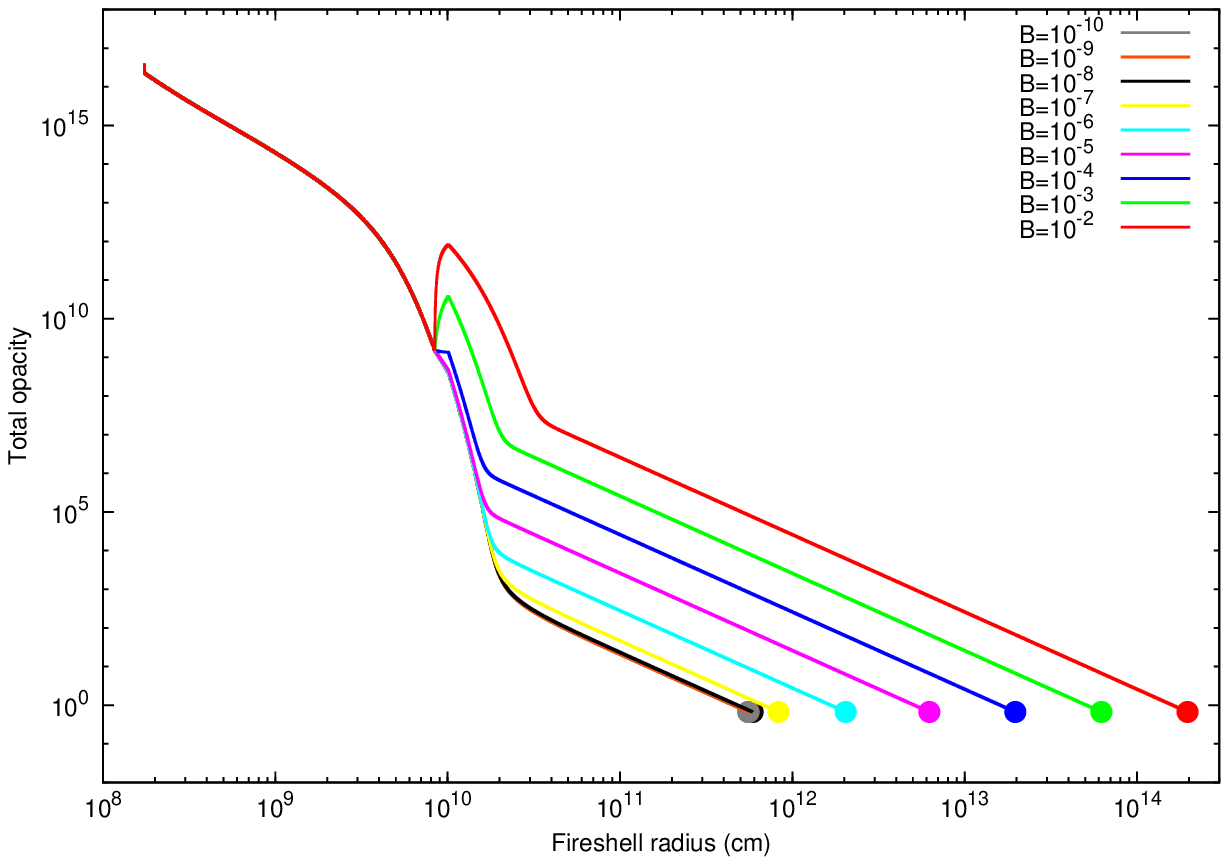}
\caption{Total opacity of the fireshell.}
\label{fig:tau}
\end{figure}

To demonstrate the richness of physical phenomena associated with gam\-ma-ray
bursts we provide below several illustrations. We calculate the temperature in the comoving and laboratory frames of the plasma as well as the Lorentz
factor with the code described in Ref.~\refcite{1999A&A...350..334R,2000A&A...359..855R} for
different values of the parameter $B$. We show our results in Fig.~\ref{fig:gamma} and \ref{fig:temperature}. From Fig.~\ref{fig:temperature} it
is clear that during early phases of the expansion the temperature decreases down
to $0.511$ MeV, as shown by the star in Fig.~\ref{fig:gamma}, and the ratio
of electron-positron pairs to photons becomes exponentially suppressed.
However, because of the accelerated expansion the apparent temperature in the
observer's frame remains almost constant, see Fig.~\ref{fig:temperature}.
Then, after collision with a baryonic remnant after which the Lorentz factor
decreases, the plasma continues to expand. At a certain moment, shown by
squares in Fig.~\ref{fig:gamma}, a departure from a thermal distribution of
electron-positron pairs occurs, due to the fact that the rate of the reaction
$\gamma\gamma\longrightarrow e^{+}e^{-}$\ becomes smaller than the expansion rate.
From that moment electron-positron pairs freeze out, analogous to what happens in the early
Universe. Finally, transparency is reached, as denoted by circles in Fig.~\ref{fig:gamma}. For high values of the baryonic loading $B>10^{-4}$ the
comoving temperature decreases in the late stages of expansion in the same way
as the observed temperature.

\begin{figure}
\includegraphics[width=\hsize]{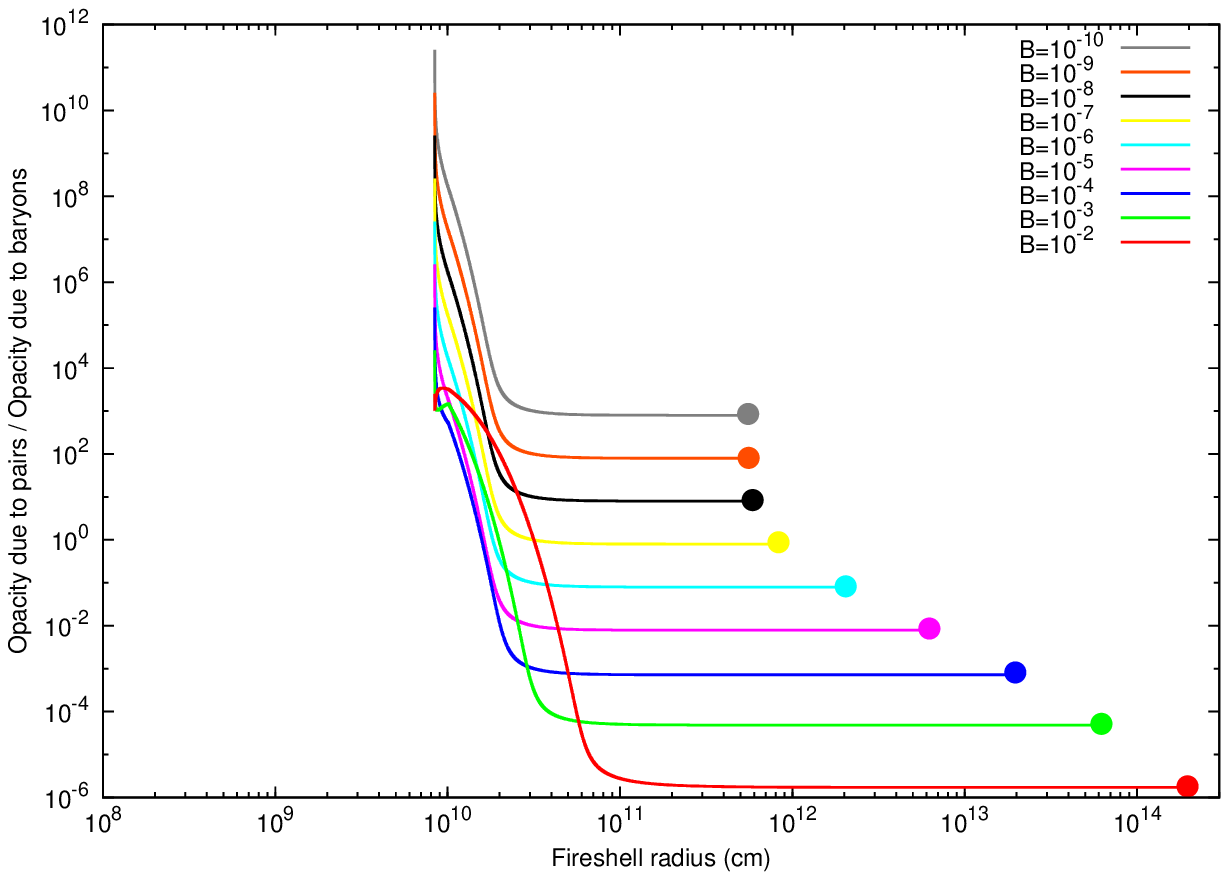}
\caption{The ratio of the opacity due to electron-positron pairs to the
opacity due to electrons associated with baryons in the fireshell.}
\label{fig:tau_ratio}
\end{figure}

Total opacity due to pair production and due to Compton scattering is shown in
Fig.~\ref{fig:tau}, while the ratio of separate contributions, i.e., opacity
due to pairs and baryons is shown in Fig.~\ref{fig:tau_ratio}. The expansion
starts with $\tau\gg1$ and the optical depth starts to decrease. Then, after
collision with the baryonic remnant containing also associated electrons, the
opacity may increase, but only for large baryonic loading $B>10^{-4}$. The
change of exponential decrease into a power law decrease seen in Fig.~\ref{fig:tau}
corresponds to the departure of distributions of electrons and positrons from
thermal ones. Finally, as one can see from Fig.~\ref{fig:tau_ratio} at the
late stage of expansion of the fireshell the opacity is dominated by pairs for
$B<10^{-7}$ and by Compton scattering for $B>10^{-7}$.

\subsection{The afterglow}

\begin{figure}  
\includegraphics[width=\hsize,clip]{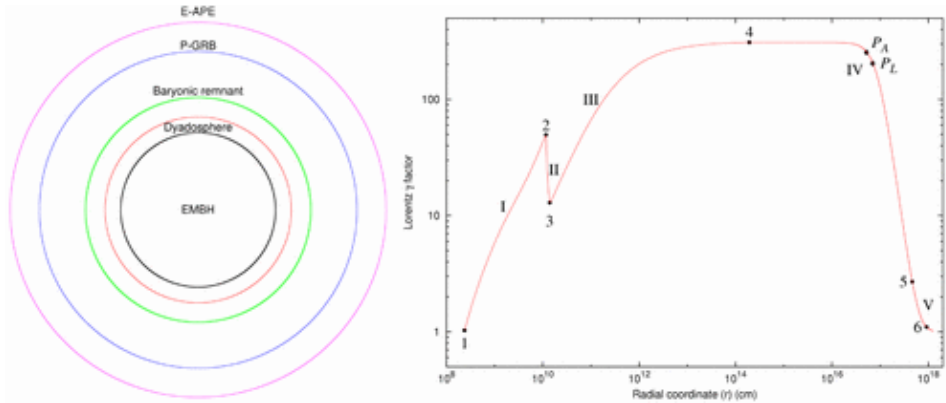} 
\caption{The GRB afterglow phase is represented here together with the optically thick phase (see Fig.~\ref{cip2}). The value of the Lorentz gamma factor is given here from the transparency point all the way to the ultrarelativisitc, relativistic and nonrelativistic regimes. Details in Ref.~\refcite{2003AIPC..668...16R}.}
\label{cip_tot} 
\end{figure}

After reaching transparency and the emission of the P-GRB, the accelerated baryonic matter (the ABM pulse) interacts with the CBM and gives rise to the afterglow (see Fig.~\ref{cip_tot}). Also in the descriptions of this last phase many differences exist between our treatment and the other ones in the current literature (see next sections).

We first look to the initial value problem. The initial conditions of the afterglow era are determined at the end of the optically thick era when the P-GRB is emitted. As recalled in the last section, the transparency condition is determined by a time of arrival $t_a^d$, a value of the gamma Lorentz factor $\gamma_\circ$, a value of the radial coordinate $r_\circ$, and the amount of baryonic matter $M_B$, all of which are only functions of the two parameters $E_{dya}$ and $B$ (see Eq.~\eqref{Bdef}).

This connection to the optically thick era is missing in the current approach in the literature which attributes the origin of the ``prompt radiation'' to an unspecified inner engine activity (see Ref.~\refcite{1999PhR...314..575P} and references therein). The initial conditions at the beginning of the afterglow era are obtained by a best fit of the later parts of the afterglow. This approach is quite unsatisfactory since the theoretical treatments currently adopted in the description of the afterglow are not appropriate\cite{2004ApJ...605L...1B,2005ApJ...620L..23B,2005ApJ...633L..13B}. The fit which uses an inappropriate theoretical treatment leads necessarily to the wrong conclusions as well as, in turn, to the determination of incorrect initial conditions.

The order of magnitude estimate usually quoted for the characteristic time scale to be expected for a burst emitted by a GRB at the moment of transparency at the end of the optically thick expansion phase is given by $\tau \sim GM/c^3$. For a $10M_\odot$ black hole this will give $\sim 10^{-3}$ s. There are reasons today not to take seriously such an order of magnitude estimate\cite{2005IJMPD..14..131R}. In any case this time is much shorter than the ones typically observed in ``prompt radiation'' of the long bursts, from a few seconds all the way to $10^2$ s. In the current literature (see e.g. Ref.~\refcite{1999PhR...314..575P} and references therein), in order to explain the ``prompt radiation'' and overcome the above difficulty it has been generally assumed that its origin should be related to a prolonged ``inner engine'' activity preceding the afterglow which is not well identified.

To us this explanation has always appeared logically inconsistent since there remain to be explained not one but two very different mechanisms, independent of each other, of similar and extremely large energetics. This approach has generated an additional very negative result: it has distracted everybody working in the field from the earlier very interesting work on the optically thick phase of GRBs.

\begin{figure}
\includegraphics[width=\hsize,clip]{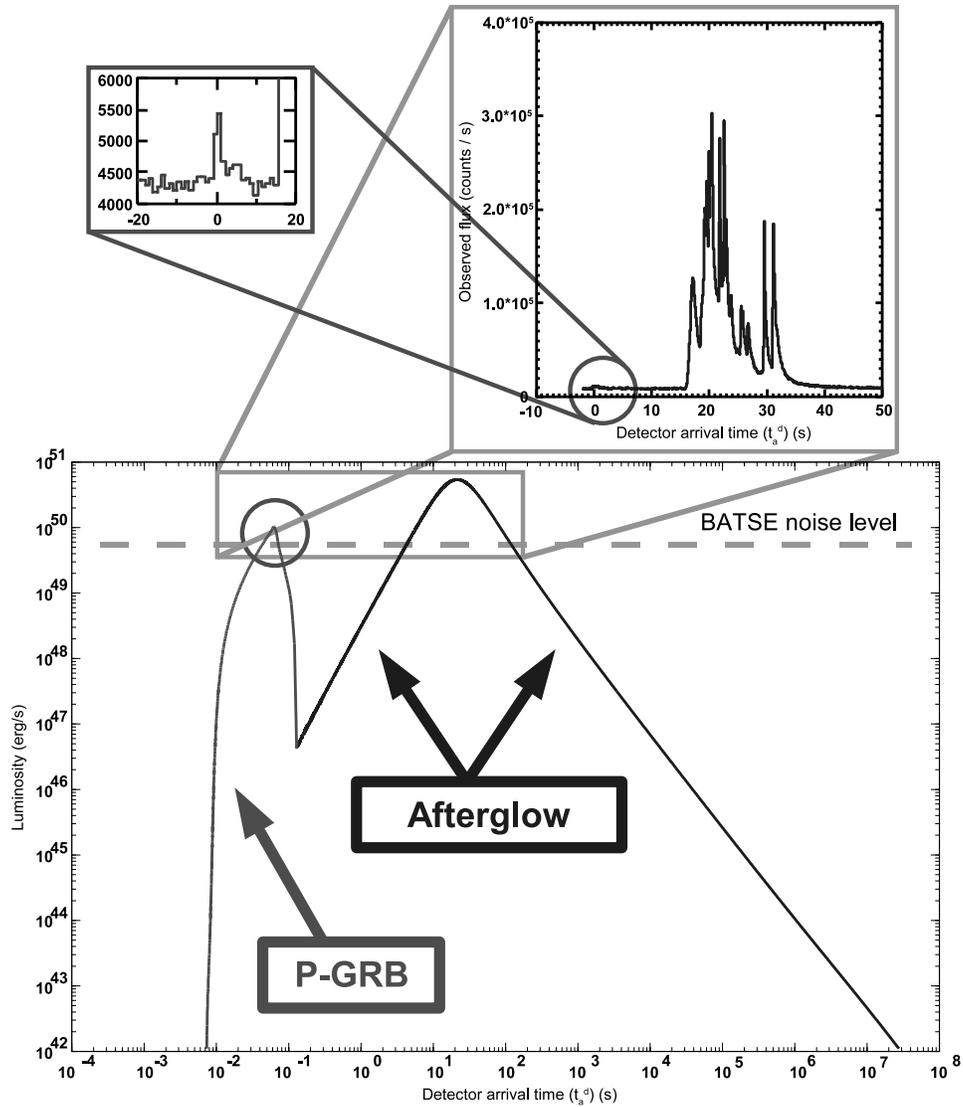}
\caption{The detailed features of GRB 991216 shown by our theoretical models are reproduced here: the P-GRB, the ``prompt radiation'' and what is generally called the afterglow. It is clear that the prompt emission observed by BATSE coincides with the extended afterglow peak emission (E-APE) and has been considered as a burst only as a consequence of the high noise threshold in the observations. The small precursor is identified with the P-GRB. Details in Ref.~\refcite{2001ApJ...555L.113R,2002ApJ...581L..19R,2003AIPC..668...16R,2005AIPC..782...42R}.}
\label{grb991216}
\end{figure}

The way out of this dichotomy in our model is different: 
{\bf 1)} indeed the optically thick phase exists and is crucial to the GRB phenomenon and terminates with a burst: the P-GRB; 
{\bf 2)} the ``prompt radiation'' follows the P-GRB; 
{\bf 3)} the ``prompt radiation'' is not a burst: it is actually the temporally extended peak emission of the afterglow (E-APE). The observed structures of the prompt radiation can all be traced back to inhomogeneities in the interstellar medium (see Fig.~\ref{grb991216} and Ref.~\refcite{2002ApJ...581L..19R}).

This approach was first tested on GRB 991216. Both the relative intensity and time separation of the P-GRB and the afterglow were duly explained (see Fig.~\ref{grb991216}) choosing a total energy of the plasma $E_{e^\pm}^{tot} = E_{dya} = 4.83\times 10^{53}$ erg and a baryon loading $B = 3.0\times 10^{-3}$ (see Ref. Ref.~\refcite{2001ApJ...555L.113R,2002ApJ...581L..19R,2003AIPC..668...16R,2005AIPC..782...42R}). Similarly, the temporal substructure in the prompt emission was explicitly shown to be related to the CBM inhomogeneities (see the following sections).

Following this early analysis and the subsequent ones on additional sources, it became clear that the CBM structure revealed by our analysis is quite different from the traditional description in the current literature. Far from considering analogies with shock wave processes developed within a fluidodynamic approach, it appears to us that the correct CBM description is a discrete one, composed of overdense ``blobs'' of typical size $\Delta R \sim 10^{14}$ cm widely spaced in underdense and inert regions.

\begin{figure}
\includegraphics[width=\hsize,clip]{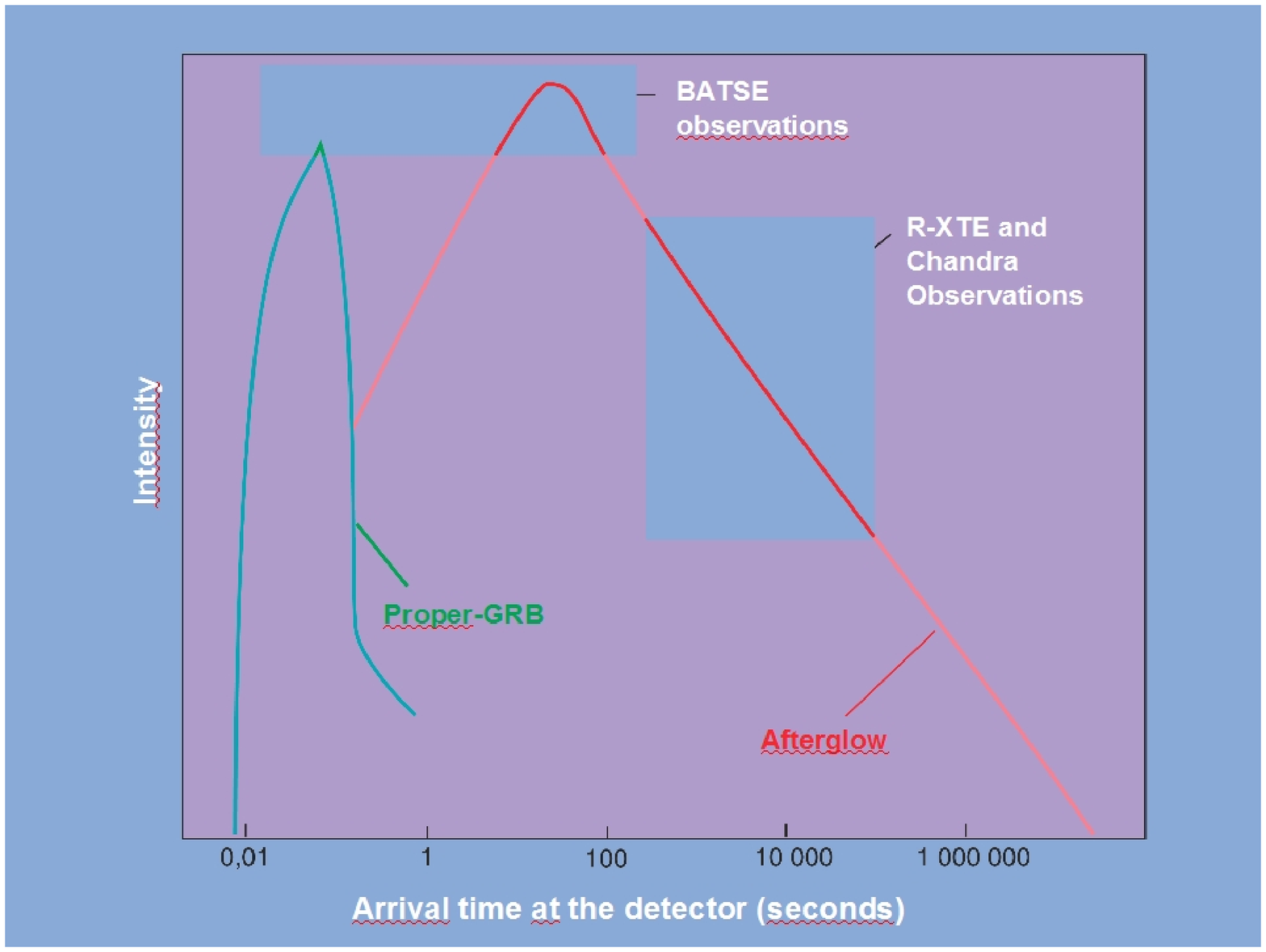}
\caption{Bolometric luminosity of P-GRB and afterglow as a function of the arrival time. Details in Ref.~\refcite{2003AIPC..668...16R}. Reproduced and adapted from Ref.~\refcite{pls} with the kind permission of the publisher.}
\label{bolum}
\end{figure}

\begin{figure}
\begin{minipage}{\hsize}
\begin{center}
\includegraphics[width=0.85\hsize,clip]{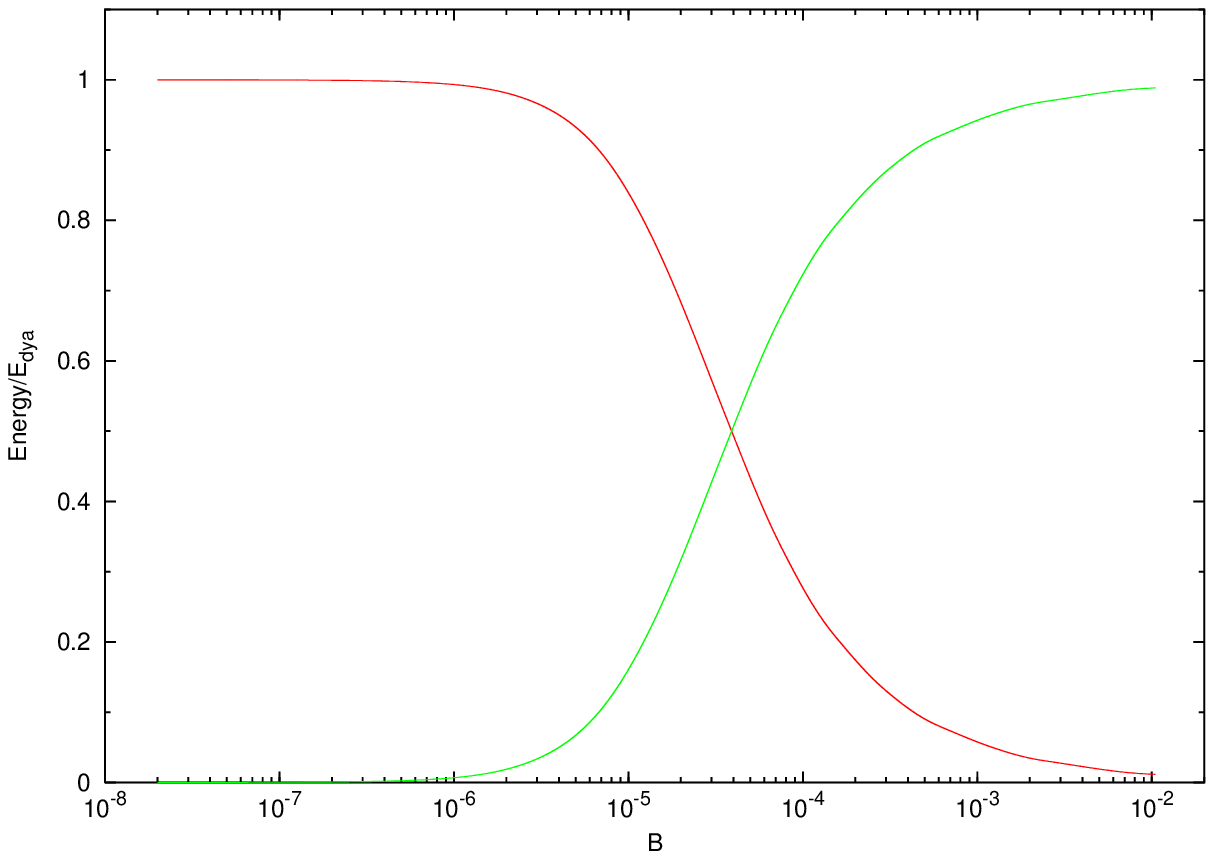}\\
\includegraphics[width=0.85\hsize,clip]{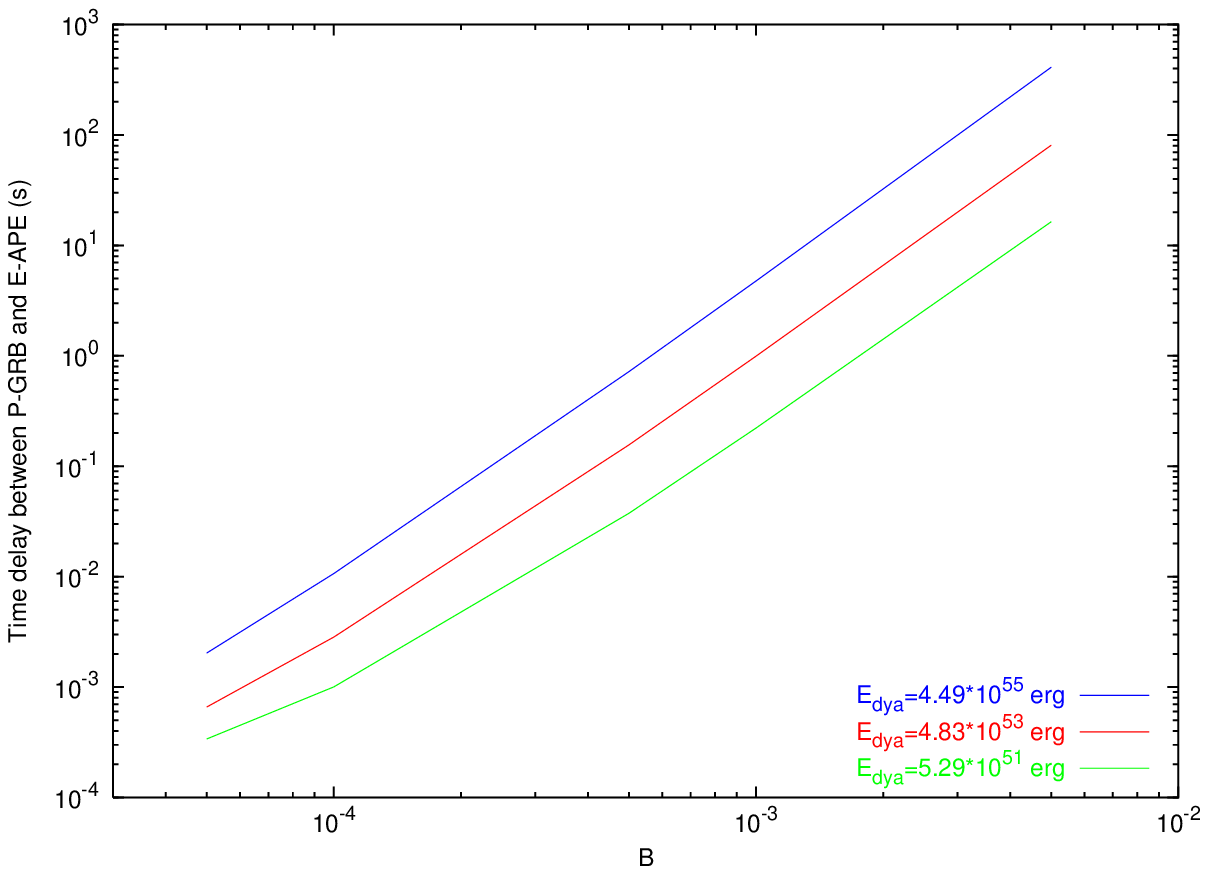}
\end{center}
\end{minipage}
\caption{{\bf Above:} The energy radiated in the P-GRB (the red line) and in the afterglow (the green line), in units of the total energy of the dyadosphere ($E_{dya}$), are plotted as functions of the $B$ parameter. {\bf Below:} The arrival time delay between the P-GRB and the peak of the afterglow is plotted as a function of the $B$ parameter for three selected values of $E_{dya}$.}
\label{bcross}
\end{figure}

We can then formulate the second paradigm, the interpretation of the burst structure (IBS) paradigm\cite{2001ApJ...555L.113R}, which covers three fundamental issues leading to the unequivocal identification of the canonical GRB structure:\\ 
{\bf a)} the existence of two different components: the P-GRB and the afterglow related by precise equations determining their relative amplitude and temporal sequence (see Fig.~\ref{bolum}, Ref.~\refcite{2003AIPC..668...16R} and next section);\\ 
{\bf b)} what in the literature has been addressed as the ``prompt emission'' and considered as a burst, in our model is not a burst at all---instead it is just the emission from the peak of the afterglow (see the clear confirmation of this result by the \emph{Swift} data of e.g. GRB 050315 in the next sections and in Ref.~\refcite{2006ApJ...645L.109R,Venezia_Orale});\\
{\bf c)} the crucial role of the parameter $B$ in determining the relative amplitude of the P-GRB to the afterglow and discriminating between the short and the long bursts (see Fig.~\ref{bcross}). Both short and long bursts arise from the same physical phenomena: the gravitational collapse to a black hole endowed with electromagnetic structure and the formation of its dyadosphere.

The fundamental diagram determining the relative intensity of the P-GRB and the afterglow as a function of the dimensionless parameter $B$ is shown in Fig.~\ref{bcross}. The main difference relates to the amount of baryonic matter engulfed by the electron-positron plasma in their optically thick phase prior to transparency. For $B < 10^{-5}$ the intensity of the P-GRB is larger and dominates the afterglow. This corresponds to the ``genuine'' short bursts \cite{2007A&A...474L..13B}. For $10^{-5} < B \le 10^{-2}$ the afterglow dominates the GRB. For $B > 10^{-2}$ we may observe a third class of ``bursts'', eventually related to a turbulent process occurring prior to transparency \cite{2000A&A...359..855R}. This third family should be characterized  by  smaller values of the Lorentz gamma factors than in the case of the short or long bursts.

\begin{figure}
\includegraphics[width=\hsize,clip]{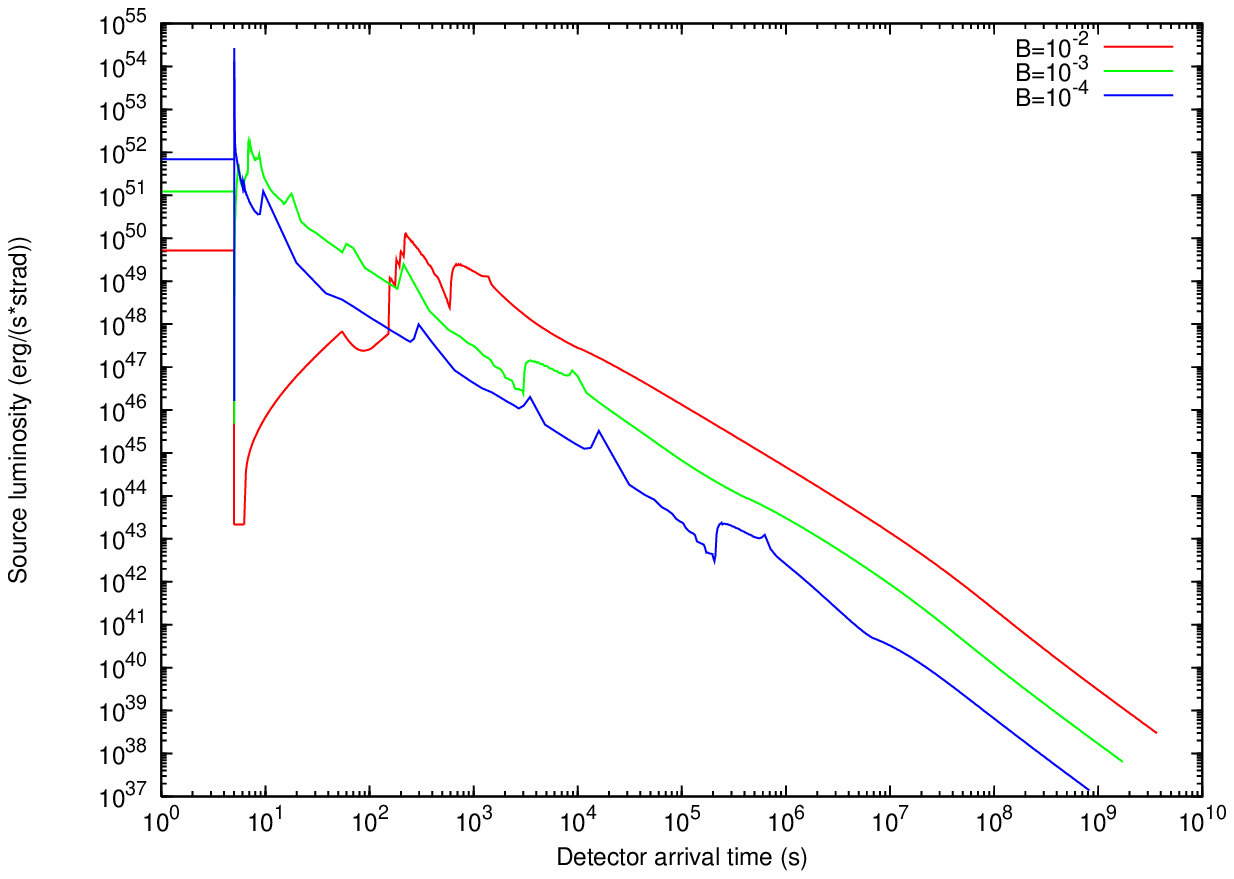}
\caption{The bolometric luminosity of a source with the same total energy and CBM distribution of GRB 991216 is represented here for selected values of the $B$ parameter, ranging from $B=10^{-2}$ to $B=10^{-4}$. The actual value for the GRB 991216 is $B=3.0\times 10^{-3}$. As expected, for smaller values of the $B$ parameter the intensity of the P-GRB increases and the total energy of the afterglow decreases. What is most remarkable is that the luminosity in the early part of the afterglow becomes very spiky and the peak luminosity actually increases.}
\label{Letizia_MultiB}
\end{figure}

Particularly enlightening for the gradual transition to the short bursts as a function of the $B$ parameter is the diagram showing how the GRB 991216 bolometric light curve would scale changing the only the value of $B$ (see Fig.~\ref{Letizia_MultiB}).

\begin{figure}
\includegraphics[width=\hsize,clip]{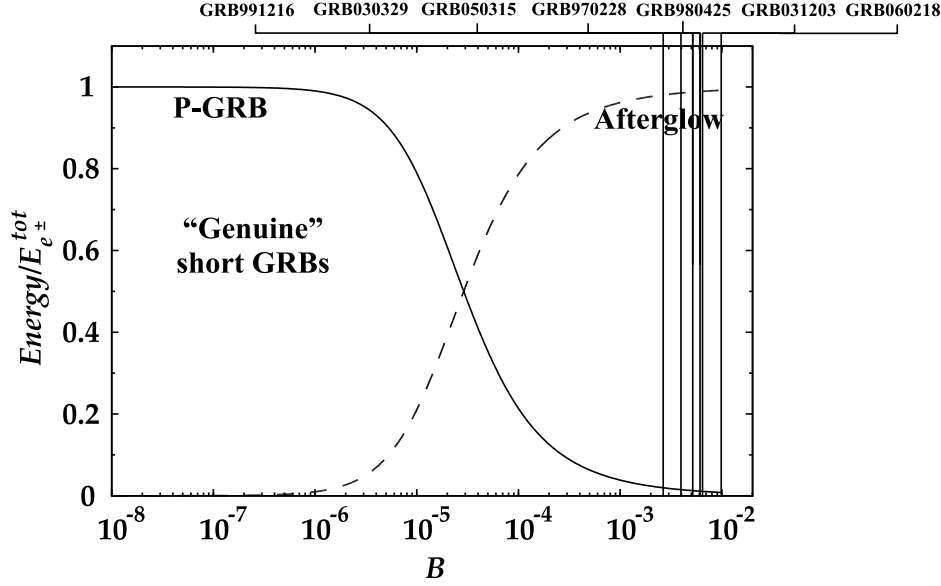}
\caption{Same as Fig.~\ref{bcross} with the values determined for selected GRBs. In order to determine the value of the $B$ parameter and the total energy we have performed the complete fit of each source. In particular, for each source we have fit the observed luminosities in selected energy bands of the entire afterglow including the prompt emission. We have verified that in each source the hard-to-soft spectral evolution is correctly fit and we have compared the theoretically computed spectral lag with the observations. Where applicable, we have also computed the relative intensity and temporal separation between the P-GRB and the peak of the afterglow and compared these values with the observed ones. The absence of spectral lag in the P-GRB is automatically verified by our model.}
\label{bcross_sorgenti}
\end{figure}

Moving from these two paradigms, and the prototypical case of GRB 991216, we have extended our analysis to a larger number of sources, such as GRB 970228\cite{2007A&A...474L..13B}, GRB 980425\cite{cospar02,Mosca_Orale}, GRB 030329\cite{2005tmgm.meet.2459B}, GRB 031203\cite{2005ApJ...634L..29B}, GRB 050315\cite{2006ApJ...645L.109R}, GRB 060218\cite{2007A&A...471L..29D} which have led to a confirmation of the validity of our canonical GRB structure (see Fig.~\ref{bcross_sorgenti}). In addition, progress has been made in our theoretical comprehension, which will be presented in the following sections.

\subsubsection{The simple model for the afterglow}

\label{aft}

When electron-positron pairs are annihilated and the transparency condition
$\tau\simeq1$ is satisfied, photons escape and the only ingredient of the
fireshell which is left is a relativistically expanding shell of protons and
electrons. The latter are not important kinematically and one can say that the
shell consists of baryons. This shell propagates in the interstellar medium
sweeping up the cold gas. The constant width approximation is good enough to
describe this process, however, we do not restrict ourselves to this
case. The only requirement is that the collision is inelastic and the energy
released in the collision is shared within the whole shell for a short time.

Consider the collision process in the lab reference frame where the CBM is
initially at rest. Assuming that the expanding baryons as well as CBM are cold,
the total energy of the shell is $E=M_{B}c^{2}\gamma=M\gamma$ with $\gamma$
being the Lorentz factor of the shell and $M$ is the rest mass of the shell
(together with its thermal energy) in units of energy. The mass-energy of the
CBM swept up within an infinitesimal time interval is $dm$. The energy released
during this process is $dE$. In our definitions $\gamma\beta=\sqrt{\gamma
^{2}-1}$. Finally, the gamma factor after the collision becomes $\gamma
+d\gamma$. Rewrite energy and momentum conservation:
\begin{align}
M\gamma+dm  &  =(M+dm+dE)(\gamma+d\gamma),\label{encon}\\
M\sqrt{\gamma^{2}-1}  &  =(M+dm+dE)\sqrt{(\gamma+d\gamma)^{2}-1}.
\label{momcon}%
\end{align}
This set of equations is equivalent to the one used in
\refcite{2003AIPC..668...16R}. From (\ref{momcon}) we get:
\begin{equation}
dE=-dm-M+M\sqrt{\frac{\gamma^{2}-1}{(\gamma+d\gamma)^{2}-1}}. \label{dE}%
\end{equation}
Substituting the last equality into (\ref{encon}) we arrive at
\begin{equation}
d\gamma=\frac{\frac{dm}{M}+\gamma(1-b)}{b},
\end{equation}
where
\begin{equation}
b=\sqrt{1+2\gamma\left(  \frac{dm}{M}\right)  +\left(  \frac{dm}{M}\right)
^{2}}.
\end{equation}
Substituting these expressions into (\ref{dE}) we find
\begin{equation}
dE=-dm-M(1-b).
\end{equation}
Since $dm$ is infinitesimally small we can expand $b$ in a series in $dm/M$
\begin{equation}
b\simeq1-\frac{dm}{M}\gamma,
\end{equation}
so we finally obtain:
\begin{align}
dE  &  =(\gamma-1)dm,\label{dyneq1}\\
d\gamma &  =-(\gamma^{2}-1)\frac{dm}{M}. \label{dyneq2}%
\end{align}
These equations are used to describe collision of the baryonic remnant of the
fireshell with the CBM \cite{1999PhR...314..575P,2000A&A...359..855R,2003AIPC..668...16R}. Note that it is a mistake to assume that $dE=0$
and use only (\ref{momcon}) to obtain a differential equation for $\gamma$ as
was done in Ref.~\refcite{1994ApJ...422..248K}. The total mass-energy swept up during
this process is given by
\begin{equation}
dM=M+dm+(1-\delta)dE-M=dm+(1-\delta)dE,
\end{equation}
where $\delta$ denotes the portion of the energy that is radiated. There are
basically two approximations which follow from this expression, namely the fully
radiative condition $\delta=1$ and the adiabatic condition $\delta=0$. With
the adiabatic condition no energy is radiated and one must assume that the
kinetic energy of the decelerating baryons is converted into radiation in
dissipation processes in shocks which are supposed to form during the
collision,\cite{1992MNRAS.258P..41R,1992ApJ...395L..83N} see also
\refcite{1999PhR...314..575P}. With the radiative condition no additional
mechanisms are required to describe the afterglow since it results from the
emission of part of the energy released during inelastic collision of
accelerated baryonic pulse with the CBM \cite{2003AIPC..668...16R}.

\subsubsection{Blandford and McKee radiative solution}

Eqs.~(\ref{dyneq1}),(\ref{dyneq2}) can be found already in Ref.~\refcite{1976PhFl...19.1130B}. However, the meaning of the quantities used there
as well as the derivation are incorrect.

Following Ref.~\refcite{1976PhFl...19.1130B} consider the relativistic
shock (blast) wave resulting from an impulsive energy release, propagating
into the homogeneous external medium with the Lorentz factor $\gamma$. Authors
assume that the shock wave sweeps up external matter. They treat this
swept-up matter as being in a thin, cold shell.

Consider the reference frame where the shock is at rest. For an observer in
this frame the external medium has relativistic Lorentz factor $\gamma$. The
shock front is a spherical massless surface. For our observer there is a flow
of matter, energy and momentum through this surface. We can calculate these
quantities using (\ref{ce}) and (\ref{connum}). Recall that the external
medium is cold so $p\ll\rho$. Thus we get
\begin{align}
\frac{dE_{tot}}{dt}  &  =-\oint_{S}T^{01}d\mathcal{S}=-\oint_{S}\rho\gamma
^{2}\beta d\mathcal{S}=-4\pi R^{2}\rho\beta\gamma^{2},\label{det}\\
\frac{d\mathbf{P}}{dt}  &  =-\oint_{S}T^{11}d\mathcal{S}=-\oint_{S}\rho
(\gamma^{2}-1)d\mathcal{S}=-4\pi R^{2}\rho(\gamma^{2}-1),\label{dp}\\
\frac{dm}{dt}  &  =-\oint_{S}\rho U^{1}d\mathcal{S}=-4\pi R^{2}\rho\beta
\gamma, \label{dm}%
\end{align}
where $\mathbf{P}$ is a radial momentum of external medium, $E_{tot}$ is its
total energy. The kinetic energy $E_{k}$ change is simply connected to its
flux:
\begin{equation}
\frac{dE_{k}}{dt}=\frac{dE_{tot}}{dt}-\frac{dm}{dt}=-4\pi R^{2}\rho\beta
\gamma(\gamma-1). \label{dek}%
\end{equation}
This formula coincides with Eq.~ (84) in Ref.~\refcite{1976PhFl...19.1130B}. The next equation (85) is of the form of our
(\ref{dyneq2}). This implies that in order to derive this result these authors
considered the full derivative of kinetic energy of the external medium with
respect to the expanding shock wave, namely:
\begin{equation}
\frac{dE_{k}}{dt}=(\gamma-1)\frac{dm}{dt}+m\frac{d\gamma}{dt}, \label{fder}%
\end{equation}
and equate these two expressions. However, if one proceeds in the same manner
with the radial momentum equation, one arrives at a strange result: they
become inconsistent. The reason is the following.

Eqs.~(\ref{det}),(\ref{dp}),(\ref{dm}) and (\ref{dek}) are of course
correct. The next step is in doubt. To write down the total derivative means
to assume that the Lorentz factor changes with time. This means that the
properties of the external medium change. However, the problem considered by
Blandford and McKee is an idealized one, namely the hydrodynamics of relativistic
shocks without consideration of any microphysics\footnote{Microphysical
interactions responsible for changes in the blast wave can be Coulumb
interactions within the plasma, ionization of external medium losses,
interactions with magnetic field and various radiative processes.}. In this
context the damping and dissipation effect cannot be incorporated directly and
the shock wave generally speaking will propagate with constant velocity.
Apart from difficulties in definition of the quantity $m$, the mass of
external medium which can be thought of as infinite, the last term in (\ref{fder})
cannot exist and therefore there is no way to get (85) from (84) in
\refcite{1976PhFl...19.1130B}.

From the physical point of view, instead of a massless shock (even if it is
present) a massive shell must be considered\footnote{In some
approximation it can be considered as the relativistic piston problem.}. The
only natural way to deal with this problem is to consider interaction of this
shell with a small shell of external medium. The problem is consequently
equivalent to the interaction of two massive particles which is obvious to treat
with the help of conservation equations instead of shock equations.

Thus, equations (\ref{dyneq1}),(\ref{dyneq2}) can be derived only from energy
and radial momentum conservation equations as was done in
\refcite{1997ApJ...490..772K} and independently in Ref.~\refcite{2000A&A...359..855R}.

\section{The ``canonical GRB'' bolometric light curve}\label{secE}

We assume that the internal energy due to kinetic collision is instantly radiated away and that the corresponding emission is isotropic. Let $\Delta \varepsilon$ be the internal energy density developed in the collision. In the comoving frame the energy per unit of volume and per solid angle is simply
\begin{equation} 
\left(\frac{dE}{dV d\Omega}\right)_{\circ}  =  \frac{\Delta \varepsilon}{4 
\pi} 
\label{dEo} 
\end{equation} 
due to the fact that the emission is isotropic in this frame. The total number of photons emitted is an invariant quantity independent of the frame used. Thus we can compute this quantity as seen by an observer in the comoving frame (which we denote with the subscript ``$\circ$'') and by an observer in the laboratory frame (which we denote with no subscripts). Doing this we find:
\begin{equation} 
\frac{dN_\gamma}{dt d \Omega d \Sigma}= \left(\frac{dN_\gamma}{dt d \Omega 
d \Sigma} \right)_{\circ} \Lambda^{-3} 
\cos \vartheta 
\, , 
\end{equation} 
where $\vartheta$ is the angle between the radial expansion velocity of a point on the fireshell surface and the line of sight, $\cos\vartheta$ comes from the projection of the elementary surface of the shell on the direction of propagation and $\Lambda = \gamma ( 1 - \beta \cos \vartheta )$ is the Doppler factor introduced in the two following differential transformation
\begin{equation} 
d \Omega_{\circ} = d \Omega \times \Lambda^{-2} 
\end{equation} 
for the solid angle transformation and 
\begin{equation} 
d t_{\circ} = d t \times \Lambda^{-1} 
\end{equation} 
for the time transformation. An extra $\Lambda$ factor comes from the energy transformation:
\begin{equation} 
E_{\circ} = E \times \Lambda\,  
\end{equation} 
(see also Ref.~\refcite{1999ApJ...512..699C}). Thus finally we obtain:
\begin{equation} 
\frac{dE}{dt d \Omega d \Sigma} = \left(\frac{dE}{dt d \Omega d \Sigma}
\right)_{\circ} \Lambda^{-4} \cos \vartheta \, . 
\end{equation} 
Doing this we clearly identify  $  \left(\frac{dE}{dt d \Omega d \Sigma} \right)_{\circ} $ 
as the energy density in the comoving frame up to a factor $\frac{v}{4\pi}$ (see Eq.~(\ref{dEo})). Then we have: 
\begin{equation} 
\frac{dE}{dt d \Omega } = \int_{shell} \frac{\Delta \varepsilon}{4 \pi} \; 
v \; \cos \vartheta \; \Lambda^{-4} \; d \Sigma\, , 
\label{fluxlab} 
\end{equation} 
where the integration in $d \Sigma$ is performed over the visible area of the ABM pulse at laboratory time $t$, namely with $0\le\vartheta\le\vartheta_{max}$ and $\vartheta_{max}$ is the boundary of the visible region defined by:
\begin{equation}
\cos\vartheta_{max} = \frac{v}{c}\, .
\label{bound}
\end{equation} 
Eq.~(\ref{fluxlab}) gives us the energy emitted toward the observer per unit solid angle and per unit laboratory time $t$ in the laboratory frame.

What we really need is the energy emitted per unit solid angle and per unit detector arrival time $t_a^d$, so we must use the complete relation between $t_a^d$ and $t$ given by:
\begin{equation} 
t_a^d = \left(1+z\right)\left[t - \frac{r}{c}\left(t\right)\cos \vartheta  + \frac{r^\star}{c}\right]\, ,
\label{ta_g} 
\end{equation} 
where $r^\star$ is the initial size of the fireshell. First we have to multiply the integrand in Eq.~(\ref{fluxlab}) by the factor $\left(dt/dt_a^d\right)$ to transform the energy density generated per unit of laboratory time $t$ into the energy density generated per unit arrival time $t_a^d$. Then we have to integrate with respect to $d \Sigma$ over the EquiTemporal Surfaces (EQTS) corresponding to arrival time $t_a^d$ instead of the ABM pulse visible area at laboratory time $t$. The analog of Eq.~(\ref{fluxlab}) for the source luminosity in detector arrival time is then:
\begin{equation} 
\frac{dE_\gamma}{dt_a^d d \Omega } = \int_{EQTS} \frac{\Delta 
\varepsilon}{4 \pi} \; v \; \cos \vartheta \; \Lambda^{-4} \; 
\frac{dt}{dt_a^d} d \Sigma\, . 
\label{fluxarr} 
\end{equation} 
It is important to note that, in the present case of GRB 991216, the Doppler factor $\Lambda^{-4}$ in Eq.~(\ref{fluxarr}) enhances the apparent luminosity of the burst compared to the intrinsic luminosity by a factor which at the peak of the afterglow is in the range between $10^{10}$ and $10^{12}$!

We are now able to reproduce in Fig.~\ref{bolum} the general behavior of the luminosity starting from the P-GRB to the latest phases of the afterglow as a function of the arrival time. It is generally agreed that the GRB afterglow originates from an ultrarelativistic shell of baryons with an initial Lorentz factor $\gamma_\circ\sim 200$--$300$ with respect to the CBM (see e.g. Ref.~\refcite{2003AIPC..668...16R,2004ApJ...605L...1B} and references therein). Using GRB 991216 as a prototype, in Ref.~\refcite{2001ApJ...555L.107R,2001ApJ...555L.113R} we have shown how from the time varying bolometric intensity of the afterglow it is possible to infer the average density $\left<n_{cbm}\right>=1$ particle/cm$^3$ of the CBM in a region of approximately $10^{17}$ cm surrounding the black hole giving rise to the GRB phenomenon.

The summary of these general results are shown in Fig.~\ref{grb991216}, where the P-GRB, the emission at the peak of the afterglow in relation to the ``prompt emission'' and the latest part of the afterglow are clearly identified for the source GRB 991216. Details in Ref.~\refcite{2003AIPC..668...16R}.

Summarizing, unlike treatments in the current literature (see e.g. Ref.~\refcite{2005RvMP...76.1143P,2006RPPh...69.2259M} and references therein), we define a ``canonical GRB'' light curve with two sharply different components (see Fig.~\ref{grb991216})\cite{2001ApJ...555L.113R,2007AIPC..910...55R,2007A&A...474L..13B}:
\begin{enumerate}
\item \textbf{The P-GRB:} it has the imprint of the black hole formation, a harder spectrum and no spectral lag. \cite{2001A&A...368..377B,2005IJMPD..14..131R}.
\item \textbf{The afterglow:} it presents a clear hard-to-soft behavior \cite{2005ApJ...634L..29B,2004IJMPD..13..843R,2006ApJ...645L.109R}; the peak of the afterglow contributes to what is usually called the ``prompt emission'' \cite{2001ApJ...555L.113R,2006ApJ...645L.109R,2007A&A...471L..29D}.
\end{enumerate}
The ratio between the total time-integrated luminosity of the P-GRB (namely, its total energy) and the corresponding one for the afterglow is the crucial quantity for the identification of a GRBs' nature. Such a ratio, as well as the temporal separation between the corresponding peaks, is a function of the $B$ parameter \cite{2001ApJ...555L.113R}.

When the P-GRB is the leading contribution to the emission and the afterglow is negligible we have a ``genuine'' short GRB \cite{2001ApJ...555L.113R}. This is the case where $B \lesssim 10^{-5}$ (see Fig.~\ref{bcross_sorgenti}): in the limit $B \to 0$ the afterglow vanishes (see Fig.~\ref{bcross_sorgenti}). In the other GRBs, with $10^{-4} \lesssim B \lesssim 10^{-2}$, the afterglow contribution is generally predominant (see Fig.~\ref{bcross_sorgenti}; for the existence of the upper limit $B \lesssim 10^{-2}$ see Ref.~\refcite{2000A&A...359..855R,2007A&A...471L..29D}). Still, this case presents two distinct possibilities:
\begin{itemize}
\item The afterglow peak luminosity is \textbf{larger} than the P-GRB peak luminosity. A clear example of this situation is GRB 991216, represented in Fig.~\ref{grb991216}.
\item The afterglow peak luminosity is \textbf{smaller} than the P-GRB one. A clear example of this situation is GRB 970228, represented in Fig.~\ref{970228_fit_prompt}.
\end{itemize}

The simultaneous occurrence of an afterglow with total time-integrated luminosity larger than the P-GRB one, but with a smaller peak luminosity, is indeed explainable in terms of a peculiarly small average value of the CBM density, compatible with a galactic halo environment, and not due to the intrinsic nature of the source (see Fig.~\ref{970228_fit_prompt})\cite{2007A&A...474L..13B}. Such a small average CBM density deflates the afterglow peak luminosity. Of course, such a deflated afterglow lasts much longer, since the total time-integrated luminosity in the afterglow is fixed by the value of the $B$ parameter (see above and Fig.~\ref{picco_n=1}). In this sense, GRBs belonging to this class are only ``fake'' short GRBs. This is GRB class identified by Ref.~\refcite{2006ApJ...643..266N}, which also GRB 060614 belongs to, and which has GRB 970228 as a prototype \cite{2007A&A...474L..13B}.

\begin{figure}
\includegraphics[width=\hsize,clip]{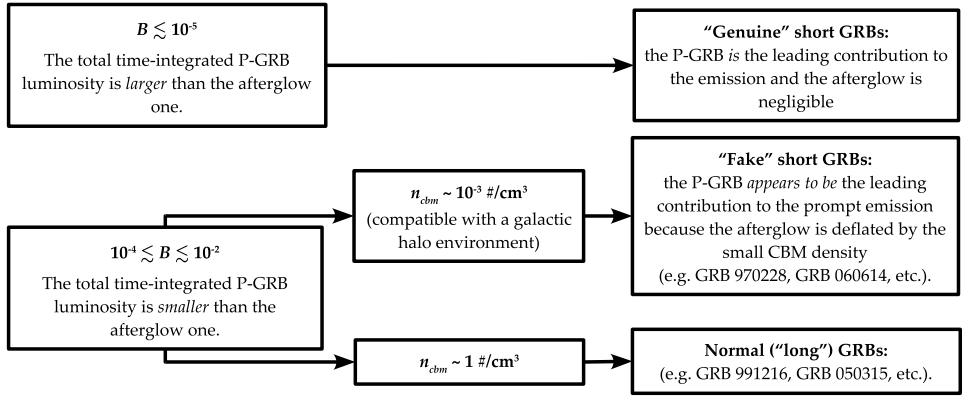}
\caption{A sketch summarizing the different possibilities predicted by the ``canonical GRB'' scenario depending on the fireshell baryon loading $B$ and the average CBM density $\langle n_{cbm} \rangle$.}
\label{canonical}
\end{figure}

Our ``canonical GRB'' scenario, therefore, especially points out the need to distinguish between ``genuine'' and ``fake'' short GRBs:
\begin{itemize}
\item The \textbf{``genuine'' short GRBs} inherit their features from an intrinsic property of their sources. The very small fireshell baryon loading, in fact, implies that the afterglow time-integrated luminosity is negligible with respect to the P-GRB one.
\item The \textbf{``fake'' short GRBs} instead inherit their features from the environment. The very small CBM density in fact implies that the afterglow peak luminosity is lower than the P-GRB one, even if the afterglow total time-integrated luminosity is higher. This deflated afterglow peak can be observed as a ``soft bump'' following the P-GRB spike, as in GRB 970228 \cite{2007A&A...474L..13B}, GRB 060614 (Caito et al., in preparation), and the sources analyzed by Ref.~\refcite{2006ApJ...643..266N}.
\end{itemize}
A sketch of the different possibilities depending on the fireshell baryon loading $B$ and the average CBM density $\langle n_{cbm} \rangle$ is given in Fig.~\ref{canonical}.

\section{The spectra of the afterglow}

In our approach we focus uniquely on the X and gamma ray radiation, which appears to be conceptually more predictable in terms of fundamental processes than the optical and radio emission. It is perfectly predictable by a set of constitutive equations, which leads to directly verifiable and very stable features in the spectral distribution of the observed GRB afterglows. In line with the observations of GRB 991216 and other GRB sources, we assume in the following that the X and gamma ray luminosity represents approximately $90$\% of the energy of the afterglow, while the optical and radio emission represents only the remaining $10$\%.

This approach differs significantly from the other ones in the current literature, where attempts are made to explain at once all the multi-wavelength emission in the radio, optical, X and gamma ray as coming from a common origin which is linked to boosted synchrotron emission. Such an approach has been shown to have a variety of difficulties \cite{2002A&A...393..409G,1998ApJ...506L..23P} and cannot anyway have the instantaneous variability needed to explain the structure in the ``prompt radiation'' in an external shock scenario, which is indeed confirmed by our model.

Here the fundamental new assumption is adopted (see also Ref.~\refcite{2004IJMPD..13..843R}) that the X and gamma ray radiation during the entire afterglow phase has a thermal spectrum in the comoving frame. The temperature is then given by:
\begin{equation} 
T_s=\left[\Delta E_{\rm int}/\left(4\pi r^2 \Delta \tau \sigma 
\mathcal{R}\right)\right]^{1/4}\, , 
\label{TdiR} 
\end{equation} 
where $\Delta E_{\rm int}$ is the internal energy developed in the collision with the CBM in a time interval $\Delta \tau$ in the co-moving frame, $\sigma$ is the Stefan-Boltzmann constant and
\begin{equation} 
\mathcal{R}=A_{eff}/A_{vis}\, , 
\label{Rdef} 
\end{equation} 
is the ratio between the ``effective emitting area'' of the ABM pulse of radius $r$ and its total visible area, which accounts for the CBM filamentary structure \cite{2005IJMPD..14...97R}. Due to the CBM inhomogeneities the ABM emitting region is in fact far from being homogeneous. In GRB 991216 such a factor is observed to be decreasing during the afterglow between: $3.01\times 10^{-8} \ge \mathcal{R} \ge 5.01 \times 10^{-12}$ \cite{2004IJMPD..13..843R}. 

\begin{figure}
\includegraphics[width=\hsize,clip]{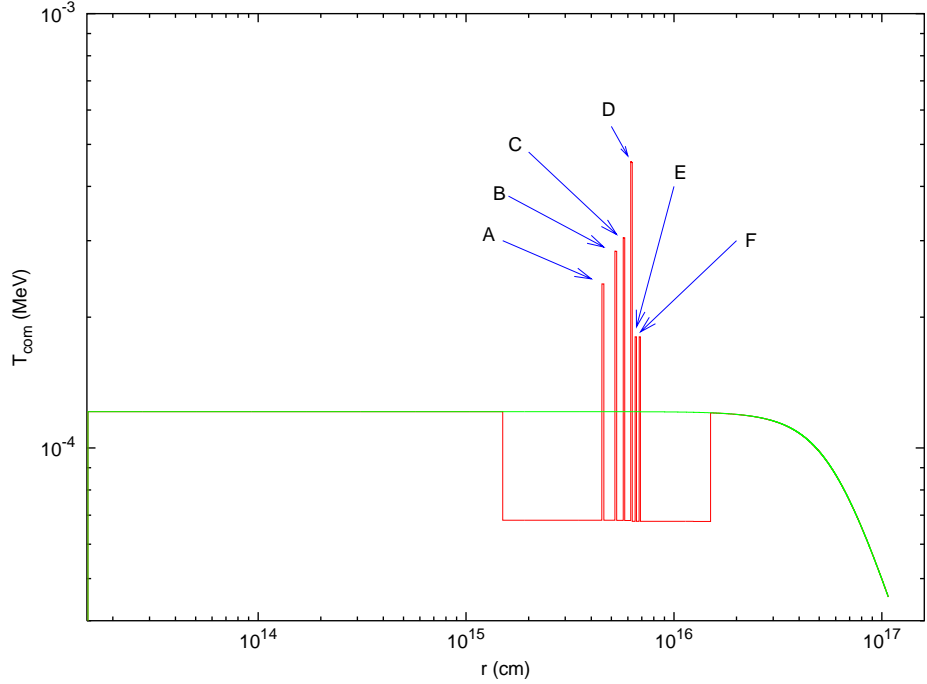}
\caption{The temperature in the comoving frame of the shock front corresponding to the density distribution with the six spikes A,B,C,D,E,F presented in Ref.~\refcite{2005IJMPD..14...97R}. The green line corresponds to an homogeneous distribution with $n_{cbm}=1$. Details in Ref.~\refcite{2005IJMPD..14...97R}.}
\label{tcom_fig}
\end{figure}

The temperature in the comoving frame corresponding to the density distribution described in Ref.~\refcite{2002ApJ...581L..19R} is shown in Fig.~\ref{tcom_fig}. 

We are now ready to evaluate the source luminosity in a given energy band. The source luminosity at a detector arrival time $t_a^d$, per unit solid angle $d\Omega$ and in the energy band $\left[\nu_1,\nu_2\right]$ is given by\cite{2003AIPC..668...16R,2004IJMPD..13..843R}: 
\begin{equation} 
\frac{dE_\gamma^{\left[\nu_1,\nu_2\right]}}{dt_a^d d \Omega } = 
\int_{EQTS} \frac{\Delta \varepsilon}{4 \pi} \; v \; \cos \vartheta \; 
\Lambda^{-4} \; \frac{dt}{dt_a^d} W\left(\nu_1,\nu_2,T_{arr}\right) d 
\Sigma\, , 
\label{fluxarrnu} 
\end{equation} 
where $\Delta \varepsilon=\Delta E_{int}/V$ is the energy density released in the interaction of the ABM pulse with the CBM inhomogeneities measured in the comoving frame, $\Lambda=\gamma(1-(v/c)\cos\vartheta)$ is the Doppler factor, $W\left(\nu_1,\nu_2,T_{arr}\right)$ is an ``effective weight'' required to evaluate only the contributions in the energy band $\left[\nu_1,\nu_2\right]$, $d\Sigma$ is the surface element of the EQTS at detector arrival time $t_a^d$ on which the integration is performed (see also Ref.~\refcite{2005IJMPD..14...97R}) and $T_{arr}$ is the observed temperature of the radiation emitted from $d\Sigma$: 
\begin{equation} 
T_{arr}=T_s/\left[\gamma 
\left(1-(v/c)\cos\vartheta\right)\left(1+z\right)\right]\, . 
\label{Tarr} 
\end{equation} 

The ``effective weight'' $W\left(\nu_1,\nu_2,T_{arr}\right)$ is given by the ratio of the integral over the given energy band of a Planckian distribution at a temperature $T_{arr}$ to the total integral $aT_{arr}^4$: 
\begin{equation} 
W\left(\nu_1,\nu_2,T_{arr}\right)=\frac{1}{aT_{arr}^4}\int_{\nu_1}^{\nu_2}\rho\left(T_{arr},\nu\right)d\left(\frac{h\nu}{c}\right)^3\, , 
\label{effweig} 
\end{equation} 
where $\rho\left(T_{arr},\nu\right)$ is the Planckian distribution at temperature $T_{arr}$: 
\begin{equation} 
\rho\left(T_{arr},\nu\right)=\left(2/h^3\right)h\nu/\left(e^{h\nu/\left(kT_{arr}\right)}-1\right) \label{rhodef} 
\end{equation} 

\section{Application to GRB 031203: the first spectral analysis}\label{sec031203}

The first verification of the above theoretical framework came from the analysis of GRB 031203. GRB 031203 was observed by IBIS, on board the INTEGRAL satellite \cite{2003GCN..2460....1M}, as well as by XMM \cite{2004ApJ...605L.101W} and Chandra \cite{2004Natur.430..648S} in the $2--10$ keV band, and by VLT \cite{2004Natur.430..648S} in the radio band. It appears as a typical long burst \cite{2004Natur.430..646S}, with a simple profile and a duration of $\approx 40$ s. The burst fluence in the $20--200$ keV band is $(2.0\pm 0.4)\times 10^{-6}$ erg/cm$^2$ \cite{2004Natur.430..646S}, and the measured redshift is $z=0.106$ \cite{2004ApJ...611..200P}. We analyze in the following the gamma ray signal received by INTEGRAL. The observations in other wavelengths, in analogy with the case of GRB 980425 \cite{2000ApJ...536..778P,2004AdSpR..34.2715R,Mosca_Orale}, could be related to the supernova event, as also suggested by Ref.~\refcite{2004Natur.430..648S}, and they will be examined elsewhere.

The INTEGRAL observations find a direct explanation in our theoretical model. We reproduce correctly the observed time variability of the prompt emission (see Fig.~\ref{fig1a})\cite{2005ApJ...634L..29B}. The radiation produced by the interaction with the CBM of the baryonic matter shell, accelerated by the fireshell, agrees with observations both for intensity and time structure.

The progress in reproducing the X and $\gamma-$ray emission as originating from a thermal spectrum in the comoving frame of the burst \cite{2004IJMPD..13..843R} leads to the characterization of the instantaneous spectral properties which are shown to drift from hard to soft during the evolution of the system. The convolution of these instantaneous spectra over the observational time scale is in very good agreement with the observed power-law spectral shape.

\subsection{The initial conditions}

The best fit of the observational data leads to a total energy of the electron-positron plasma $E_{e^\pm}^{tot}=1.85\times10^{50}$ erg. Assuming a black hole mass $M=10M_{\odot}$, we then have a black hole charge to mass ratio $\xi=6.8\times 10^{-3}$; the plasma is created between the radii $r_1=2.95\times10^6$ cm and $r_2=2.81\times10^7$ cm with an initial temperature $T=1.52$ MeV and a total number of pairs $N_{e^\pm}=2.98\times10^{55}$. The amount of baryonic matter in the remnant is $B = 7.4\times10^{-3}$.

After the transparency point and the P-GRB emission, the initial Lorentz gamma factor of the accelerated baryons is $\gamma_\circ=132.8$ at an arrival time at the detector $t^d_a=8.14\times 10^{-3}$ s and a distance from the Black Hole $r_\circ=6.02\times 10^{12}$ cm. The CBM parameters are: $<n_{cbm}>=0.3$ particle/$cm^3$ and $<\mathcal{R}>=7.81\times 10^{-9}$.

\subsection{The GRB luminosity in fixed energy bands}\label{par3}

The aim of our model is to derive from first principles both the luminosity in selected energy bands and the time resolved/integrated spectra. We recall that the luminosity in selected energy bands is evaluated integrating over the EQTSs \cite{2004IJMPD..13..843R} the energy density released in the interaction of the accelerated baryons with the CBM measured in the co-moving frame, duly boosted in the observer frame. The radiation viewed in the comoving frame of the accelerated baryonic matter is assumed to have a thermal spectrum and to be produced by the interaction of the CBM with the front of the expanding baryonic shell.

In order to evaluate the contributions in the band $[\nu_1,\nu_2]$ we have to multiply the bolometric luminosity by an ``effective weight'' $W(\nu_1,\nu_2,T_{arr})$, where $T_{arr}$ is the observed temperature. $W(\nu_1,\nu_2,T_{arr})$ is given by the ratio of the integral over the given energy band of a Planckian distribution at temperature $T_{arr}$ to the total integral $aT_{arr}^4$ \cite{2004IJMPD..13..843R}. The resulting expression for the emitted luminosity is Eq.~(\ref{fluxarrnu}).

\subsection{The ``prompt emission''}

\begin{figure}
\includegraphics[width=\hsize,clip]{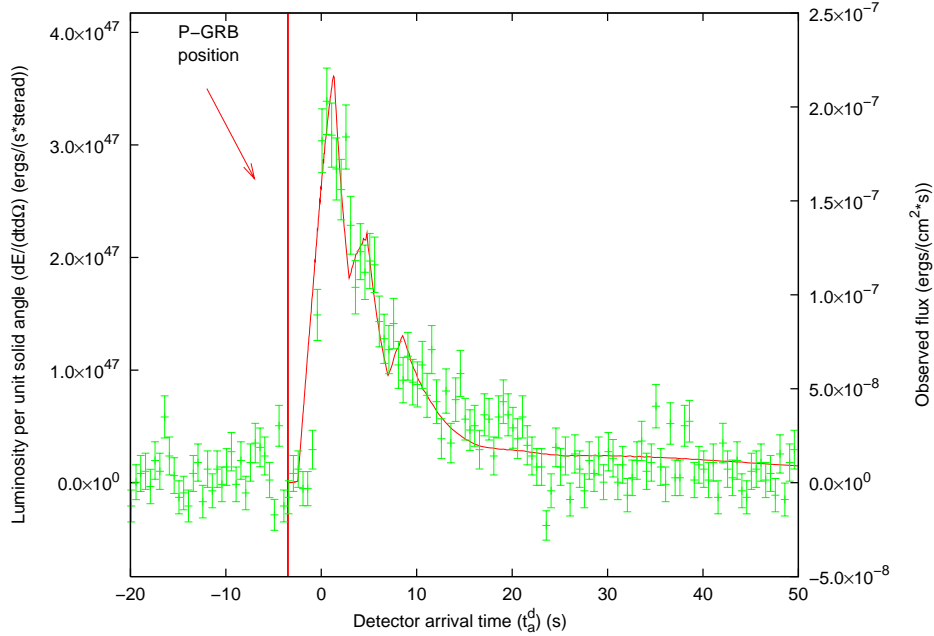}
\caption{Theoretically simulated light curve of the GRB 031203 prompt emission in the $20-200$ keV energy band (solid red line) is compared with the observed data (green points) from Ref.~\refcite{2004Natur.430..646S}. The vertical bold red line indicates the time position of the P-GRB.}
\label{fig1a}
\end{figure}

In order to compare our theoretical prediction with the observations, it is important to notice that there is a shift between the initial time of the GRB event and the moment in which the satellite instrument has been triggered. In fact, in our model the GRB emission starts at the transparency point when the P-GRB is emitted. If the P-GRB is under the threshold of the instrument, the trigger starts a few seconds later with respect to the real beginning of the event. Therefore it is crucial, in the theoretical analysis, to estimate and take this time delay into proper account. In the present case it results in $\Delta t^d_a=3.5$ s (see the bold red line in Fig.~\ref{fig1a}). In what follows, the detector arrival time is referred to the onset of the instrument.

The structure of the prompt emission of GRB 031203, which is a single peak with a slow decay, is reproduced assuming an CBM which does not have a constant density but instead several density spikes with $<n_{cbm}>=0.16$ particle/cm$^3$. Such density spikes corresponding to the main peak are modeled as three spherical shells with width $\Delta$ and density contrast $\Delta n/n$: we adopted for the first peak $\Delta=3.0\times10^{15}$ cm and $\Delta n/n=8$, for the second peak $\Delta=1.0\times10^{15}$ cm and $\Delta n/n=1.5$ and for the third one $\Delta=7.0\times10^{14}$ cm and $\Delta n/n=1$. To describe the details of the CBM filamentary structure we would require intensity vs.\ time information with an arbitrarily high resolving power. With the finite resolution of the INTEGRAL instrument, we can only describe the average density distribution compatible with the given accuracy. Only structures at scales of $10^{15}$ cm can be identified. Smaller structures would need a stronger signal and/or a smaller time resolution of the detector. The three clouds here considered are necessary and sufficient to reproduce the observed light curve: a smaller number would not fit the data, while a larger number is unnecessary and would be indeterminable.

The result (see Fig.~\ref{fig1a}) shows a good agreement with the light curve reported by Ref.~\refcite{2004Natur.430..646S}, and it provides further evidence for the possibility of reproducing light curves with a complex time variability through CBM inhomogeneities \cite{2002ApJ...581L..19R,2003AIPC..668...16R,2005AIPC..782...42R}.

\subsection{The instantaneous spectrum}\label{inst}

\begin{figure}
\includegraphics[width=\hsize,clip]{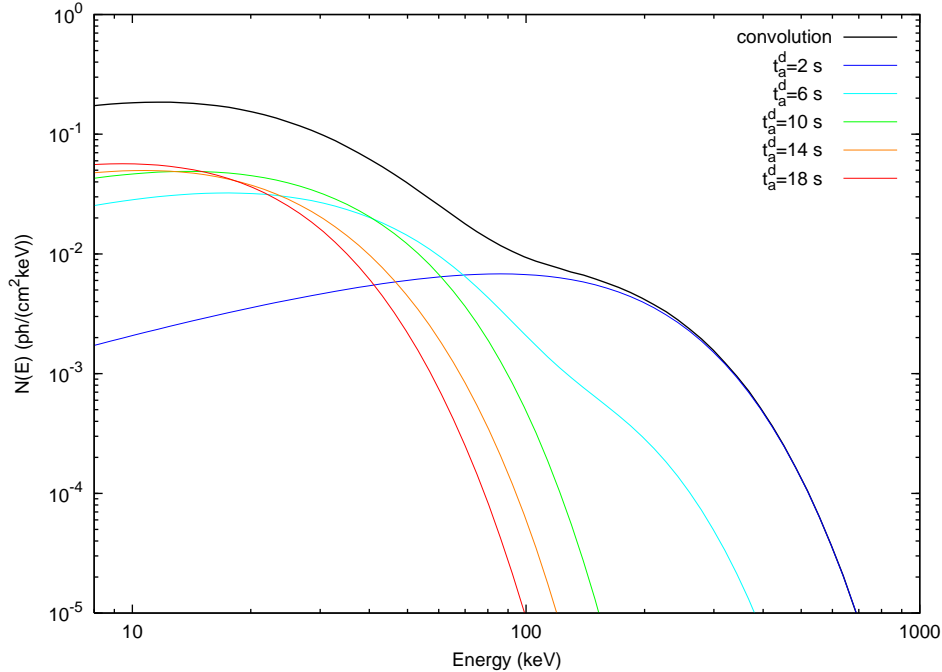}
\caption{Five different theoretically predicted instantaneous photon number spectra $N(E)$ for $t_a^d=2$, $6$, $10$, $14$, $18$ s are represented here (colored curves) together with their own temporal convolution (black bold curve). The shapes of the instantaneous spectra are not blackbodies due to the spatial convolution over the EQTS (see text).}
\label{fig2a}
\end{figure}

In addition to the the luminosity in fixed energy bands we can also derive the instantaneous photon number spectrum $N(E)$. In Fig.~\ref{fig2a} are shown samples of time-resolved spectra for five different values of the arrival time which cover the whole duration of the event.

It is clear from this picture that, although the spectrum in the comoving frame of the expanding pulse is thermal, the shape of the final spectrum in the laboratory frame is clearly not thermal. In fact, as explained in Ref.~\refcite{2004IJMPD..13..843R}, each single instantaneous spectrum is the result of an integration of hundreds of thermal spectra over the corresponding EQTS. This calculation produces a nonthermal instantaneous spectrum in the observer frame (see Fig.~\ref{fig2a}).

\begin{figure}
\includegraphics[width=\hsize,clip]{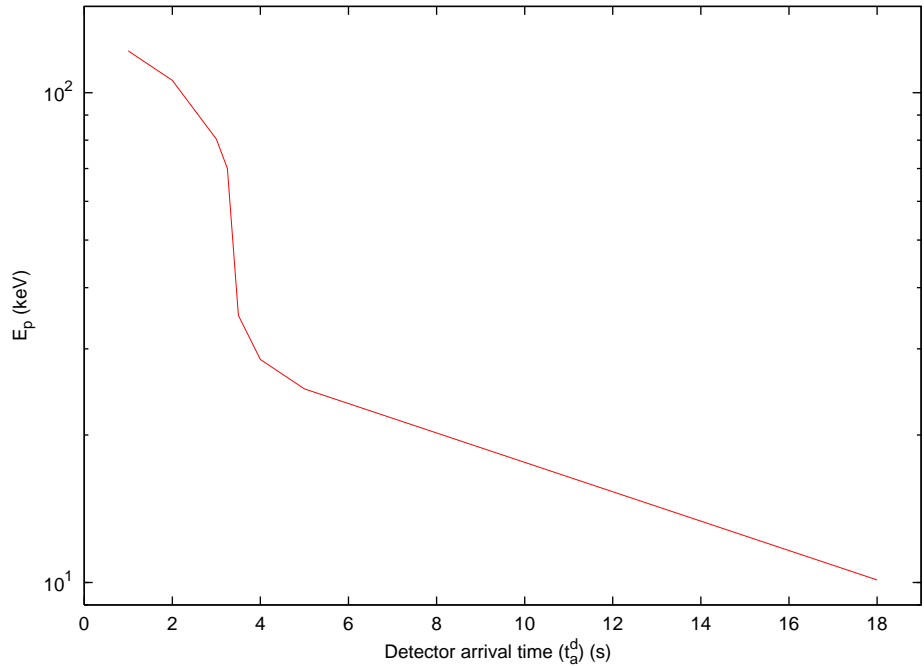}
\caption{The energy of the peak of the instantaneous photon number spectrum $N(E)$ is here represented as a function of the arrival time during the ``prompt emission'' phase. The clear hard to soft behavior is shown.}
\label{fig3a}
\end{figure}

Another distinguishing feature of the GRBs spectra which is also present in these instantaneous spectra, as shown in Fig.~\ref{fig2a}, is the hard to soft transition during the evolution of the event \cite{1997ApJ...479L..39C,1999PhR...314..575P,2000ApJS..127...59F,2002A&A...393..409G}. In fact the peak of the energy distributions $E_p$ drift monotonically to softer frequencies with time (see Fig.~\ref{fig3a}). This feature explains the change in the power-law low energy spectral index $\alpha$ \cite{1993ApJ...413..281B} which at the beginning of the prompt emission of the burst ($t_a^d=2$ s) is $\alpha=0.75$, and progressively decreases for later times (see Fig.~\ref{fig2a}). In this way the link between $E_p$ and $\alpha$ identified by Ref.~\refcite{1997ApJ...479L..39C} is explicitly shown. This theoretically predicted evolution of the spectral index during the event unfortunately cannot be detected in this particular burst by INTEGRAL because of the insufficient quality of the data (poor photon statistics, see Ref.~\refcite{2004Natur.430..646S}).

\subsection{The time-integrated spectrum: comparison with the observed data}

The time-integrated observed GRB spectra show a clear power-law behavior. Within a different framework Shakura, Sunyaev and Zel'dovich (see e.g. Ref.~\refcite{1983ASPRv...2..189P} and references therein) argued that it is possible to obtain such power-law spectra from a convolution of many non-power-law instantaneous spectra evolving in time. This result was recalled and applied to GRBs by Blinnikov et al.\cite{1999ARep...43..739B} assuming for the instantaneous spectra a thermal shape with a temperature changing with time. They showed that the integration of such energy distributions over the observation time gives a typical power-law shape possibly consistent with GRB spectra.

Our specific quantitative model is more complicated than the one considered by Blinnikov et al.\cite{1999ARep...43..739B}: the instantaneous spectrum here is not a black body. Each instantaneous spectrum is obtained by an integration over the corresponding EQTS: it is itself a convolution, weighted by appropriate Lorentz and Doppler factors, of $\sim 10^6$ thermal spectra with variable temperature. Therefore, the time-integrated spectra are not plain convolutions of thermal spectra: they are convolutions of convolutions of thermal spectra (see Fig.~\ref{fig2a}).

\begin{figure}
\includegraphics[width=\hsize,clip]{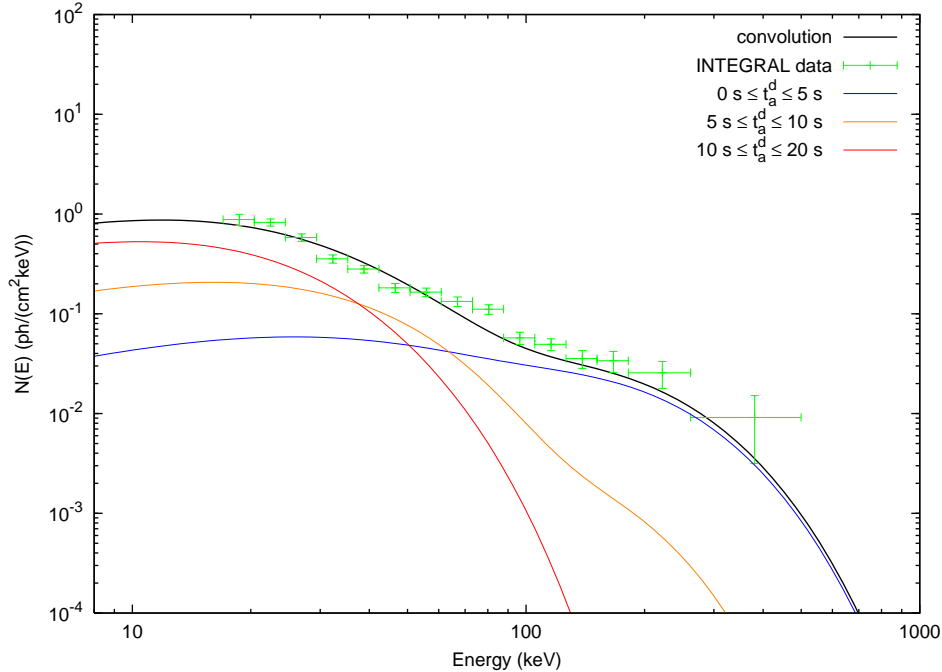}
\caption{Three theoretically predicted time-integrated photon number spectra $N(E)$ are here represented for $0 \le t_a^d \le 5$ s, $5 \le t_a^d \le 10$ s and $10 \le t_a^d \le 20$ s (colored curves). The hard to soft behavior presented in Fig.~\ref{fig3a} is confirmed. Moreover, the theoretically predicted time-integrated photon number spectrum $N(E)$ corresponding to the first $20$ s of the ``prompt emission'' (black bold curve) is compared with the data observed by INTEGRAL (green points)\cite{2004Natur.430..646S,saz2}. This curve is obtained as a convolution of 108 instantaneous spectra, which are enough to get a good agreement with the observed data.}
\label{fig4a}
\end{figure}

The simple power-law shape of the integrated spectrum is more evident if we sum tens of instantaneous spectra, as in Fig.~\ref{fig4a}. In this case we divided the prompt emission in three different time interval, and for each one we integrated on time the energy distribution. The resulting three time-integrated spectra have a clear nonthermal behavior, and still show the characteristic hard to soft transition.

Finally, we integrated the photon number spectrum $N(E)$ over the entire duration of the prompt event (see again Fig.~\ref{fig4a}): in this way we obtain a typical nonthermal power-law spectrum which turns out to be in good agreement with the INTEGRAL data \cite{2004Natur.430..646S,saz2} and gives clear evidence of the possibility that the observed GRBs spectra originate from a thermal emission.

The precise knowledge we have here acquired on GRB 031203 helps in clarifying the overall astrophysical system GRB 031203 - SN 2003lw - the $2-10$ keV XMM and Chandra data (see sections \ref{031203} and \ref{secURCA}, where the late $2--10$ keV XMM and Chandra data are also discussed).

\section{Application to GRB 050315: the first complete light curve fitting}\label{secJ}

The fit of GRB 991216 was specially important in showing a good agreement between the bolometric luminosity predicted by our theory and the observations, evidencing the clear separation between the P-GRB and the afterglow. The INTEGRAL observations of GRB 031203 have been crucial in leading to the confirmation of our theoretical approach for the spectral shape in the prompt emission of GRBs. It has been the welcome consequence of the Swift satellite to have the possibility of receiving high quality data continuously spanning from the early part of the afterglow (the prompt emission) to the late phases of the afterglow, leading to the emergence of a standard GRB structure. Our first analysis of the Swift data came with GRB 050315.

GRB 050315 \cite{2006ApJ...638..920V} had been triggered and located by the BAT instrument \cite{2004SPIE.5165..175B,2005SSRv..120..143B} on board the {\em Swift} satellite \cite{2004ApJ...611.1005G} at 2005-March-15 20:59:42 UT \cite{2005GCN..3094....1P}. The narrow field instrument XRT \cite{2004SPIE.5165..201B,2005SSRv..120..165B} began observations $\sim 80$ s after the BAT trigger, one of the earliest XRT observations yet made, and continued to detect the source for $\sim 10$ days \cite{2006ApJ...638..920V}. The spectroscopic redshift has been found to be $z = 1.949$ \cite{2005GCN..3101....1K}.

We present here the results of the fit of the \emph{Swift} data of this source in $5$ energy bands in the framework of our theoretical model, pointing out a new step toward the uniqueness of the explanation of the overall GRB structure. We first recall the essential features of our theoretical model; then we fit the GRB 050315 observations by both the BAT and XRT instruments; we also present the instantaneous spectra for selected values of the detector arrival time ranging from $60$ s (i.e., during the so called ``prompt emission'') all the way to $3.0\times 10^4$ s (i.e., the latest afterglow phases).

\subsection{The fit of the observations}\label{050315_fit}

The best fit of the observational data leads to a total energy of the black hole dyadosphere, generating the $e^\pm$ plasma, $E_{e^\pm}^{tot}= 1.46\times 10^{53}$ erg (the observational \emph{Swift} $E_{iso}$ is $> 2.62\times 10^{52}$ erg)\cite{2006ApJ...638..920V}, so that the plasma is created between the radii $r_1 = 5.88\times 10^6$ cm and $r_2 = 1.74 \times 10^8$ cm with an initial temperature $T = 2.05 MeV$ and a total number of pairs $N_{e^+e^-} = 7.93\times 10^{57}$. The second parameter of the theory, the amount $M_B$ of baryonic matter in the plasma, is found to be such that $B \equiv M_Bc^2/E_{dya} = 4.55 \times 10^{-3}$. The transparency point and the P-GRB emission occurs then with an initial Lorentz gamma factor of the accelerated baryons $\gamma_\circ = 217.81$ at a distance $r = 1.32 \times 10^{14}$ cm from the black hole.

\subsubsection{The BAT data}

\begin{figure}
\begin{minipage}{\hsize}
\begin{center}
\includegraphics[width=0.49\hsize,clip]{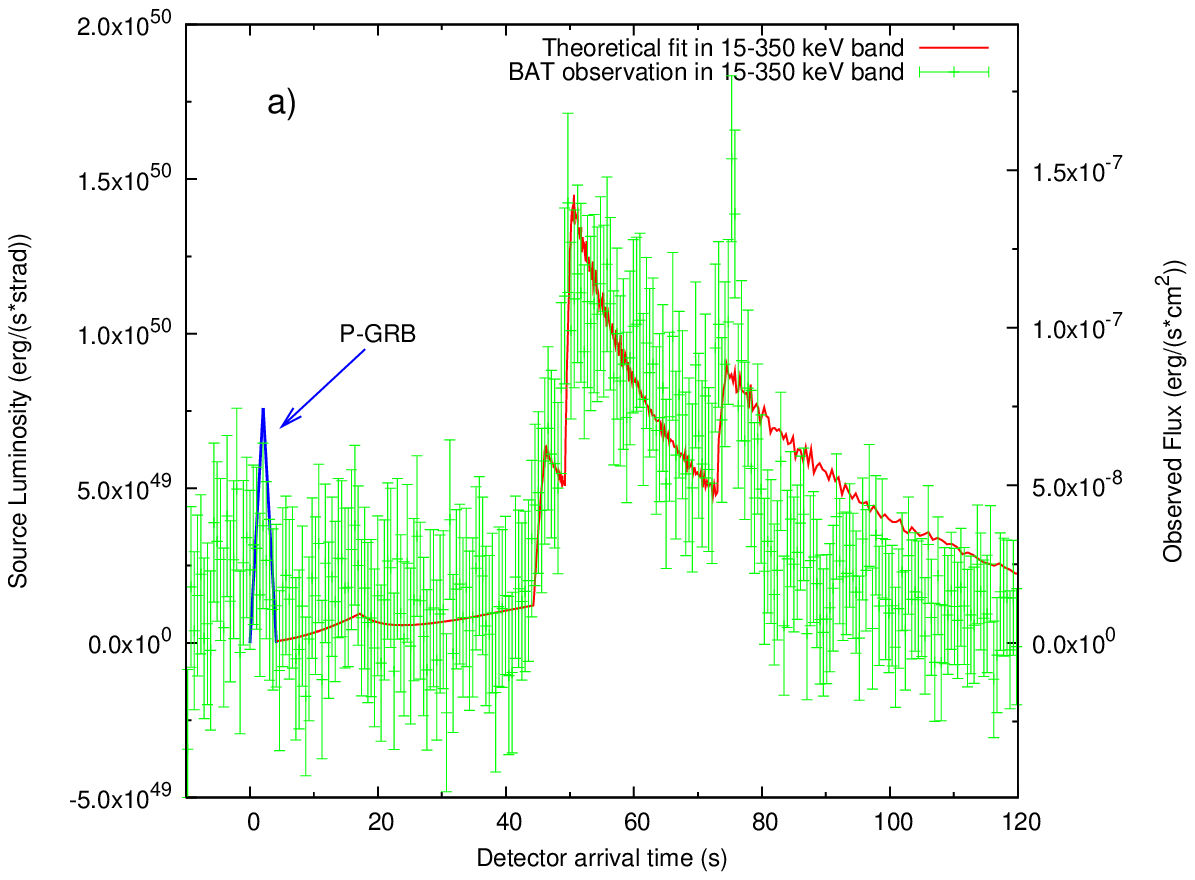}
\includegraphics[width=0.49\hsize,clip]{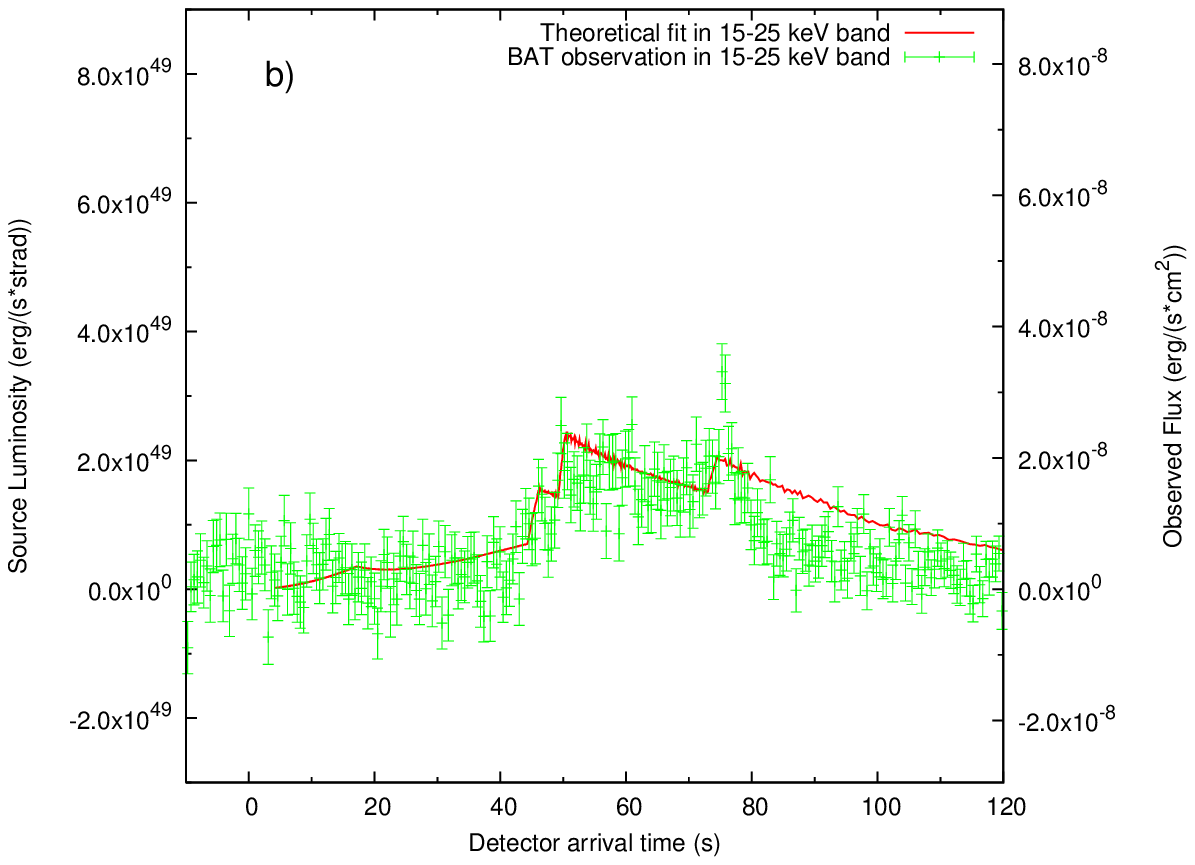}\\
\includegraphics[width=0.49\hsize,clip]{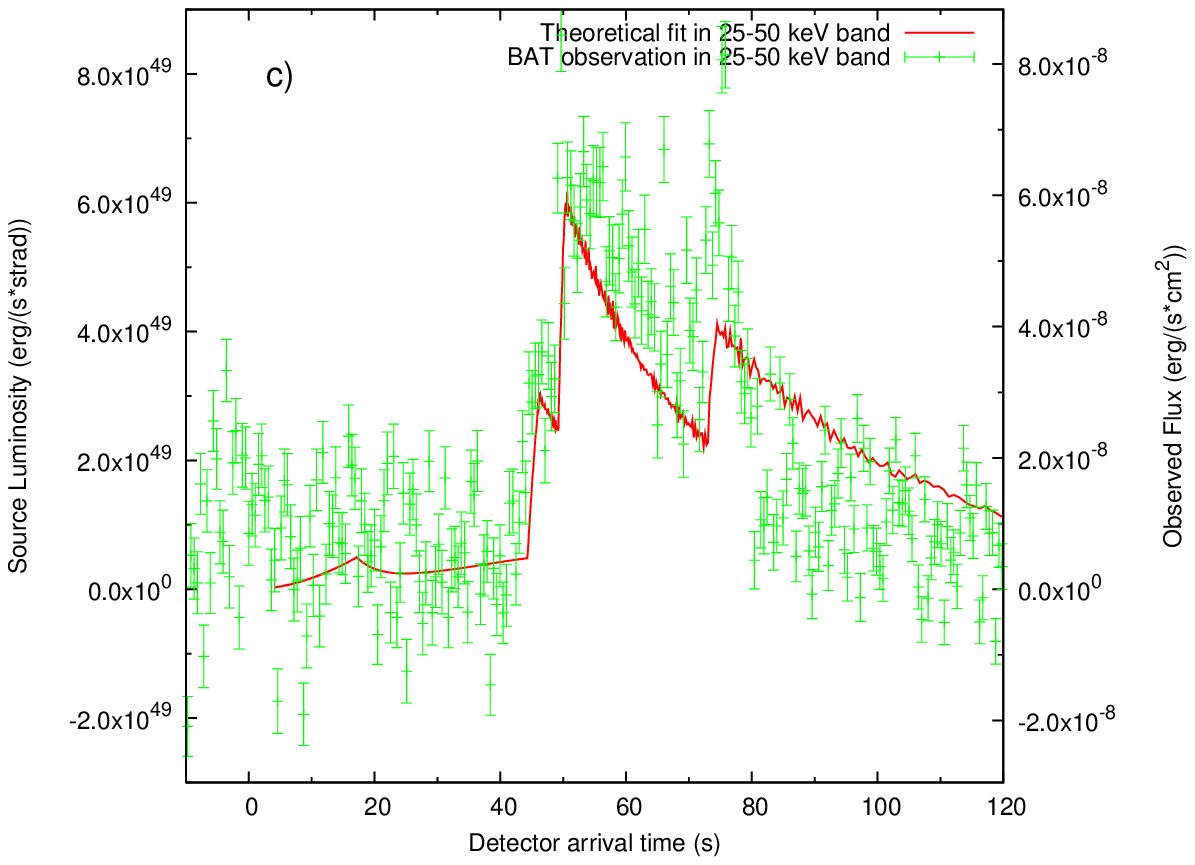}
\includegraphics[width=0.49\hsize,clip]{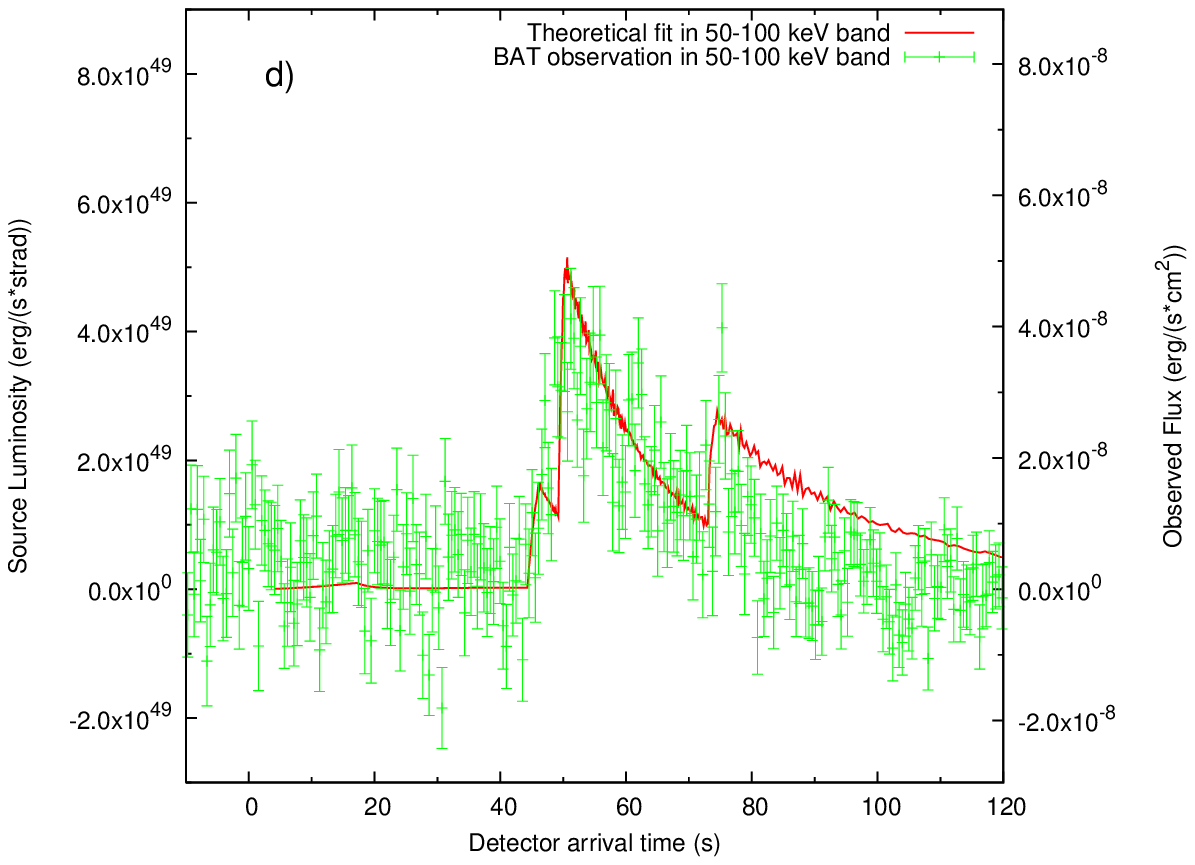}
\end{center}
\end{minipage}
\caption{Our theoretical fit (red line) of the BAT observations (green points) of GRB 050315 in the $15$--$350$ keV (a), $15$--$25$ keV (b), $25$--$50$ keV (c), $50$--$100$ keV (d) energy bands \cite{2006ApJ...638..920V}. The blue line in panel (a) represents our theoretical prediction for the intensity and temporal position of the P-GRB.}
\label{tot}
\label{050315_fit_prompt}
\end{figure}

In Fig.~\ref{tot} we represent our theoretical fit of the BAT observations in the three energy channels $15$--$25$ keV, $25$--$50$ keV and $50$--$100$ keV and in the whole $15$--$350$ keV energy band.

We have already recalled how in our model the GRB emission starts at the transparency point when the P-GRB is emitted; this instant of time is often different from the moment in which the satellite instrument triggers, due to the fact that sometimes the P-GRB is under the instrumental noise threshold or comparable with it. In order to compare our theoretical predictions with the observations, it is important to estimate and take into account this time shift. In the present case of GRB 050315 there has been observed \cite{2006ApJ...638..920V} a possible precursor before the trigger. Such a precursor is indeed in agreement with our theoretically predicted P-GRB, both in its isotropic energy emitted (which we theoretically predict to be $E_{P-GRB} = 1.98 \times 10^{51}$ erg) and its temporal separation from the peak of the afterglow (which we theoretically predicted to be $\Delta t^d_a = 51$ s). In Fig.~\ref{tot}a the blue line shows our theoretical prediction for the P-GRB in agreement with the observations.

After the P-GRB emission, all the observed radiation is produced by the interaction of the expanding baryonic shell with the interstellar medium. In order to reproduce the complex time variability of the light curve of the prompt emission as well as of the afterglow, we describe the CBM filamentary structure, for simplicity, as a sequence of overdense spherical regions separated by much less dense regions. Such overdense regions are nonhomogeneously filled, leading to an effective emitting area $A_{eff}$ determined by the dimensionless parameter ${\cal R}$ \cite{2004IJMPD..13..843R,2005IJMPD..14...97R}. Clearly, in order to describe any detailed structure of the time variability an authentic three-dimensional representation of the CBM structure would be needed. However, this finer description would not change the substantial agreement of the model with the observational data. Anyway, in the ``prompt emission'' phase, the small angular size of the source visible area due to the relativistic beaming makes such a spherical approximation an excellent one (see also for details Ref.~\refcite{2002ApJ...581L..19R}).

The structure of the ``prompt emission'' has been reproduced assuming three overdense spherical CBM regions with width $\Delta$ and density contrast $\Delta n/\langle n\rangle$: we chose for the first region, at $r = 4.15\times 10^{16}$ cm, $\Delta = 1.5\times 10^{15}$ cm and $\Delta n/\langle n\rangle = 5.17$, for the second region, at $r = 4.53\times 10^{16}$ cm, $\Delta = 7.0\times 10^{14}$ cm and $\Delta n/\langle n\rangle = 36.0$ and for the third region, at $r = 5.62\times 10^{16}$ cm, $\Delta = 5.0\times 10^{14}$ cm and $\Delta n/\langle n\rangle = 85.4$. The CBM mean density during this phase is $\left\langle n_{cbm} \right\rangle=0.81$ particles/cm$^3$ and $\left\langle {\cal R} \right\rangle = 1.4 \times 10^{-7}$. With this choice of the density mask we obtain agreement with the observed light curve, as shown in Fig.~\ref{tot}. A small discrepancy occurs in coincidence with the last peak: this is due to the fact that at this stage the source visible area due to the relativistic beaming is comparable with the size of the clouds, therefore the spherical shell approximation should be properly modified by a detailed analysis of a full three-dimensional treatment of the CBM filamentary structure. Such a topic is currently under investigation (see also for details Ref.~\refcite{2002ApJ...581L..19R}). Fig.~\ref{tot} also shows the theoretical fit of the light curves in the three BAT energy channels in which the GRB has been detected ($15$--$25$ keV in Fig.~\ref{tot}b, $25$--$50$ keV in Fig.~\ref{tot}c, $50$--$100$ keV in Fig.~\ref{tot}d).

\subsubsection{The XRT data}

\begin{figure}
\includegraphics[width=\hsize,clip]{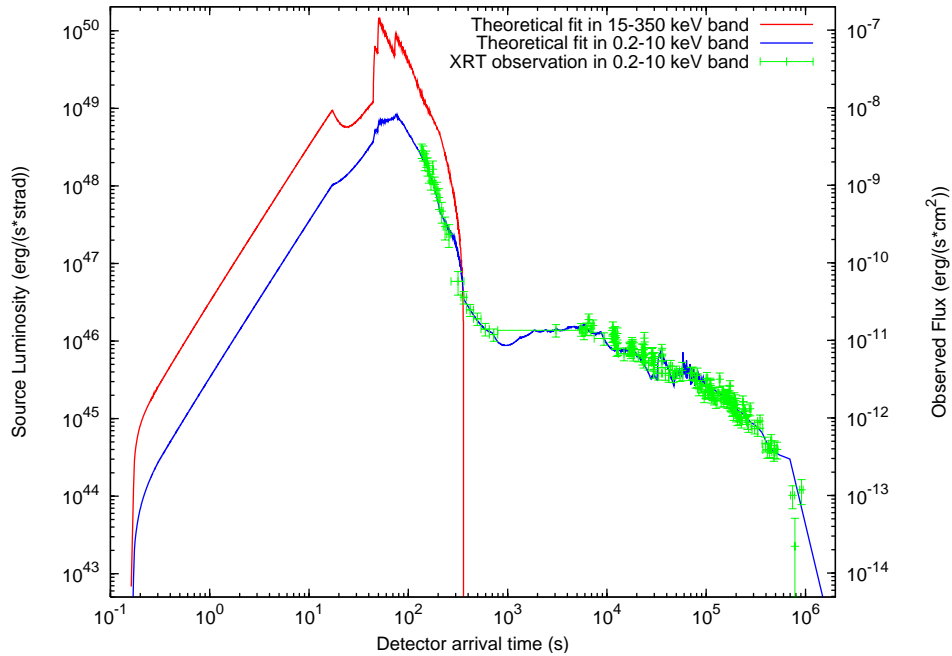}
\caption{Our theoretical fit (blue line) of the XRT observations (green points) of GRB 050315 in the $0.2$--$10$ keV energy band \cite{2006ApJ...638..920V}. The theoretical fit of the BAT observations (see Fig.~\ref{tot}a) in the $15$--$350$ keV energy band is also represented (red line).}
\label{global}
\label{050315_fit_aft}
\end{figure}

The same analysis can be applied to explain the features of the XRT light curve in the afterglow phase. It has been recently pointed out \cite{2006ApJ...642..389N} that almost all the GRBs observed by {\em Swift} show a ``canonical behavior'': an initial very steep decay followed by a shallow decay and finally a steeper decay. In order to explain these features many different approaches have been proposed \cite{2006RPPh...69.2259M,2006ApJ...642..389N,2006MNRAS.366.1357P,2006ApJ...642..354Z}. In our treatment these behaviors are automatically described by the same mechanism responsible for the prompt emission described above: the baryonic shell expands in a CBM region, between $r = 9.00\times 10^{16}$ cm and $r = 5.50\times 10^{18}$ cm, which is at significantly lower density ($\left\langle n_{cbm} \right\rangle=4.76 \times 10^{-4}$ particles/cm$^3$, $\left\langle {\cal R} \right\rangle = 7.0 \times 10^{-6}$) then the one corresponding to the prompt emission, and this produces a slower decrease of the velocity of the baryons with a consequently longer duration of the afterglow emission. The initial steep decay of the observed flux is due to the smaller number of collisions with the CBM. In Fig.~\ref{global} is represented our theoretical fit of the XRT data, together with the theoretically computed $15$--$350$ keV light curve of Fig.~\ref{tot}a (without the BAT observational data in order not to overwhelm the picture too much).

What is impressive is that no different scenarios need to be advocated in order to explain the features of the light curves: both the prompt and the afterglow emission are just due to the thermal radiation in the comoving frame produced by inelastic collisions with the CBM properly boosted by the relativistic transformations over the EQTSs.

\subsection{The instantaneous spectrum}\label{spectra}

\begin{figure}
\includegraphics[width=\hsize,clip]{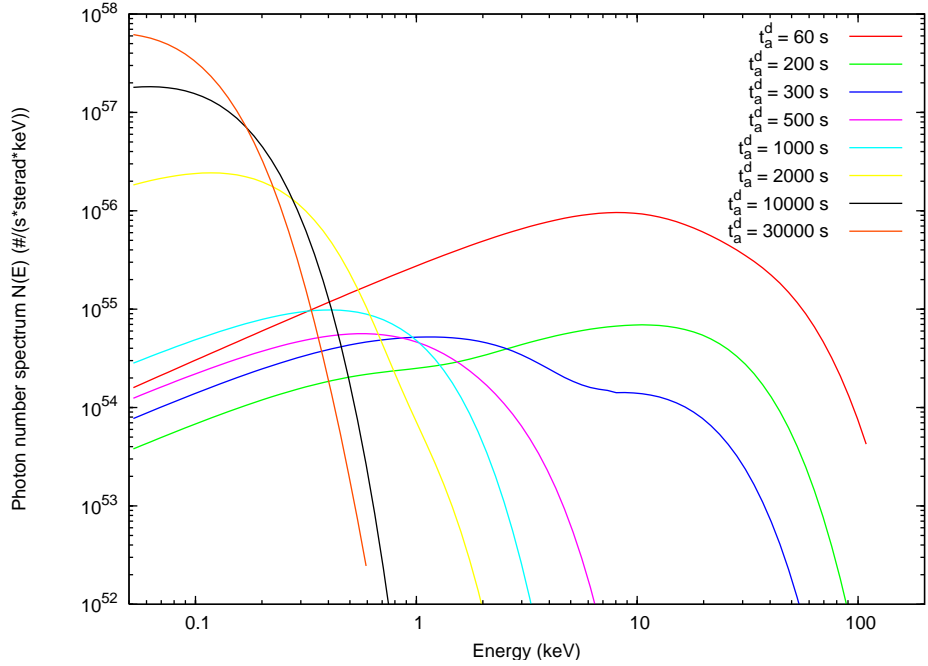}
\caption{Eight theoretically predicted instantaneous photon number spectra $N(E)$ are represented here for different values of the arrival time (colored curves). The hard to soft behavior is confirmed.}
\label{spettro}
\end{figure}

In addition to the the luminosity in fixed energy bands we can derive also the instantaneous photon number spectrum $N(E)$ starting from the same assumptions. In Fig.~\ref{spettro} are shown samples of time-resolved spectra for eight different values of the arrival time which cover the entire duration of the event. It is clear from this picture that although the spectrum in the co-moving frame of the expanding pulse is thermal, the shape of the final spectrum in the laboratory frame is clearly nonthermal. In fact, as we have recalled and explained in Ref.~\refcite{2004IJMPD..13..843R}, each single instantaneous spectrum is the result of an integration of thousands of thermal spectra over the corresponding EQTS. This calculation produces a nonthermal instantaneous spectrum in the observer frame (see Fig.~\ref{spettro}).

A distinguishing feature of the GRB spectra which is also present in these instantaneous spectra is the hard to soft transition during the evolution of the event \cite{1997ApJ...479L..39C,1999PhR...314..575P,2000ApJS..127...59F,2002A&A...393..409G}. In fact the peak of the energy distribution $E_p$ drifts monotonically to softer frequencies with time. This feature is linked to the change in the power-law low energy spectral index $\alpha$ \cite{1993ApJ...413..281B}, so the correlation between $\alpha$ and $E_p$ \cite{1997ApJ...479L..39C} is explicitly shown.

It is important to stress that there is no difference in the nature of the spectrum during the prompt and the afterglow phases: the observed energy distribution changes from hard to soft, with continuity, from the ``prompt emission'' all the way to the latest phases of the afterglow.

\section{Problems with the definition of ``long'' GRBs evidenced by GRB 05015}

\begin{figure}
\includegraphics[width=\hsize,clip]{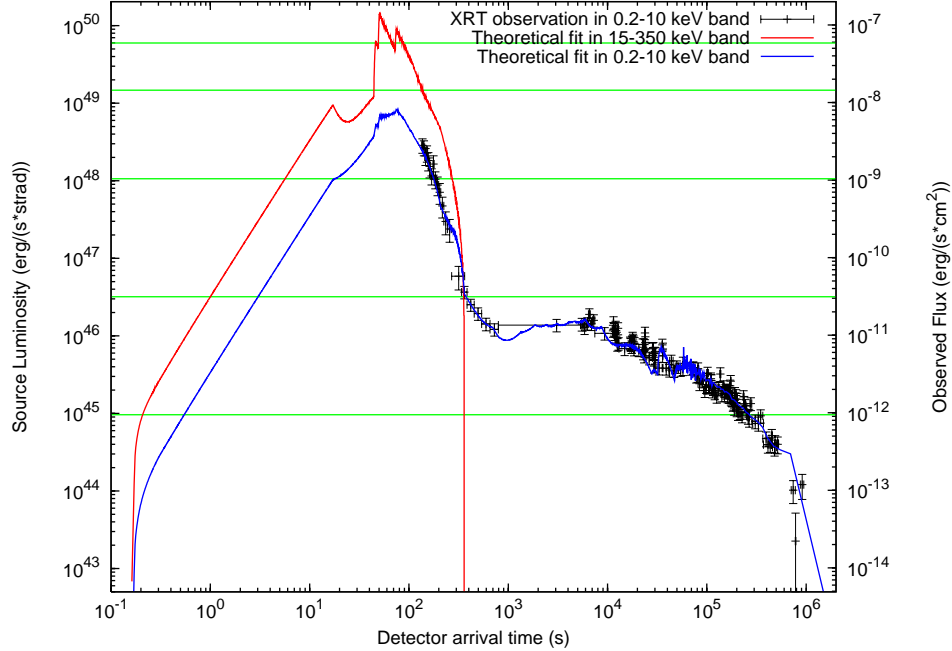}
\caption{Same as Fig.~\ref{global}. The horizontal green lines correspond to different possible instrumental thresholds. It is clear that long GRB durations are just functions of the observational threshold.}
\label{global_th}
\end{figure}

The confirmation by \emph{Swift} of our prediction of the overall afterglow structure, and especially the coincidence of the ``prompt emission'' with the peak of the afterglow, opens a new problematic in the definition of the long GRBs. It is clear in fact that the identification of the ``prompt emission'' in the current GRB literature is not at all intrinsic to the phenomenon but is merely due to the threshold of the instruments used in the observations (e.g. BATSE in the $50$--$300$ keV energy range, or BeppoSAX GRBM in $40$--$700$ keV, or \emph{Swift} BAT in $15$--$350$ keV). As it is clear from Fig.~\ref{global_th}, there is no natural way to identify in the source a special extension of the peak of the afterglow that is not the one purely defined by the experimental threshold. It is clear, therefore, that long GRBs, as defined until today, are just the peak of the afterglow and there is no way, as explained above, to define their ``prompt emission'' duration as a characteristic signature of the source. As the \emph{Swift} observations show, the duration of the long GRBs has to coincide with the duration of the entire afterglow. A Kouveliotou-Tavani plot of the long GRBs, done following our interpretation which is clearly supported by the recent \emph{Swift} data (see Fig.~\ref{global_th}), will present enormous dispersion on the temporal axis.

We recall that in our theory both ``short'' and ``long'' GRBs originate from the same process of black hole formation. The major difference between the two is the value of the baryon loading parameter $B$ (see Fig.~\ref{bcross}). In the limit of small baryon loading, all the plasma energy is emitted at the transparency in the P-GRB, with negligible afterglow observed flux. For higher values of the baryon loading, the relative energy content of the P-GRB with respect to the afterglow diminishes (see Ref.~\refcite{2005AIPC..782...42R} and references therein).

\section{Application to GRB 060218: the first evidence of the critical value of the baryon loading $B\sim 10^{-2}$}\label{sec060218}

GRB 060218 triggered the BAT instrument of {\em Swift} on 18 February 2006 at 03:36:02 UT and has a $T_{90} = (2100 \pm 100)$ s \cite{2006GCN..4775....1C}. The XRT instrument \cite{2006GCN..4776....1K,2006GCN..4775....1C} began observations $\sim 153$ s after the BAT trigger and continued for $\sim 12.3$ days \cite{2006GCN..4822....1S}. The source is characterized by a flat gamma ray light curve and a soft spectrum \cite{2006GCN..4780....1B}. It has an X-ray light curve with a long, slow rise and gradual decline and it is considered an X-Ray Flash (XRF) since its peak energy occurs at $E_p=4.9^{+0.4}_{-0.3}$ keV \cite{2006Natur.442.1008C}. It has been observed by the \emph{Chandra} satellite on February 26.78 and March 7.55 UT ($t\simeq 8.8$ and $17.4$ days) for $20$ and $30$ ks respectively \cite{2006Natur.442.1014S}. The spectroscopic redshift has been found to be $z=0.033$ \cite{2006A&A...454..503S,2006ApJ...643L..99M}. The corresponding isotropic equivalent energy is $E_{iso}=(1.9\pm 0.1)\times 10^{49}$ erg \cite{2006GCN..4822....1S} which sets this GRB as a low luminous one, consistent with most of the GRBs associated with supernovas \cite{2007ApJ...662.1111L,2006ApJ...645L.113C,2007ApJ...657L..73G}.

GRB 060218 is associated with SN2006aj whose expansion velocity is $v\sim 0.1c$ \cite{2006Natur.442.1011P,2006GCN..4809....1F,2006GCN..4804....1S,2006ApJ...645L.113C}. The host galaxy of SN2006aj is a low luminosity, metal poor star-forming dwarf galaxy \cite{2007AIPC..924..120F} with an irregular morphology \cite{2007A&A...464..529W} similar to that of other GRBs associated with supernovas \cite{2006ApJ...645L..21M,2006A&A...454..503S}.

\subsection{The fit of the observed data}\label{fit}

\begin{figure}
\includegraphics[width=\hsize,clip]{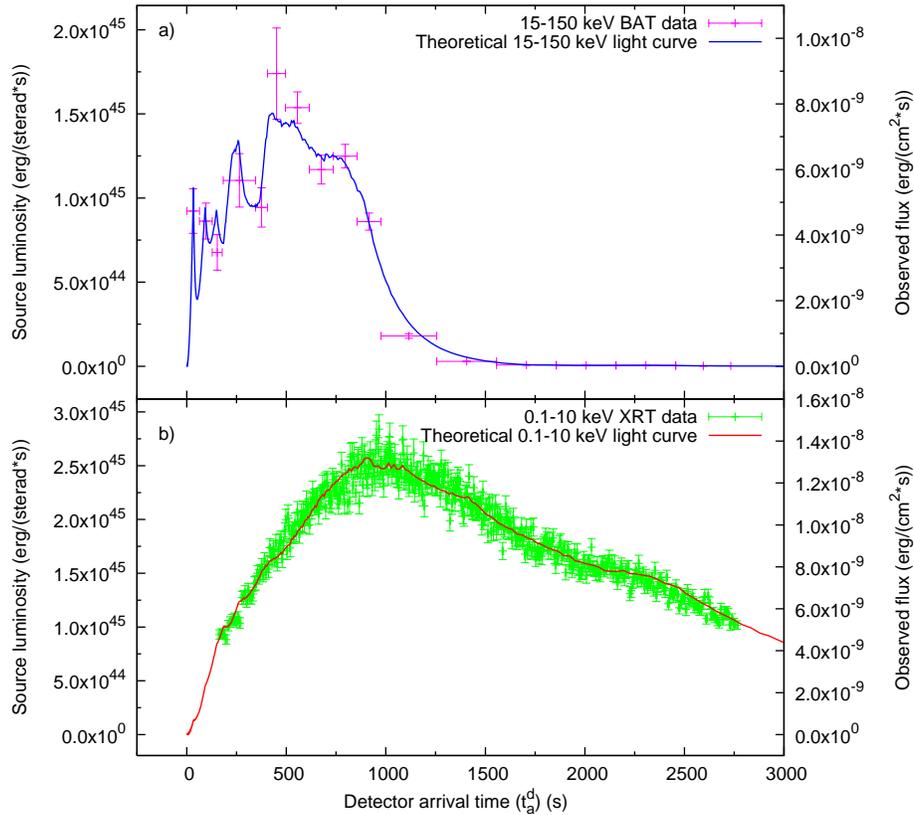}
\caption{GRB 060218 prompt emission: a) our theoretical fit (blue line) of the BAT observations in the $15$--$150$ keV energy band (pink points); b) our theoretical fit (red line) of the XRT observations in the $0.3$--$10$ keV energy band (green points)\cite{2006Natur.442.1008C}.}
\label{060218_tot}
\end{figure}

In this section we present the fit of our fireshell model to the observed data (see Figs.~\ref{060218_tot}, \ref{global2}). The fit leads to a total energy of the $e^\pm$ plasma $E_{e^\pm}^{tot}= 2.32\times 10^{50}$ erg, with an initial temperature $T = 1.86$ MeV and a total number of pairs $N_{e^\pm} = 1.79\times 10^{55}$. The second parameter of the theory, $B = 1.0 \times 10^{-2}$, is the highest value ever observed and is close to the limit for the stability of the adiabatic optically thick acceleration phase of the fireshell (for further details see Ref.~\refcite{2000A&A...359..855R}). The Lorentz gamma factor obtained solving the fireshell equations of motion \cite{2005ApJ...620L..23B,2005ApJ...633L..13B} is $\gamma_\circ=99.2$ at the beginning of the afterglow phase at a distance from the progenitor $r_\circ=7.82\times 10^{12}$ cm. It is much larger than $\gamma \sim 5$ estimated by Ref.~\refcite{2007ApJ...654..385K} and Ref.~\refcite{2007ApJ...659.1420T}.

In Fig.~\ref{060218_tot} we show the afterglow light curves fitting the prompt emission both in the BAT ($15$--$150$ keV) and in the XRT ($0.3$--$10$ keV) energy ranges, as expected in our ``canonical GRB'' scenario \cite{2007A&A...471L..29D}. Initially the two luminosities are comparable to each other, but for a detector arrival time $t_a^d > 1000$ s the XRT curves becomes dominant. The displacement between the peaks of these two light curves leads to a theoretically estimated spectral lag greater than $500$ s in perfect agreement with the observations \cite{2006ApJ...653L..81L}. We obtain that the bolometric luminosity in this early part coincides with the sum of the BAT and XRT light curves (see Fig.~\ref{global2}) and the luminosity in the other energy ranges is negligible.

We recall that at $t_a^d \sim 10^4$ s there is a sudden enhancement in the radio luminosity and there is an optical luminosity dominated by the SN2006aj emission \cite{2006Natur.442.1008C,2006Natur.442.1014S,2006JCAP...09..013F}. Although our analysis addresses only the BAT and XRT observations, for $r > 10^{18}$ cm corresponding to $t_a^d > 10^4$ s the fit of the XRT data implies two new features: \textbf{1)} a sudden increase of the ${\cal R}$ factor from ${\cal R} = 1.0\times 10^{-11}$ to ${\cal R} = 1.6\times 10^{-6}$, corresponding to a significantly more homogeneous effective CBM distribution (see Fig.\ref{060218_global}b); \textbf{2)} an XRT luminosity much smaller than the bolometric one (see Fig.~\ref{global2}). These theoretical predictions may account for the energetics of the enhancement of the radio and possibly optical and UV luminosities. Therefore, we identify two different regimes in the afterglow, one for $t_a^d < 10^4$ s and the other for $t_a^d > 10^4$ s. Nevertheless, there is a unifying feature: the determined effective CBM density decreases with the distance $r$ monotonically and continuously through both these two regimes from $n_{cbm} = 1$ particle/cm$^3$ at $r = r_\circ$ to $n_{cbm} = 10^{-6}$ particle/cm$^3$ at $r = 6.0 \times 10^{18}$ cm: $n_{cbm} \propto r^{-\alpha}$, with $1.0 \lesssim \alpha \lesssim 1.7$ (see Fig.~\ref{060218_global}a).

Our assumption of spherical symmetry is supported by the observations which set for GRB 060218 an opening beaming angle larger than $\sim 37^\circ$ \cite{2007ApJ...662.1111L,2006Natur.442.1008C,2006Natur.442.1014S,2007ApJ...657L..73G}.

\subsection{The procedure of the fit\label{procedure} }

The arrival time of each photon at the detector depends on the entire previous history of the fireshell \cite{2001ApJ...555L.107R}. Moreover, all the observables depends on the EQTS \cite{2004ApJ...605L...1B,2005ApJ...620L..23B} which in turn depend crucially on the equations of motion of the fireshell. The CBM engulfment has to be computed self-consistently through the entire dynamical evolution of the fireshell and not separately at each point. Any change in the CBM distribution strongly influences the entire dynamical evolution of the fireshell and, due to the EQTS structure, produces observable effects up to a much later time. For example if we change the density mask at a certain distance from the black hole we modify the shape of the lightcurve and consequently the evolution changes at larger radii corresponding to later times. 
Anyway the change of the density is not the only problem to face in the fitting of the source, in fact first of all we have to choose the energy in order to have a Lorentz gamma factor sufficiently high to fit the entire GRB.  
In order to show the sensitivity of the fitting procedure I also present two examples of fits with the same value of $B$ and a different value of $E_{e^\pm}^{tot}$.

\begin{figure}
  \includegraphics[width=\hsize,clip]{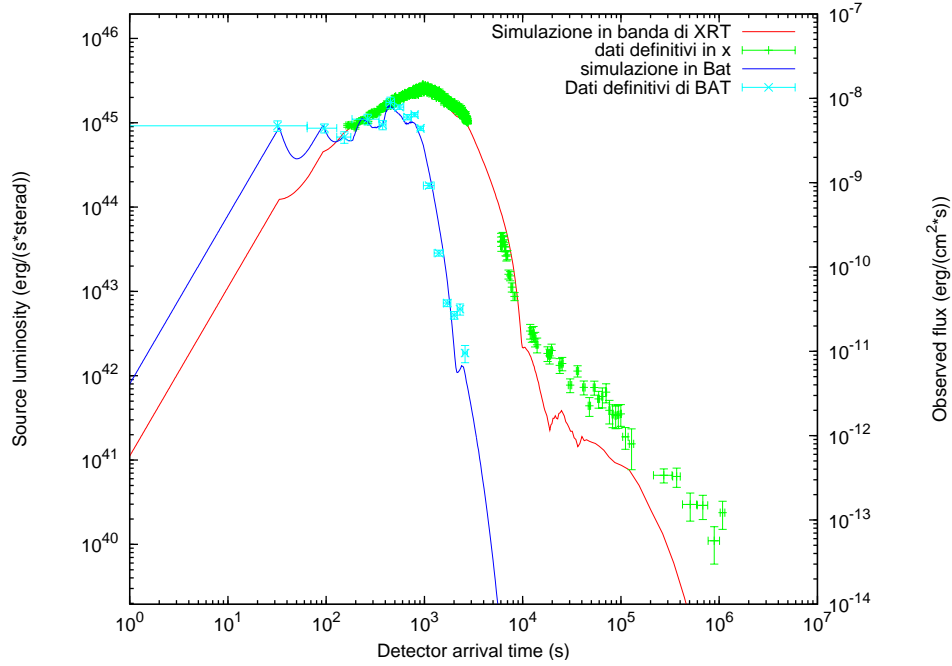}
  \caption{GRB 060218 light curves with $E_{e^\pm}^{tot}= 1.36\times 10^{50}$ erg: our theoretical fit (blue line) of the $15$--$150$ keV BAT observations (pink points), our theoretical fit (red line) of the $0.3$--$10$ keV XRT observations (green points) and the $0.3$--$10$ keV \textit{Chandra} observations (black points) are represented together with our theoretically computed bolometric luminosity (black line) (Data from: Ref.~\refcite{2006Natur.442.1008C,2006Natur.442.1014S}).}
  \label{060218bolometricasotto}
\end{figure}

The first example has an $E_{e^\pm}^{tot}$ = $1.36\times 10^{50}$ erg . This fit was unsuccessful as we see from the Fig.~\ref {060218bolometricasotto}, because the bolometric lightcurve is under the XRT peak of the afterglow. This means that the value of the energy chosen is too small to fit any data points after the peak of the afterglow. So we have to increase the value of the energy to a have a better fit. In fact the parameter values have been found with various attempts in order to obtain the best fit. 

\begin{figure}
  \includegraphics[width=\hsize,clip]{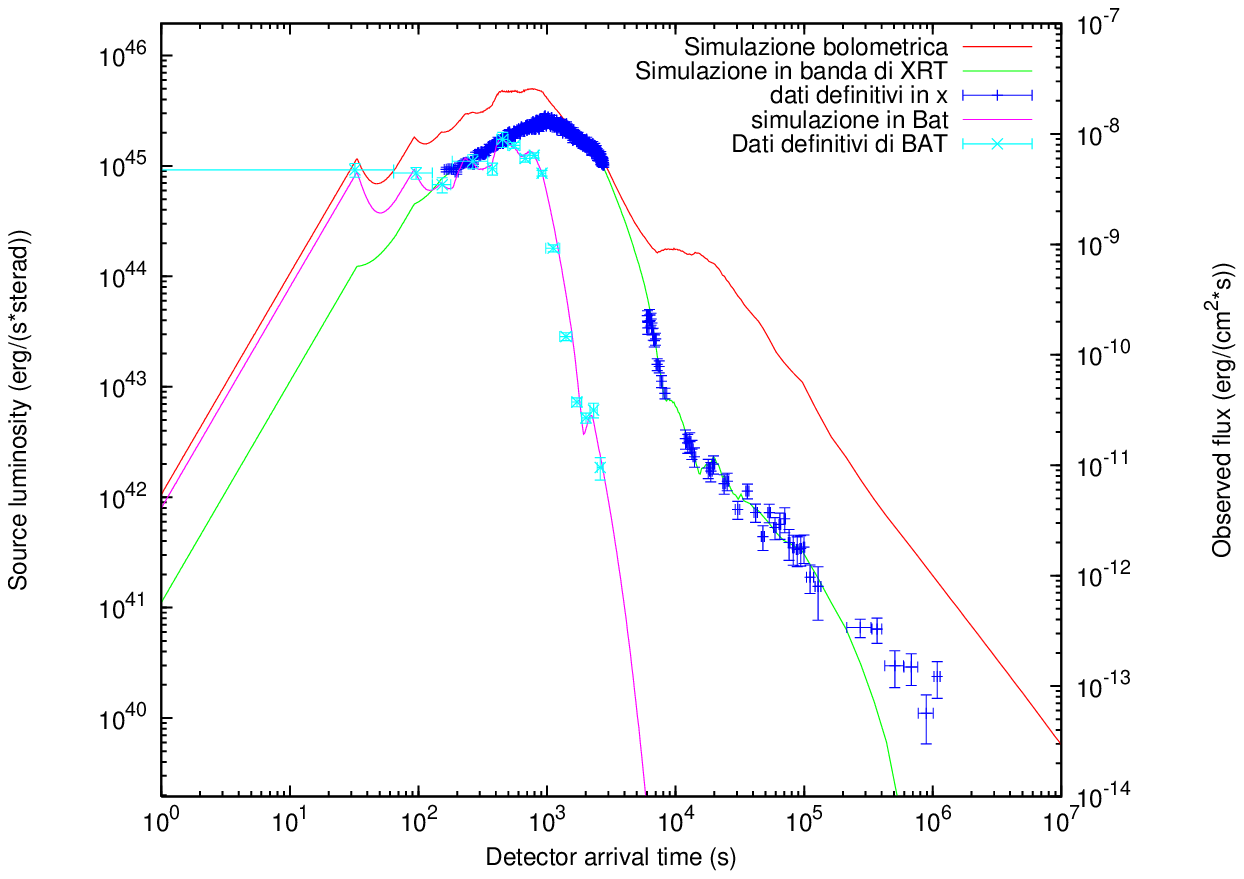}
  \caption{GRB 060218 light curves with $E_{e^\pm}^{tot}= 1.61\times 10^{50}$ erg: our theoretical fit (blue line) of the $15$--$150$ keV BAT observations (pink points), our theoretical fit (red line) of the $0.3$--$10$ keV XRT observations (green points) and the $0.3$--$10$ keV \textit{Chandra} observations (black points) are represented together with our theoretically computed bolometric luminosity (black line). Data from: Ref.~\refcite{2006Natur.442.1008C,2006Natur.442.1014S}.}
  \label{060218senzaultimipunti}
\end{figure}

\begin{figure}
\includegraphics[width=\hsize,clip]{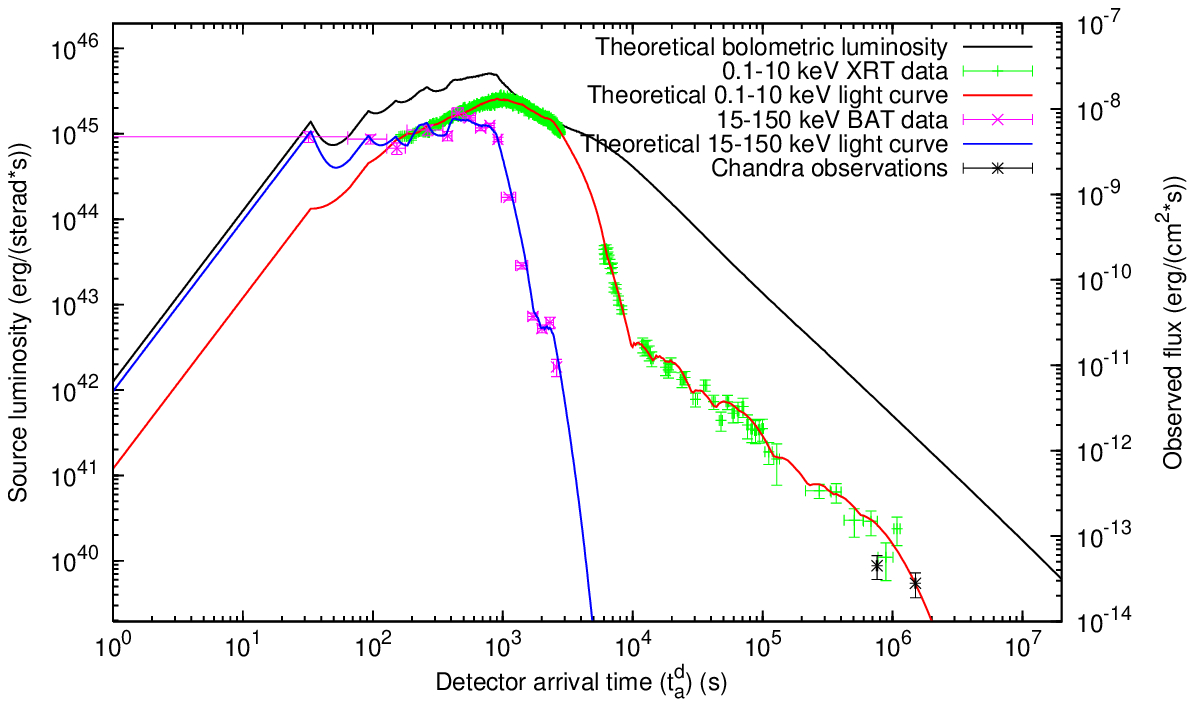}
\caption{GRB 060218 complete light curves: our theoretical fit (blue line) of the $15$--$150$ keV BAT observations (pink points), our theoretical fit (red line) of the $0.3$--$10$ keV XRT observations (green points) and the $0.3$--$10$ keV \textit{Chandra} observations (black points)\cite{2006Natur.442.1008C,2006Natur.442.1014S} are represented together with our theoretically computed bolometric luminosity (black line). In this case we have $E_{e^\pm}^{tot}= 2.32\times 10^{50}$ ergs.}
\label{global2}
\end{figure}

The second example is characterized by $E_{e^\pm}^{tot}= 1.61\times 10^{50}$ erg and the all the data are fit except for the last point from $2.0\times 10^{2}$s to the end (see Fig.~\ref{060218senzaultimipunti}). We attempt to fit these last points trying to diminish the $R$ values in order to enhance the energy emission, but again the low value of the Lorentz gamma factor that in this case is $3$ prevents the fireshell from expanding. So again in this case the value of the energy chosen is too small, but it is better than the previous attempt. In this case we increased the energy value by 24\%, but it is not enough so we decide to increase 16\%.

So the final fit is characterized by the $B=1.0\times10^{-2}$ and by the $E_{e^\pm}^{tot}= 2.32\times 10^{50}$ erg. With this value of the energy we are able to fit all the experimental points.

\subsection{The fireshell fragmentation}\label{sec4}

\begin{figure}
\includegraphics[width=\hsize,clip]{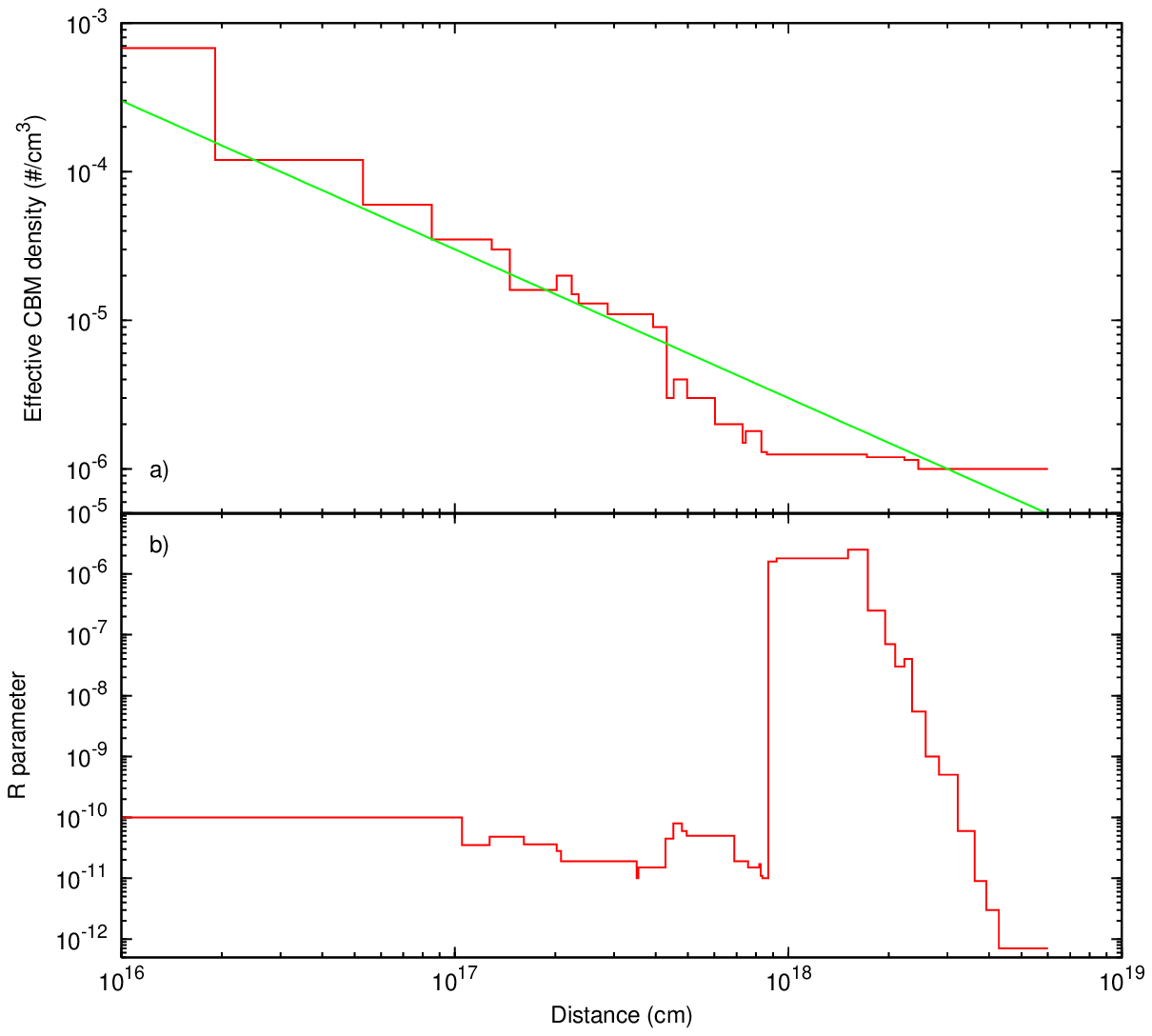}
\caption{The CBM distribution parameters: a) the effective CBM number density (red line) monotonically decreases with the distance $r$ following Eq.~(\ref{nismact}) (green line); b) the ${\cal R}$ parameter vs.\ distance.}
\label{060218_global}
\end{figure}

GRB 060218 presents different peculiarities: the extremely long $T_{90}$, the very low effective CBM density decreasing with the distance and the largest possible value of $B=10^{-2}$. These peculiarities appear to be correlated. Following Ref.~\refcite{Mosca_Orale}, we propose that in the present case the fireshell is fragmented. This implies that the surface of the fireshell does not increase any longer like $r^2$ but like $r^\beta$ with $\beta < 2$. Consequently, the effective CBM density $n_{cbm}$ is linked to the actual one $n_{cbm}^{act}$ by:
\begin{equation}
n_{cbm} = {\cal R}_{shell} n_{cbm}^{act}\, , \quad \mathrm{with} \quad {\cal R}_{shell} \equiv \left(r^\star/r\right)^\alpha\, ,
\label{nismact}
\end{equation}
where $r^\star$ is the starting radius at which the fragmentation occurs and $\alpha = 2 - \beta$ (see Fig.~\ref{060218_global}a). For $r^\star = r_\circ$ we have $n_{cbm}^{act}=1$ particles/cm$^3$, as expected for a ``canonical GRB'' \cite{2007AIPC..910...55R} and in agreement with the apparent absence of a massive stellar wind in the CBM \cite{2006Natur.442.1014S,2006JCAP...09..013F,2007MNRAS.375..240L}.

The ${\cal R}$ parameter defined in Eq.~(\ref{060218_Rdef}) has to take into account both the effect of the fireshell fragmentation (${\cal R}_{shell}$) and the effective CBM porosity (${\cal R}_{cbm}$):
\begin{equation}
{\cal R} \equiv {\cal R}_{shell} \times {\cal R}_{cbm}\, .
\label{060218_Rdef}
\end{equation}

The phenomenon of the clumpiness of the ejecta, whose measure is the filling factor, is an aspect well known in astrophysics.  For example, in the case of Novae the filling factor has been measured to be in the range $10^{-2}$--$10^{-5}$ \cite{2006A&A...459..875E}. Such a filling factor coincides, in our case, with ${\cal R}_{shell}$.

\subsection{Binaries as progenitors of GRB-supernova systems}\label{binary}

The majority of the existing models in the literature appeal to a single astrophysical phenomenon to explain both the GRB and the SN (``collapsar'', see e.g. Ref.~\refcite{2006ARA&A..44..507W}). On the contrary, a distinguishing feature of our theoretical approach is to distinguish between the supernova and the GRB process. The GRB is assumed to occur during the formation process of a black hole. The supernova is assumed to lead to the formation of a neutron star (NS) or to a complete disruptive explosion without remnants and in no way to the formation of a black hole. In the case of SN2006aj the formation of such a NS has been actually inferred by Ref.~\refcite{2007ApJ...658L...5M} because of the large amount of $^{58}$Ni ($0.05 M_\odot$). Moreover the significantly small initial mass of the supernova progenitor star $M \approx 20 M_\odot$ is expected to form a NS rather than a black hole when its core collapses \cite{2007ApJ...658L...5M,2007AIPC..924..120F,2006Natur.442.1018M,2007astro.ph..2472N}. In order to fulfill both the above requirement, we assume that the progenitor of the GRB and the supernova consists of a binary system formed by a NS close to its critical mass collapsing to a black hole, and a companion star evolved out of the main sequence originating the supernova. The temporal coincidence between the GRB and the supernova phenomenon is explained in terms of the concept of ``induced'' gravitational collapse \cite{2001ApJ...555L.117R,Mosca_Orale}. There is also the distinct possibility of observing the young born NS out of the supernova (see e.g., Ref.~\refcite{Mosca_Orale} and references therein).

It has been often proposed that GRBs associated with Ib/c supernovas, at a smaller redshift $0.0085 < z < 0.168$ (see e.g. Ref.~\refcite{2006AIPC..836..367D} and references therein), form a different class, less luminous and possibly much more numerous than the high luminosity GRBs at higher redshift \cite{2006Natur.442.1011P,2004Natur.430..648S,2007ApJ...658L...5M,2006AIPC..836..367D}. Therefore they have been proposed to originate from a separate class of progenitors \cite{2007ApJ...662.1111L,2006ApJ...645L.113C}. In our model this is explained by the nature of the progenitor system leading to the formation of the black hole with the smallest possible mass: the one formed by the collapse of a just overcritical NS \cite{Mosca_Orale}.

The recent observation of GRB 060614 at $z=0.125$ without an associated supernova \cite{2006Natur.444.1050D,2007A&A...470..105M} gives strong support to our scenario, alternative to the collapsar model. Also in this case the progenitor of the GRB appears to be a binary system composed of two NSs or a NS and a white dwarf.

\subsection{Conclusions on GRB 060218}

GRB 060218 presents a variety of peculiarities, including its extremely large $T_{90}$ and its classification as an XRF. Nevertheless, a crucial point of our analysis is that we have successfully applied to this source our ``canonical GRB'' scenario.

Within our model there is no need for inserting GRB 060218 in a new class of GRBs, such as the XRFs, alternative to the ``canonical'' ones. This same point recently received strong observational support in the case of GRB 060218 \cite{2006ApJ...653L..81L} and a consensus by other models in the literature \cite{2007ApJ...654..385K}.

The anomalously long $T_{90}$ led us to infer a monotonic decrease in the CBM effective density giving the first clear evidence for the fragmentation of the fireshell, which indeed was predicted for values of the baryon loading $B > 10^{-2}$. For GRB 060218 there is no need within our model for a new or unidentified source such as a magnetar or a collapsar.

GRB 060218 is the first GRB associated with a supernova with complete coverage of data from the onset all the way up to $\sim 10^6$ s. This fact offers an unprecedented opportunity to verify theoretical models on such a GRB class. For example, GRB 060218 fulfills the Ref.~\refcite{2002A&A...390...81A} relation unlike other sources in its same class. This is particularly significant, since GRB 060218 is the only source in such a class to have an excellent data coverage without gaps. We are currently examining if the missing data in the other sources of such a class may have a prominent role in their non-fulfillment of the Ref.~\refcite{2002A&A...390...81A} relation\cite{2006MNRAS.372.1699G}.

\section{Application to GRB 970228: the appearance of ``fake'' short GRBs}\label{sec970228}

GRB 970228 was detected by the Gamma-Ray Burst Monitor (GRBM, $40$--$700$ keV) and Wide Field Cameras (WFC, $2$--$26$ keV) on board BeppoSAX on February $28.123620$ UT \cite{1998ApJ...493L..67F}. The burst prompt emission is characterized by an initial $5$ s strong pulse followed, after $30$ s, by a set of three additional pulses of decreasing intensity \cite{1998ApJ...493L..67F}. Eight hours after the initial detection, the NFIs on board BeppoSAX were pointed at the burst location for a first target of opportunity observation and a new X-ray source was detected in the GRB error box: this is the first ``afterglow'' ever detected \cite{1997Natur.387..783C}. A fading optical transient has been identified in a position consistent with the X-ray transient \cite{1997Natur.386..686V}, coincident with a faint galaxy with redshift $z=0.695$ \cite{2001ApJ...554..678B}. Further observations by the Hubble Space Telescope clearly showed that the optical counterpart was located in the outskirts of a late-type galaxy with an irregular morphology \cite{1997Natur.387R.476S}.

The BeppoSAX observations of GRB 970228 prompt emission revealed a discontinuity in the spectral index between the end of the first pulse and the beginning of the three additional ones \cite{1997Natur.387..783C,1998ApJ...493L..67F,2000ApJS..127...59F}. The spectrum during the first $3$ s of the second pulse is significantly harder than during the last part of the first pulse \cite{1998ApJ...493L..67F,2000ApJS..127...59F}, while the spectrum of the last three pulses appear to be consistent with the late X-ray afterglow \cite{1998ApJ...493L..67F,2000ApJS..127...59F}. This was soon recognized by Ref.~\refcite{1998ApJ...493L..67F,2000ApJS..127...59F} as pointing to an emission mechanism producing the X-ray afterglow already taking place after the first pulse.

The simultaneous occurrence of an afterglow with total time-integrated luminosity larger than the P-GRB one, but with a smaller peak luminosity, is indeed explainable in terms of a peculiarly small average value of the CBM density and not due to the intrinsic nature of the source. In this sense, GRBs belonging to this class are only ``fake'' short GRBs. We show that GRB 970228 is a very clear example of this situation. We identify the initial spikelike emission with the P-GRB, and the late soft bump with the peak of the afterglow. GRB 970228 shares the same morphology and observational features with the sources analyzed by Ref.~\refcite{2006ApJ...643..266N} as well as with e.g. GRB 050709 \cite{2005Natur.437..855V}, GRB 050724 \cite{2006A&A...454..113C} and GRB 060614 \cite{2006Natur.444.1044G}. Therefore, we propose GRB 970228 as a prototype for this new GRB class.

\subsection{The analysis of GRB 970228 prompt emission}\label{theo}

\begin{figure}
\includegraphics[width=\hsize,clip]{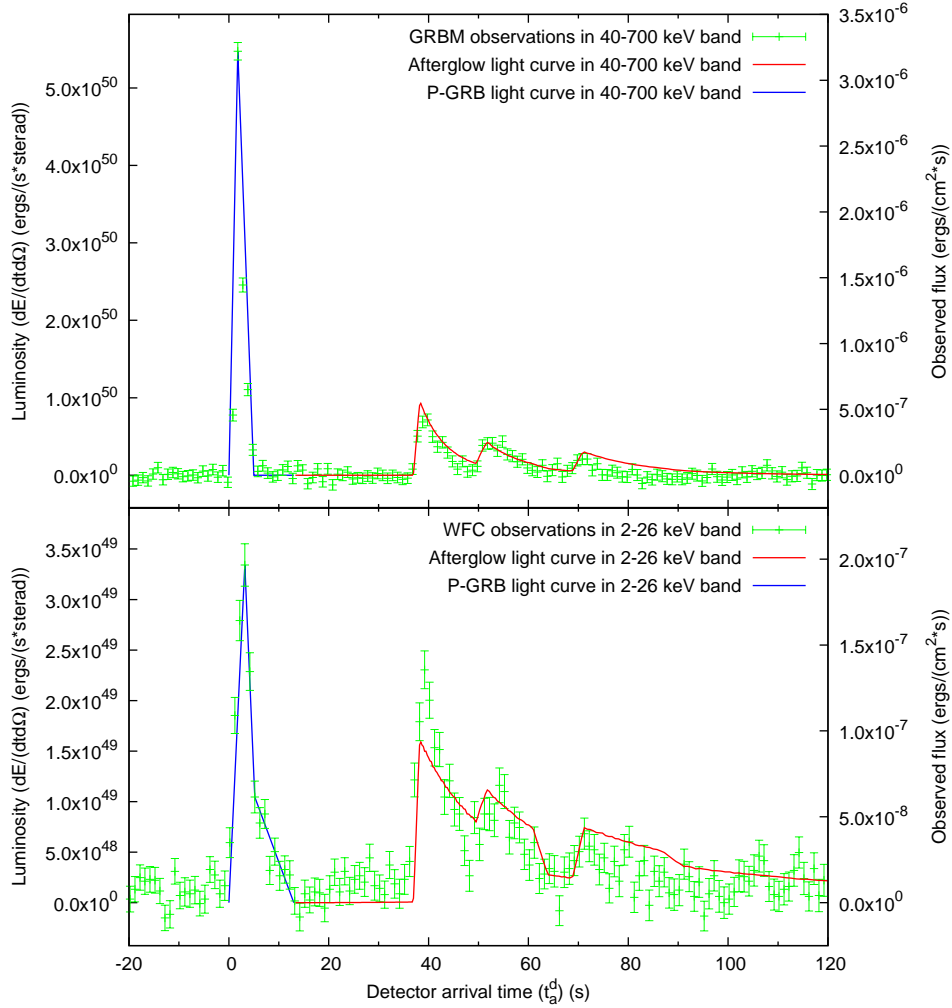}
\caption{The ``canonical GRB'' light curve theoretically computed for the prompt emission of GRB 970228. BeppoSAX GRBM ($40$--$700$ keV, above) and WFC ($2$--$26$ keV, below) light curves (data points) are compared with the afterglow peak theoretical ones (solid lines). The onset of the afterglow coincides with the end of the P-GRB (represented qualitatively by the dotted lines). For this source we have $B\simeq 5.0\times 10^{-3}$ and $\langle n_{cbm} \rangle \sim 10^{-3}$ particles/cm$^3$.}
\label{970228_fit_prompt}
\end{figure}

\begin{figure}
\includegraphics[width=\hsize,clip]{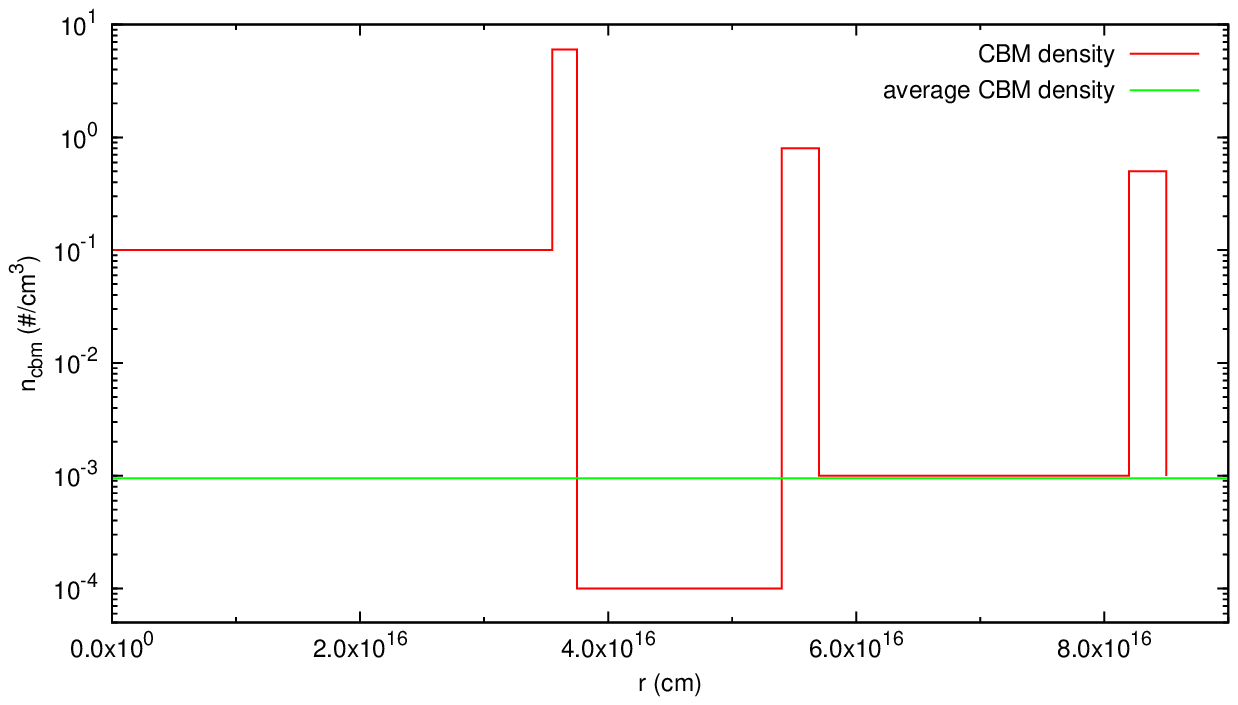}
\caption{The CBM density profile we assumed to reproduce the last three pulses of the GRB 970228 prompt emission (red line), together with its average value $\langle n_{cbm} \rangle = 9.5\times 10^{-4}$ particles/cm$^3$ (green line).}
\label{mask}
\end{figure}

In Fig.~\ref{970228_fit_prompt} we present the theoretical fit of BeppoSAX GRBM ($40$--$700$ keV) and WFC ($2$--$26$ keV) light curves of GRB 970228 prompt emission \cite{1998ApJ...493L..67F}. Within our ``canonical GRB'' scenario we identify the first main pulse with the P-GRB and the three additional pulses with the afterglow peak emission, consistent with the above mentioned observations by Ref.~\refcite{1997Natur.387..783C} and Ref.~\refcite{1998ApJ...493L..67F}.  The last three such pulses have been reproduced assuming three overdense spherical CBM regions (see Fig.~\ref{mask}) with very good agreement (see Fig.~\ref{970228_fit_prompt}).

We therefore obtain for the two parameters characterizing the source in our model $E_{e^\pm}^{tot}=1.45\times 10^{54}$ erg and $B = 5.0\times 10^{-3}$. This implies an initial $e^\pm$ plasma created between the radii $r_1 = 3.52\times10^7$ cm and $r_2 = 4.87\times10^8$ cm with a total number of $e^{\pm}$ pairs $N_{e^\pm} = 1.6\times 10^{59}$ and an initial temperature $T = 1.7$ MeV. The theoretically estimated total isotropic energy emitted in the P-GRB is $E_{P-GRB}=1.1\% E_{e^\pm}^{tot}=1.54 \times 10^{52}$ erg, in excellent agreement with the one observed in the first main pulse ($E_{P-GRB}^{obs} \sim 1.5 \times 10^{52}$ erg in $2-700$ keV energy band, see Fig.~\ref{970228_fit_prompt}), as expected due to their identification. After the transparency point at $r_0 = 4.37\times 10^{14}$ cm from the progenitor, the initial Lorentz gamma factor of the fireshell is $\gamma_0 = 199$. On average, during the afterglow peak emission phase we have for the CBM $\langle {\cal R} \rangle = 1.5\times 10^{-7}$ and $\langle n_{cbm} \rangle = 9.5\times 10^{-4}$ particles/cm$^3$. This very low average value for the CBM density is compatible with the observed occurrence of GRB 970228 in its host galaxy's halo \cite{1997Natur.387R.476S,1997Natur.386..686V,2006MNRAS.367L..42P} and it is crucial in explaining the light curve behavior.

The values of $E_{e^\pm}^{tot}$ and $B$ we determined are univocally fixed by two tight constraints. The first one is the total energy emitted by the source all the way up to the latest afterglow phases (i.e., up to $\sim 10^6$ s). The second one is the ratio between the total time-integrated luminosity of the P-GRB and the corresponding one of the entire afterglow (i.e., up to $\sim 10^6$ s). In particular, in GRB 970228 such a ratio turns out to be $\sim 1.1\%$ (see Fig.~\ref{bcross_sorgenti}). However, the P-GRB peak luminosity actually turns out to be much more intense than the afterglow one (see Fig.~\ref{970228_fit_prompt}). This is due to the very low average value of the CBM density $\langle n_{cbm} \rangle = 9.5\times 10^{-4}$ particles/cm$^3$, which produces a less intense afterglow emission. Since the afterglow total time-integrated luminosity is fixed, such a less intense emission lasts longer than what we would expect for an average density $\langle n_{cbm} \rangle \sim 1$ particles/cm$^3$.

\subsection{Rescaling the CBM density}\label{rescale}

\begin{figure}
\includegraphics[width=\hsize,clip]{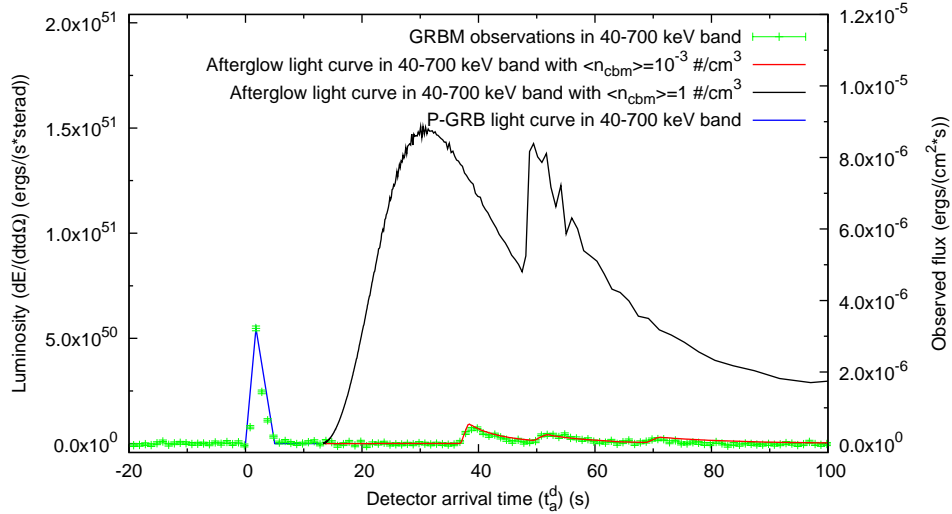}
\caption{The theoretical fit of the BeppoSAX GRBM observations (red line, see Fig.~\ref{970228_fit_prompt}) is compared with the afterglow light curve in the $40$--$700$ keV energy band obtained rescaling the CBM density to $\langle n_{cbm} \rangle = 1$ particle/cm$^3$ keeping constant its shape and the values of the fundamental parameters of the theory $E_{e^\pm}^{tot}$ and $B$ (black line). The P-GRB duration and luminosity (blue line), depending only on $E_{e^\pm}^{tot}$ and $B$, are not affected by this process of rescaling the CBM density.}
\label{picco_n=1}
\end{figure}

We present now an explicit example in order to probe the crucial role of the average CBM density in explaining the relative intensities of the P-GRB and of the afterglow peak in GRB 970228. We keep fixed the basic parameters of the source, namely the total energy $E_{e^\pm}^{tot}$ and the baryon loading $B$, therefore keeping fixed the P-GRB and the afterglow total time-integrated luminosities. Then we rescale the CBM density profile given in Fig.~\ref{mask} by a constant numerical factor in order to raise its average value to the standard one $\langle n_{cbm} \rangle = 1$ particle/cm$^3$. We then compute the corresponding light curve, shown in Fig.~\ref{picco_n=1}.

We notice a clear enhancement of the afterglow peak luminosity with respect to the P-GRB one in comparison with the fit of the observational data presented in Fig.~\ref{970228_fit_prompt}. The two light curves actually cross at $t_a^d \simeq 1.8\times 10^4$ s since their total time-integrated luminosities must be the same. The GRB ``rescaled'' to $\langle n_{cbm} \rangle = 1$ particle/cm$^3$ appears to be totally similar to, e.g., GRB 050315 \cite{2006ApJ...645L.109R} and GRB 991216 \cite{2003AIPC..668...16R,2004IJMPD..13..843R,2005AIPC..782...42R}.

It is appropriate to emphasize that, although the two underlying CBM density profiles differ by a constant numerical factor, the two afterglow light curves in Fig.~\ref{picco_n=1} do not. This is because the absolute value of the CBM density at each point affects in a nonlinear way all the following evolution of the fireshell due to the feedback on its dynamics \cite{2005ApJ...633L..13B}. Moreover, the shape of the surfaces of equal arrival time of the photons at the detector (EQTS) is strongly elongated along the line of sight \cite{2005ApJ...620L..23B}. Therefore photons coming from the same CBM density region are observed over a very long arrival time interval.

\subsection{GRB 970228 and the Amati relation}\label{amati_rel}

We turn now to the ``Amati relation'' \cite{2002A&A...390...81A,2006MNRAS.372..233A} between the isotropic equivalent energy emitted in the prompt emission $E_{iso}$ and the peak energy of the corresponding time-integrated spectrum $E_{p,i}$ in the source rest frame. It has been shown by Ref.~\refcite{2002A&A...390...81A,2006MNRAS.372..233A} that this correlation holds for almost all the ``long'' GRBs which have a redshift and an $E_{p,i}$ measured, but not for the ones classified as ``short'' \cite{2006MNRAS.372..233A}. If we focus on the ``fake'' short GRBs, namely the GRBs belonging to this new class, at least in one case (GRB 050724)\cite{2006A&A...454..113C} it has been shown that the correlation is recovered if also the extended emission is considered \cite{amatiIK}. 

It clearly follows from our treatment that for the ``canonical GRBs'' with large values of the baryon loading and high $\left\langle n_{cbm}\right\rangle$, which presumably are most of the GRBs for which the correlation holds, the leading contribution to the prompt emission is the afterglow peak emission. The case of the ``fake'' short GRBs is completely different: it is crucial to consider separately the two components since the P-GRB contribution to the prompt emission in this case is significant.

\begin{figure}
\includegraphics[width=\hsize,clip]{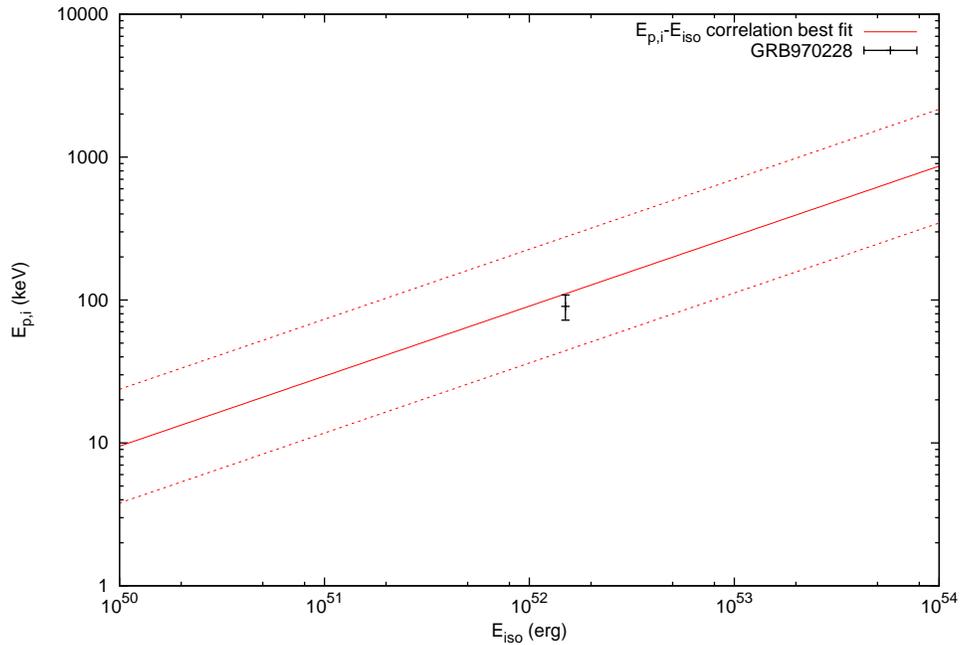}
\caption{The estimated values for $E_{p,i}$ and $E_{iso}$ obtained by our analysis (black dot) compared with the ``Amati relation'' \cite{2002A&A...390...81A}: the solid line is the best fitting power law \cite{2006MNRAS.372..233A} and the dashed lines delimit the region corresponding to a vertical logarithmic deviation of $0.4$ \cite{2006MNRAS.372..233A}. The uncertainty in the theoretical estimated value for $E_{p,i}$ has been assumed conservatively as $20\%$.}
\label{amati}
\end{figure}

To test this scenario, we evaluated from our fit of GRB 970228 $E_{iso}$ and $E_{p,i}$ only for the afterglow peak emission component, i.e., from $t_a^d= 37$ s to $t_a^d= 81.6$ s. We found an isotropic energy emitted in the $2$--$400$ keV energy band $E_{iso}=1.5 \times 10^{52}$ erg, and $E_{p,i}=90.3$ keV. As it is clearly shown in Fig.~\ref{amati}, the sole afterglow component of GRB 970228 prompt emission is in perfect agreement with the Amati relation. If this behavior is confirmed for other GRBs belonging to this new class, this will reinforce our identification of the ``fake'' short GRBs. This result will also provide a theoretical explanation for the apparent absence of such a correlation for the initial spikelike component in the different nature of the P-GRB.

\subsection{Conclusions on GRB 970228}\label{concl}

We conclude that GRB 970228 is a ``canonical GRB'' with a large value of the baryon loading quite near to the maximum $B \sim 10^{-2}$ (see Fig.~\ref{bcross_sorgenti}). The difference with e.g. GRB 050315 \cite{2006ApJ...645L.109R} or GRB 991216 \cite{2003AIPC..668...16R,2004IJMPD..13..843R, 2005AIPC..782...42R} is the low average value of the CBM density $\langle n_{cbm} \rangle \sim 10^{-3}$ particles/cm$^3$ which deflates the afterglow peak luminosity. Hence, the predominance of the P-GRB, coincident with the initial spikelike emission, over the afterglow is just apparent: $98.9\%$ of the total time-integrated luminosity is indeed in the afterglow component. Such a low average CBM density is consistent with the occurrence of GRB 970228 in the galactic halo of its host galaxy \cite{1997Natur.387R.476S,1997Natur.386..686V}, where lower CBM densities have to be expected \cite{2006MNRAS.367L..42P}.

We propose GRB 970228 as the prototype for the new class of GRBs comprising GRB 060614 and the GRBs analyzed by Norris \& Bonnell\cite{2006ApJ...643..266N}. We naturally explain the hardness and the absence of spectral lag in the initial spikelike emission with the physics of the P-GRB originating from the gravitational collapse leading to the black hole formation. The hard-to-soft behavior in the afterglow is also naturally explained by the physics of the relativistic fireshell interacting with the CBM, clearly evidenced in GRB 031203 \cite{2005ApJ...634L..29B} and in GRB 050315 \cite{2006ApJ...645L.109R}. Also justified is the applicability of the Amati relation to the sole afterglow component \cite{2006MNRAS.372..233A,amatiIK}.

This class of GRBs with $z \sim 0.4$ appears to be nearer than the other GRBs detected by \emph{Swift} ($z \sim 2.3$)\cite{2006astro.ph.10408G}. This may be explained by the afterglow peak luminosity deflation. The absence of a jet break in those afterglows has been pointed out \cite{2006A&A...454..113C,2006A&A...454L.123W}, consistently with our spherically symmetric approach. Their association with non-star-forming host galaxies appears to be consistent with the merging of a compact object binary \cite{2005Natur.438..994B,2005Natur.437..845F}. It is here appropriate, however, to caution on this conclusion, since the association of GRB 060614 and GRB 970228 with the explosion of massive stars is not excluded \cite{2006Natur.444.1050D,2000ApJ...536..185G}.

Most of the sources of this class appear indeed not to be related to bright ``Hypernovae'', to be in the outskirts of their host galaxies \cite{2005Natur.437..845F} and a consistent fraction of them are in galaxy clusters with CBM densities $\langle n_{cbm} \rangle \sim 10^{-3}$ particles/cm$^3$ \cite{2003ApJ...586..135L,2007ApJ...660..496B}. This suggests a spiraling out binary nature of their progenitor systems \cite{KMG11} made of neutron stars and/or white dwarfs leading to a black hole formation.

Moreover, we verified the applicability of the Amati relation to the sole afterglow component in GRB 970228 prompt emission, in analogy with what happens for some of the GRBs belonging to this new class. In fact it has been shown by Ref.~\refcite{2006MNRAS.372..233A,amatiIK} that the ``fake'' short GRBs do not fulfill the $E_{p,i}$--$E_{iso}$ correlation when the sole spikelike emission is considered, while they do if the long soft bump is included. Since the spikelike emission and the soft bump contributions are comparable, it is natural to expect that the soft bump alone will fulfill the correlation as well.

Within our ``canonical GRB'' scenario the sharp distinction between the P-GRB and the afterglow provide a natural explanation for the observational features of the two contributions. We naturally explain the hardness and the absence of spectral lag in the initial spikelike emission with the physics of the P-GRB originating from the gravitational collapse leading to the black hole formation. The hard-to-soft behavior in the afterglow is also naturally explained by the physics of the relativistic fireshell interacting with the CBM, clearly evidenced in GRB 031203 \cite{2005ApJ...634L..29B} and in GRB 050315 \cite{2006ApJ...645L.109R}. Therefore, we expect naturally that the $E_{p,i}$--$E_{iso}$ correlation holds only for the afterglow component and not for the P-GRB. Actually we find that the correlation is recovered for the afterglow peak emission of GRB 970228.

In the original work of Ref.~\refcite{2002A&A...390...81A,2006MNRAS.372..233A} only the prompt emission is considered and not the late afterglow one. In our theoretical approach the afterglow peak emission contributes to the prompt emission and continues up to the latest GRB emission. Hence, the meaningful procedure within our model to recover the Amati relation is to look at a correlation between the total isotropic energy and the peak of the time-integrated spectrum of the whole afterglow. A first attempt to obtain such a correlation has already been performed using GRB 050315 as a template, giving very satisfactory results (...).

\section{The GRB-Supernova Time Sequence (GSTS) paradigm: the concept of induced gravitational collapse}\label{sec3par}

\begin{table}
\tbl{}
{
\begin{tabular}{lcccccrlcc}
\toprule
GRB/SN & $E_{e^\pm}^{tot}$ & ${E_{SN}^{bolom}}^g$ & ${E_{SN}^{kin}}^h$ & $E_{URCA}^i$ & $B$ & $\gamma_\circ$ & $z^j$ & $S_X/S_\gamma^k$ &  $\begin{array}{c}\left\langle  n_{cbm} \right\rangle\\(\mathrm{\#}/\mathrm{cm}^3)\end{array}$\\
\colrule
060218/2006aj & $1.8\times 10^{50}$  & $9.2\times 10^{48}$ & $2.0\times10^{51}$ & $?$ & $1.0\times 10^{-2}$  & $99$ & $0.033$ & $3.54$(XRF) & $1.0$\\ 
980425/1998bw$^a$ & $1.2\times 10^{48}$ & $2.3\times 10^{49}$ &  $1.0\times 10^{52}$ & $3\times 10^{48}$ & $7.7\times 10^{-3}$  & $124$     & $0.0085$ & $0.58$ (XRR) & $2.5\times 10^{-2}$ \\
031203/2003lw$^b$ & $1.8\times 10^{50}$  & $3.1\times 10^{49}$ & $1.5\times10^{52}$ &  $2\times10^{49}$ & $7.4\times 10^{-3}$  & $133$     & $0.105$ & $0.49$(XRR/XRF) & $0.3$\\
030329/2003dh$^c$ & $2.1\times 10^{52}$  & $1.8\times 10^{49}$ & $8.0\times10^{51}$ & $3\times 10^{48}$ & $4.8\times 10^{-3}$  & $206$     & $0.168$   & $0.56$(XRR)     & $1.0$  \\ 
050315$^d$        & $1.5\times 10^{53}$ & & & & $4.5\times 10^{-3}$  & $217$     & $1.949$     & $1.58$(XRF) &   $0.8$\\ 
970228/?$^e$      & $1.4\times 10^{54}$ & & & & $5.0\times 10^{-3}$  & $326$     & $0.695$      & GRB &        $1.0\times 10^{-3}$\\ 
991216$^f$        & $4.8\times 10^{53}$ & & & & $2.7\times 10^{-3}$  & $340$     & $1.0$ & GRB & $3.0$\\ 
\botrule
\end{tabular}
}
\begin{tabnote}
see: a) Ref.~\refcite{Mosca_Orale}; b) Ref.~\refcite{2005ApJ...634L..29B}; c) Ref.~\refcite{2004AIPC..727..312B}; d) Ref.~\refcite{2006ApJ...645L.109R}; e) Ref.~\refcite{2007A&A...474L..13B}; f) Ref.~\refcite{2005AIPC..782...42R}; g) see Ref.~\refcite{2007ApJ...654..385K}; h) Mazzali, P., private communication at MG11 meeting in Berlin, July 2006; i) evaluated fitting the URCAs with a power law followed by an exponentially decaying part; j) respectively Ref.~\refcite{2006ApJ...643L..99M}, Ref.~\refcite{1998Natur.395..670G}, Ref.~\refcite{2004ApJ...611..200P}, Ref.~\refcite{2003ApJ...599.1223G}, Ref.~\refcite{2005GCN..3101....1K}, Ref.~\refcite{2001GCN..1152....1I}, Ref.~\refcite{2001ApJ...554..678B}, Ref.~\refcite{2001GCN..1147....1P}; k) respectively Ref.~\refcite{2006GCN..4776....1K}, Ref.~\refcite{2006GCN..4822....1S}, XRR is considered in Ref.~\refcite{2006GCN..4776....1K}, while XRF as suggested by Ref.~\refcite{2004ApJ...605L.101W}, Ref.~\refcite{2006GCN..4776....1K}, Ref.~\refcite{2006ApJ...638..920V}, Ref.~\refcite{2001ApJ...554..678B}.
\end{tabnote}
\label{tab1}
\end{table}

Following the result of Ref.~\refcite{1998Natur.395..670G} who discovered the temporal coincidence of GRB 980425 and SN 1998bw, the association of other nearby GRBs with Type Ib/c SNe has been spectroscopically confirmed (see Tab.~\ref{tab1}). The approaches in the current literature have attempted to explain both the supernova and the GRB as two aspects of the same astrophysical phenomenon. Hence, GRBs have been assumed to originate from a specially strong supernova process, a hypernova or a collapsar (see e.g. Ref.~\refcite{1998ApJ...494L..45P,1998Natur.395..663K,1998Natur.395..672I,2006ARA&A..44..507W} and references therein). Both these possibilities imply very dense and strongly wind-like CBM structure.

\begin{figure}
\includegraphics[width=\hsize,clip]{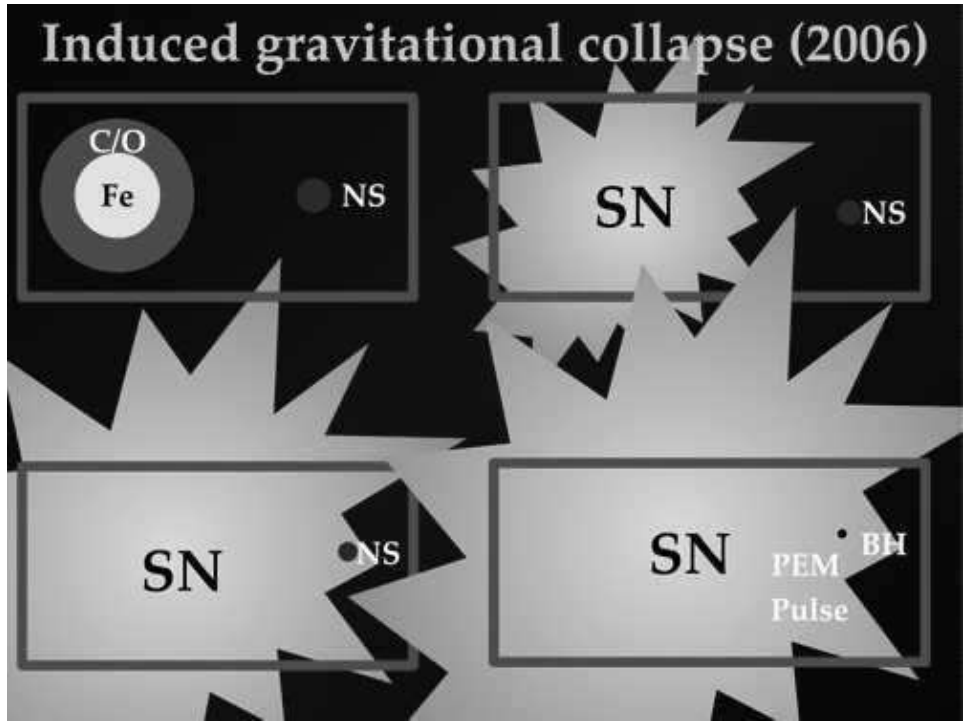}
\caption{A possible process of gravitational collapse to a black hole ``induced'' by the Ib/c supernova on a companion neutron star in a close binary system.}
\label{IndColl06}
\end{figure}

In our model we assumed that the GRB consistently originates from the gravitational collapse to a black hole. The supernova follows instead the complex pattern of the final evolution of a massive star, possibly leading to a neutron star or to a complete explosion but never to a black hole. The temporal coincidence of the two phenomena, the supernova explosion and the GRB, have then to be explained by the novel concept of ``induced gravitational collapse'', introduced in Ref.~\refcite{2001ApJ...555L.117R}. We have to recognize that still today we do not have a precise description of how this process of ``induced gravitational collapse'' occurs. At this stage, it is more a framework to be implemented by additional theoretical work and observations. Two different possible scenarios have been outlined. In the first version \cite{2001ApJ...555L.117R} we have considered the possibility that the GRBs may have caused the trigger of the supernova event. For the occurrence of this scenario, the companion star had to be in a very special phase of its thermonuclear evolution and three different possibilities were considered:
\begin{enumerate}
\item A white dwarf, close to its critical mass. In this case, the GRB may implode the star enough to ignite thermonuclear burning.
\item The GRB enhances in an iron-silicon core the capture of the electrons on the iron nuclei and consequently decreases the Fermi energy of the core, leading to the onset of gravitational instability.
\item The pressure waves of the GRB may trigger a massive and instantaneous nuclear burning process leading to the collapse.
\end{enumerate}
More recently \cite{Mosca_Orale}, a quite different possibility has been envisaged: the supernova, originating from a very evolved core, undergoes explosion in presence of a companion neutron star with a mass close to its critical one. The supernova blast wave may then trigger the collapse of the companion neutron star to a black hole and the emission of the GRB (see Fig.~\ref{IndColl06}). It is clear that, in both scenarios, the GRB and the supernova occur in a binary system.

There are many reasons to propose this concept of ``induced gravitational collapse'':
\begin{enumerate}
\item The fact that GRBs occur from the gravitational collapse to a black hole.
\item The fact that CBM density for the occurrence of GRBs is inferred from the analysis of the afterglow to be on the order of $1$ particle/cm$^3$ (see Tab.~\ref{tab1}) except for few cases (see e.g. sections \ref{secGRB-SN} and \ref{sec970228}). This implies that the process of collapse has occurred in a region of space filled with a very little amount of baryonic matter. The only significant contribution to the baryonic matter component in this process is the one represented by the fireshell baryon loading, which is anyway constrained by the inequality $B \le 10^{-2}$.
\item The fact that the energetics of the GRBs associated with supernovas appears to be particularly weak is consistent with the energy originating from the gravitational collapse to the smallest possible black hole: the one with mass $M$ just over the neutron star critical mass.
\end{enumerate}
There are also at work very clearly selection effects among the association between supernovas and GRBs:
\begin{enumerate}
\item Many type Ib/c supernovas exist without an associated GRB \cite{2007ApJ...657L..73G}.
\item Some GRBs do not show the presence of an associated supernova, although they are close enough for the supernova to be observed \cite{2006Natur.444.1050D}.
\item The presence in all observed GRB-supernova systems of an URCA source, a peculiar late time X-ray emission. These URCA sources have been identified and presented for the first time at the Tenth Marcel Grossmann meeting held in Rio de Janeiro (Brazil) in the Village of Urca, and named consequently. They appear to be one of the most novel issues still to be understood about GRBs. We will return on these aspects in the section \ref{secURCA}.
\end{enumerate}

The issue of triggering the gravitational collapse instability induced by the GRB on the progenitor star of the supernova or, vice versa, by the supernova on the progenitor star of the GRB needs accurate timing. The occurrence of new nuclear physics and/or relativistic phenomena is very likely. The general relativistic instability induced on a nearby star by the formation of a black hole needs some very basic new developments.

Only a very preliminary work exists on this subject, by Jim Wilson and his collaborators \cite{2005tmgm.meet.1802M}. The reason for the complexity in answering such a question is simply stated: unlike the majority of theoretical work on black holes and binary X-ray sources, which deals mainly with one-body black hole solutions in the Newtonian field of a companion star, we now have to address a many-body problem in general relativity. We are starting in these days to reconsider, in this framework, some classic work by Ref.~\refcite{f21,1973PhRvD...8.3259H,1947PhRv...72..390M,p47,1973PhRvD...7.2874P,2007PhRvD..75d4012B,2007PhLA..360..515B} which may lead to a new understanding of general relativistic effects in these many-body systems. This is a welcome effect of GRBs on the conceptual development of general relativity.

\section{Some unexplained features in the GRB-supernova association: the URCA sources}\label{secGRB-SN}

Models of GRBs based on a single source (the ``collapsar'') generating both the supernova and the GRB abounds in the literature \cite{2006ARA&A..44..507W}. Since the two phenomena are qualitatively very different, in our approach we have emphasized the concept of induced gravitational collapse, which occurs strictly in a binary system. The supernova originates from a star evolved out of the main sequence and the GRB from the collapse to a black hole. The concept of induced collapse implies at least two alternative scenarios. In the first, the GRB triggers a supernova explosion in the very last phase of the thermonuclear evolution of a companion star \cite{2001ApJ...555L.117R}. In the second, the early phases of the SN induce gravitational collapse of a companion neutron star to a black hole. Of course, in absence of a supernova, there is also the possibility that the collapse to a black hole, generating the GRB, occurs in a single star system or in the final collapse of a binary neutron star system. Still, in such a case there is also the possibility that the black hole progenitor is represented by a binary system composed by a white dwarf and/or a neutron star and/or a black hole in various combinations. What is most remarkable is that, following the ``uniqueness of the black hole'' \cite{RuKerr}, all these collapses lead to a common GRB independently of the nature of their progenitors.

Having obtained success in the fit of GRB 991216, as well as of GRB 031203 and GRB 050315 (see sections \ref{sec031203} and \ref{secJ}), we turn to the application of our theoretical analysis to the GRBs associated with supernovas. We start with GRB 980425 / SN 1998bw. We must, however, be cautious about the validity of this fit. From the available data of BeppoSAX, BATSE, XMM and Chandra, only the data of the prompt emission ($t_a^d < 10^2$ s) and of the latest afterglow phases ($t_a^d > 10^5$ s all the way to more than $10^8$ s!) were available. Our fit refers only to the prompt emission, as usually interpreted as the peak of the afterglow. The fit, therefore, represents an underestimate of the GRB 980425 total energy and in this sense it is not surprising that it does not fit the Ref.~\refcite{2002A&A...390...81A} relation. The latest afterglow emission, the URCA-1 emission, presents a different problematic which we will shortly address (see below).

\subsection{GRB 980425 / SN 1998bw / URCA-1}\label{980425}

\begin{figure}
\includegraphics[width=\hsize,clip]{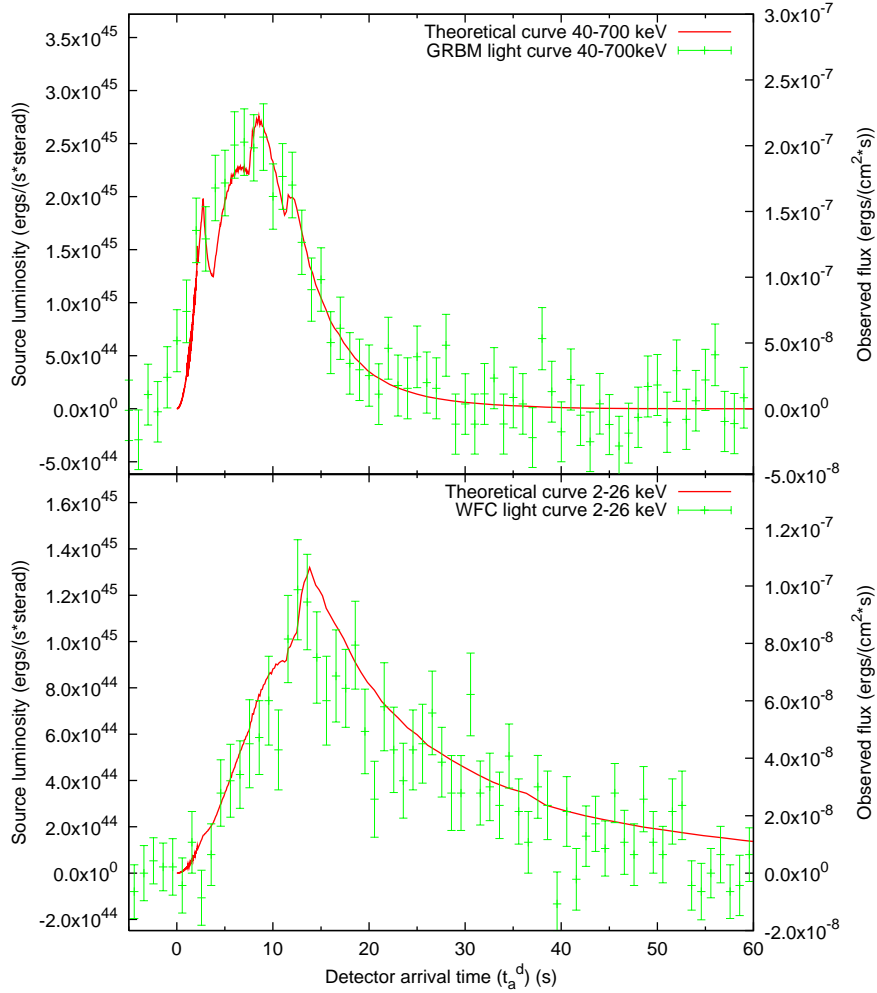}
\caption{Theoretical light curves of GRB 980425 prompt emission in the $40$--$700$ keV and $2$--$26$ keV energy bands (red line), compared with the observed data respectively from Beppo-SAX GRBM and WFC \cite{2000ApJ...536..778P,2000ApJS..127...59F}.}
\label{980425_picco}
\end{figure}

\begin{figure}
\includegraphics[width=\hsize,clip]{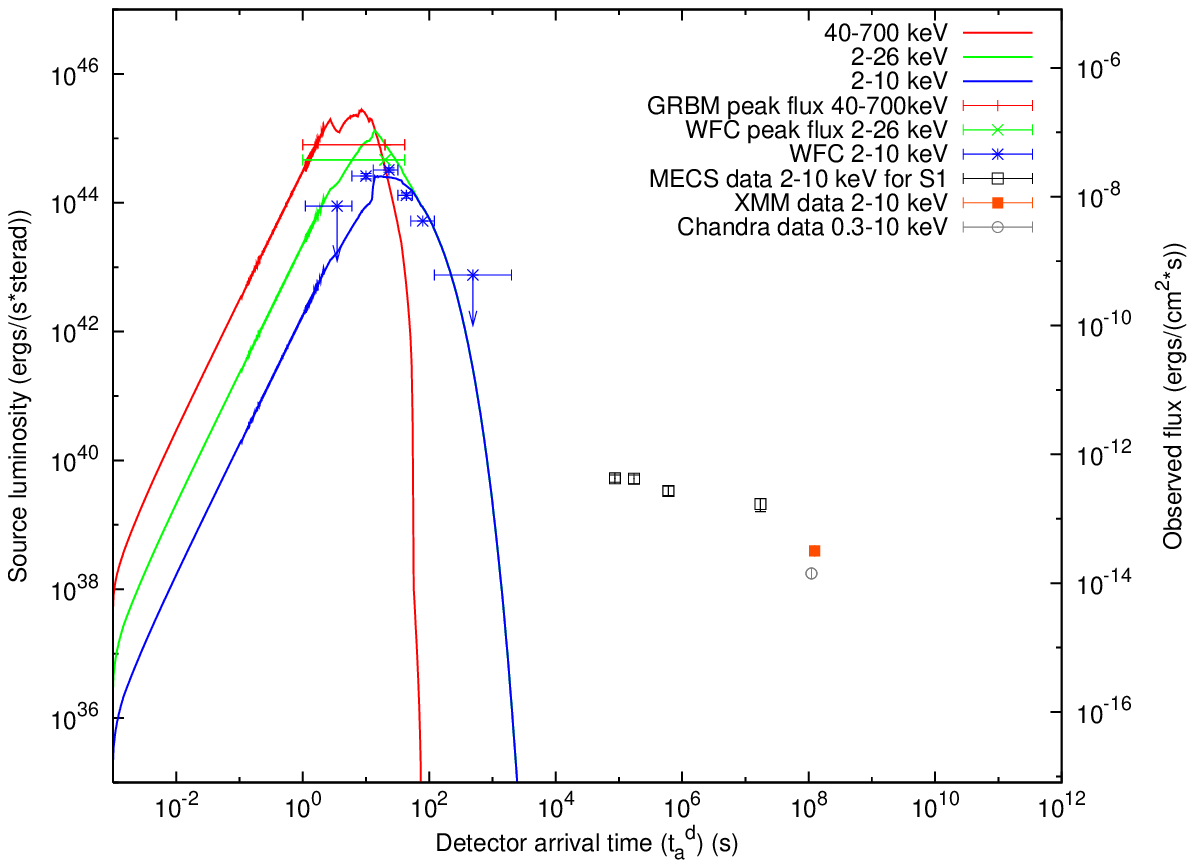}
\caption{Theoretical light curves of GRB 980425 in the $40$--$700$ keV (red line), $2$--$26$ keV (green line), $2$--$10$ keV (blue line) energy bands, represented together with URCA-1 observational data. All observations are by BeppoSAX \cite{2000ApJ...536..778P}, with the exception of the last two URCA-1 points, which is observed by XMM and Chandra \cite{2004AdSpR..34.2711P,2004ApJ...608..872K}.}
\label{980425_global}
\end{figure}

The best fit of the observational data of GRB 980425 \cite{2000ApJ...536..778P,2000ApJS..127...59F} leads to $E_{e^\pm}^{tot}=1.2\times10^{48}$ erg and $B = 7.7\times10^{-3}$. This implies an initial $e^\pm$ plasma with $N_{e^+e^-} = 3.6\times10^{53}$ and with an initial temperature $T = 1.2$ MeV. After the transparency point, the initial Lorentz gamma factor of the accelerated baryons is $\gamma_\circ = 124$. The variability of the luminosity, due to the inhomogeneities of the CBM, is characterized by a density contrast $\delta n / n \sim 10^{-1}$ on a length scale of $\Delta \sim 10^{14}$ cm. We determine the effective CBM parameters to be: $\langle n_{cbm} \rangle = 2.5\times 10^{-2}$ particle/$cm^3$ and $\langle \mathcal{R} \rangle = 1.2\times 10^{-8}$.

In Fig.~\ref{980425_picco} we test our specific theoretical assumptions comparing our theoretically computed light curves in the $40$--$700$ and $2$--$26$ keV energy bands with the observations by the BeppoSAX GRBM and WFC during the first $60$ s of data \cite{2000ApJ...536..778P,2000ApJS..127...59F}. The agreement between observations and theoretical predictions in Fig.~\ref{980425_picco} is very satisfactory.

In Fig.~\ref{980425_global} we summarize some of the problematics implicit in the old pre-\emph{Swift} era: data are missing in the crucial time interval between $60$ s and $10^5$ s, when the BeppoSAX NFI starts to point the GRB 980425 location. In this region we have assumed, for the effective CBM parameters, constant values inferred by the last observational data. Currently we are relaxing this condition, also in view of the interesting paper by Ref.~\refcite{2006MNRAS.372.1699G}. 

The follow-up of GRB 980425 with BeppoSAX NFI 10 hours, one week and 6 months after the event revealed the presence of an X-ray source consistent with SN1998bw \cite{2000ApJ...536..778P}, confirmed also by observations by XMM \cite{2004AdSpR..34.2711P} and Chandra \cite{2004ApJ...608..872K}. The S1 X-ray light curve shows a decay much slower than usual X-ray GRB afterglows \cite{2000ApJ...536..778P}. We then address to this peculiar X-ray emission as ``URCA-1'' (see section \ref{sec3par} and the following sections). In Fig.~\ref{urca123+GRB_full}A we represent the URCA-1 observations \cite{2000ApJ...536..778P,2004AdSpR..34.2711P,2004ApJ...608..872K}. The separation between the light curves of GRB 980425 in the $2$--$700$ keV energy band, of SN 1998bw in the optical band \cite{2007astro.ph..2472N,2006Natur.442.1011P}, and of the above mentioned URCA-1 observations is evident.

\subsection{GRB 030329 / SN 2003dh / URCA-2}\label{030329}

For GRB 030329 we have obtained \cite{2004AIPC..727..312B,2005tmgm.meet.2459B,Mosca_Orale} a total energy $E_{e^\pm}^{tot}=2.12\times10^{52}$ erg and a baryon loading $B = 4.8\times10^{-3}$. This implies an initial $e^\pm$ plasma with $N_{e^+e^-}=1.1\times10^{57}$ and with an initial temperature $T=2.1$ MeV. After the transparency point, the initial Lorentz gamma factor of the accelerated baryons is $\gamma_\circ = 206$. The effective CBM parameters are $\langle n_{cbm} \rangle = 2.0$ particle/$cm^3$ and $\langle \mathcal{R} \rangle = 2.8\times 10^{-9}$, with a density contrast $\delta n / n \sim 10$ on a length scale of $\Delta \sim 10^{14}$ cm. The resulting fit of the observations, both of the prompt phase and of the afterglow have been presented in Ref.~\refcite{2004AIPC..727..312B,2005tmgm.meet.2459B}. We compare in Fig.~\ref{urca123+GRB_full}B the light curves of GRB 030329 in the $2$--$400$ keV energy band, of SN 2003dh in the optical band \cite{2007astro.ph..2472N,2006Natur.442.1011P} and of the possible URCA-2 emission observed by XMM-EPIC in $2$--$10$ keV energy band \cite{2003A&A...409..983T,2004A&A...423..861T}.

\subsection{GRB 031203 / SN 2003lw / URCA-3}\label{031203}

\begin{figure}
\centering
\includegraphics[width=0.6\hsize,clip]{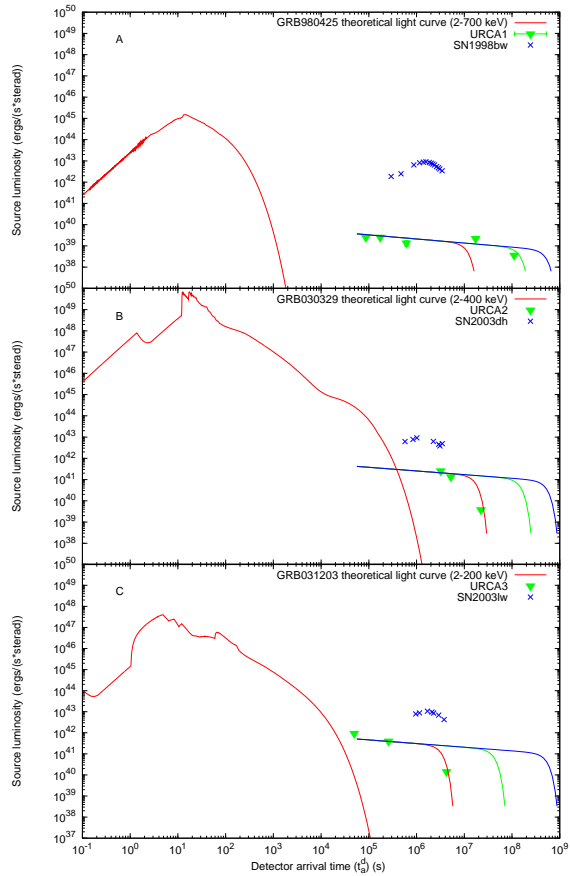}
\caption{Theoretically computed light curves of GRB 980425 in the $2$--$700$ keV band (A), of GRB 030329 in the $2$--$400$ keV band (B) and of GRB 031203 in the $2$--$200$ keV band (C) are represented, together with the URCA observational data and qualitative representative curves for their emission, fit with a power law followed by an exponentially decaying part. The luminosity of the supernovas in the $3000-24000$ {\AA} are also represented \cite{2007astro.ph..2472N,2006Natur.442.1011P}.}
\label{urca123+GRB_full}
\end{figure}

We will show in section \ref{sec031203} the detailed analysis of GRB 031203 which leads to a total energy $E_{e^\pm}^{tot}=1.85\times10^{50}$ erg and to a baryon loading $B = 7.4\times10^{-3}$. This implies an initial $e^\pm$ plasma with $N_{e^+e^-}=3.0\times 10^{55}$ and with an initial temperature $T=1.5$ MeV. After the transparency point, the initial Lorentz gamma factor of the accelerated baryons is $\gamma_\circ = 132$. The effective CBM parameters are $\langle n_{cbm} \rangle = 1.6\times 10^{-1}$ particle/$cm^3$ and $\langle \mathcal{R} \rangle = 3.7\times 10^{-9}$, with a density contrast $\delta n / n \sim 10$ on a length scale of $\Delta \sim 10^{15}$ cm. In Fig.~\ref{urca123+GRB_full}C we compare the light curves of GRB 031203 in the $2$--$200$ keV energy band, of SN 2003lw in the optical band \cite{2007astro.ph..2472N,2006Natur.442.1011P} and of the possible URCA-3 emission observed by XMM-EPIC in the $0.2$--$10$ keV energy band \cite{2004ApJ...605L.101W} and by Chandra in the $2$--$10$ keV energy band \cite{2004Natur.430..648S}.

\subsection{The GRB / SN / URCA connection}

\begin{table}
\tbl{}
{
\begin{tabular}{ccccccccccc}
\toprule
GRB & $\begin{array}{c}E_{e^\pm}^{tot}\\ \mathrm{(erg)}\end{array}$ & $B$ & $\gamma_0$ & $\begin{array}{c}E_{SN}^{bolom}\\ \mathrm{(erg)^a}\end{array}$ & $\begin{array}{c}E_{SN}^{kin}\\ \mathrm{(erg)^b}\end{array}$ & $\begin{array}{c}E_{URCA}\\ \mathrm{(erg)^c}\end{array}$ & $\displaystyle\frac{E_{e^\pm}^{tot}}{E_{URCA}}$ & $\displaystyle\frac{E_{SN}^{kin}}{E_{URCA}}$ & $\begin{array}{c}R_{NS}\\ \mathrm{(km)^d}\end{array}$ & $z^e$ \\
\colrule
980425 & $1.2\times 10^{48}$ & $7.7\times10^{-3}$ & $124$ & $2.3\times 10^{49}$ & $1.0\times 10^{52}$ & $3\times 10^{48}$ & $0.4$ & $1.7\times10^{4}$ & $ 8$ & $0.0085$\\
030329 & $2.1\times 10^{52}$ & $4.8\times10^{-3}$ & $206$ & $1.8\times 10^{49}$ & $8.0\times10^{51}$ & $3\times10^{49}$ & $6\times 10^{2}$ & $1.2\times10^{3}$ & $14$ & $0.1685$\\
031203 & $1.8\times 10^{50}$ & $7.4\times10^{-3}$ & $133$ & $3.1\times 10^{49}$ & $1.5\times10^{52}$ & $2\times10^{49}$ & $8.2$ & $3.0\times10^{3}$ & $20$ & $0.105$\\
060218 & $1.8\times 10^{50}$ & $1.0\times10^{-2}$ & $99$ & $9.2\times 10^{48}$ & $2.0\times10^{51}$ & $?$ & $?$ & $?$ & $?$ & $0.033$\\
\botrule
\end{tabular}
}
\begin{tabnote}
a) see Ref.~\refcite{2007ApJ...654..385K}; b) Mazzali, P., private communication at MG11 meeting in Berlin, July 2006; c) evaluated fitting the URCAs with a power law followed by an exponentially decaying part; d) evaluated assuming a mass of the neutron star $M=1.5 M_\odot$ and $T \sim 5$--$7$ keV in the source rest frame; e) see Ref.~\refcite{1998Natur.395..670G,2003ApJ...599.1223G,2004ApJ...611..200P,2006ApJ...643L..99M}.
\end{tabnote}
\label{tabella}
\end{table}

In Tab. \ref{tabella} we summarize the representative parameters of the above three GRB-supernova systems together with GRB 060218-SN 2006aj, including the very large kinetic energy observed in all supernovas \cite{mazzaliVen}. Some general conclusions on these weak GRBs at low redshift, associated to SN Ib/c, can be established on the grounds of our analysis:\\
{\bf 1)} From the detailed fit of their light curves, as well as their accurate spectral analysis, it follows that all the above GRB sources originate consistently from the formation of a black hole. This result extends to this low-energy GRB class at small cosmological redshift the applicability of our model, which now spans over a range of energy of six orders of magnitude from $10^{48}$ to $10^{54}$ ergs \cite{2003AIPC..668...16R,2004AdSpR..34.2715R,Mosca_Orale,2004AIPC..727..312B,2005tmgm.meet.2459B,2005ApJ...634L..29B,2006ApJ...645L.109R}. Distinctive of this class is the very high value of the baryon loading which in one case (GRB 060218)\cite{2007A&A...471L..29D} is very close to the maximum limit compatible with the dynamical stability of the adiabatic optically thick acceleration phase of the GRBs \cite{2000A&A...359..855R}. Correspondingly, the maximum Lorentz gamma factors are systematically smaller than the ones of the more energetic GRBs at large cosmological distances. This in turn implies the smoothness of the observed light curves in the so-called ``prompt phase''. The only exception to this is the case of GRB 030329.\\
{\bf 2)} The accurate fits of the GRBs allow us to also infer some general properties of the CBM. While the size of the clumps of the inhomogeneities is $\Delta \approx 10^{14}$ cm, the effective CBM average density is consistently smaller than in the case of more energetic GRBs: we have in fact $\langle n_{cbm} \rangle$ in the range between $\sim 10^{-6}$ particle/$cm^3$ (GRB 060218) and $\sim 10^{-1}$ particle/$cm^3$ (GRB 031203), while only in the case of GRB 030329 it is $\sim 2$ particle/$cm^3$.\\
{\bf 3)} Still within their weakness these four GRB sources present a large variability in their total energy: a factor $10^4$ between GRB 980425 and GRB 030329. Remarkably, the supernova emissions both in their very high kinetic energy and in their bolometric energy appear to be almost constant respectively $10^{52}$ erg and $10^{49}$ erg. The URCAs present also a remarkably steady behavior around a ``standard luminosity'' and a typical temporal evolution. The weakness in the energetics of GRB 980425 and GRB 031203, and the sizes of their dyadospheres, suggest that they originate from the formation of the smallest possible black hole, just over the critical mass of the neutron star (see Fig.~\ref{IndColl06}).

\subsection{URCA-1, URCA-2 and URCA-3}\label{secURCA}

We turn to the search for the nature of URCA-1, URCA-2 and URCA-3. These systems are not yet understood and may have an important role in the comprehension of the astrophysical scenario of GRB sources. It is important to perform additional observations in order to verify if the URCA sources are related to the black hole originating the GRB phenomenon or to the supernova. Even a single observation of an URCA source with a GRB in absence of a would prove their relation with the black hole formation. Such a result is today theoretically unexpected and would open new problematics in relativistic astrophysics and in the physics of black holes. Alternatively, even a single observation of an URCA source during the early expansion phase of a Type Ib/c supernova in absence of a GRB would prove the early expansion phases of the supernova remnants. In the case that none of such two conditions are fulfilled, then the URCA sources must be related to the GRBs occurring in presence of a supernova. In such a case, one of the possibilities would be that for the first time we are observing a newly born neutron star out of the supernova phenomenon unveiled by the GRB. This last possibility would offer new fundamental information about the outcome of the gravitational collapse, and especially about the equations of state at supranuclear densities and about a variety of fundamental issues of relativistic astrophysics of neutron stars.

The names of ``URCA-1'' and ``URCA-2'' for the peculiar late X-ray emission of GRB 980425 and GRB 030329 were given in the occasion of the Tenth Marcel Grossmann meeting held in Rio de Janeiro (Brazil) in the Village of Urca \cite{2005tmgm.meet..369R}. Their identification was made at that time and presented at that meeting. However, there are additional reasons for the choice of these names. Another important physical phenomenon was indeed introduced in 1941 in the same Village of Urca by George Gamow and Mario Schoenberg \cite{1941PhRv...59..539G}. The need for a rapid cooling process due to neutrino anti-neutrino emission in the process of gravitational collapse leading to the formation of a neutron star was there considered for the first time. It was Gamow who named this cooling as ``Urca process'' \cite{GamowBook-MyWorldlines}. Since then, a systematic analysis of the theory of neutron star cooling was advanced by Ref.~\refcite{1964PhDT........34T,1979PhR....56..237T,1966CaJPh..44.1863T,2002ApJ...571L.143T,1978pans.proc..448C}. The coming of age of X-ray observatories such as Einstein (1978-1981), EXOSAT (1983-1986), ROSAT (1990-1998), and the contemporary missions of Chandra and XMM-Newton since 1999 dramatically presented an observational situation establishing very embarrassing and stringent upper limits to the surface temperature of neutron stars in well known historical supernova remnants \cite{1987ApJ...313..718R}. For some remnants, notably SN 1006 and the Tycho supernova, the upper limits to the surface temperatures were significantly lower than the temperatures given by standard cooling times \cite{1987ApJ...313..718R}. Much of the theoretical work has been mainly directed, therefore, to find theoretical arguments in order to explain such low surface temperature $T_s \sim 0.5$--$1.0\times 10^6$ K --- embarrassingly low, when compared to the initial hot ($\sim 10^{11}$ K) birth of a neutron star in a supernova explosion \cite{1987ApJ...313..718R}. Some important contributions in this research have been presented by Ref.~\refcite{1988ApJ...329..339V,1991ApJS...75..449V,1986ApJ...307..178B,1994ApJ...425..802L,2004ARA&A..42..169Y}. The youngest neutron star to be searched for thermal emission has been the pulsar PSR J0205+6449 in 3C 58 \cite{2004ARA&A..42..169Y}, which is $820$ years old! Ref.~\refcite{2005esns.conf..117T} reported evidence for the detection of thermal emission from the crab nebula pulsar which is, again, $951$ years old.

URCA-1, URCA-2 and URCA-3 may explore a totally different regime: the X-ray emission possibly from a recently born neutron star in the first days -- months of its existence. The thermal emission from the young neutron star surface would in principle give information on the equations of state in the core at supranuclear densities and on the detailed mechanism of the formation of the neutron star itself with the related neutrino emission. It is also possible that the neutron star is initially fast rotating and its early emission could be dominated by the magnetospheric emission or by accretion processes from the remnant which would overshadow the thermal emission. A periodic signal related to the neutron star rotational period should in principle be observable in a close enough GRB-supernova system. In order to attract attention to this problematic, we have given in Tab.~\ref{tabella} an estimate of the corresponding neutron star radius for URCA-1, URCA-2 and URCA-3. It has been pointed out \cite{2000ApJ...536..778P} the different spectral properties between the GRBs and the URCAs. It would be also interesting to compare and contrast the spectra of all URCAs in order to evidence any analogy among them. Observations of a powerful URCA source on time scales of $0.1$--$10$ seconds would be highly desirable.

\section{Conclusions}

GRBs are giving the first clear evidence for the extraction of energy from black holes during the last phases of their formation process. This new form of energy is unprecedented in the Universe, both for its magnitude and its very high efficiency in transforming matter into radiation, which reaches the $50\%$ limit while the nuclear energy reaches efficiency of $2-3\%$ only. These sources, with their energy of $10^{54}$ ergs/pulse, dwarf the corresponding nuclear energy events with their energy of $\sim 10^{22}$ ergs/pulse.

We have shown how the quest for understanding the initial conditions leading to this new gravitational electromagnetic phenomenon has originated a new inquiry into the properties of nuclear matter in bulk which sheds new light on two classical physical problems: the heavy nuclei and neutron stars. Such an analysis follows a new conceptual unified approach of important heuristic content. It may well lead to the solution of yet unsolved problems in physics and astrophysics and on much different scales, such as the emission of the remnant in supernova phenomena on a macroscopic scale or the limits on the dimension of the constituent of elementary particles on a microscopic scale.

The richness of the experimental data obtained from GRBs, especially thanks to the Swift and INTEGRAL satellites, has been exemplified in a selected number of sources. This allowed the development and for the first time the testing of the theory of ultrarelativistic collisions of baryonic matter with a Lorentz gamma factor up to $\gamma\sim 300$ and involving scales extending up to a few light years. The interpretation of these observations needed the development of a highly self-consistent theoretical framework, which has been tested with high accuracy analyzing both spectra and luminosities in selected energy bands on unprecedented time scales ranging from few milliseconds all the way to $10^6$ seconds. In turn, the CBM structure has been analyzed using GRB light curves as its tomographic image. From all these analyses a ``canonical'' GRB scenario is emerging, quite independent on their enormous energy ranging from $10^{48}$ erg to $10^{54}$ erg. Such a ``canonical'' GRB scenario promises to be relevant for the future use of GRBs as cosmological standard candles.

As the comprehension of the GRB phenomenon progresses, so the astrophysical setting where they originate is further clarified. We have given evidence for the GRB observation originating in active star formation region, with $n_{cbm}\sim 1$ particle/cm$^3$, as well as in galactic halos, with $n_{cbm}\sim 10^{-3}$ particles/cm$^3$. There is also mounting evidence that all observed GRBs originate from a variety of binary systems. There are, in this respect, at least three different systems: 1) GRBs associated with supernova originating from an initial binary system formed by a massive star and a neutron star and evolving to a neutron star, 2) originating from the supernova event, and 3) a black hole, originating from the gravitational collapse of the neutron star. The occurrence of these GRBs is strictly linked to the process of induced gravitational collapse. These supernova-related GRBs are therefore the less energetic ones, being formed by the smallest possible black holes just over the critical mass of a neutron star. There are also much more energetic phenomena occurring both in the galactic halos (e.g. GRB 970228) and in star-forming regions (e.g. GRB 050315), which must definitely originate from the collapse of binary systems formed by two neutron stars or a white dwarf and a neutron star or two intermediate mass black holes.

\begin{figure}
\includegraphics[width=\hsize,clip]{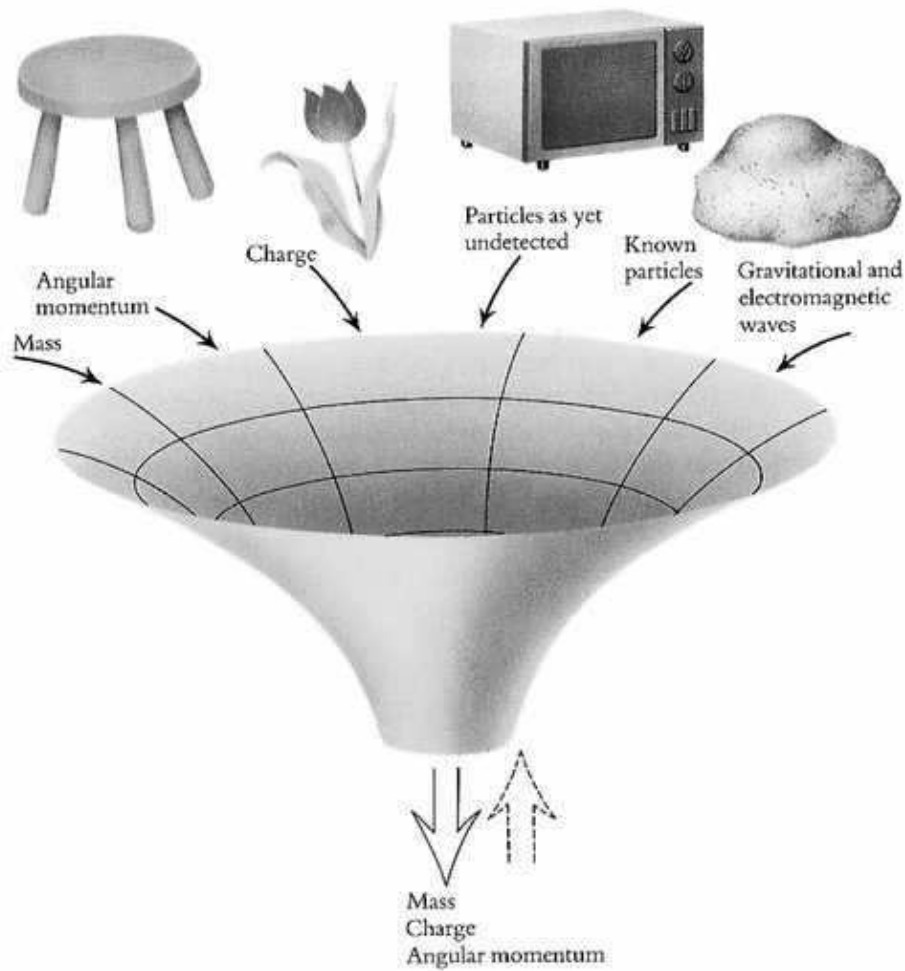}
\caption{The uniqueness of the black hole. Reproduced from the paper ``Introducing the black hole''\cite{1971PhT....24a..30R}.}
\label{grbuniq1}
\end{figure}

\begin{figure}
\includegraphics[width=\hsize,clip]{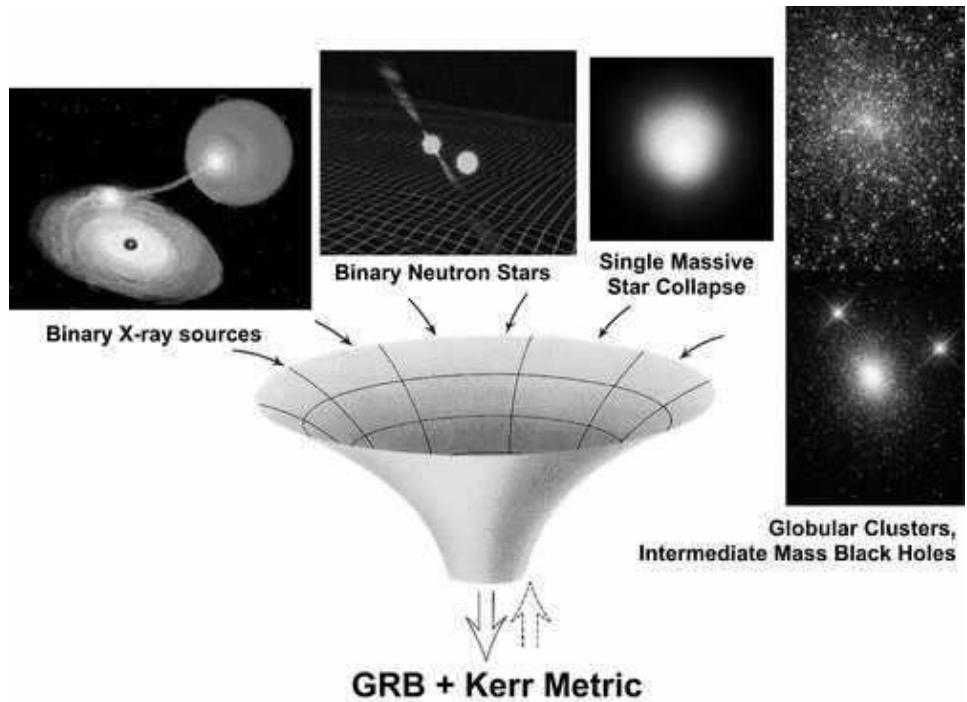}
\caption{The uniqueness of the black hole picture updated for GRB connection.}
\label{grbuniq2}
\end{figure}

What would appear to be {\em a priori} very surprising is that such a large variety of initial conditions leads to a ``canonical'' GRB scenario consistent over a range of energies spanning $6$ orders of magnitude. Again, the crucial explanation for this is due to the uniqueness of the black hole, which was represented in a thought-provoking form in the paper ``Introducing the black hole''\cite{1971PhT....24a..30R} which we reproduce here in Fig.~\ref{grbuniq1} and we also represent in its updated version for the GRB connection in Fig.~\ref{grbuniq2}.

\begin{thebibliography}{999}

\bibitem{2007AIPC..910...55R}
R.~{Ruffini}, M.~G. {Bernardini}, C.~L. {Bianco}, L.~{Caito}, P.~{Chardonnet},
  M.~G. {Dainotti}, F.~{Fraschetti}, R.~{Guida}, M.~{Rotondo},
  G.~{Vereshchagin}, L.~{Vitagliano} and S.-S. {Xue}, The blackholic energy and
  the canonical gamma-ray burst, in {\em XIIth Brazilian School of Cosmology
  and Gravitation\/},  eds. M.~{Novello} and S.~E. {Perez Bergliaffa}, American
  Institute of Physics Conference Series, Vol.~910 (June 2007).
  
\bibitem{1936PhRv...50..899A}
E.~Amaldi, E.~Fermi, {\em Physical Review}, {\bf 50}, 899 (1936).

\bibitem{1939NW.....27...11H}
O.~Hahn, F.~Strassmann, {\em Naturwissenschaften}, {\bf 27}, 11 (1939).

\bibitem{mf39}
L.~Meitner, O.R.~Frisch, {\em Nature}, {\bf 143}, 239 (1939).

\bibitem{1952AmJPh..20..536F}
E.~Fermi, {\em American Journal of Physics}, {\bf 20}, 536 (1952).

\bibitem{1949PhRv...75.1561F}
R.~P.~Feynman, N.~Metropolis and E.~Teller, {\em Physical Review}, {\bf 75}, 1561 (1949).

\bibitem{OppenheimerCase}
U.~A.~E. Commission, {\em A Nation's Security: the Case of Dr. J. Robert Oppenheimer} (Secker \& Warburg, 1955).

\bibitem{WignerBook}
G.~Birkhoff and E.~Wigner, {\em Nuclear Reactor Theory} (American Mathematical Society, 1961).

\bibitem{StalinBomb}
D.~Holloway, {\em Stalin and the Bomb} (Yale University Press, 1996).

\bibitem{1929ZPhy...53..157K}
O. Klein, \emph{Zeitschrift fur Physik}, \textbf{53}, 157 (1929).

\bibitem{1931ZPhy...69..742S}
F. Sauter, \emph{Zeitschrift fur Physik}, \textbf{69}, 742 (1931).

\bibitem{1935NW.....23..246E}
H. Euler, B. Kockel, \emph{Naturwissenschaften}, \textbf{23}, 246 (1935).

\bibitem{1936ZPhy...98..714H}
W. Heisenberg, H. Euler, \emph{Zeitschrift fur Physik}, \textbf{98}, 714 (1936).

\bibitem{w36}
V. Weisskopf, \emph{Kgl. Danske Viden. Selskab}, \textbf{14}, 6 (1936).

\bibitem{1934ZPhy...89...27W}
V. Weisskopf, \emph{Zeitschrift fur Physik}, \textbf{89}, 27 (1934).

\bibitem{1971PhT....24a..30R}
R. Ruffini, J.A. Wheeler, \emph{Physics Today}, \textbf{24}, 30 (1971).

\bibitem{s51}
J. Schwinger, \emph{Phys. Rev.}, \textbf{82}, 664 (1951).

\bibitem{s54a}
J. Schwinger, \emph{Phys. Rev.}, \textbf{93}, 615 (1954).

\bibitem{s54b}
J. Schwinger, \emph{Phys. Rev.}, \textbf{94}, 1362 (1954).

\bibitem{Dir30}
P.A.M. Dirac, \emph{Proceedings of the Cambridge Philos. Soc.}, \textbf{26}, 361 (1930).

\bibitem{Dir33}
P.A.M. Dirac, \emph{Rep. at Inst. of Phys. Solvay}, 203 (1933).

\bibitem{demourwkb}
T. Damour, \emph{Proceedings of the I Marcel Grossmann Meeting on General Relativity}, ed. R. Ruffini, North-Holland Publishing Company, p. 459 (1975).

\bibitem{gamow-book}
G. Gamow, \emph{Constitution of Atomic Nuclei and Radioactivity} (Clarendon Press, 1931).

\bibitem{landaumedium}
L.D. Landau, E.M. Lifshitz, \emph{The Electrodynamics of Continuum Medium} (Pergamon Press, 1975).

\bibitem{dirac1934}
P.A.M. Dirac, \emph{Proceedings of the Cambridge Philos. Soc.}, \textbf{30}, 150 (1934).

\bibitem{heisenberg1934}
W. Heisenberg, \emph{Zeitschrift fur Physik}, \textbf{90}, 209 (1934).

\bibitem{LanLif75a}
L.D. Landau, E.M. Lifshitz, \emph{Non-Relativistic Quantum Mechanics} (Pergamon Press, 1975).

\bibitem{LanLif75b}
L.D. Landau, E.M. Lifshitz, \emph{Relativistic Quantum Theory} (Pergamon Press, 1975).

\bibitem{z4a}
P.A.M. Dirac, \emph{Proceedings Roy. Soc.}, \textbf{117}, 610 (1928).

\bibitem{z4b}
P.A.M. Dirac, \emph{Proceedings Roy. Soc.}, \textbf{118}, 34 (1928)1.

\bibitem{z5}
P.A.M. Dirac, \emph{Principles of Quantum Mechanics} (Clarendon Press, 1958).

\bibitem{z7}
W. Gordon, \emph{Zeitschrift fur Physik}, \textbf{48}, 11 (1928).

\bibitem{z6}
C.G. Darwin, \emph{Proceedings Roy. Soc. A}, \textbf{118}, 654 (1928).

\bibitem{Sommerfeld-X}
A. Sommerfeld, \emph{Atombau und Spektrallinien} (Braunschweig, F. Vieweg 1922).

\bibitem{grc98}
W. Greiner, J. Reinhardt, in \emph{Quantum Aspects of Beam Physics, Proc. 15$^{th}$ Advanced ICFA Beam Dynamics Workshop}, ed. P. Chen, World Scientific (1998).

\bibitem{g1a}
I. Pomeranchuk, J. Smordinsky, \emph{J. Phys. USSR}, \textbf{9}, 97 (1945).

\bibitem{g1b}
N. Case, \emph{Phys. Rev.}, \textbf{80}, 797 (1950).

\bibitem{g1c}
F.G. Werner, J.A. Wheeler, \emph{Phys. Rev.}, \textbf{109}, 126 (1958).

\bibitem{g1d}
V. Vorankov, N.N. Kolesinkov, \emph{Sov. Phys. JETP}, \textbf{12}, 136 (1961).

\bibitem{z10}
V.S. Popov, \emph{Yad. Fiz.}, \textbf{12}, 429 (1970) [\emph{Sov. J. Nucl. Phys.}, \textbf{12} 235 (1971)].

\bibitem{z11a}
V.S. Popov, \emph{Zhetf Pis. Red.}, \textbf{11}, 254 (1970) [\emph{JETP Lett.}, \textbf{11}, 162 (1970)].

\bibitem{z11b}
V.S. Popov, \emph{Zh. Eksp. Theor Fiz.}, \textbf{59}, 965 (1970) [\emph{Sov. Phys. JEPT}, \textbf{32}, 526 (1971)].

\bibitem{z12}
V.S. Popov, \emph{Zh. Eksp. Theor Fiz.}, \textbf{60}, 1228 (1971) [\emph{Sov. Phys. JEPT}, \textbf{33}, 665 (1971)].

\bibitem{z}
Ya.B. Zel'dovich, V.S. Popov, \emph{Sov. Phys. USPEKHI}, \textbf{14}, 673 (1972).

\bibitem{popov2001}
V.S. Popov, \emph{Yad. Fiz.}, \textbf{64}, 421 (2001) [\emph{Phys. Atomic Nuclei}, \textbf{64}, 367 (2001)].

\bibitem{GZ69}
S.S. Ger\v{s}tein, Ya.B. Zel'dovich, \emph{Lett. Nuovo Cimento}, \textbf{1}, 835 (1969).

\bibitem{GZ70}
S.S. Ger\v{s}tein, Ya.B. Zel'dovich, \emph{Zh. Eksp. Theor. Fiz.}, \textbf{57}, 654 (1969) [\emph{Sov. Phys. JETP}, \textbf{30}, 358 (1970)].

\bibitem{z65}
V.S. Popov, \emph{Yad. Fiz.}, \textbf{14}, 458 (1971) [\emph{Sov. J. Nucl. Phys.}, \textbf{14}, 257 (1972)].

\bibitem{z20}
Ya.B. Zel'dovich, E.M. Rabinovich, \emph{Zh. Eksp. Theor. Fiz.}, \textbf{37}, 1296 (1959) [\emph{Sov. Phys. JETP}, \textbf{10}, 924 (1960)].

\bibitem{z21}
\'E.\'E. Shnol', \emph{Teor. Mat. Fiz.}, \textbf{4}, 239 (1970).

\bibitem{z22}
A.I. Baz', Ya.B. Zel'dovich, A.M. Peremolov, \emph{Scattering, reactions, and Decays in Relativistic Quantum mechanics, 2nd edition} (Nauka, 1971).

\bibitem{z23}
A.B. Migdal, A.M. Peremolov, V.S. Popov, \emph{Yad. Fis.}, \textbf{14}, 874 (1971) [\emph{Sov. J. Nucl. Phys.}, \textbf{14}, 488 (1972)].

\bibitem{z24}
A.M. Peremolov, V.S. Popov, \emph{Zh. Eksp. Teor. Fiz.}, \textbf{61}, 1743 (1971) [\emph{Sov. Phys. JETP}, \textbf{34}, 928 (1972)].

\bibitem{Greinerbook}
W. Greiner, J. Reinhardt, \emph{Quantum Electrodynamics} (Springer-Verlag, 1992).

\bibitem{g3a}
B. M\"uller, \emph{Ann. Rev. Nucl. Sci.}, \textbf{26}, 351 (1976).

\bibitem{g3b}
J. Reinhardt, W. Greiner, \emph{Rep. Prog. Phys.}, \textbf{40}, 219 (1977).

\bibitem{g3c}
J. Rafelski, L. F\"ulcher, A. Klein, \emph{Phys. Rep. C}, \textbf{38}, 227 (1978).

\bibitem{g3d}
S.J. Brodsky, P.J. Mohr, in \emph{Structure and Collisions of Ions and Atoms}, ed. I.A. Sellin, Springer-Verlag (1978).

\bibitem{grc98-5a}
M. Gyulassy, \emph{Nucl. Phys. A}, \textbf{244}, 497 (1975).

\bibitem{grc98-5b}
G.A. Rinker, L. Wilets, \emph{Nucl. Phys. A}, \textbf{12}, 748 (1975).

\bibitem{grc98-6}
G. Soff, P. Schl\"uter, B. M\"uller, W. Greiner, \emph{Phys. Rev. Lett.}, \textbf{48}, 1465 (1982).

\bibitem{g}
J.S. Greenberg, W. Greiner, \emph{Physics Today}, \textbf{35}, 24 (1982).

\bibitem{greiner1972a}
B. M\"uller, J. Rafelski, W. Greiner, \emph{Z. Phys.}, \textbf{257}, 62 (1972).

\bibitem{greiner1972b}
B. M\"uller, J. Rafelski, W. Greiner, \emph{Z. Phys.}, \textbf{257}, 183 (1972).

\bibitem{g5}
J. Rafelski, B. M\"uller, W. Greiner, \emph{Z. Phys. A}, \textbf{285}, 49 (1978).

\bibitem{g6}
J. Reinhardt, U. M\"uller, B. M\"uller, W. Greiner, \emph{Z. Phys. A}, \textbf{303}, 173 (1981).

\bibitem{grc98-30}
O. Graf, J. Reinhardt, B. M\"uller, W. Greiner, G. Soff, \emph{Phys. Rev. Lett.}, \textbf{61}, 2831 (1981).

\bibitem{g10}
C. Kozhuharov, \emph{Electronic and Atomic Collisions}, ed. S. Datz, North Holland, (1981).

\bibitem{g11}
P. Vincent, K.H. Heinig, J.-U. J\"ager, K.-H. Kaun, H. Richler, H. Woittenek, in \emph{Proceedings of NASI Conference on the Quantum Electrodynamics of Strong Fields}, ed. W. Greiner, Plenum, New York (1981).

\bibitem{g7a}
P. Kienle, H. Backe, H. Bokemeyer, in \emph{Proceedings of NASI Conference on the Quantum Electrodynamics of Strong Fields}, ed. W. Greiner, Plenum, New York (1981).

\bibitem{g7b}
J.S. Greenberg, \emph{Electronic and Atomic Collisions}, eds. N. Oda and K. Takayanagi, North Holland (1980).

\bibitem{g7c}
P. Kienle, \emph{Atomic Physics}, eds. D. Kleppner and F. Pipkin, Plenum, New York, Vol. 7 (1981).

\bibitem{g8}
H. Backe et al., \emph{Phys. Rev. Lett.}, \textbf{40}, 1443 (1978).

\bibitem{g9}
C.Kozhuharov et al., \emph{Phys. Rev. Lett.}, \textbf{42}, 376 (1979).

\bibitem{g18-22a}
U. M\"uller, J. Reinhardt, T. de Reus, P. Schl\"uter, G. Soff, K.H. Wietschorke, B. M\"uller, in \emph{Proceedings of NASI Conference on the Quantum Electrodynamics of Strong Fields}, Ed. W. Greiner, Plenum, New York (1981).

\bibitem{g18-22b}
H. Bokemeyer et al., in \emph{Proceedings of NASI Conference on the Quantum Electrodynamics of Strong Fields}, Ed. W. Greiner, Plenum, New York (1981).

\bibitem{g18-22c}
H. Backe et al., in \emph{Proceedings of NASI Conference on the Quantum Electrodynamics of Strong Fields}, Ed. W. Greiner, Plenum, New York (1981).

\bibitem{zexp1}
J. Schweppe et al., \emph{Phys. Rev. Lett.}, \textbf{77}, 2838 (1995).

\bibitem{G...96}
R. Ganz et al., \emph{Phys. Lett. B}, \textbf{389}, 4 (1996).

\bibitem{L...97}
U. Leinberger et al., \emph{Phys. Lett. B}, \textbf{394}, 16 (1997).

\bibitem{zexp4}
U. Leinberger et al., \emph{Eur. Phys. Journ. A}, \textbf{1}, 249 (1998).

\bibitem{H...98}
S. Heinz et al., \emph{Eur. Phys. J. A}, \textbf{1}, 27 (1998).

\bibitem{A...95}
I. Ahmad et al., \emph{Phys. Rev. Lett.}, \textbf{75}, 2658 (1995).

\bibitem{grc98-52b}
I. Ahmad et al., \emph{Phys. Rev. Lett.}, \textbf{78}, 618 (1997).

\bibitem{grc98-53a}
T.E. Cowan, J.S. Greenberg, \emph{Phys. Rev. Lett.}, \textbf{77}, 2838 (1995).

\bibitem{grc98-53b}
I. Ahmad et al., Reply to the Comment, \emph{Phys. Rev. Lett.}, \textbf{77}, 2839 (1995).

\bibitem{ll2}
L.D. Landau, E.M. Lifshitz, \emph{The classical theory of fields} (Pergamon Press, 1975).

\bibitem{cr71}
D. Christodoulou, R. Ruffini, \emph{Phys. Rev. D}, \textbf{4}, 3552 (1971).

\bibitem{dr12z}
Ya.B. Zel'dovich, \emph{Zh. Eksp. Theor. Fiz.}, \textbf{62}, 2076 (1972) [\emph{Sov.~Phys.~JEPT}, \textbf{35}, 1085 (1972)].

\bibitem{dr12s}
A.A. Starobinsky, \emph{Zh. Eksp. Theor. Fiz.}, \textbf{64}, 48 (1973) [\emph{Sov. Phys. JEPT}, \textbf{37}, 28 (1973)].

\bibitem{dr12unruh}
W.G. Unruh, \emph{Phys. Rev. D}, \textbf{14}, 870 (1976).

\bibitem{dr12mg1}
N. Deruelle, in \emph{Proceedings of the first Marcel Grossmann meeting}, ed. R. Ruffini, North Holland, Amsterdam (1977).

\bibitem{dr13zaumen}
W.T. Zaumen, Nature, 247, 530 (1974).

\bibitem{dr13gibbons}
G.W. Gibbons, \emph{Commun. Math. Phys.}, \textbf{44}, 245 (1975).

\bibitem{dr75}
T. Damour, R. Ruffini, \emph{Phys. Rev. Lett.}, \textbf{35}, 463 (1975).

\bibitem{r70}
R. Ruffini, On the energetics of Black Holes, in \emph{Black Holes - Les astres occlus}, eds. C. and B.S. De Witt, Gordon and Breach, New York (1973).

\bibitem{dr15}
B. Carter, \emph{Commun. Math. Phys.}, \textbf{10}, 280 (1968).

\bibitem{mtw73}
C.W. Misner, K.S. Thorne, J.A. Wheeler, \emph{Gravitation} (Freeman, 1973).

\bibitem{c97}
B. Carter, \emph{Proceedings of the Eighth Marcel Grossmann Meeting on General Relativity}, eds. T. Piran, R. Ruffini, World Scientific, p. 136 (1997).

\bibitem{bcjr02}
D. Bini, C. Cherubini, R. Jantzen, R. Ruffini, \emph{Prog. Theor. Phys.}, \textbf{107}, 967 (2002).

\bibitem{ruKl}
R. Ruffini, in \emph{Fluctuating Paths and Fields}, eds. W. Janke, A. Pelster, H.-J. Schmidt, M. Bachmann, World Scientific (2001).

\bibitem{1997Natur.387..783C}
E.~{Costa}, F.~{Frontera}, J.~{Heise}, M.~{Feroci}, J.~{in't Zand}, F.~{Fiore},
  M.~N. {Cinti}, D.~{Dal Fiume}, L.~{Nicastro}, M.~{Orlandini}, E.~{Palazzi},
  M.~{Rapisarda\#}, G.~{Zavattini}, R.~{Jager}, A.~{Parmar}, A.~{Owens},
  S.~{Molendi}, G.~{Cusumano}, M.~C. {Maccarone}, S.~{Giarrusso}, A.~{Coletta},
  L.~A. {Antonelli}, P.~{Giommi}, J.~M. {Muller}, L.~{Piro} and R.~C. {Butler},
  {\em Nature} {\bf 387}, 783 (June 1997).

\bibitem{rukyoto}
R. Ruffini, in \emph{Black Holes and High Energy Astrophysics}, ed. H. Sato and N. Sugiyama, Universal Academic Press, Vol. 167 (1998).

\bibitem{prx02}
G. Preparata, R. Ruffini, S.-S. Xue, \emph{J. Korean Phys. Soc.}, \textbf{42}, S99 (2003).

\bibitem{prx98}
G. Preparata, R. Ruffini, S.-S. Xue, \emph{Astron. Astrophys.}, \textbf{338}, L87 (1998).

\bibitem{1999A&A...350..334R}
R.~{Ruffini}, J.~D. {Salmonson}, J.~R. {Wilson} and S.-S. {Xue}, {\em Astronomy
  \& Astrophysics} {\bf 350}, 334 (October 1999).

\bibitem{2000A&A...359..855R}
R.~{Ruffini}, J.~D. {Salmonson}, J.~R. {Wilson} and S.-S. {Xue}, {\em Astronomy
  \& Astrophysics} {\bf 359}, 855 (July 2000).

\bibitem{r78}
R. Ruffini, in {\em Physics and Astrophysics of Neutron Stars and Black Holes}, ed. R. Giacconi, R. Ruffini, North Holland, Amsterdam (1978).

\bibitem{prx01}
G. Preparata, R. Ruffini, S.-S. Xue, in {\em Proceedings of the III ICRA Network Workshop and VI Italo-Korean Meeting}, eds. C.Cherubini and R. Ruffini, SIF (2001).

\bibitem{arv07}
A.G. Aksenov, R. Ruffini, G.V. Vereshchagin, \emph{Phys. Rev. Lett.}, \textbf{99}, 125003 (2007).

\bibitem{crv02}
C. Cherubini, R. Ruffini, L. Vitagliano, \emph{Phys. Lett. B}, \textbf{545}, 226 (2002).

\bibitem{rv02a}
R. Ruffini, L. Vitagliano, \emph{Phys. Lett. B}, \textbf{545}, 233 (2002).

\bibitem{rv02b}
R. Ruffini, L. Vitagliano, \emph{Int. Journ. Mod. Phys. D}, \textbf{12}, 121 (2003).

\bibitem{rvx03sep}
R. Ruffini, L. Vitagliano, S.-S. Xue, \emph{Phys. Lett. B}, \textbf{573}, 33 (2003).

\bibitem{Jorge_IC4}
C.~Cherubini, A.~Geralico, J.A.~Rueda Hernandez, and R.~Ruffini, On the ``Dyadotorus'' of the Kerr-Newman Spacetime, in {\em Relativistic Astrophysics}, eds. C.L..~{Bianco} and S.-S. {Xue}, American Institute of Physics Conference Series, Vol.~966 (February 2008).

\bibitem{damour}
T. Damour, \emph{Proceedings of the Second Marcel Grossmann Meeting of General Relativity}, 587 (1982).

\bibitem{damourotros}
T. Damour, R.S. Hanni, R. Ruffini, J.R. Wilson, \emph{\prd}, \textbf{17}, 1518 (1978).

\bibitem{n70}
N.B. Narozhnyi, A.I. Nikishov, \emph{Sov. J. Nucl. Phys.}, \textbf{11}, 596 (1970).

\bibitem{PR71}
V.S. Popov, T.I. Rozhdestvenskaya, \emph{JETP Lett.}, \textbf{14}, 177 (1971).

\bibitem{S...83}
J. Schweppe, et al., \emph{Phys. Rev. Lett.}, \textbf{51}, 2261 (1983).

\bibitem{B...95}
R. B\"{a}r, et al., \emph{Nucl. Phys. A}, \textbf{583}, 237 (1995).

\bibitem{B...97}
D.L. Burke, et al., \emph{Phys. Rev. Lett.}, \textbf{79}, 1626 (1997).

\bibitem{R01}
A. Ringwald, \emph{Phys. Lett. B}, \textbf{510}, 107 (2001).

\bibitem{AHRSV01}
R. Alkofer, M.B. Hecht, C.D. Roberts, S.M. Schmidt, D.V. Vinnik, \emph{Phys. Rev. Lett.}, \textbf{87}, 193902 (2001).

\bibitem{RSV02}
C.D. Roberts, S.M. Schmidt, D.V. Vinnik, \emph{Phys. Rev. Lett.}, \textbf{89}, 153901 (2002).

\bibitem{2001ApJ...555L.107R}
R.~{Ruffini}, C.~L. {Bianco}, F.~{Fraschetti}, S.-S. {Xue} and P.~{Chardonnet},
  {\em Astrophysical Journal} {\bf 555}, L107 (July 2001).

\bibitem{2001ApJ...555L.113R}
R.~{Ruffini}, C.~L. {Bianco}, F.~{Fraschetti}, S.-S. {Xue} and P.~{Chardonnet},
  {\em Astrophysical Journal} {\bf 555}, L113 (July 2001).

\bibitem{2001ApJ...555L.117R}
R.~{Ruffini}, C.~L. {Bianco}, F.~{Fraschetti}, S.-S. {Xue} and P.~{Chardonnet},
  {\em Astrophysical Journal} {\bf 555}, L117 (July 2001).

\bibitem{2002ApJ...581L..19R}
R.~{Ruffini}, C.~L. {Bianco}, P.~{Chardonnet}, F.~{Fraschetti} and S.-S. {Xue},
  {\em Astrophysical Journal} {\bf 581}, L19 (December 2002).

\bibitem{rvx03}
R. Ruffini, L. Vitagliano, S.-S. Xue, in \emph{Quantum Aspects of Beam Physics, 28th Advanced IFCA Beam Dynamics Workshop}, ed. P. Chen, World Scientific, Singapore (2003), \url{astro-ph/0304306}.

\bibitem{liebsimon}
E. Lieb, B. Simon, \emph{Phys. Rev. Lett.}, \textbf{31}, 681 (1973).

\bibitem{ruffinistella81}
R. Ruffini, L. Stella, \emph{Phys. Lett. B}, \textbf{102}, 442 (1981).

\bibitem{muller72}
B. M\"uller, H. Peitz, J. Rafelski, W. Greiner, \emph{Phys. Rev. Lett.}, \textbf{28}, 1235 (1972).
 
\bibitem{muller75}
B. M\"uller, J. Rafelski, \emph{Phys. Rev. Lett.}, \textbf{34}, 349 (1975).

\bibitem{migdal76}
A.B. Migdal, D.N. Voskresenskii, V.S. Popov, \emph{JETP Letters}, \textbf{24}, 186 (1976).

\bibitem{ruffinistella80}
J. Ferreirinho, R. Ruffini, L. Stella, \emph{Phys. Lett. B}, \textbf{91}, 314 (1980).

\bibitem{segrebook}
E. Segr\'e, ``Nuclei e Particle'', Second Edition, W.A. Benjamin (1977).

\bibitem{ov39}
J.R. Oppenheimer, R. Volkoff, \emph{Phys. Rev.}, \textbf{55}, 374 (1939).

\bibitem{ruffiniphd}
R. Ruffini, S. Bonazzola, \emph{Phys. Rev. D}, \textbf{187}, 1767 (1969).

\bibitem{sato1970}
H.A. Bethe, G. B\"orner, K. Sato, \emph{\aap}, \textbf{7}, 279 (1970).

\bibitem{cameron1970}
A.G.W. Cameron, \emph{Annual Review of Astronomy and Astrophysics}, \textbf{8}, 179 (1970).

\bibitem{Thomas}
L.H. Thomas, \emph{Proc. Cambridge Phil. Society}, \textbf{23}, 542 (1927).

\bibitem{Fermi}
E. Fermi, \emph{Rend. Accad. Lincei}, \textbf{6}, 602 (1927).

\bibitem{scorza28}
G. Scorza-Dragoni, \emph{Rend. Accad. Lincei}, \textbf{8}, 301 (1928).

\bibitem{scorza29}
G. Scorza-Dragoni, \emph{Rend. Accad. Lincei}, \textbf{9}, 378 (1929).

\bibitem{sommerfeld}
A. Sommerfeld, \emph{Z. Physik}, \textbf{78}, 283 (1932).

\bibitem{Miranda}
C. Miranda, \emph{Mem. Accad. Italia}, \textbf{5}, 285 (1934).

\bibitem{Lieb}
E. Lieb, \emph{Rev. Mod. Physics}, \textbf{53}, 603 (1981).

\bibitem{Spruch}
L. Spruch, \emph{Rev. Mod. Physics}, \textbf{63}, 151 (1991).

\bibitem{rrx06}
R. Ruffini, M. Rotondo, S.-S. Xue, \emph{Int. J. Mod. Phys. D}, \textbf{16}, 1 (2007).

\bibitem{e05}
A. Einstein, \emph{Ann. Phys. (Germany)}, \textbf{17}, 891 (1905).

\bibitem{2003AIPC..668...16R}
R.~{Ruffini}, C.~L. {Bianco}, P.~{Chardonnet}, F.~{Fraschetti}, L.~{Vitagliano}
  and S.-S. {Xue}, New perspectives in physics and astrophysics from the
  theoretical understanding of gamma-ray bursts, in {\em Cosmology and
  Gravitation\/},  eds. M.~{Novello} and S.~E. {Perez Bergliaffa}, American
  Institute of Physics Conference Series, Vol.~668 (June 2003).

\bibitem{1999PhR...314..575P}
T.~{Piran}, {\em Physics Reports} {\bf 314}, 575 (June 1999).

\bibitem{p00}
T. Piran, \emph{Phys. Rep.}, \textbf{333-334}, 529 (2000).

\bibitem{rm94}
M.J. Rees, P. M\'esz\'aros, \emph{\apjl}, \textbf{430}, L93 (1994).

\bibitem{px94}
B. Paczy\'nski, \& Xu, G. (1994), \apj, 427, 708.

\bibitem{sp97}
R. Sari, T. Piran, \emph{\apj}, \textbf{485}, 270 (1997).

\bibitem{f99}
E.E. Fenimore, \emph{\apj}, \textbf{518}, 375 (1999).

\bibitem{fcrsyn99}
E.E. Fenimore, C. Cooper, E. Ramirez-Ruiz, M.C. Sumner, A. Yoshida, M. Namiki, \emph{\apj}, \textbf{512}, 683 (1999).

\bibitem{1993MNRAS.263..861P}
T.~{Piran}, A.~{Shemi} and R.~{Narayan}, {\em Monthly Notices of the Royal
  Astronomical Society} {\bf 263}, p. 861 (August 1993).

\bibitem{bm95} 
G.S. Bisnovatyi-Kogan, M.V.A. Murzina, \emph{\prd}, \textbf{52}, 4380 (1995).

\bibitem{1993ApJ...415..181M}
P.~{Meszaros}, P.~{Laguna} and M.~J. {Rees}, {\em Astrophysical Journal} {\bf
  415}, 181 (September 1993).

\bibitem{2007arXiv0705.2411B}
C.~L. {Bianco}, R.~{Ruffini}, G.~{Vereshchagin} and S.-S. {Xue}, {\em J.Korean
  Phys.Soc.} {\bf 49}, 722  (2006).

\bibitem{1990ApJ...365L..55S}
A.~{Shemi} and T.~{Piran}, {\em Astrophysical Journal} {\bf 365}, L55 (December
  1990).

\bibitem{Vereshchagin07}
G.~V. {Vereshchagin}, Pair plasma and gamma ray bursts, PhD thesis, IRAP PhD,
  (2007).

\bibitem{1948PhRv...74..328T}
A.~H. {Taub}, {\em Physical Review} {\bf 74}, 328 (August 1948).

\bibitem{1976PhFl...19.1130B}
R.~D. {Blandford} and C.~F. {McKee}, {\em Physics of Fluids} {\bf 19}, 1130
  (August 1976).

\bibitem{1992MNRAS.258P..41R}
M.~J. {Rees} and P.~{Meszaros}, {\em \mnras} {\bf 258}, 41P (September 1992).

\bibitem{1992ApJ...395L..83N}
R.~{Narayan}, B.~{Paczynski} and T.~{Piran}, {\em Astrophysical Journal} {\bf
  395}, L83 (August 1992).

\bibitem{1994ApJ...422..248K}
J.~I. {Katz}, {\em Astrophysical Journal} {\bf 422}, 248 (February 1994).

\bibitem{1987per..book.....L}
L.~D. {Landau} and E.~M. {Lifshits}, {\em Fluid Mechanics (Course of
  Theoretical Physics)} (Pergamon, New York, 1987).

\bibitem{2005ApJ...635..516N}
E.~{Nakar}, T.~{Piran} and R.~{Sari}, {\em Astrophysical Journal} {\bf 635},
  516 (December 2005).

\bibitem{1998MNRAS.300.1158G}
O.~M. {Grimsrud} and I.~{Wasserman}, {\em Monthly Notices of the Royal
  Astronomical Society} {\bf 300}, 1158 (November 1998).

\bibitem{2004ApJ...605L...1B}
C.~L. {Bianco} and R.~{Ruffini}, {\em Astrophysical Journal} {\bf 605}, L1
  (April 2004).

\bibitem{2005ApJ...620L..23B}
C.~L. {Bianco} and R.~{Ruffini}, {\em Astrophysical Journal} {\bf 620}, L23
  (February 2005).

\bibitem{2005ApJ...633L..13B}
C.~L. {Bianco} and R.~{Ruffini}, {\em Astrophysical Journal} {\bf 633}, L13
  (November 2005).

\bibitem{2005IJMPD..14..131R}
R.~{Ruffini}, F.~{Fraschetti}, L.~{Vitagliano} and S.-S. {Xue}, {\em
  International Journal of Modern Physics D} {\bf 14}, 131  (2005).

\bibitem{2005AIPC..782...42R}
R.~{Ruffini}, M.~G. {Bernardini}, C.~L. {Bianco}, P.~{Chardonnet},
  F.~{Fraschetti}, V.~{Gurzadyan}, L.~{Vitagliano} and S.-S. {Xue}, The
  blackholic energy: long and short gamma-ray bursts (new perspectives in
  physics and astrophysics from the theoretical understanding of gamma-ray
  bursts, ii), in {\em XIth Brazilian School of Cosmology and Gravitation\/},
  eds. M.~{Novello} and S.~E. {Perez Bergliaffa}, American Institute of Physics
  Conference Series, Vol.~782 (August 2005).

\bibitem{pls}
R. Ruffini, P. Chardonnet, C.L. Bianco, S.-S. Xue, F. Fraschetti, \emph{Pour la science}, \textbf{294}, 26 (2002).

\bibitem{2006ApJ...645L.109R}
R.~{Ruffini}, M.~G. {Bernardini}, C.~L. {Bianco}, P.~{Chardonnet},
  F.~{Fraschetti}, R.~{Guida} and S.-S. {Xue}, {\em Astrophysical Journal} {\bf
  645}, L109 (July 2006).

\bibitem{Venezia_Orale}
R. Ruffini, M.G. Bernardini, C.L. Bianco, P. Chardonnet, F. Fraschetti, R. Guida, S.-S. Xue, \emph{Il Nuovo Cimento B}, \textbf{121}, 1367 (2007).

\bibitem{2007A&A...474L..13B}
M.~G. {Bernardini}, C.~L. {Bianco}, L.~{Caito}, M.~G. {Dainotti}, R.~{Guida}
  and R.~{Ruffini}, {\em Astronomy \& Astrophysics} {\bf 474}, L13 (October
  2007).

\bibitem{cospar02}
R. Ruffini, C.L. Bianco, P. Chardonnet, F. Fraschetti, S.-S. Xue, \emph{Adv. Sp. Res.}, \textbf{34}, 2715 (2004).

\bibitem{Mosca_Orale}
R. Ruffini, M.G. Bernardini, C.L. Bianco, L. Caito, P. Chardonnet, M.G. Dainotti, F. Fraschetti, R. Guida, G. Vereshchagin, S.-S. Xue, \emph{ESA Special Publication}, \textbf{SP-622}, 561 (2007).

\bibitem{2005tmgm.meet.2459B}
M.~G. {Bernardini}, C.~L. {Bianco}, R.~{Ruffini}, S.-S. {Xue}, P.~{Chardonnet}
  and F.~{Fraschetti}, General features of grb 030329 in the embh model, in
  {\em The Tenth Marcel Grossmann Meeting. On recent developments in
  theoretical and experimental general relativity, gravitation and relativistic
  field theories\/},  eds. M.~{Novello}, S.~{Perez Bergliaffa} and R.~{Ruffini}
  (Singapore: World Scientific, January 2005).

\bibitem{2005ApJ...634L..29B}
M.~G. {Bernardini}, C.~L. {Bianco}, P.~{Chardonnet}, F.~{Fraschetti},
  R.~{Ruffini} and S.-S. {Xue}, {\em Astrophysical Journal} {\bf 634}, L29
  (November 2005).

\bibitem{2007A&A...471L..29D}
M.~G. {Dainotti}, M.~G. {Bernardini}, C.~L. {Bianco}, L.~{Caito}, R.~{Guida}
  and R.~{Ruffini}, {\em Astronomy \& Astrophysics} {\bf 471}, L29 (August
  2007).

\bibitem{1997ApJ...490..772K}
J.~I. {Katz} and T.~{Piran}, {\em \apj} {\bf 490}, 772 (December 1997).

\bibitem{1999ApJ...512..699C}
J.~{Chiang} and C.~D. {Dermer}, {\em Astrophysical Journal} {\bf 512}, 699
  (February 1999).

\bibitem{2005RvMP...76.1143P}
T.~{Piran}, {\em Reviews of Modern Physics} {\bf 76}, 1143 (January 2005).

\bibitem{2006RPPh...69.2259M}
P.~{Meszaros}, {\em Reports of Progress in Physics} {\bf 69}, 2259  (2006).

\bibitem{2001A&A...368..377B}
C.~L. {Bianco}, R.~{Ruffini} and S.-S. {Xue}, {\em Astronomy \& Astrophysics}
  {\bf 368}, 377 (March 2001).

\bibitem{2004IJMPD..13..843R}
R.~{Ruffini}, C.~L. {Bianco}, S.-S. {Xue}, P.~{Chardonnet}, F.~{Fraschetti} and
  V.~{Gurzadyan}, {\em International Journal of Modern Physics D} {\bf 13}, 843
   (2004).

\bibitem{2006ApJ...643..266N}
J.~P. {Norris} and J.~T. {Bonnell}, {\em Astrophysical Journal} {\bf 643}, 266
  (May 2006).

\bibitem{2002A&A...393..409G}
G.~{Ghirlanda}, A.~{Celotti} and G.~{Ghisellini}, {\em Astronomy \&
  Astrophysics} {\bf 393}, 409 (October 2002).

\bibitem{1998ApJ...506L..23P}
R.~D. {Preece}, M.~S. {Briggs}, R.~S. {Mallozzi}, G.~N. {Pendleton}, W.~S.
  {Paciesas} and D.~L. {Band}, {\em Astrophysical Journal} {\bf 506}, L23
  (October 1998).

\bibitem{2005IJMPD..14...97R}
R.~{Ruffini}, C.~L. {Bianco}, S.-S. {Xue}, P.~{Chardonnet}, F.~{Fraschetti} and
  V.~{Gurzadyan}, {\em International Journal of Modern Physics D} {\bf 14}, 97
  (2005).

\bibitem{2003GCN..2460....1M}
S.~{Mereghetti} and D.~{Gotz}, {\em GRB Coordinates Network (GCN)} {\bf 2460}
  (2003).

\bibitem{2004ApJ...605L.101W}
D.~{Watson}, J.~{Hjorth}, A.~{Levan}, P.~{Jakobsson}, P.~T. {O'Brien}, J.~P.
  {Osborne}, K.~{Pedersen}, J.~N. {Reeves}, J.~A. {Tedds}, S.~A. {Vaughan},
  M.~J. {Ward} and R.~{Willingale}, {\em Astrophysical Journal} {\bf 605}, L101
  (April 2004).

\bibitem{2004Natur.430..648S}
A.~M. {Soderberg}, S.~R. {Kulkarni}, E.~{Berger}, D.~W. {Fox}, M.~{Sako}, D.~A.
  {Frail}, A.~{Gal-Yam}, D.~S. {Moon}, S.~B. {Cenko}, S.~A. {Yost}, M.~M.
  {Phillips}, S.~E. {Persson}, W.~L. {Freedman}, P.~{Wyatt}, R.~{Jayawardhana}
  and D.~{Paulson}, {\em Nature} {\bf 430}, 648 (August 2004).

\bibitem{2004Natur.430..646S}
S.~Y. {Sazonov}, A.~A. {Lutovinov} and R.~A. {Sunyaev}, {\em Nature} {\bf 430},
  646 (August 2004).

\bibitem{2004ApJ...611..200P}
J.~X. {Prochaska}, J.~S. {Bloom}, H.-W. {Chen}, K.~C. {Hurley}, J.~{Melbourne},
  A.~{Dressler}, J.~R. {Graham}, D.~J. {Osip} and W.~D. {Vacca}, {\em
  Astrophysical Journal} {\bf 611}, 200 (August 2004).

\bibitem{2000ApJ...536..778P}
E.~{Pian}, L.~{Amati}, L.~A. {Antonelli}, R.~C. {Butler}, E.~{Costa},
  G.~{Cusumano}, J.~{Danziger}, M.~{Feroci}, F.~{Fiore}, F.~{Frontera},
  P.~{Giommi}, N.~{Masetti}, J.~M. {Muller}, L.~{Nicastro}, T.~{Oosterbroek},
  M.~{Orlandini}, A.~{Owens}, E.~{Palazzi}, A.~{Parmar}, L.~{Piro}, J.~J.~M.
  {in't Zand}, A.~{Castro-Tirado}, A.~{Coletta}, D.~{Dal Fiume}, S.~{Del
  Sordo}, J.~{Heise}, P.~{Soffitta} and V.~{Torroni}, {\em Astrophysical
  Journal} {\bf 536}, 778 (June 2000).

\bibitem{2004AdSpR..34.2715R}
R.~{Ruffini}, M.~G. {Bernardini}, C.~L. {Bianco}, P.~{Chardonnet},
  F.~{Fraschetti} and S.-S. {Xue}, {\em Advances in Space Research} {\bf 34},
  2715  (2004).

\bibitem{1997ApJ...479L..39C}
A.~{Crider}, E.~P. {Liang}, I.~A. {Smith}, R.~D. {Preece}, M.~S. {Briggs},
  G.~N. {Pendleton}, W.~S. {Paciesas}, D.~L. {Band} and J.~L. {Matteson}, {\em
  Astrophysical Journal} {\bf 479}, p. L39 (April 1997).

\bibitem{2000ApJS..127...59F}
F.~{Frontera}, L.~{Amati}, E.~{Costa}, J.~M. {Muller}, E.~{Pian}, L.~{Piro},
  P.~{Soffitta}, M.~{Tavani}, A.~{Castro-Tirado}, D.~{Dal Fiume}, M.~{Feroci},
  J.~{Heise}, N.~{Masetti}, L.~{Nicastro}, M.~{Orlandini}, E.~{Palazzi} and
  R.~{Sari}, {\em Astrophysical Journal Supplement Series} {\bf 127}, 59 (March
  2000).

\bibitem{1993ApJ...413..281B}
D.~{Band}, J.~{Matteson}, L.~{Ford}, B.~{Schaefer}, D.~{Palmer},
  B.~{Teegarden}, T.~{Cline}, M.~{Briggs}, W.~{Paciesas}, G.~{Pendleton},
  G.~{Fishman}, C.~{Kouveliotou}, C.~{Meegan}, R.~{Wilson} and P.~{Lestrade},
  {\em Astrophysical Journal} {\bf 413}, 281 (August 1993).

\bibitem{1983ASPRv...2..189P}
L.~A. {Pozdniakov}, I.~M. {Sobol} and R.~A. {Siuniaev}, {\em Astrophysics and
  Space Physics Reviews} {\bf 2}, 189  (1983).

\bibitem{1999ARep...43..739B}
S.~I. {Blinnikov}, A.~V. {Kozyreva} and I.~E. {Panchenko}, {\em Astronomy
  Reports} {\bf 43}, 739 (November 1999).

\bibitem{saz2}
S.~{Sazonov}, A.~{Lutovinov} and R.~{Sunyaev}, \emph{Private communication} (2004).

\bibitem{2006ApJ...638..920V}
S.~{Vaughan}, M.~R. {Goad}, A.~P. {Beardmore}, P.~T. {O'Brien}, J.~P.
  {Osborne}, K.~L. {Page}, S.~D. {Barthelmy}, D.~N. {Burrows}, S.~{Campana},
  J.~K. {Cannizzo}, M.~{Capalbi}, G.~{Chincarini}, J.~R. {Cummings},
  G.~{Cusumano}, P.~{Giommi}, O.~{Godet}, J.~E. {Hill}, S.~{Kobayashi},
  P.~{Kumar}, V.~{La Parola}, A.~{Levan}, V.~{Mangano}, P.~{M{\'e}sz{\'a}ros},
  A.~{Moretti}, D.~C. {Morris}, J.~A. {Nousek}, C.~{Pagani}, D.~M. {Palmer},
  J.~L. {Racusin}, P.~{Romano}, G.~{Tagliaferri}, B.~{Zhang} and N.~{Gehrels},
  {\em Astrophysical Journal} {\bf 638}, 920 (February 2006).

\bibitem{2004SPIE.5165..175B}
S.~D. {Barthelmy}, Burst alert telescope (bat) on the swift midex mission, in
  {\em X-Ray and Gamma-Ray Instrumentation for Astronomy XIII.\/},  eds. K.~A.
  {Flanagan} and O.~H.~W. {Siegmund}, Society of Photo-Optical Instrumentation
  Engineers (SPIE) Conference Proceedings, Vol.~5165 (February 2004).

\bibitem{2005SSRv..120..143B}
S.~D. {Barthelmy}, L.~M. {Barbier}, J.~R. {Cummings}, E.~E. {Fenimore},
  N.~{Gehrels}, D.~{Hullinger}, H.~A. {Krimm}, C.~B. {Markwardt}, D.~M.
  {Palmer}, A.~{Parsons}, G.~{Sato}, M.~{Suzuki}, T.~{Takahashi}, M.~{Tashiro}
  and J.~{Tueller}, {\em Space Science Reviews} {\bf 120}, 143 (October 2005).

\bibitem{2004ApJ...611.1005G}
N.~{Gehrels}, G.~{Chincarini}, P.~{Giommi}, K.~O. {Mason}, J.~A. {Nousek},
  A.~A. {Wells}, N.~E. {White}, S.~D. {Barthelmy}, D.~N. {Burrows}, L.~R.
  {Cominsky}, K.~C. {Hurley}, F.~E. {Marshall}, P.~{M{\'e}sz{\'a}ros}, P.~W.~A.
  {Roming}, L.~{Angelini}, L.~M. {Barbier}, T.~{Belloni}, S.~{Campana}, P.~A.
  {Caraveo}, M.~M. {Chester}, O.~{Citterio}, T.~L. {Cline}, M.~S. {Cropper},
  J.~R. {Cummings}, A.~J. {Dean}, E.~D. {Feigelson}, E.~E. {Fenimore}, D.~A.
  {Frail}, A.~S. {Fruchter}, G.~P. {Garmire}, K.~{Gendreau}, G.~{Ghisellini},
  J.~{Greiner}, J.~E. {Hill}, S.~D. {Hunsberger}, H.~A. {Krimm}, S.~R.
  {Kulkarni}, P.~{Kumar}, F.~{Lebrun}, N.~M. {Lloyd-Ronning}, C.~B.
  {Markwardt}, B.~J. {Mattson}, R.~F. {Mushotzky}, J.~P. {Norris},
  J.~{Osborne}, B.~{Paczynski}, D.~M. {Palmer}, H.-S. {Park}, A.~M. {Parsons},
  J.~{Paul}, M.~J. {Rees}, C.~S. {Reynolds}, J.~E. {Rhoads}, T.~P. {Sasseen},
  B.~E. {Schaefer}, A.~T. {Short}, A.~P. {Smale}, I.~A. {Smith}, L.~{Stella},
  G.~{Tagliaferri}, T.~{Takahashi}, M.~{Tashiro}, L.~K. {Townsley},
  J.~{Tueller}, M.~J.~L. {Turner}, M.~{Vietri}, W.~{Voges}, M.~J. {Ward},
  R.~{Willingale}, F.~M. {Zerbi} and W.~W. {Zhang}, {\em Astrophysical Journal}
  {\bf 611}, 1005 (August 2004).

\bibitem{2005GCN..3094....1P}
A.~{Parsons}, S.~{Barthelmy}, L.~{Barbier}, J.~{Cummings}, E.~{Fenimore},
  R.~{Fink}, N.~{Gehrels}, S.~{Holland}, D.~{Hullinger}, K.~{Hurley},
  H.~{Krimm}, C.~{Markwardt}, D.~{Palmer}, S.~{Piranomonte}, T.~{Sakamoto},
  G.~{Sato}, A.~{Smale}, M.~{Suzuki} and J.~{Tueller}, {\em GRB Coordinates
  Network (GCN)} {\bf 3094}  (2005).

\bibitem{2004SPIE.5165..201B}
D.~N. {Burrows}, J.~E. {Hill}, J.~A. {Nousek}, A.~A. {Wells}, G.~{Chincarini},
  A.~F. {Abbey}, A.~P. {Beardmore}, J.~{Bosworth}, H.~W. {Br{\"a}uninger},
  W.~{Burkert}, S.~{Campana}, M.~{Capalbi}, W.~{Chang}, O.~{Citterio}, M.~J.
  {Freyberg}, P.~{Giommi}, G.~D. {Hartner}, R.~{Killough}, B.~{Kittle},
  R.~{Klar}, C.~{Mangels}, M.~{McMeekin}, B.~J. {Miles}, A.~{Moretti},
  K.~{Mori}, D.~C. {Morris}, K.~{Mukerjee}, J.~P. {Osborne}, A.~D.~T. {Short},
  G.~{Tagliaferri}, F.~{Tamburelli}, D.~J. {Watson}, R.~{Willingale} and M.~E.
  {Zugger}, The swift x-ray telescope, in {\em X-Ray and Gamma-Ray
  Instrumentation for Astronomy XIII\/},  eds. K.~A. {Flanagan} and O.~H.~W.
  {Siegmund}, Society of Photo-Optical Instrumentation Engineers (SPIE)
  Conference Proceedings, Vol.~5165 (February 2004).

\bibitem{2005SSRv..120..165B}
D.~N. {Burrows}, J.~E. {Hill}, J.~A. {Nousek}, J.~A. {Kennea}, A.~{Wells},
  J.~P. {Osborne}, A.~F. {Abbey}, A.~{Beardmore}, K.~{Mukerjee}, A.~D.~T.
  {Short}, G.~{Chincarini}, S.~{Campana}, O.~{Citterio}, A.~{Moretti},
  C.~{Pagani}, G.~{Tagliaferri}, P.~{Giommi}, M.~{Capalbi}, F.~{Tamburelli},
  L.~{Angelini}, G.~{Cusumano}, H.~W. {Br{\"a}uninger}, W.~{Burkert} and G.~D.
  {Hartner}, {\em Space Science Reviews} {\bf 120}, 165 (October 2005).

\bibitem{2005GCN..3101....1K}
D.~{Kelson} and E.~{Berger}, {\em GRB Coordinates Network (GCN)} {\bf 3101}
  (2005).

\bibitem{2006ApJ...642..389N}
J.~A. {Nousek}, C.~{Kouveliotou}, D.~{Grupe}, K.~L. {Page}, J.~{Granot},
  E.~{Ramirez-Ruiz}, S.~K. {Patel}, D.~N. {Burrows}, V.~{Mangano},
  S.~{Barthelmy}, A.~P. {Beardmore}, S.~{Campana}, M.~{Capalbi},
  G.~{Chincarini}, G.~{Cusumano}, A.~D. {Falcone}, N.~{Gehrels}, P.~{Giommi},
  M.~R. {Goad}, O.~{Godet}, C.~P. {Hurkett}, J.~A. {Kennea}, A.~{Moretti},
  P.~T. {O'Brien}, J.~P. {Osborne}, P.~{Romano}, G.~{Tagliaferri} and A.~A.
  {Wells}, {\em Astrophysical Journal} {\bf 642}, 389 (May 2006).

\bibitem{2006MNRAS.366.1357P}
A.~{Panaitescu}, P.~{M{\'e}sz{\'a}ros}, N.~{Gehrels}, D.~{Burrows} and
  J.~{Nousek}, {\em Monthly Notices of the Royal Astronomical Society} {\bf
  366}, 1357 (March 2006).

\bibitem{2006ApJ...642..354Z}
B.~{Zhang}, Y.~Z. {Fan}, J.~{Dyks}, S.~{Kobayashi}, P.~{M{\'e}sz{\'a}ros},
  D.~N. {Burrows}, J.~A. {Nousek} and N.~{Gehrels}, {\em Astrophysical Journal}
  {\bf 642}, 354 (May 2006).

\bibitem{2006GCN..4775....1C}
G.~{Cusumano}, S.~{Barthelmy}, N.~{Gehrels}, S.~{Hunsberger}, S.~{Immler},
  F.~{Marshall}, D.~{Palmer} and T.~{Sakamoto}, {\em GRB Coordinates Network
  (GCN)} {\bf 4775}  (2006).

\bibitem{2006GCN..4776....1K}
J.~A. {Kennea}, D.~N. {Burrows}, G.~{Cusumano} and G.~{Tagliaferri}, {\em GRB
  Coordinates Network (GCN)} {\bf 4776}  (2006).

\bibitem{2006GCN..4822....1S}
T.~{Sakamoto}, L.~{Barbier}, S.~{Barthelmy}, J.~{Cummings}, E.~{Fenimore},
  N.~{Gehrels}, D.~{Hullinger}, H.~{Krimm}, C.~{Markwardt}, D.~{Palmer},
  A.~{Parsons}, G.~{Sato} and J.~{Tueller}, {\em GRB Coordinates Network (GCN)}
  {\bf 4822}  (2006).

\bibitem{2006GCN..4780....1B}
L.~{Barbier}, S.~{Barthelmy}, J.~{Cummings}, G.~{Cusumano}, E.~{Fenimore},
  N.~{Gehrels}, D.~{Hullinger}, H.~{Krimm}, C.~{Markwardt}, D.~{Palmer},
  A.~{Parsons}, T.~{Sakamoto}, G.~{Sato} and J.~{Tueller}, {\em GRB Coordinates
  Network (GCN)} {\bf 4780}  (2006).

\bibitem{2006Natur.442.1008C}
S.~{Campana}, V.~{Mangano}, A.~J. {Blustin}, P.~{Brown}, D.~N. {Burrows},
  G.~{Chincarini}, J.~R. {Cummings}, G.~{Cusumano}, M.~{Della Valle},
  D.~{Malesani}, P.~{M{\'e}sz{\'a}ros}, J.~A. {Nousek}, M.~{Page},
  T.~{Sakamoto}, E.~{Waxman}, B.~{Zhang}, Z.~G. {Dai}, N.~{Gehrels},
  S.~{Immler}, F.~E. {Marshall}, K.~O. {Mason}, A.~{Moretti}, P.~T. {O'Brien},
  J.~P. {Osborne}, K.~L. {Page}, P.~{Romano}, P.~W.~A. {Roming},
  G.~{Tagliaferri}, L.~R. {Cominsky}, P.~{Giommi}, O.~{Godet}, J.~A. {Kennea},
  H.~{Krimm}, L.~{Angelini}, S.~D. {Barthelmy}, P.~T. {Boyd}, D.~M. {Palmer},
  A.~A. {Wells} and N.~E. {White}, {\em Nature} {\bf 442}, 1008 (August 2006).

\bibitem{2006Natur.442.1014S}
A.~M. {Soderberg}, S.~R. {Kulkarni}, E.~{Nakar}, E.~{Berger}, P.~B. {Cameron},
  D.~B. {Fox}, D.~{Frail}, A.~{Gal-Yam}, R.~{Sari}, S.~B. {Cenko},
  M.~{Kasliwal}, R.~A. {Chevalier}, T.~{Piran}, P.~A. {Price}, B.~P. {Schmidt},
  G.~{Pooley}, D.-S. {Moon}, B.~E. {Penprase}, E.~{Ofek}, A.~{Rau},
  N.~{Gehrels}, J.~A. {Nousek}, D.~N. {Burrows}, S.~E. {Persson} and P.~J.
  {McCarthy}, {\em Nature} {\bf 442}, 1014 (August 2006).

\bibitem{2006A&A...454..503S}
J.~{Sollerman}, A.~O. {Jaunsen}, J.~P.~U. {Fynbo}, J.~{Hjorth}, P.~{Jakobsson},
  M.~{Stritzinger}, C.~{F{\'e}ron}, P.~{Laursen}, J.-E. {Ovaldsen}, J.~{Selj},
  C.~C. {Th{\"o}ne}, D.~{Xu}, T.~{Davis}, J.~{Gorosabel}, D.~{Watson},
  R.~{Duro}, I.~{Ilyin}, B.~L. {Jensen}, N.~{Lysfjord}, T.~{Marquart}, T.~B.
  {Nielsen}, J.~{N{\"a}r{\"a}nen}, H.~E. {Schwarz}, S.~{Walch}, M.~{Wold} and
  G.~{{\"O}stlin}, {\em Astronomy \& Astrophysics} {\bf 454}, 503 (August
  2006).

\bibitem{2006ApJ...643L..99M}
N.~{Mirabal}, J.~P. {Halpern}, D.~{An}, J.~R. {Thorstensen} and D.~M.
  {Terndrup}, {\em Astrophysical Journal} {\bf 643}, L99 (June 2006).

\bibitem{2007ApJ...662.1111L}
E.~{Liang}, B.~{Zhang}, F.~{Virgili} and Z.~G. {Dai}, {\em Astrophysical
  Journal} {\bf 662}, 1111 (June 2007).

\bibitem{2006ApJ...645L.113C}
B.~E. {Cobb}, C.~D. {Bailyn}, P.~G. {van Dokkum} and P.~{Natarajan}, {\em
  Astrophysical Journal} {\bf 645}, L113 (July 2006).

\bibitem{2007ApJ...657L..73G}
D.~{Guetta} and M.~{Della Valle}, {\em Astrophysical Journal} {\bf 657}, L73
  (March 2007).

\bibitem{2006Natur.442.1011P}
E.~{Pian}, P.~A. {Mazzali}, N.~{Masetti}, P.~{Ferrero}, S.~{Klose},
  E.~{Palazzi}, E.~{Ramirez-Ruiz}, S.~E. {Woosley}, C.~{Kouveliotou},
  J.~{Deng}, A.~V. {Filippenko}, R.~J. {Foley}, J.~P.~U. {Fynbo}, D.~A. {Kann},
  W.~{Li}, J.~{Hjorth}, K.~{Nomoto}, F.~{Patat}, D.~N. {Sauer}, J.~{Sollerman},
  P.~M. {Vreeswijk}, E.~W. {Guenther}, A.~{Levan}, P.~{O'Brien}, N.~R.
  {Tanvir}, R.~A.~M.~J. {Wijers}, C.~{Dumas}, O.~{Hainaut}, D.~S. {Wong},
  D.~{Baade}, L.~{Wang}, L.~{Amati}, E.~{Cappellaro}, A.~J. {Castro-Tirado},
  S.~{Ellison}, F.~{Frontera}, A.~S. {Fruchter}, J.~{Greiner}, K.~{Kawabata},
  C.~{Ledoux}, K.~{Maeda}, P.~{M{\o}ller}, L.~{Nicastro}, E.~{Rol} and
  R.~{Starling}, {\em Nature} {\bf 442}, 1011 (August 2006).

\bibitem{2006GCN..4809....1F}
T.~A. {Fatkhullin}, V.~V. {Sokolov}, A.~V. {Moiseev}, S.~{Guziy} and A.~J.
  {Castro-Tirado}, {\em GRB Coordinates Network (GCN)} {\bf 4809}  (2006).

\bibitem{2006GCN..4804....1S}
A.~M. {Soderberg}, E.~{Berger} and B.~P. {Schmidt}, {\em GRB Coordinates
  Network (GCN)} {\bf 4804}  (2006).

\bibitem{2007AIPC..924..120F}
P.~{Ferrero}, E.~{Palazzi}, E.~{Pian} and S.~{Savaglio}, Optical observations
  of grb 060218/sn 2006aj and its host galaxy, in {\em American Institute of
  Physics Conference Series\/}, , American Institute of Physics Conference
  Series Vol.~924 (August 2007).

\bibitem{2007A&A...464..529W}
K.~{Wiersema}, S.~{Savaglio}, P.~M. {Vreeswijk}, S.~L. {Ellison}, C.~{Ledoux},
  S.-C. {Yoon}, P.~{M{\o}ller}, J.~{Sollerman}, J.~P.~U. {Fynbo}, E.~{Pian},
  R.~L.~C. {Starling} and R.~A.~M.~J. {Wijers}, {\em Astronomy \& Astrophysics}
  {\bf 464}, 529 (March 2007).

\bibitem{2006ApJ...645L..21M}
M.~{Modjaz}, K.~Z. {Stanek}, P.~M. {Garnavich}, P.~{Berlind}, S.~{Blondin},
  W.~{Brown}, M.~{Calkins}, P.~{Challis}, A.~M. {Diamond-Stanic}, H.~{Hao},
  M.~{Hicken}, R.~P. {Kirshner} and J.~L. {Prieto}, {\em Astrophysical Journal}
  {\bf 645}, L21 (July 2006).

\bibitem{2007ApJ...654..385K}
Y.~{Kaneko}, E.~{Ramirez-Ruiz}, J.~{Granot}, C.~{Kouveliotou}, S.~E. {Woosley},
  S.~K. {Patel}, E.~{Rol}, J.~J.~M.~i. {Zand}, A.~J. {van der Horst},
  R.~A.~M.~J. {Wijers} and R.~{Strom}, {\em Astrophysical Journal} {\bf 654},
  385 (January 2007).

\bibitem{2007ApJ...659.1420T}
K.~{Toma}, K.~{Ioka}, T.~{Sakamoto} and T.~{Nakamura}, {\em Astrophysical
  Journal} {\bf 659}, 1420 (April 2007).

\bibitem{2006ApJ...653L..81L}
E.-W. {Liang}, B.-B. {Zhang}, M.~{Stamatikos}, B.~{Zhang}, J.~{Norris},
  N.~{Gehrels}, J.~{Zhang} and Z.~G. {Dai}, {\em Astrophysical Journal} {\bf
  653}, L81 (December 2006).

\bibitem{2006JCAP...09..013F}
Y.-Z. {Fan}, T.~{Piran} and D.~{Xu}, {\em Journal of Cosmology and
  Astro-Particle Physics} {\bf 9}, p.~13 (September 2006).

\bibitem{2007MNRAS.375..240L}
L.-X. {Li}, {\em Monthly Notices of the Royal Astronomical Society} {\bf 375},
  240 (February 2007).

\bibitem{2006A&A...459..875E}
A.~{Ederoclite}, E.~{Mason}, M.~{Della Valle}, R.~{Gilmozzi}, R.~E. {Williams},
  L.~{Germany}, I.~{Saviane}, F.~{Matteucci}, B.~E. {Schaefer}, F.~{Walter},
  R.~J. {Rudy}, D.~{Lynch}, S.~{Mazuk}, C.~C. {Venturini}, R.~C. {Puetter},
  R.~B. {Perry}, W.~{Liller} and A.~{Rotter}, {\em Astronomy \& Astrophysics}
  {\bf 459}, 875 (December 2006).

\bibitem{2006ARA&A..44..507W}
S.~E. {Woosley} and J.~S. {Bloom}, {\em Annual Review of Astronomy and
  Astrophysics} {\bf 44}, 507 (September 2006).

\bibitem{2007ApJ...658L...5M}
K.~{Maeda}, K.~{Kawabata}, M.~{Tanaka}, K.~{Nomoto}, N.~{Tominaga},
  T.~{Hattori}, T.~{Minezaki}, T.~{Kuroda}, T.~{Suzuki}, J.~{Deng}, P.~A.
  {Mazzali} and E.~{Pian}, {\em Astrophysical Journal} {\bf 658}, L5 (March
  2007).

\bibitem{2006Natur.442.1018M}
P.~A. {Mazzali}, J.~{Deng}, K.~{Nomoto}, D.~N. {Sauer}, E.~{Pian},
  N.~{Tominaga}, M.~{Tanaka}, K.~{Maeda} and A.~V. {Filippenko}, {\em Nature}
  {\bf 442}, 1018 (August 2006).

\bibitem{2007astro.ph..2472N}
K.~{Nomoto}, N.~{Tominaga}, M.~{Tanaka}, K.~{Maeda}, T.~{Suzuki}, J.~S. {Deng}
  and P.~A. {Mazzali}, {\em ArXiv:astro-ph/0702472}  (February 2007).

\bibitem{2006AIPC..836..367D}
M.~{Della Valle}, Supernova and grb connection: Observations and questions, in
  {\em Gamma-Ray Bursts in the Swift Era\/},  eds. S.~S. {Holt}, N.~{Gehrels}
  and J.~A. {Nousek}, American Institute of Physics Conference Series, Vol.~836
  (May 2006).

\bibitem{2006Natur.444.1050D}
M.~{Della Valle}, G.~{Chincarini}, N.~{Panagia}, G.~{Tagliaferri},
  D.~{Malesani}, V.~{Testa}, D.~{Fugazza}, S.~{Campana}, S.~{Covino},
  V.~{Mangano}, L.~A. {Antonelli}, P.~{D'Avanzo}, K.~{Hurley}, I.~F. {Mirabel},
  L.~J. {Pellizza}, S.~{Piranomonte} and L.~{Stella}, {\em Nature} {\bf 444},
  1050 (December 2006).

\bibitem{2007A&A...470..105M}
V.~{Mangano}, S.~T. {Holland}, D.~{Malesani}, E.~{Troja}, G.~{Chincarini},
  B.~{Zhang}, V.~{La Parola}, P.~J. {Brown}, D.~N. {Burrows}, S.~{Campana},
  M.~{Capalbi}, G.~{Cusumano}, M.~{Della Valle}, N.~{Gehrels}, P.~{Giommi},
  D.~{Grupe}, C.~{Guidorzi}, T.~{Mineo}, A.~{Moretti}, J.~P. {Osborne}, S.~B.
  {Pandey}, M.~{Perri}, P.~{Romano}, P.~W.~A. {Roming} and G.~{Tagliaferri},
  {\em Astronomy \& Astrophysics} {\bf 470}, 105 (July 2007).

\bibitem{2002A&A...390...81A}
L.~{Amati}, F.~{Frontera}, M.~{Tavani}, J.~J.~M. {in't Zand}, A.~{Antonelli},
  E.~{Costa}, M.~{Feroci}, C.~{Guidorzi}, J.~{Heise}, N.~{Masetti},
  E.~{Montanari}, L.~{Nicastro}, E.~{Palazzi}, E.~{Pian}, L.~{Piro} and
  P.~{Soffitta}, {\em Astronomy \& Astrophysics} {\bf 390}, 81 (July 2002).

\bibitem{2006MNRAS.372.1699G}
G.~{Ghisellini}, G.~{Ghirlanda}, S.~{Mereghetti}, Z.~{Bosnjak}, F.~{Tavecchio}
  and C.~{Firmani}, {\em Monthly Notices of the Royal Astronomical Society}
  {\bf 372}, 1699 (November 2006).

\bibitem{1998ApJ...493L..67F}
F.~{Frontera}, E.~{Costa}, L.~{Piro}, J.~M. {Muller}, L.~{Amati}, M.~{Feroci},
  F.~{Fiore}, G.~{Pizzichini}, M.~{Tavani}, A.~{Castro-Tirado}, G.~{Cusumano},
  D.~{dal Fiume}, J.~{Heise}, K.~{Hurley}, L.~{Nicastro}, M.~{Orlandini},
  A.~{Owens}, E.~{Palazzi}, A.~N. {Parmar}, J.~{in 't Zand} and G.~{Zavattini},
  {\em Astrophysical Journal} {\bf 493}, p. L67 (February 1998).

\bibitem{1997Natur.386..686V}
J.~{van Paradijs}, P.~J. {Groot}, T.~{Galama}, C.~{Kouveliotou}, R.~G. {Strom},
  J.~{Telting}, R.~G.~M. {Rutten}, G.~J. {Fishman}, C.~A. {Meegan},
  M.~{Pettini}, N.~{Tanvir}, J.~{Bloom}, H.~{Pedersen}, H.~U.
  {N{\o}rdgaard-Nielsen}, M.~{Linden-V{\o}rnle}, J.~{Melnick}, G.~{van der
  Steene}, M.~{Bremer}, R.~{Naber}, J.~{Heise}, J.~{in't Zand}, E.~{Costa},
  M.~{Feroci}, L.~{Piro}, F.~{Frontera}, G.~{Zavattini}, L.~{Nicastro},
  E.~{Palazzi}, K.~{Bennet}, L.~{Hanlon} and A.~{Parmar}, {\em Nature} {\bf
  386}, 686 (April 1997).

\bibitem{2001ApJ...554..678B}
J.~S. {Bloom}, S.~G. {Djorgovski} and S.~R. {Kulkarni}, {\em Astrophysical
  Journal} {\bf 554}, 678 (June 2001).

\bibitem{1997Natur.387R.476S}
K.~C. {Sahu}, M.~{Livio}, L.~{Petro}, F.~D. {Macchetto}, J.~{van Paradijs},
  C.~{Kouveliotou}, G.~J. {Fishman}, C.~A. {Meegan}, P.~J. {Groot} and
  T.~{Galama}, {\em Nature} {\bf 387}, 476 (May 1997).

\bibitem{2005Natur.437..855V}
J.~S. {Villasenor}, D.~Q. {Lamb}, G.~R. {Ricker}, J.-L. {Atteia}, N.~{Kawai},
  N.~{Butler}, Y.~{Nakagawa}, J.~G. {Jernigan}, M.~{Boer}, G.~B. {Crew}, T.~Q.
  {Donaghy}, J.~{Doty}, E.~E. {Fenimore}, M.~{Galassi}, C.~{Graziani},
  K.~{Hurley}, A.~{Levine}, F.~{Martel}, M.~{Matsuoka}, J.-F. {Olive},
  G.~{Prigozhin}, T.~{Sakamoto}, Y.~{Shirasaki}, M.~{Suzuki}, T.~{Tamagawa},
  R.~{Vanderspek}, S.~E. {Woosley}, A.~{Yoshida}, J.~{Braga}, R.~{Manchanda},
  G.~{Pizzichini}, K.~{Takagishi} and M.~{Yamauchi}, {\em Nature} {\bf 437},
  855 (October 2005).

\bibitem{2006A&A...454..113C}
S.~{Campana}, G.~{Tagliaferri}, D.~{Lazzati}, G.~{Chincarini}, S.~{Covino},
  K.~{Page}, P.~{Romano}, A.~{Moretti}, G.~{Cusumano}, V.~{Mangano},
  T.~{Mineo}, V.~{La Parola}, P.~{Giommi}, M.~{Perri}, M.~{Capalbi},
  B.~{Zhang}, S.~{Barthelmy}, J.~{Cummings}, T.~{Sakamoto}, D.~N. {Burrows},
  J.~A. {Kennea}, J.~A. {Nousek}, J.~P. {Osborne}, P.~T. {O'Brien}, O.~{Godet}
  and N.~{Gehrels}, {\em Astronomy \& Astrophysics} {\bf 454}, 113 (July 2006).

\bibitem{2006Natur.444.1044G}
N.~{Gehrels}, J.~P. {Norris}, S.~D. {Barthelmy}, J.~{Granot}, Y.~{Kaneko},
  C.~{Kouveliotou}, C.~B. {Markwardt}, P.~{M{\'e}sz{\'a}ros}, E.~{Nakar}, J.~A.
  {Nousek}, P.~T. {O'Brien}, M.~{Page}, D.~M. {Palmer}, A.~M. {Parsons},
  P.~W.~A. {Roming}, T.~{Sakamoto}, C.~L. {Sarazin}, P.~{Schady},
  M.~{Stamatikos} and S.~E. {Woosley}, {\em Nature} {\bf 444}, 1044 (December
  2006).

\bibitem{2006MNRAS.367L..42P}
A.~{Panaitescu}, {\em Monthly Notices of the Royal Astronomical Society} {\bf
  367}, L42 (March 2006).

\bibitem{2006MNRAS.372..233A}
L.~{Amati}, {\em Monthly Notices of the Royal Astronomical Society} {\bf 372},
  233 (October 2006).

\bibitem{amatiIK}
L.~{Amati}. talk presented at the congress ``10$^{th}$ Italian-Korean
  Meeting'', Pescara, Italy, June 25-29,  (2007).

\bibitem{2006astro.ph.10408G}
D.~{Guetta}, {\em ArXiv:astro-ph/0610408}  (October 2006).

\bibitem{2006A&A...454L.123W}
D.~{Watson}, J.~{Hjorth}, P.~{Jakobsson}, D.~{Xu}, J.~P.~U. {Fynbo},
  J.~{Sollerman}, C.~C. {Th{\"o}ne} and K.~{Pedersen}, {\em Astronomy \&
  Astrophysics} {\bf 454}, L123 (August 2006).

\bibitem{2005Natur.438..994B}
S.~D. {Barthelmy}, G.~{Chincarini}, D.~N. {Burrows}, N.~{Gehrels}, S.~{Covino},
  A.~{Moretti}, P.~{Romano}, P.~T. {O'Brien}, C.~L. {Sarazin},
  C.~{Kouveliotou}, M.~{Goad}, S.~{Vaughan}, G.~{Tagliaferri}, B.~{Zhang},
  L.~A. {Antonelli}, S.~{Campana}, J.~R. {Cummings}, P.~{D'Avanzo}, M.~B.
  {Davies}, P.~{Giommi}, D.~{Grupe}, Y.~{Kaneko}, J.~A. {Kennea}, A.~{King},
  S.~{Kobayashi}, A.~{Melandri}, P.~{Meszaros}, J.~A. {Nousek}, S.~{Patel},
  T.~{Sakamoto} and R.~A.~M.~J. {Wijers}, {\em Nature} {\bf 438}, 994 (December
  2005).

\bibitem{2005Natur.437..845F}
D.~B. {Fox}, D.~A. {Frail}, P.~A. {Price}, S.~R. {Kulkarni}, E.~{Berger},
  T.~{Piran}, A.~M. {Soderberg}, S.~B. {Cenko}, P.~B. {Cameron}, A.~{Gal-Yam},
  M.~M. {Kasliwal}, D.-S. {Moon}, F.~A. {Harrison}, E.~{Nakar}, B.~P.
  {Schmidt}, B.~{Penprase}, R.~A. {Chevalier}, P.~{Kumar}, K.~{Roth},
  D.~{Watson}, B.~L. {Lee}, S.~{Shectman}, M.~M. {Phillips}, M.~{Roth}, P.~J.
  {McCarthy}, M.~{Rauch}, L.~{Cowie}, B.~A. {Peterson}, J.~{Rich}, N.~{Kawai},
  K.~{Aoki}, G.~{Kosugi}, T.~{Totani}, H.-S. {Park}, A.~{MacFadyen} and K.~C.
  {Hurley}, {\em Nature} {\bf 437}, 845 (October 2005).

\bibitem{2000ApJ...536..185G}
T.~J. {Galama}, N.~{Tanvir}, P.~M. {Vreeswijk}, R.~A.~M.~J. {Wijers}, P.~J.
  {Groot}, E.~{Rol}, J.~{van Paradijs}, C.~{Kouveliotou}, A.~S. {Fruchter},
  N.~{Masetti}, H.~{Pedersen}, B.~{Margon}, E.~W. {Deutsch}, M.~{Metzger},
  L.~{Armus}, S.~{Klose} and B.~{Stecklum}, {\em Astrophysical Journal} {\bf
  536}, 185 (June 2000).

\bibitem{2003ApJ...586..135L}
A.~D. {Lewis}, D.~A. {Buote} and J.~T. {Stocke}, {\em Astrophysical Journal}
  {\bf 586}, 135 (March 2003).

\bibitem{2007ApJ...660..496B}
E.~{Berger}, M.-S. {Shin}, J.~S. {Mulchaey} and T.~E. {Jeltema}, {\em
  Astrophysical Journal} {\bf 660}, 496 (May 2007).

\bibitem{KMG11}
M.~{Kramer}, in {\em The Eleventh Marcel Grossmann Meeting.\/},  eds. R.~T.
  {Jantzen}, H.~{Kleinert} and R.~{Ruffini} (Singapore: World Scientific, in
  press).

\bibitem{2004AIPC..727..312B}
M.~G. {Bernardini}, C.~L. {Bianco}, P.~{Chardonnet}, F.~{Fraschetti},
  R.~{Ruffini} and S.-S. {Xue}, A new astrophysical ''triptych'':
  Grb030329/sn2003dh/urca-2, in {\em Gamma-Ray Bursts: 30 Years of
  Discovery\/},  eds. E.~{Fenimore} and M.~{Galassi}, American Institute of
  Physics Conference Series, Vol.~727 (September 2004).

\bibitem{1998Natur.395..670G}
T.~J. {Galama}, P.~M. {Vreeswijk}, J.~{van Paradijs}, C.~{Kouveliotou},
  T.~{Augusteijn}, H.~{B{\"o}hnhardt}, J.~P. {Brewer}, V.~{Doublier}, J.-F.
  {Gonzalez}, B.~{Leibundgut}, C.~{Lidman}, O.~R. {Hainaut}, F.~{Patat},
  J.~{Heise}, J.~{in't Zand}, K.~{Hurley}, P.~J. {Groot}, R.~G. {Strom}, P.~A.
  {Mazzali}, K.~{Iwamoto}, K.~{Nomoto}, H.~{Umeda}, T.~{Nakamura}, T.~R.
  {Young}, T.~{Suzuki}, T.~{Shigeyama}, T.~{Koshut}, M.~{Kippen},
  C.~{Robinson}, P.~{de Wildt}, R.~A.~M.~J. {Wijers}, N.~{Tanvir},
  J.~{Greiner}, E.~{Pian}, E.~{Palazzi}, F.~{Frontera}, N.~{Masetti},
  L.~{Nicastro}, M.~{Feroci}, E.~{Costa}, L.~{Piro}, B.~A. {Peterson},
  C.~{Tinney}, B.~{Boyle}, R.~{Cannon}, R.~{Stathakis}, E.~{Sadler}, M.~C.
  {Begam} and P.~{Ianna}, {\em Nature} {\bf 395}, 670 (October 1998).

\bibitem{2003ApJ...599.1223G}
J.~{Greiner}, S.~{Klose}, M.~{Salvato}, A.~{Zeh}, R.~{Schwarz}, D.~H.
  {Hartmann}, N.~{Masetti}, B.~{Stecklum}, G.~{Lamer}, N.~{Lodieu}, R.~D.
  {Scholz}, C.~{Sterken}, J.~{Gorosabel}, I.~{Burud}, J.~{Rhoads},
  I.~{Mitrofanov}, M.~{Litvak}, A.~{Sanin}, V.~{Grinkov}, M.~I. {Andersen},
  J.~M. {Castro Cer{\'o}n}, A.~J. {Castro-Tirado}, A.~{Fruchter}, J.~U.
  {Fynbo}, J.~{Hjorth}, L.~{Kaper}, C.~{Kouveliotou}, E.~{Palazzi}, E.~{Pian},
  E.~{Rol}, N.~R. {Tanvir}, P.~M. {Vreeswijk}, R.~A.~M.~J. {Wijers} and E.~{van
  den Heuvel}, {\em Astrophysical Journal} {\bf 599}, 1223 (December 2003).

\bibitem{2001GCN..1152....1I}
L.~{Infante}, P.~M. {Garnavich}, K.~Z. {Stanek} and L.~{Wyrzykowski}, {\em GRB
  Coordinates Network (GCN)} {\bf 1152}  (2001).

\bibitem{2001GCN..1147....1P}
L.~{Piro}, {\em GRB Coordinates Network (GCN)} {\bf 1147}  (2001).

\bibitem{1998ApJ...494L..45P}
B.~{Paczynski}, {\em Astrophysical Journal} {\bf 494}, p. L45 (February 1998).

\bibitem{1998Natur.395..663K}
S.~R. {Kulkarni}, D.~A. {Frail}, M.~H. {Wieringa}, R.~D. {Ekers}, E.~M.
  {Sadler}, R.~M. {Wark}, J.~L. {Higdon}, E.~S. {Phinney} and J.~S. {Bloom},
  {\em Nature} {\bf 395}, 663 (October 1998).

\bibitem{1998Natur.395..672I}
K.~{Iwamoto}, P.~A. {Mazzali}, K.~{Nomoto}, H.~{Umeda}, T.~{Nakamura},
  F.~{Patat}, I.~J. {Danziger}, T.~R. {Young}, T.~{Suzuki}, T.~{Shigeyama},
  T.~{Augusteijn}, V.~{Doublier}, J.-F. {Gonzalez}, H.~{Boehnhardt},
  J.~{Brewer}, O.~R. {Hainaut}, C.~{Lidman}, B.~{Leibundgut}, E.~{Cappellaro},
  M.~{Turatto}, T.~J. {Galama}, P.~M. {Vreeswijk}, C.~{Kouveliotou}, J.~{van
  Paradijs}, E.~{Pian}, E.~{Palazzi} and F.~{Frontera}, {\em Nature} {\bf 395},
  672 (October 1998).

\bibitem{2005tmgm.meet.1802M}
G.~J. {Mathews} and J.~R. {Wilson}, Relativistic induced compression of neutron
  stars and white dwarfs, in {\em The Tenth Marcel Grossmann Meeting. On recent
  developments in theoretical and experimental general relativity, gravitation
  and relativistic field theories\/},  eds. M.~{Novello}, S.~{Perez Bergliaffa}
  and R.~{Ruffini} (Singapore: World Scientific, January 2005).

\bibitem{f21}
E.~{Fermi}, {\em Il Nuovo Cimento} {\bf 22}, 176  (1921).

\bibitem{1973PhRvD...8.3259H}
R.~S. {Hanni} and R.~{Ruffini}, {\em Physical Review D} {\bf 8}, 3259  (1973).

\bibitem{1947PhRv...72..390M}
S.~D. {Majumdar}, {\em Physical Review} {\bf 72}, 390 (September 1947).

\bibitem{p47}
A.~{Papapetrou}, {\em Proceedings of the Royal Irish Academy} {\bf 51}, p. 191
  (1945).

\bibitem{1973PhRvD...7.2874P}
L.~{Parker}, R.~{Ruffini} and D.~{Wilkins}, {\em Physical Review D} {\bf 7},
  2874 (May 1973).

\bibitem{2007PhRvD..75d4012B}
D.~{Bini}, A.~{Geralico} and R.~{Ruffini}, {\em Physical Review D} {\bf 75}, p.
  044012 (February 2007).

\bibitem{2007PhLA..360..515B}
D.~{Bini}, A.~{Geralico} and R.~{Ruffini}, {\em Physics Letters A} {\bf 360},
  515 (January 2007).

\bibitem{RuKerr}
R.~{Ruffini}, The ergosphere and dyadosphere of black holes, in {\em The Kerr
  Spacetime\/},  eds. D.~L. {Wiltshire}, M.~{Visser} and S.~{Scott} (Cambridge
  University Press, in press, in press).

\bibitem{2004AdSpR..34.2711P}
E.~{Pian}, P.~{Giommi}, L.~{Amati}, E.~{Costa}, J.~{Danziger}, M.~{Feroci},
  M.~T. {Fiocchi}, F.~{Frontera}, C.~{Kouveliotou}, N.~{Masetti},
  L.~{Nicastro}, E.~{Palazzi}, L.~{Piro}, M.~{Tavani} and J.~J.~M. {in 'T
  Zand}, {\em Advances in Space Research} {\bf 34}, 2711  (2004).

\bibitem{2004ApJ...608..872K}
C.~{Kouveliotou}, S.~E. {Woosley}, S.~K. {Patel}, A.~{Levan}, R.~{Blandford},
  E.~{Ramirez-Ruiz}, R.~A.~M.~J. {Wijers}, M.~C. {Weisskopf}, A.~{Tennant},
  E.~{Pian} and P.~{Giommi}, {\em Astrophysical Journal} {\bf 608}, 872 (June
  2004).

\bibitem{2003A&A...409..983T}
A.~{Tiengo}, S.~{Mereghetti}, G.~{Ghisellini}, E.~{Rossi}, G.~{Ghirlanda} and
  N.~{Schartel}, {\em Astronomy \& Astrophysics} {\bf 409}, 983 (October 2003).

\bibitem{2004A&A...423..861T}
A.~{Tiengo}, S.~{Mereghetti}, G.~{Ghisellini}, F.~{Tavecchio} and
  G.~{Ghirlanda}, {\em Astronomy \& Astrophysics} {\bf 423}, 861 (September
  2004).

\bibitem{mazzaliVen}
P.~{Mazzali}. talk presented at the congress ``Swift and GRBs: unveiling the
  relativistic universe'', Venice, Italy, June 5-9,  (2006).

\bibitem{2005tmgm.meet..369R}
R.~{Ruffini}, M.~G. {Bernardini}, C.~L. {Bianco}, L.~{Vitagliano}, S.-S. {Xue},
  P.~{Chardonnet}, F.~{Fraschetti} and V.~{Gurzadyan}, Black hole physics and
  astrophysics: The grb-supernova connection and urca-1 - urca-2, in {\em The
  Tenth Marcel Grossmann Meeting. On recent developments in theoretical and
  experimental general relativity, gravitation and relativistic field
  theories\/},  eds. M.~{Novello}, S.~{Perez Bergliaffa} and R.~{Ruffini}
  (Singapore: World Scientific, January 2005).

\bibitem{1941PhRv...59..539G}
G.~{Gamow} and M.~{Schoenberg}, {\em Physical Review} {\bf 59}, 539 (April
  1941).

\bibitem{GamowBook-MyWorldlines}
G.~{Gamow}, {\em My wordlines - an informal autobiography} (New York: Viking
  press, 1970).

\bibitem{1964PhDT........34T}
S.~{Tsuruta}, PhD thesis, , Columbia Univ., (1964), (1964).

\bibitem{1979PhR....56..237T}
S.~{Tsuruta}, {\em Physics Reports} {\bf 56}, 237  (1979).

\bibitem{1966CaJPh..44.1863T}
S.~{Tsuruta} and A.~G.~W. {Cameron}, {\em Canadian Journal of Physics} {\bf
  44}, p. 1863  (1966).

\bibitem{2002ApJ...571L.143T}
S.~{Tsuruta}, M.~A. {Teter}, T.~{Takatsuka}, T.~{Tatsumi} and R.~{Tamagaki},
  {\em Astrophysical Journal} {\bf 571}, L143 (June 2002).

\bibitem{1978pans.proc..448C}
V.~{Canuto}, Neutron stars, in {\em Physics and Astrophysics of Neutron Stars
  and Black Holes\/},  eds. R.~{Giacconi} and R.~{Ruffini} (1978).

\bibitem{1987ApJ...313..718R}
R.~W. {Romani}, {\em Astrophysical Journal} {\bf 313}, 718 (February 1987).

\bibitem{1988ApJ...329..339V}
K.~A. {van Riper}, {\em Astrophysical Journal} {\bf 329}, 339 (June 1988).

\bibitem{1991ApJS...75..449V}
K.~A. {van Riper}, {\em Astrophysical Journal Supplement Series} {\bf 75}, 449
  (February 1991).

\bibitem{1986ApJ...307..178B}
A.~{Burrows} and J.~M. {Lattimer}, {\em Astrophysical Journal} {\bf 307}, 178
  (August 1986).

\bibitem{1994ApJ...425..802L}
J.~M. {Lattimer}, K.~A. {van Riper}, M.~{Prakash} and M.~{Prakash}, {\em
  Astrophysical Journal} {\bf 425}, 802 (April 1994).

\bibitem{2004ARA&A..42..169Y}
D.~G. {Yakovlev} and C.~J. {Pethick}, {\em Annual Review of Astronomy and
  Astrophysics} {\bf 42}, 169 (September 2004).

\bibitem{2005esns.conf..117T}
J.~E. {Tr{\"u}mper}, Observations of cooling neutron stars, in {\em NATO ASIB
  Proc. 210: The Electromagnetic Spectrum of Neutron Stars\/},  eds.
  A.~{Baykal}, S.~K. {Yerli}, S.~C. {Inam} and S.~{Grebenev} (January 2005).

\end{thebibliography}
\end{document}